\documentstyle[sprocl,epsfig,epsf]{article}

\def\be{\begin{equation}}
\def\ee{\end{equation}}
\def\bea{\begin{eqnarray}}
\def\eea{\end{eqnarray}}

\newcommand{\nn}{\nonumber}

\newcommand{\alpS}{\alpha_{S}}

\def\bmag#1{{|{\mathbf #1}|}}
\def\bmat#1{{{\mathbf #1}}}

\def\Dsl{\hbox{/\kern-.6000em D}} 

\def\vev#1{\left\langle{#1}\right\rangle}
\def\dsl{\,\raise.15ex\hbox{/}\mkern-13.5mu D}
\def\bsigma{\mbox{\boldmath $\sigma$}}
\def\lqcd{\Lambda_{\rm QCD}}
\def\psip#1{\psi_{\mathbf{#1}}}
\def\chip#1{\chi_{\mathbf{#1}}}
\def\bsigma{\mbox{\boldmath $\sigma$}}

\def\abs#1{\left| #1 \right|}
\def\ltap{\ \raise.3ex\hbox{$<$\kern-.75em\lower1ex\hbox{$\sim$}}\ }
\def\gtap{\ \raise.3ex\hbox{$>$\kern-.75em\lower1ex\hbox{$\sim$}}\ }
\def\QQbar{Q\bar{Q}}
\def\LQCD{{\cal L}_{\rm QCD}}
\def\LNRQCD{{\cal L}_{\rm NRQCD}}
\def\LpNRQCD{{\cal L}_{\rm pNRQCD}}

\def\OMIT#1{}
\def\Od#1#2{\mbox{\boldmath $O$}^\dagger_{\mbox{\scriptsize\boldmath $#1$},#2}}

\def\msb{{\overline{\rm MS}}}

\def\ms{$\overline{\rm MS}$ }

\def\O#1#2{\mbox{\boldmath $O$}_{\mbox{\scriptsize $\mathbf #1$},#2}}
\newcommand{\bmk}{\mathbf k}
\newcommand{\bmp}{\mathbf p}
\newcommand{\bmq}{\mathbf q}
\newcommand{\bmr}{\mathbf r}
\newcommand{\bmA}{\mathbf A}
\newcommand{\bmE}{\mathbf E}
\newcommand{\bmP}{\mathbf P}
\newcommand{\bmD}{\mathbf D}
\newcommand{\bmS}{\mathbf S}
\newcommand{\bmsigma}{\mathbf \bsigma}
\newcommand{\bmnabla}{\mathbf \nabla}
\newcommand{\bmx}{\mathbf x}
\newcommand{\bmy}{\mathbf y}
\def\lsim{\mathrel{\raise.3ex\hbox{$<$\kern-.75em\lower1ex\hbox{$\sim$}}}}
\def\gsim{\mathrel{\raise.3ex\hbox{$>$\kern-.75em\lower1ex\hbox{$\sim$}}}}
\def\Li2{{\rm Li}_2}

\def\bfsigma{\mbox{\boldmath $\sigma$}}
\def\bfnabla{\mbox{\boldmath $\nabla$}}

\def\citebk#1{\mbox{[\hspace{0.9mm}\raisebox{-1.85mm}[0mm][0mm]
  {\Large\cite{#1}}\hspace{-0.1mm}]}}
\def\citebkcap#1{\mbox{[\hspace{0.8mm}\raisebox{-1.5mm}[0mm][0mm]
  {\large\cite{#1}}\hspace{-0.2mm}]}}
\def\citebkcapx#1{\mbox{[\hspace{0.8mm}\raisebox{-1.7mm}[0mm][0mm]
  {\large\cite{#1}}\hspace{-0.2mm}]}}

\begin{document}

\sloppy

%
%
%
%
%

\newpage

\pagestyle{myheadings}  
\markboth{\small \em Handbook of QCD / Volume 4}{\small \em 
 Heavy Quarkonium Dynamics
}

\thispagestyle{empty}
\begin{flushright}
{\bf MPI-PhT/2002-14}\\
{\bf hep-ph/0204299}\\
{\bf April 2002}\\
\end{flushright}
\vspace{.5cm}
\title{HEAVY QUARKONIUM DYNAMICS\mbox{\hspace{1mm}}\footnote{
To be published in ``At the Frontier of Particle
Physics/Handbook of QCD, Volume 4'', edited by M. Shifman (World Scientific, 
Singapore)}}

\author{ANDRE H. HOANG}

\address{Max-Planck-Institut f\"ur Physik\\
(Werner-Heisenberg-Institut)\\
F\"ohringer Ring 6, 80805 M\"unchen, Germany}

\maketitle\abstracts{ 
An introduction to the
recent developments in perturbative heavy quarkonium physics 
is given. Covered are the effective theories NRQCD, pNRQCD and
vNRQCD, the threshold expansion, the role of renormalons,
and applications to bottom and top quark physics.
}

\vspace{0.1cm}

\tableofcontents

\newpage

\section{Introduction} 
\label{sectionintroduction}

The study of non-relativistic two-body systems consisting of a heavy
quark-antiquark ($\QQbar$) pair has a long tradition since the
discovery of the $J/\psi$ and the work of Appelquist and
Politzer,\cite{Appelquist1} which had shown that, owing to asymptotic
freedom, non-relativistic quantum mechanics should apply to a good
approximation to heavy $\QQbar$ systems. Thus, heavy quarkonium
systems should be predominantly Coulomb-like bound states sharing many
features of positronium and an ideal tool studying the interplay of
perturbative and non-perturbative aspects of quantum chromodynamics
(QCD).  

The realization of this idea has proven a difficult task.
Not even to the first approximation does
the Balmer spectrum of QED give a good quantitative
description of the charmonium or bottomonium spectrum. It was found
that perturbation theory completely fails to describe the long
distance part of the heavy quark potential. Subsequently, potential
models emerged,\cite{Buchmuller1} which incorporated a linear rising
potential at long distances in accordance to the idea of quark
confinement\,\cite{Wilson1} and a Coulomb-like
potential at short distances in accordance to perturbation
theory. Potential models were highly successful in describing the
observed spectra and transition rates, but they were not derived from
first  principles in QCD. As such they cannot be used for quantitative
tests of QCD.

On the other hand, it was found that non-perturbative effects in heavy
quarkonia, where the quark mass $m$ is sufficiently large, such that
$m\gg mv\gg mv^2\gg\lqcd$, $v$ being the quark velocity and $\lqcd$
the typical hadronization scale, can be described by an operator
product expansion in terms of local and gauge-invariant quark and gluon
condensates.\cite{Shifman1,Voloshin1,Leutwyler1} Only the top-antitop
quark system truly belongs into this category. For bottomonia, the
ground states and, maybe, low radial excitations might belong to this
category too, but mesons containing charm do certainly not. However,
even for the perturbative contributions for systems in this 
category, it was not straightforward to systematically
determine higher orders corrections, either due
to ultraviolet (UV) divergences in approaches starting from the
non-relativistic approximation, or due to the amount of technical
difficulties in the relativistic approach, the Bethe-Salpeter
formalism.\cite{Bethe1}

A new conceptual approach using an effective field
theory was proposed by Caswell and Lepage\,\cite{Caswell1} and
Bodwin, Braaten and Lepage.\,\cite{Bodwin1} These authors found that,
for the description of quarkonia, QCD can be reformulated in terms of
an effective 
non-renormalizable Lagrangian built from quark and gluonic field
operators describing fluctuations below $m$, whereas hard effects
associated with the scale $m$ are incorporated in the coefficients of
the operators. The theory is called ``non-relativistic QCD'' (NRQCD).
At present, NRQCD is one of the standard instruments to describe
quarkonium production and decay properties at collider 
experiments. In NRQCD, 
production and decay of quarkonia is described as a sum of terms each
of which is a product of an process-dependent hard coefficient and a
universal operator matrix element. The hard coefficient can be
computed perturbatively, whereas the matrix element has to be
determined from fits to experimental data.
Since I will not touch this aspect of quarkonium physics in this
review I refer to Refs.\,\citebk{NRQCDreviews}. 
NRQCD also made lattice computations of quarkonium systems possible,
since by integrating out hard fluctuations the required lattice size
became small enough to be manageable for present day computing
power.\,\cite{Latticereviews} 

For the computation of the perturbative contributions of heavy
quarkonium systems with $m\gg mv\gg mv^2\gg\lqcd$, NRQCD provided a
prescription to deal with the UV divergences that arise in
relativistic corrections to the non-relativistic Schr\"odinger
equation.\cite{Hoang1} However, NRQCD is not yet a satisfactory
theory for systematic perturbative computations of heavy quarkonium
properties, since no separation of non-relativistic fluctuations is
carried out. The consequences are that NRQCD becomes inconsistent in 
dimensional regularization, that there is no consistent
power counting in $v$, and that it is not possible to sum large QCD
logarithms of $v$. Consistent computations are
only possible in cutoff schemes, which makes analytic QCD calculations
complicated and practically impossible at higher orders.

These deficiencies of NRQCD were addressed in two new effective
theories called ``potential NRQCD'' (pNRQCD)\,\cite{Pineda1} and
``velocity NRQCD'' (vNRQCD).\cite{Luke1} The new effective theories are
more complicated than NRQCD and based on a complete separation of the
modes that fluctuate in the various momentum regions that exist for
$m\gg mv\gg mv^2\gg\lqcd$. Both theories are formulated in dimensional 
regularization. The separation is achieved by a strict expansion in
energy and momentum components that are small in a given region.
For the computation of an asymptotic expansion in $v$ of
Feynman diagrams involving non-relativistic $\QQbar$ pairs this method
has been called the ``threshold expansion''.\cite{Beneke1}
   
In this review I give an introduction to the perturbative aspects of
heavy quarkonium systems with $m\gg mv\gg mv^2\gg\lqcd$ using
effective theory methods. 
The progress on the conceptual as well as on
the calculational level in the last few years has been remarkable.
This period was (at least for me) an exciting time, since 
the transition from having methods, which worked well
but would need to be extended for new applications, to
having theories has been achieved. Certainly, the development is not
completed, but the path to go on is by far clearer now than let's say
five years ago. Although the application of the new developments is
restricted to top and some aspects of bottom physics, the new
knowledge will certainly help also for a better quantitative
understanding of systems where $\lqcd$ is of order $mv^2$ or larger.

The review is divided into two parts. Sections\
\ref{sectionNRQCD}--\ref{sectionquarkmass} cover the conceptual developments
and Secs.\ \ref{sectionQQbarproduction}--\ref{sectionspectrum} the
practical applications. Each section has been written in a
self-contained way.
In Sec.\ \ref{sectionNRQCD} the perturbative
aspects of NRQCD are reviewed. In particular, it is shown why NRQCD fails
to be consistent when dimensional regularization is used. As mentioned
before, the phenomenological application of the NRQCD factorization
formalism to charmonium and bottomonium production and decay is not
covered. In Sec.\ \ref{sectionthresholdexpansion} the method of
regions and the threshold expansion are discussed. An introduction
to pNRQCD and vNRQCD is given in Secs.\ \ref{sectionpNRQCD} and
\ref{sectionvNRQCD}, respectively. Since the concepts used in the
construction of both theories have some subtle differences and since
both theories appear not to be equivalent, I have devoted
Sec.\ \ref{sectionvNRQCDvspNRQCD} to a comparison of
pNRQCD and vNRQCD to visualize the differences.
The important role of quark mass definitions in perturbative
computations of heavy quarkonium properties is reviewed in Sec.\
\ref{sectionquarkmass}. In Sec.\ \ref{sectionQQbarproduction} it is
shown, how the total heavy quark production cross section at
next-to-next-to-leading order in fixed order and
renormalization-group-improved perturbation theory is computed, and
the differences between both types of perturbative expansions is
discussed. Section\ \ref{sectionQQbarproduction} covers the
application of these computations to top quark pair production close to
threshold at the Linear Collider and their impact on the prospect of
measurements of the top quark mass and other top quark properties. In 
this section I concentrate on applications of the total cross
section and leave many other interesting aspects unmentioned.
The applications of the total cross section calculations in bottom mass
extractions from non-relativistic sum rules for $\Upsilon$ mesons are
reviewed in Sec.\ \ref{sectionsumrules}. Sec.\
\ref{sectionspectrum} discusses applications of perturbative
computations of the quarkonium spectrum. Conclusions and an outlook
are given in Sec.\ \ref{sectionconclusions}.

\vspace{1cm}

\section{Effective Theories I: NRQCD}
\label{sectionNRQCD}

The effective theory NRQCD was proposed by
Caswell and Lepage\,\cite{Caswell1} and Bodwin, Braaten and
Lepage\,\cite{Bodwin1} 
to describe non-relativistic
$\QQbar$ dynamics and quarkonium production and decay properties. 
At present it is one of the major instruments for the theoretical
analysis of quarkonium production and decay data at collider
experiments.

\subsection{Basic Ideas}

The basic idea behind the construction of NRQCD is that all hard
fluctuations with frequencies of order $m$ are integrated out. The
effective theory is then 
built from fields for the heavy quark and the light degrees of
freedom, which describe non-relativistic fluctuations below the hard
scale. The various non-relativistic scales $mv$, $mv^2$ and $\lqcd$
are not discriminated. The theory is regulated by a momentum space
cutoff $\Lambda$ that is of order $m$. 
Dimensional regularization leads to inconsistencies 
in NRQCD if the power counting of
Ref.~\citebk{Bodwin1} is employed.\cite{Manohar1} 
Therefore, a consistent renormalization group 
scaling of NRQCD operators is difficult to construct.
From NRQCD an effective Schr\"odinger formalism can be
constructed\,\cite{Caswell1} that contains instantaneous $\QQbar$
potentials.

\subsection{Cutoff Regularization}
\label{subsectionnrqcdcutoff}

The NRQED Lagrangian reads\,\cite{Caswell1,Bodwin1}
\begin{eqnarray}
\lefteqn{
{\cal L}_{\rm NRQCD} 
\, = \,
\psi^{\dagger} \bigg\{ i D_0
+ \, {{\bf D}^2\over 2 m} + \, c_1 {{\bf D}^4\over 8 m^3}
+ c_2\, g {{\bf \bfsigma \cdot B} \over 2 m}
}
\nonumber
\\ && 
+ c_3 \, g { \left({\bf D \cdot E} - {\bf E \cdot D} \right) \over 8 m^2}
+ i c_4 \, g { {\bf \bfsigma \cdot \left(D \times E -E \times D\right)}
    \over 8 m^2} +\ldots \,\bigg\} \psi
+ (\psi \to \chi )
\nonumber
\\ && 
- \,\frac{d_1\,g^2}{4\,m^2}\,
  (\psi^\dagger{\mbox{\boldmath $\sigma$}}\sigma_2\chi^*)\,
  (\chi^T\sigma_2{\mbox{\boldmath $\sigma$}}\psi)
\nonumber
\\ && 
+ \frac{d_2\,g^2}{3\,m^4}\,
  \frac{1}{2}\Big[\,
  (\psi^\dagger{\mbox{\boldmath $\sigma$}}\sigma_2\chi^*)\,
  (\chi^T\sigma_2{\mbox{\boldmath $\sigma$}}
    (-\mbox{$\frac{i}{2}$}
  {\stackrel{\leftrightarrow}{\mbox{\boldmath $D$}}})^2\psi)
  +\mbox{h.c.}\,\Big] + \ldots
\nonumber
\\ && 
-\frac{1}{4}G^{\mu\nu}G_{\mu \nu}
\,,
\label{NRQCDLagrangian1}
\end{eqnarray}
where $\psi$ and $\chi$ are two-component quark and antiquark Pauli
spinors; $i D_0=i\partial_0
-gA_0$ and $i{\bmD}=i{\bmnabla}+g{\bmA}$
are the time and space components
of the gauge covariant derivative $D_\mu$, 
$E^i = G^{0 i}$ and $B^i = \frac{1}{2}\epsilon^{i j k} G^{j k}$ 
are the chromo-electric and chromo-magnetic
components of the gluon field strength tensor, 
$i g G^{\mu\nu}=\left[D^\mu,D^\nu\right]$.
Operators involving light quark are not displayed, color indices are
suppressed, and $m$ is the heavy quark pole mass.
At Born level the effective Lagrangian is obtained by simply expanding
the vertices of full QCD in the non-relativistic limit.
Beyond Born level the short-distance coefficients $c_i$ and $d_i$
encode the effects from momenta of order $m$, which are integrated
out. They are determined from matching NRQCD amplitudes to QCD
amplitudes.

From the above Lagrangian one may in principle derive explicit Feynman
rules and carry out bound state computations. Questions concerning the
power counting are relevant only in so far as one needs
to include a sufficient number of operators for a computation at a
certain order. In particular, according to the power counting rules in
Ref.~\citebk{Bodwin1},
at leading order one has to include the
quark kinetic term $\frac{{\bmD}^2}{2m}$ because
$D_0\sim\frac{{\bmD}^2}{2m}\sim mv^2$, and the correct leading order
quark propagator reads 
$$\frac{i}{k_0-\frac{{\bmk}^2}{2m}+i\delta}\,.$$ 
This is the approach of lattice calculations, where the dynamics is 
computed in a non-perturbative manner. In a cutoff scheme a
discrimination between different non-relativistic 
fluctuations with soft ($\sim m v$) or ultrasoft ($\sim m v^2$)
frequencies is not mandatory conceptually, since the cutoff
automatically includes only the relevant fluctuations. 
For analytic bound state calculations, however, the present formalism
is still too cumbersome to be used in practical calculations because the
contributions from the Coulomb potential exchange 
need to be summed to all orders, while subleading terms can be treated
as perturbations. The NRQCD Feynman rules do not yet provide a systematic
distinction of the various contributions.
Therefore, to carry out analytic bound state calculations, the NRQCD
framework needs to be rewritten in 
terms of a Schr\"odinger theory which contains potentials and
radiation effects and which can be used to carry out
perturbation theory for bound states.\cite{Caswell1}
In fact all analytic NRQCD bound state calculations were
carried out in such a Schr\"odinger theory. 

The conceptually most developed approach to such a Schr\"odinger theory was 
suggested by Labelle.\cite{Labelle1} The most efficient gauge for
this program  is the Coulomb gauge. In the Coulomb gauge, the
longitudinal gluon (the time component of the vector
potential) has an energy-independent propagator, 
$\frac{i}{{\bmk}^2}$. This means that the
interaction associated with the exchange of a longitudinal gluon
corresponds to an instantaneous potential. The power counting of
diagrams containing instantaneous potentials is particularly simple,
because an instantaneous propagator has no particle pole (and no
$i\delta$-prescription) and the scale of ${\bmk}$ is
just set by the average three-momentum of the quarks 
$\sim mv$ in the bound state. 
So one readily obtains the well known result that the Coulomb
potential is the leading order interaction.
On the other hand, the transverse gluon (the spatial component
of the vector potential) has an energy-dependent  propagator of the
form  
$$\frac{i}{(k_0^2 - {\bmk}^2+i\delta)}
\left(\delta^{ij}-\frac{k^ik^j}{{\bmk}^2}\right)\,.$$ 
In this case,
the propagator has a particle pole, and $k_0$ and 
${\bmk}$ can be of order $m v$ (soft) or $mv^2$ (ultrasoft). Also
the regime with  $k_0\sim mv^2\ll {\bmk}\sim mv$ is 
described by the
propagator. This has the consequence   
that NRQCD diagrams containing transverse gluons involve
contributions from soft and ultrasoft scales and, therefore, do not
contribute to a unique order in $v$.
In a cutoff regularization scheme this is a priori not problematic
for the consistency of the theory, but it makes analytic calculations
unnecessarily complicated, because there is no transparent 
power counting. Therefore one does not gain too much in separating only
the hard fluctuations.  Labelle
suggested to also separate explicitly the soft and ultrasoft fluctuations. 
A diagram containing a quark-antiquark pair and a transverse gluon
contains a loop integral of the generic form  
\begin{eqnarray}
\int \frac{d^4k}{
[k_0+\frac{E}{2}-\frac{({\bmk}+{\bmp})^2}{2m}+i\delta]
[k_0-\frac{E}{2}+\frac{({\bmk}+{\bmp})^2}{2m}-i\delta]
[k_0^2-{\bmk}^2+i\delta]}
\,,
\label{Labelleexample}
\end{eqnarray}
where $E\sim mv^2$ and ${\bmp}\sim mv$ stand for some external
center-of-mass 
energies and momenta.
Carrying out the $k_0$-integration by contours, which picks up the
gluon and quark poles, one obtains 
terms of the form
\begin{eqnarray}
\int \frac{d^3{\bmk}}{|{\bmk}|\,
[\frac{E}{2}-\frac{({\bmk}+{\bmp})^2}{2m}-|{\bmk}|]}
\,,
\label{transverseprop}
\end{eqnarray}
where I have dropped a factor 
$2i\pi/(E-\frac{({\bmk}+{\bmp})^2}{m}+i\delta)$ for
the quark-antiquark propagation.
From this, one can see that different contributions arise
depending on whether the scale of ${\bmk}$ is set by
${\bmp}\sim mv$ or by  
$E\sim{\bmp}^2/m \sim mv^2$. The effects associated
with the latter scale are the retardation effects, which are for
example
responsible for the Lamb shift in hydrogen.\cite{Labelle2} The
contributions  
from both momentum regions will not contribute to the same order in
$v$, i.e.\ the power counting is broken.
Labelle\,\cite{Labelle1} suggested to reformulate NRQCD in such a way that the  
contributions associated with the different scales are coming from
separate diagrams. 
This is achieved by Taylor-expanding the
NRQCD diagrams containing Eq.~(\ref{transverseprop}) 
around ${\bmk} \sim mv$ and around 
${\bmk} \sim mv^2$. The prescription anticipated one
of the crucial ingredients  of the threshold expansion 
(Sec.\ \ref{sectionthresholdexpansion}) and the effective theory
vNRQCD (Sec.\ \ref{sectionvNRQCD}). It is a generalization of the
multipole expansion of vertices in ${\bmp}^2/m$ and by now
generally called ``multipole expansion'' in the literature. 
One finds that the lowest order term of the expansion around 
${\bmk}\sim mv$ gives a contribution of order 
\begin{equation}
\int d^3{\bmk}\,
{1 \over |{\bmk}|^2}\, \sim {(mv)^3 \over (mv)^2}
\sim mv
\,.
\label{labellesoft}
\end{equation}
It originates from the region $k_0\sim\frac{{\bmk}^2}{m}\sim mv^2$
in Eq.\ (\ref{Labelleexample}).
Higher order terms in the expansion
also contain contributions from the region $k_0\sim{\bmk}\sim mv$.
The lowest order term of the expansion around 
${\bmk} \sim mv^2$ gives 
\begin{equation}
\int d^3{\bmk}\, 
\frac{1}{|{\bmk}|\, [\frac{E}{2}-\frac{{\bmp}^2}{2m}-|{\bmk}|]} 
\sim  {(mv^2)^3 \over (mv^2)^2} \sim mv^2
\,.
\label{labelleusoft}
\end{equation}
This contributions comes from the region $k_0\sim{\bmk}\sim mv^2$.
The results from Eqs.~(\ref{labellesoft}) and (\ref{labelleusoft}) 
show that at leading order the transverse photon propagator reduces to
an instantaneous potential, which simply corresponds to approximating
the transverse photon propagator $1/(k_0^2-\mbox{\boldmath $k$}^2)$ by 
$-1/\mbox{\boldmath $k$}^2$.
Due to the $v$-suppression of the couplings of a
transverse gluon to heavy quarks, the potentials from the
transverse gluons are of order $v^2$ with respect to the Coulomb
potential.
The prescription of Labelle allows the construction of
a Schr\"odinger theory where the leading order binding of the $\QQbar$
pair is described by the well known non-relativistic Schr\"odinger
equation and where higher order corrections such as
$v$-suppressed potentials and retardation effects can be computed
separately as perturbations.   

All analytic bound state calculations that were based on  NRQED or
NRQCD have essentially
used the prescription proposed by Labelle.
No explicit discussion of QED calculations
shall be given at this place. For a number of original results in the
NRQED framework on the muonium hyperfine
splitting and on positronium, see e.g.\ Refs.~\citebk{Kinoshita1,Hoang2,Hill1}.  
A major application of this formalism in QCD was the determination of
the  total cross section of $\QQbar$ pairs close to threshold 
at next-to-next-to-leading order (NNLO) in the fixed order 
non-relativistic expansion, which is discussed in Secs.\
\ref{sectionquarkmass}--\ref{sectionsumrules}.

Although a cutoff regularization is in principle feasible, it has a
number of disadvantages, particularly in QCD.
In analytic QCD bound state calculations the cutoff breaks gauge
invariance at intermediate steps. This problem does not occur on the
lattice, because there NRQCD can be defined in a manifest
gauge-invariant way. In addition, in analytic QCD bound state
calculations a cutoff introduces also subtle scheme-dependences at
intermediate steps, because the results for diagrams in the effective
theory depend on the routing of loop momenta (see
e.g.\ Ref.~\citebk{Labelle3}). 

\begin{figure}[t!] 
\begin{center}
\leavevmode
\epsfxsize=2.cm
\leavevmode
\epsffile[240 440 440 480]{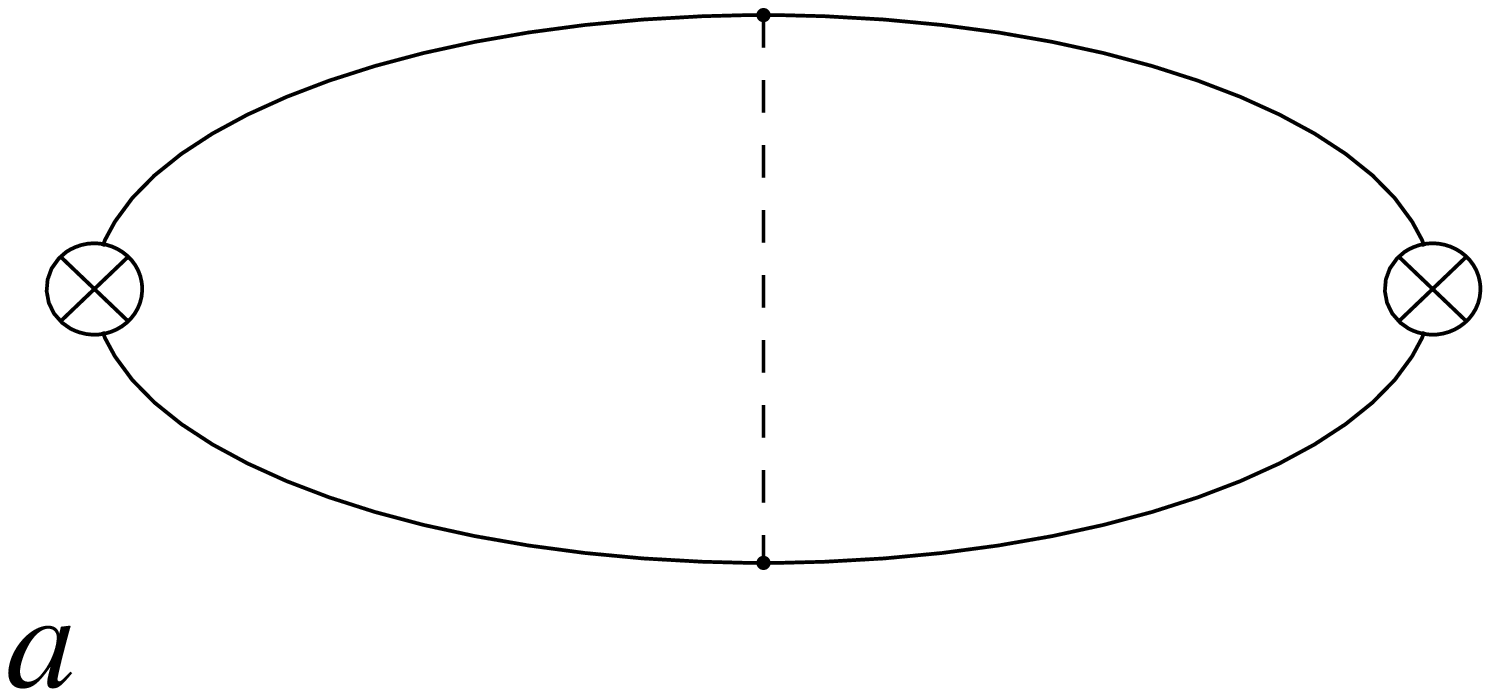}
\hspace{4cm}
\leavevmode
\epsfxsize=2.cm
\leavevmode
\epsffile[240 440 440 480]{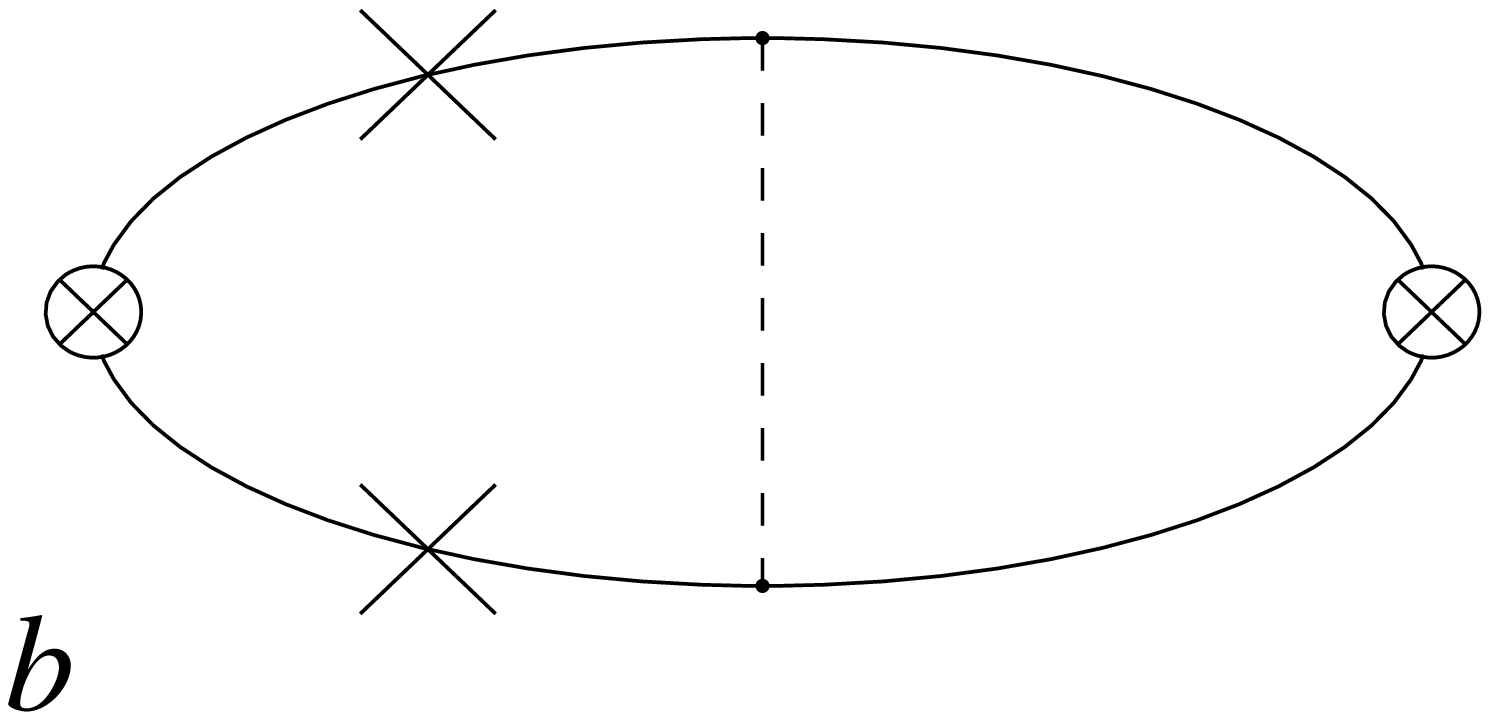}
%
%
\vskip  1.5cm
 \caption{\label{fignrqcdvector}   
The two-loop NRQCD vector current correlator with the exchange of one
Coulomb gluon without relativistic corrections (a) and with the
kinetic energy correction indicated by the crosses (b). 
}
 \end{center}
\end{figure}

A problem with both analytic and lattice
computations is that the power counting is never manifest with a
cutoff prescription, regardless whether all relevant momentum regions
are separated or not. (In the latter case, however, the effects of
power counting breaking are more severe.) The reason is that the
cutoff $\Lambda\sim m$ arises as an additional scale in the diagrams
of the effective theory. This introduces a sensitivity to the hard
scale $m$, when there are linear, quadratic, etc. UV divergences.
This means that an operator that contributes for the first time,
let's say, at NNLO according to the non-relativistic power counting
can lead to lower order contributions, if $\Lambda\sim m$. Such terms
are compensated by NLO matching corrections of coefficients in the
effective Lagrangian. In practice this means that whenever additional
higher dimensional operators of the effective theory are included to
increase the order of approximation, the matching conditions of all
coefficients, i.e.\ also of the coefficients of lower dimensional
operators, obtain additional lower order corrections. This is a feature
well known in NRQCD lattice computations, see also
Ref.~\citebk{Lepage1}. 
To illustrate this issue consider the two-loop NRQCD vector
current correlator at LO in Fig.\ \ref{fignrqcdvector}a,
which contains the exchange of a longitudinal Coulomb gluon.
The power counting in Labelle's scheme indicates that the dominant
contributions to the absorptive
part of the diagram is of order $\alpha_s v^0$,
\begin{eqnarray}
\mbox{Im$[$Fig.\ \ref{fignrqcdvector}a$]$}  & = & 
\frac{3\,N_c\,C_F\,\alpha_s\,m^2}{2\, \pi^2}\,\bigg[\,
\frac{\pi^2}{2}-\frac{4\,|{\bmp}|}{\Lambda} + 
{\cal{O}}\left(\frac{|{\bmp}|^2}{m^3}\right)
\,\bigg]
\,,
\end{eqnarray}
where $\pm{\bmp}$ is the three-momentum of the produced quarks, and
$N_c=3$, $C_F=4/3$.
Now consider the kinetic energy corrections in the same two-loop
diagram in Fig.\ \ref{fignrqcdvector}b
caused by the $\frac{{\bmD}^4}{8m^3}$ terms in the NRQCD
Lagrangian. According to Labelle's scheme the dominant contribution to
the absorptive part
is of order $\alpha_s v^2$, but the actual result reads
\begin{eqnarray}
\mbox{Im$[$Fig.\ \ref{fignrqcdvector}b$]$}
& = & 
\frac{3\,N_c\,C_F\,\alpha_s\,m^2}{4\, \pi^2}\,\bigg[\, 
\frac{\Lambda\,|{\bmp}|}{m^2} +
\frac{{\bmp}^2\,\pi^2}{2\,m^2} +  
{\cal{O}}\left( \frac{|{\bmp}|^3}{m^3} \right)
\,\bigg]
\,,
\end{eqnarray}
and contributes at order $\alpha_s v$ for $\Lambda\sim m$ because of a
linear UV divergence.

\subsection{Dimensional Regularization}
\label{subsectionNRQCDdimreg}

It has become common practice in analytic QCD computations to use
dimensional regularization to regulate UV and IR
divergences. Dimensional regularization maintains gauge invariance and
provides a comfortable framework to determine the anomalous dimensions
of the operators in  effective theories. In addition, power counting
breaking effects as mentioned at the end of the previous section do
not arise, because power divergences are automatically set to zero. 
However, it turns out that the NRQCD Lagrangian cannot be defined
in dimensional regularization, if the non-relativistic power counting
rule $D_0\sim\frac{{\bmD}^2}{m}$ is maintained. Manohar\,\cite{Manohar1}
pointed out that the resulting effective theory contains spurious
particle poles at the hard scale, which lead to the breakdown of the
power counting and make a consistent matching to QCD or even the
determination of anomalous dimensions impossible.
As an example, consider the one-loop selfenergy of a quark due to a
gluon. Taking the external quark in its rest frame
the selfenergy has the generic form
\begin{eqnarray}
\int \frac{d^Dk}{(k_0^2-{\bmk}^2+i\delta)
(k_0 - \frac{{\bmk}^2}{2m}+i\delta)}
\,.
\label{selfenergyunexpanded}
\end{eqnarray}
Carrying out the $k_0$ integration by contours 
one finds a term proportional to 
\begin{eqnarray}
\int \frac{d^{D-1}k}{|{\bmk}|^2(|{\bmk}|+2m)}
\,,
\end{eqnarray}
which is non-zero and dominated by hard momenta $k\sim m$.  
The same feature arises at higher loop orders and for
$\QQbar$ scattering diagrams. Thus NRQCD does not properly
describe non-relativistic degrees of freedom, if it is defined in
dimensional regularization and if the non-relativistic power counting
$D_0\sim\frac{{\bmD}^2}{2m}\sim mv^2$
is imposed.

Manohar pointed out that the problem can be resolved for the single
quark sector (i.e.\ all operators bilinear in the quark
field and with arbitrary number of light fields) by treating the
kinetic energy term $\frac{{\bmD}^2}{2m}$ as a perturbation. In this
case the selfenergy has the form
\begin{eqnarray}
\int \frac{d^Dk}{(k_0^2-{\bmk}^2+i\delta)}\,
\bigg\{\,
\frac{1}{(k_0+i\delta)}
+\frac{{\bmk}^2}{2m(k_0+i\delta)^2}+\ldots
\, \bigg\}
\,,
\label{selfenergyexpanded}
\end{eqnarray}
and vanishes at any order in the expansion.
A sensitivity to the hard scale does not arise. It is important to
realize that the integrals in Eqs.\ (\ref{selfenergyunexpanded}) and 
(\ref{selfenergyexpanded}) are not equivalent, even if the terms in the
$1/m$ expansion in Eq.~(\ref{selfenergyexpanded}) are summed up. This
is because in dimensional
regularization, expansion and integration do not commute, in contrast
to a cutoff scheme. 
Therefore the single quark sector of NRQCD can only be consistently
defined if the $1/m$ counting of heavy quark effective 
theory\,\cite{HQETpapers} (HQET) 
is adopted. A comparable 
prescription for the four-quark sector of NRQCD was not found.
However, Grinstein and Rothstein\,\cite{Grinstein1} pointed out that
Labelle's multi-pole expansion can be understood as a generalization of
Manohar's prescription (see also Ref.~\citebk{Luke2}). Thus, Manohar's
prescription follows from the 
fact that in the multipole expansion the combination of propagators in
Eq.\ (\ref{selfenergyunexpanded}) does simply not exist. When keeping
the form $1/(k_0^2-{\bmk}^2)$ for the gluon propagator one has
$k_0\sim{\bmk}\ll m$ and the quark propagator has to be expanded in
$1/m$. On the other hand, when keeping the form
$1/(k_0-\frac{{\bmk}^2}{2m})$ for the quark propagator one has
$k_0\ll{\bmk}\ll m$ and one has to expand the gluon propagator in
$k_0$. The latter contribution is a scaleless integral and vanishes to
all orders in the expansion.

\vspace{1cm}

\section{Method of Regions and Threshold Expansion} 
\label{sectionthresholdexpansion}

In bound state calculations one frequently needs a small-velocity
expansion of Feynman diagrams involving a heavy quark pair close to
threshold. The threshold expansion is a
prescription developed by Beneke and Smirnov\,\cite{Beneke1} to carry out
such an expansion for loop diagrams in dimensional
regularization. The prescription of the threshold expansion is based
on the more general ``method of regions'' and fully
appreciated for the first time the meaning of the multipole expansion
in dimensional regularization. 

\subsection{Basic Idea}

The problems in constructing an asymptotic expansion and the basic
idea of the method of regions can be illustrated with a simple 
example. Consider the one-dimensional integral
\begin{eqnarray}
f(a) & = & \int\limits_{-\infty}^{+\infty}\! dk\, 
\frac{|\arctan(k)|}{(k^2+a^2)^2}
\,,
\nonumber
\end{eqnarray}
which one might take as a simplified version of a diagram with 
propagator-like factors of $1/(k^2+a^2)$ and a non-trivial numerator
structure. The task is to expand $f(a)$ in $a\ll 1$. Here it is
impossible to naively expand $f$ before integration, because each term
is ``IR divergent'' at $k=0$. Altogether there are two relevant
integration regimes one has to consider, the ``hard'' regimes where
$k\sim 1$ and the ``soft'' regime where $k\sim a$. In a cutoff scheme
one deals with this situation by introducing a cutoff $\Lambda$ with
$a\ll\Lambda\ll 1$. The cutoff separates the full integration range
into a ``soft'' regime $|k|<\Lambda$ and a ``hard'' regime
$|k|>\Lambda$. In the ``soft'' regime we have $k\sim a\ll 1$, and we can
expand the numerator. In the ``hard'' regime we have $k\gg a$ and we can
expand in $a$, 
\begin{eqnarray}
\begin{array}{lclcl}
\mbox{``hard''} & \mbox{:} & \displaystyle
\int_{|k|>\Lambda}\! dk\, \frac{|\arctan(k)|}{k^4} + \ldots
& = &  \displaystyle
\frac{1}{\Lambda^2}-\frac{2}{9}+\frac{2}{3}\ln\Lambda + \ldots
\,,
\\[4mm]
\mbox{``soft''} & \mbox{:} &  \displaystyle
\int_{|k|<\Lambda}\! dk\, \frac{|k-\frac{k^3}{3}+\ldots|}
{(k^2+a^2)^2}
& = & \displaystyle
\frac{1}{a^2}-\frac{1}{\Lambda^2}+
\frac{1}{3}\,\bigg(1+2\ln\Big(\frac{a}{\Lambda}\Big)\bigg)+\ldots
\,.
\end{array}
\nonumber
\end{eqnarray}
Carrying out the Taylor expansions before the integration considerably
simplified the computation, but it is not really necessary for the
method to work because expansion and integration commute in the
presence of the cutoff $\Lambda$. Adding together the ``hard'' and the
``soft'' contributions we readily find the correct asymptotic
expansion of the integral $f$, 
\begin{eqnarray}
f(a) & = & \frac{1}{a^2}+\frac{1}{9}+
\frac{2}{3}\,\ln a + \ldots
\,.
\nonumber
\end{eqnarray}
The $\Lambda$-dependent divergent terms cancel in the sum. The
logarithmic divergent terms are special because their existence is
still visible in the final sum through the non-analytical $\ln a$
term. Thus the ``large'' logarithmic $\ln a$ terms originate from
logarithmic divergences at the borders of the regions. 
If we would consider a similar situation for the computation of a
complicated multi-loop Feynman diagram with a non-trivial structure of
relevant regions in QCD, the method described above would be
in principle feasible, but quite uncomfortable and cumbersome due to the
existence of additional cutoff scales. In QCD this also leads to the
breakdown of gauge invariance in intermediate steps of the computation.

A more economic (but also less physical) regularization method is to use
an analytic regularization method. 
In QCD computations the most common choice is to use 
dimensional regularization in the \ms scheme, where the
four-dimensional integration measure is continued to $D=4-2\epsilon$
dimensions, and $\epsilon$ is an arbitrarily small complex
parameter. In our simple example we continue the integration measure
to $\bar D=1-2\epsilon$ dimensions,
$$\int_{-\infty}^{+\infty} dk\to \tilde\mu^{2\epsilon}\int d^{\bar D}k
=\frac{\Omega(\bar D)}{\tilde\mu^{-2\epsilon}}\int_0^\infty dk
k^{-2\epsilon}\,,$$ where 
$$\Omega(\bar D)=\frac{2\,\pi^{\frac{\bar D}{2}}}{\Gamma(\frac{\bar D}{2})}$$ 
is the
$\bar D$-dimensional angular integral and 
$$\tilde\mu=\mu\left(e^{\gamma_E+\ln 4\pi}\right)^{\frac{1}{2}}\,,$$ 
$\gamma_E=0.577216\ldots$.
This is just one specific regularization method
of an infinite number of possibilities to regulate
the integral analytically.
The separation of the ``hard'' and the ``soft'' regions is
not implemented through a restriction to integration regions, but
by {\it strictly} Taylor expanding out the hierarchies in the regions.
For the contributions from the ``hard'' and the ``soft'' regions one
now finds
\begin{eqnarray}
\begin{array}{lclcl}
\mbox{``hard''} & \mbox{:} & \displaystyle
\frac{\Omega(\bar D)}{\tilde\mu^{-2\epsilon}}\int_0^\infty\!\! dk\, 
\frac{\arctan(k)}{k^{4+2\epsilon}} + \ldots
& = &  \displaystyle
\frac{1}{3\epsilon}-\frac{2}{9}+\frac{2}{3}\ln\mu + \ldots
\,,
\\[4mm]
\mbox{``soft''} & \mbox{:} &  \displaystyle
\frac{\Omega(\bar D)}{\tilde\mu^{-2\epsilon}}\int_0^\infty\!\! dk\, 
\frac{k-\frac{k^3}{3}+\ldots}
{k^{2\epsilon}(k^2+a^2)^2}
& = & \displaystyle
\frac{1}{a^2}-\frac{1}{3\epsilon}+
\frac{1}{3}\bigg(1+2\ln\Big(\frac{a}{\mu}\Big)\bigg)+\ldots
\,.
\end{array}
\nonumber
\end{eqnarray}
The sum of the terms gives again the asymptotic expansion of $f$ and
the $\mu$-dependent terms cancel. 
Linearly divergent contributions do
not arise at all, whereas the large logarithmic terms $\ln a$ are
associated with the $1/\epsilon$ singularities and $\ln\mu$ terms that
arise from divergences at the boundaries of the regions. The method
also works if the original integral is divergent itself for
$\epsilon\to 0$. In this case the outcome of the method agrees with
the result of the full calculation in the limit $\epsilon\to 0$
supplemented by an expansion in $a$. Choices for the analytic
regularization scheme that are different from the one used in the
simple example above lead to different intermediate results for the
``hard'' and the ``soft'' contributions, but to the same final result, if
the original integral is finite. 
The method breaks down if an expansion it not carried out in a strict
way. In particular, nothing is gained keeping some
contributions unexpanded with the intention to sum up certain terms.

It is an important feature of this method that each term in the
expansion only contributes to a single power of $a$ and that the order
of $a$ to which each term contributes can be easily determined before
integration by considering the $a$-scaling of $k$ in the two regions.
In the ``hard'' regime we have $k\sim 1$ and $dk\sim 1$, and we just
expand in powers of $a$ up to the required order in $a$. In the ``soft''
regime we have $k\sim a$ and $dk\sim a$, and $1/(k^2+a^2)\sim
a^{-2}$ for the propagator-like term. Thus the leading term in
the expansion in the soft region is of order $a^{-2}$, etc..

\subsection{Threshold Expansion and Regions}

The prescription for the threshold expansion goes along the lines of
the previous simple example. In Ref.~\citebk{Beneke1} four different
loop momentum regions were identified to be relevant by an analysis of
Feynman diagrams in full QCD involving non-relativistic $\QQbar$ pairs
close to threshold, 
\begin{eqnarray}
\begin{array}{lcrcl}
\mbox{hard}&\mbox{:}&\quad(k^0,\bmk)&\sim&(m,m)\,,\\[2mm]
\mbox{soft}&\mbox{:}&(k^0,\bmk)&\sim&(mv,mv)\,,\\[2mm]
\mbox{potential}&\mbox{:}&(k^0,\bmk)&\sim&(mv^2,mv)\,,\\[2mm]
\mbox{ultrasoft}&\mbox{:}&(k^0,\bmk)&\sim&(mv^2,mv^2)\,.
\end{array}
\label{momentumregions}
\end{eqnarray}
The energy and three-momentum components of the loop momentum in the
potential region scale differently with $v$, because Lorentz-invariance
is broken for $\QQbar$ pairs close to threshold. The structure of the
regions is most transparent if the $\QQbar$ pair is in the
center-of-mass frame.

The small velocity expansion of a full QCD diagram, which involves a
non-relativistic $\QQbar$ pair, is obtained by
writing it as a sum of 
terms that arise from dividing each loop into all possible
regions. The separation of the regions is achieved by strictly
expanding loop momenta according to the hierarchy in the various
regions. All terms from all regions that arise from this procedure are
integrated over the full $D$-dimensional space using dimensional
regularization in let's say the \ms scheme. 
All scaleless integrals have to be set to zero. 
The explicit expansion of energies and three-momenta
that are small is essential to make the method work, because the
expansion is the (only) instrument that separates the regions. 

It is one of the amazing features of dimensional regularization that
this heuristic
prescription, which involves subtle cancellations of
IR and UV divergences in neighboring regions, appears to work. This
``method of regions'' is quite general and
has been used earlier for other problems, such as the large mass
or the large energy expansion.\cite{Smirnov1} Also extended and
modified versions of the threshold 
expansion for different kinematic situations and particle content
have been devised.\cite{Czarnecki1,Czarnecki2}
A mathematical and more rigorously defined formulation of the method in
terms of so called R-operations has been developed earlier by 
Smirnov\cite{Smirnov2} (see also Ref.~\citebk{Smirnovnewbook}). 
Some recent formal considerations pointing out potential problems 
can be found in Ref.~\citebk{Smirnov3}.
Up to now  the method has not been proven mathematically to work in
general for arbitrary diagrams with on-shell
external particles,\cite{Smirnov3} but no counter example has been
found in cases where the expansion of the complete result was
available. This should be kept in mind, since for many multi-loop
calculations an asymptotic expansion using the method of
regions seems to be the only way to tackle the problem in the first
place. A mathematical proof, however, exists for off-shell
external particles.\cite{Smirnov2} It should also be noted that  
it is in principle not excluded that regions other than in Eq.\
(\ref{momentumregions}) could become relevant in some cases. 
In such a case the new regions would simply have to be included
without affecting the method itself.
Interesting examples (but not dealing with non-relativistic $\QQbar$
pairs), where the number of
regions increases indefinitely at higher loop orders,
have been reported in Ref.~\citebk{Smirnov3}.

In actual calculations, the threshold expansion has been applied
correctly, if each integral only contributes to a single power in
$v$. The order of $v$ to which a certain term in the threshold
expansion contributes, can be determined by power counting rules that
are obtained from the $v$-scaling of the regions in Eq.\
(\ref{momentumregions}) similar to the simple example discussed in the
previous section. The feature of the power counting makes the
threshold expansion method quite economic, because all terms needed
for a certain order in $v$ can be identified unambiguously before
doing any integrations. For example, the integration measure $d^4k$ in
the hard, soft, potential and ultrasoft regions scale $v^0$, $v^4$,
$v^5$ and $v^8$, respectively. 

The threshold expansion had a significant influence on the development
of non-relativistic effective theories. Its features, such as
the existence of the different separable momentum regions, the
existence of power counting rules and the requirement to 
``multipole-expand'' in
small momenta in dimensional regularization 
also play an important role in effective theories. The
present use of language, for 
example calling a gluon propagating in the soft region a ``soft gluon'',
or a quark propagating in the potential region a ``potential quark'',
etc., was finally established with the threshold expansion.
Since the time it has been devised, a number of calculations, which
were important for matching calculations in effective theories for
non-relativistic $\QQbar$ pairs, have been carried out using the
threshold expansion, see e.g.
Refs.~\citebk{Czarnecki3,Kniehl1,Czarnecki4}. 
However, the threshold expansion is a prescription to obtain the
asymptotic expansion of Feynman diagrams and not an effective theory
itself, because it includes on-shell as well as off-shell modes. 
(An attempt to formulate the threshold expansion
in form of an effective Lagrangian containing fields
for the off-shell soft heavy quarks has been given in 
Ref.~\citebk{Griesshammer1}.) 
In addition, the concept of renormalization is not implemented into
the threshold expansion.  
Therefore the summation of large logarithmic terms is impossible. 
It should also be noted that the 
definition of the regions in Eq.\ (\ref{momentumregions}) is not
necessarily equivalent to the corresponding regions in effective
theories due to the freedom in the choice of the operator basis.
For example, the contributions obtained from the hard momentum regime
in the threshold expansion are in general not equivalent to the hard matching
conditions of operators obtained in an effective theory.\,\cite{Hoang3}
However, an effective theory has to be able to reproduce the sum of 
contributions from the different regions of Eq.\
(\ref{momentumregions}).

\subsection{Example: One-Loop Vertex Diagram}

\begin{figure}[ht] 
\begin{center}
\leavevmode
\epsfxsize=2.3cm
\leavevmode
\epsffile[240 510 450 680]{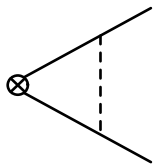}
\end{center}
%
%
\vskip  0cm
 \caption{\label{figscalabox}   
One-loop scalar vertex diagram with external massive lines close to
threshold. Solid lines are massive, and the dashed line is massless.
All lines represent scalars.
}
\end{figure}

As a simple example for the application of the threshold expansion let us
consider the one-loop vertex diagram in Fig.\ \ref{figscalabox},
which describes production of two massive scalars
(solid lines) with momenta $p_{1/2}=(p_0,\pm{\bmp})$ and the
subsequent exchange of a massless scalar (dashed line). 
The presentation given here is slightly different to the one of
Ref.~\citebk{Beneke1} with respect to the interpretation of the soft
and ultrasoft contributions.
Close to threshold we have $m E\equiv p_0^2-m^2\sim  
{\bmp}^2 \sim (mv)^2$. We assume that the produced massive scalars are
on-shell, i.e.\ we have the relation $m E = {\bmp}^2$. The on-shell
relation is, however, not essential for the threshold expansion as
long as the $v$-counting of the external energies and three-momenta is 
well-defined. The full loop integral reads ($D=4-2\epsilon$)
\begin{eqnarray}
\lefteqn{
\int\frac{d^Dk}{(2\pi)^D}\,
\frac{1}{[(k+p_1)^2-m^2][(k-p_2)^2-m^2]
[k^2]}
}
\nonumber\\ & = &\!\!
\int\!\!\!\frac{d^Dk}{(2\pi)^D}
\frac{1}{[(k_0+p_0)^2-({\bmk}\!+\!{\bmp})^2-m^2]
         [(k_0-p_0)^2-({\bmk}\!+\!{\bmp})^2-m^2]
[k_0^2-{\bmk}^2]}.
\nonumber\\
\label{fulloneloop}
\end{eqnarray}
The
renormalization scale $\mu$ is set to $1$ for simplicity and an
conventional overall factor $(4\pi)^{-\epsilon}$ is understood. 
The usual 
$+i\delta$ prescription in all propagators is implied.
In the following we determine the respective leading order
contributions from the different regions.

In the hard regime we have to expand the propagators in ${\bmp}$,
${\bmp}^\prime$ and $mE$. To leading order we obtain,
\begin{eqnarray}
& &
\int\frac{d^Dk}{(2\pi)^D}\,
\frac{1}{[k^2+2k_0p_0+i\delta]\,[k^2-2k_0p_0+i\delta]\,[k^2+i\delta]}
\nonumber\\
&  & \,\,=\,
- \frac{i}{32\,\pi^2\, (p_0^2)^{1+\epsilon}}\,
\frac{\Gamma(\epsilon)}{1+2\epsilon}
\,.\mbox{\hspace{2cm}}
\label{thresholdexpansionhard}
\end{eqnarray}
The result is of order $m^{-2}$, which can be obtained before
integration from the scaling $d^4k\sim m^4$ and $k_0\sim{\bmk}\sim
p_0\sim m$.

In the potential regime we have to expand the propagators in
$k_0^2$.  To leading order we obtain,
\begin{eqnarray} &&
\int\!\!\frac{d^Dk}{(2\pi)^D}\,
\frac{1}{[+2k_0p_0-({\bmk}\!+\!{\bmp})^2+m E+i\delta]
[-2k_0p_0-({\bmk}\!+\!{\bmp})^2+m E+i\delta]
[-{\bmk}^2]}
\,.
\nonumber\\
\end{eqnarray}
For on-shell external particles the result reads
\begin{equation}
\frac{i}{32\,p_0\,(-mE-i\delta)^{\frac{1}{2}+\epsilon}}\,
\frac{\Gamma(\epsilon+\mbox{$\frac{1}{2}$})}{2\,\pi^{\frac{3}{2}}\,\epsilon}
\,.
\end{equation}
It is of order $m^{-2}v^{-1}$, which can be obtained before
integration from the scaling $d^4k\sim m^4v^5$, $k_0\sim mv^2$ and 
${\bmk}\sim{\bmp}\sim mv$.
The massless propagator does not have an $i\delta$ prescription
since it does not depend on $k_0$ and cannot develop a particle pole.

In the soft regime the $k_0p_0$ terms in the massive propagators are
dominant. The leading term is of order 
$(mv)^4(m^2 v)^{-2}(m^2v^2)^{-1}=m^{-2}$ and reads
\begin{eqnarray} &&
\int\frac{d^Dk}{(2\pi)^D}\,
\frac{1}{[+2k_0p_0+i\delta]
[-2k_0p_0+i\delta]
[k_0^2-{\bmk}^2+i\delta]}
\,.
\label{thresholdexpansionsoft}
\end{eqnarray}
The integral is scaleless and proportional to
$(\frac{1}{\epsilon_{\rm UV}}-\frac{1}{\epsilon_{\rm IR}})$, 
and there is no contribution from the soft
region. However, the integral contains a pinch-singularity at $k_0=0$
and is in fact ill-defined.
However, the poles at $k_0=\pm
i\delta$ in the complex $k_0$-plane are unphysical because the
massive lines cannot become on-shell in the soft region.
Thus, the integral can be defined by removing the $i\delta$'s from
the massive propagators with the prescription that the $k_0$ is
carried out first by contours.

Finally, in the ultrasoft regime we have to expand the  
massive propagators in $k^2$ and ${\bmk}\cdot{\bmp}$.
The leading term is of order 
$(mv^2)^4(m^2v^2)^{-2}(m^2v^4)^{-1}=m^{-2}$ and reads
\begin{eqnarray} &&
\int\frac{d^Dk}{(2\pi)^D}\,
\frac{1}{[+ 2k_0p_0 + mE  - {\bmp}^2 + i\delta]
[- 2k_0p_0 + mE - {\bmp}^2 + i\delta]
[k_0^2-{\bmk}^2+i\delta]}
\,.
\nonumber\\
\label{thresholdexpansionusoft}
\end{eqnarray}
Interestingly, the ultrasoft contribution is the same as the soft one
in Eq.~(\ref{thresholdexpansionsoft})
for on-shell external momenta. However, soft and ultrasoft
contributions are not equivalent, because for off-shell external
momenta the soft contribution in Eq.~(\ref{thresholdexpansionsoft})
has the pinch-singularity, the ultrasoft contribution, on the other
hand, does not. In the off-shell case 
Eq.~(\ref{thresholdexpansionusoft}) gives an UV divergence
$\propto\frac{1}{\epsilon_{\rm UV}}$. In the on-shell case there is an
additional IR divergence and the overall contribution is zero.
The complete one-loop result for Eq.~(\ref{fulloneloop})
for on-shell momenta can be 
determined analytically and reads  
$$\frac{i}{16\pi^2}(-mE-i\delta)^{-1-\epsilon}\,
{}_2F_1\left(\frac{1}{2},1+\epsilon,\frac{3}{2},
\frac{p_0}{mE+i\delta}\right)\,
\frac{\Gamma(1+\epsilon)}{2\epsilon}\,,$$ where
${}_2F_1$ is the hypergeometric function. An expansion in $E$ yields
the contributions determined with the threshold expansion.

\vspace{1cm}

\section{Effective Theories II: Potential NRQCD}
\label{sectionpNRQCD}

The effective theory pNRQCD, first proposed by Pineda and
Soto,\cite{Pineda1} has been devised for the description of
non-relativistic $\QQbar$ systems where $m\gg mv\gg mv^2\gg \lqcd$ as
well as for cases where $mv^2$ is of order or small than $\lqcd$. 
In particular, pNRQCD is proposed to describe
dynamic $\QQbar$ pairs, which are relevant for physical application,
as well as static $\QQbar$ systems, which, for example, can be studied 
non-perturbatively on the lattice. 
In pNRQCD the separation of the momentum regions is achieved in two
steps and dimensional regularization is used.
In this presentation I will mainly discuss the case $m\gg
mv\gg mv^2\gg \lqcd$, because it has the most solid level of
understanding.

\subsection{Basic Idea}

The basic idea behind the construction of pNRQCD is that 
the quark and gluon field components that are off-shell in the
hard, soft, potential and ultrasoft momentum regions can be integrated
out in two steps. To be more precise, pNRQCD is the theory that
results from this two-step procedure. Starting from full QCD, first
the quark and gluonic off-shell degrees of freedom in the hard region
are integrated out at the scale $\mu=m$. The resulting theory for
$\mu<m$ is equal to NRQCD (Sec.~\ref{sectionNRQCD}) for bilinear 
quark terms and, for simplicity, also called NRQCD in this context. 
This theory is then
scaled down to the soft scale $\mu=mv$, where in a second step the
quark and gluonic degrees of freedom in the soft regime and
the gluonic degrees of freedom in the potential regime are integrated
out. The theory that results for $\mu<mv$ is called
``potential'' NRQCD, because the four-quark interactions that are
generated through the matching procedure are
the potentials in the Schr\"odinger perturbation theory that can be
derived from pNRQCD. 
The theory pNRQCD is then scaled down to $\mu=mv^2$, where matrix
elements are calculated. At the scale $\mu=mv^2$ matrix elements
should be free of any large logarithmic terms, and all logarithms
should be resummed into Wilson coefficients. 
Graphically the scheme is as follows:
\begin{equation}
\label{pNRQCDscheme}
\begin{array}{ccc}
\LQCD
 && (\mu > m)\\[2mm]
\Downarrow\\[2mm]
\LNRQCD
&& (m > \mu > mv)\\[2mm]
\Downarrow\\[2mm]
\LpNRQCD
&\hspace{0.6cm}
& (mv > \mu > mv^2)
\end{array}
\end{equation}

\subsection{Matching, Running and Power Counting in NRQCD}

The intermediate theory, NRQCD, 
is constructed in close analogy to the effective theory proposed by 
Caswell and Lepage\,\cite{Caswell1} and
Bodwin, Braaten and Lepage\,\cite{Bodwin1} discussed in
Sec.~\ref{sectionNRQCD}. This means in particular that the
gluon field describes all fluctuations of frequencies below $m$.
To avoid a sensitivity of NRQCD loop
diagrams to the hard scale (Sec.~\ref{subsectionNRQCDdimreg}), 
the non-relativistic power counting is
abandoned and a strict $1/m$ counting is carried out everywhere.
Thus the theory is defined in a static expansion,
and the bilinear quark sector of NRQCD is equivalent to
HQET.\cite{Manohar1}
The $1/m$ counting is not only applied to the bilinear quark sector
but, in particular, also to interactions between quarks and
antiquarks such as four quark operators. Therefore, the intermediate
theory in the scheme of Eq.\ (\ref{pNRQCDscheme}) is not equivalent to
the NRQCD theory originally proposed by Caswell and
Lepage\,\cite{Caswell1} (even when supplemented by Manohar's
prescription\,\cite{Manohar1} for
the single quark sector) but rather a specific extension of
it. Nevertheless, for simplicity we will call the intermediate theory
of Eq.~(\ref{pNRQCDscheme}) also ``NRQCD''. Up to order $1/m^2$ the 
Lagrangian has the form
\begin{eqnarray}
\lefteqn{
{\cal L}_{\rm NRQCD} 
\, = \,
\psi^{\dagger} \bigg\{ i D_0
+ \, c_k {{\bf D}^2\over 2 m} + \, c_4{{\bf D}^4\over 8 m^3}
+ c_F\, g {{\bf \bfsigma \cdot B} \over 2 m}
}
\nonumber
\\ && 
+ c_D \, g { \left({\bf D \cdot E} - {\bf E \cdot D} \right) \over 8 m^2}
+ i c_S \, g { {\bf \bfsigma \cdot \left(D \times E -E \times D\right)}
    \over 8 m^2} + \ldots \bigg\} \psi
+ (\psi \to \chi )
\nonumber
\\ && 
+  {d_{ss} \over m^2} \psi^{\dag} \psi \chi^{\dag} \chi
+ {d_{sv} \over m^2} \psi^{\dag} {\bfsigma} \psi
                         \chi^{\dag} {\bfsigma} \chi
\nonumber
\\ && 
+
  {d_{vs} \over m^2} \psi^{\dag} {\rm T}^a \psi
                         \chi^{\dag} {\rm T}^a \chi
+
  {d_{vv} \over m^2} \psi^{\dag} {\rm T}^a {\bfsigma} \psi
                         \chi^{\dag} {\rm T}^a {\bfsigma} \chi
\nonumber
\\ && 
-\frac{1}{4}G^{\mu\nu}G_{\mu \nu}
\,,
\label{NRQCDLagrangian2}
\end{eqnarray}
where $\psi$ is the Pauli spinor that annihilates the massive quark,
$\chi$ is the Pauli spinor that creates the massive antiquark, and $$i
D_0=i\partial_0 -gA_0\,,\quad i{\bf D}=i\bfnabla+g{\bf A}\,,
\quad i g G^{\mu\nu}=\left[D^\mu,D^\nu\right]\,.$$
The convention for the four quark operators has been taken from
Refs.~\citebk{Pineda2,Pineda3} and $m$ is the pole mass of the heavy quark.
Operators involving light quarks are not displayed. In the following I
take the convention that, when I mention gluons, I refer to light
quarks as well. It is important that the overall 
$1/m$-counting for $\QQbar$ scattering treats the kinetic energy
term $\frac{\bmat{D}^2}{m}$ as a perturbation and, at the same time,
removes the potential region from the theory, because the hierarchy
$$D_0\sim k^0 \gg \frac{\bmat{D}^2}{m}\sim \frac{\bmag{k}^2}{2m}$$ is
enforced.  
This can be seen from the fact that in dimensional regularization an
expansion of the quark propagator in $1/m$, 
\begin{equation}
\frac{i}{k^0-\frac{{\bmk}^2}{2m}+i\delta} = 
\frac{i}{k^0+i\delta} 
+ \frac{i}{(k^0+i\delta)^2}\frac{{\bmk}^2}{2m}
+ \ldots
\,,
\end{equation}
sets $k^0 \sim \bmag{k}$ of order or smaller than $mv$. 
It is claimed in Refs.~\citebk{Pineda1,Pineda2,Brambilla1} that the
removal of the potential region does not affect the matching and
running, or any physical prediction for dynamical quarks 
made with pNRQCD. In particular,
it is stated that the potential region is still contained in the
theory, but hidden in a subtle way (see Sec.~\ref{subsectionopenissues}). 
For the bilinear quark sector the matching procedure is carried out
using the prescription of Manohar.\cite{Manohar1} For amplitudes
describing quark-antiquark interactions and also for currents describing
$\QQbar$ production the matching to QCD is carried out 
on-shell and at threshold, i.e.\ the four-momenta of the external
quarks in the full QCD amplitudes need to be set to $k=(m,0)$. The
idea is that in this kinematic situation all NRQCD loop
computations give exactly zero, because the integrals are scaleless.
For the loop computations in full QCD only the hard contributions 
should be left over. At the one-loop level
this can be seen from the vertex computation 
discussed in Sec.~\ref{sectionthresholdexpansion} using the
threshold expansion.
Setting the external momenta for the massive lines 
in the hard loop, Eq.~(\ref{thresholdexpansionhard}), to
$p_{1,2}=(m,0)$ we obtain
\begin{eqnarray}
&&
\int\frac{d^Dk}{(2\pi)^D}\,
\frac{1}{[k^2+2k_0 m+i\delta][k^2-2k_0 m+i\delta][k^2+i\delta]}
\nonumber\\ &&
 = 
- \frac{i}{32\,\pi^2\,(m^2)^{1+\epsilon}}\,
\frac{\Gamma(\epsilon)}{1+2\epsilon}
\,.\,\,\,\,\,
\end{eqnarray}
On the other hand, the potential, soft and ultrasoft loop
contributions are scaleless integrals that vanish. The $1/m$ expansion
is not carried out for those integrals since they are part of the complete
computation. This prescription should hold for
any number of loops, because any connected potential, soft and
ultrasoft loop integral in the threshold expansion for 
$\QQbar$ scattering diagrams is zero in this particular kinematic
situation.
To obtain the matching conditions for operators that contain
covariant derivatives
a suitable number of derivatives with respect to the external momentum
should be applied before setting the external quark momenta to  
$(m,0)$. Matching appears to be impossible for any other choice of
external momenta due to the $1/m$ expansion in NRQCD.

The matching conditions at $\mu=m$ for the two-quark operators are
equivalent to the results in HQET, and for
the four-quark operators they were derived in Ref.\ \citebk{Pineda2}
(see also Ref.\ \citebk{Pineda3}). 
The RG equations for the coefficients of the two-quark sector 
are again equivalent to those of HQET.\cite{Bauer1} The RG equations
at one-loop 
for the four-quark operators and their
solution for the coefficients can be found
in Ref.~\citebk{Pineda3}.

\subsection{Matching, Running and Power Counting in pNRQCD}
\label{subsectionpNRQCDrunning}

\begin{figure}[t] 
\begin{center}
\leavevmode
\epsfxsize=3.2cm
\leavevmode
\epsffile[100 480 470 720]{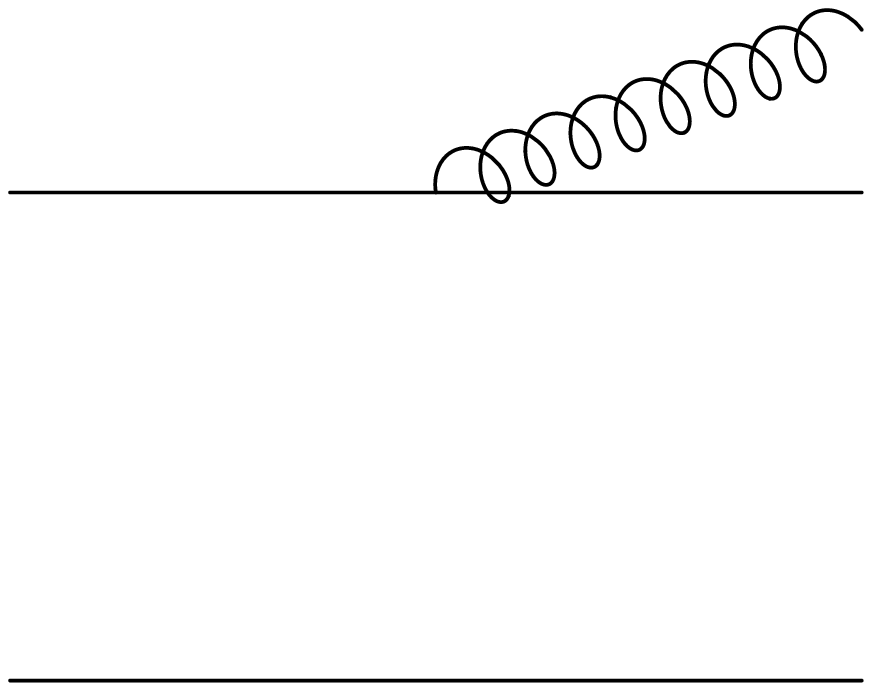}
\leavevmode
\epsfxsize=3.2cm
\leavevmode
\epsffile[100 480 470 720]{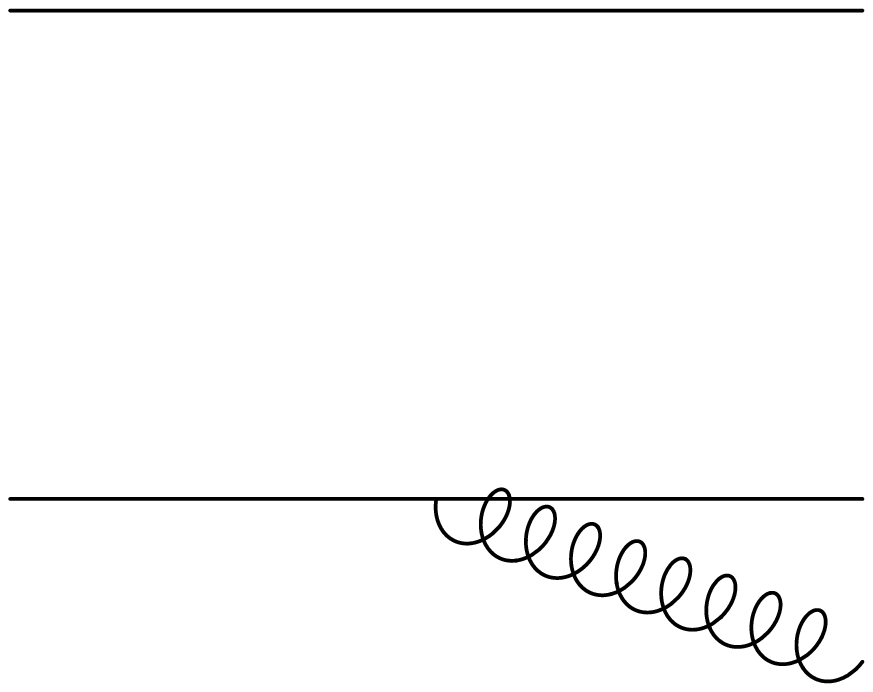}
\leavevmode
\epsfxsize=3.2cm
\leavevmode
\epsffile[100 480 470 720]{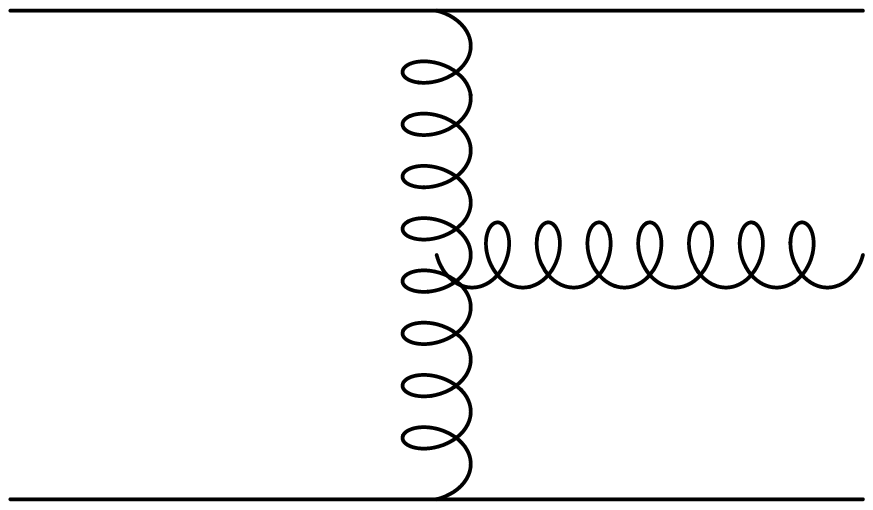}
 \end{center}
%
%
\vskip  0cm
 \caption{\label{figpnrqcdradiation}   
NRQCD diagrams that contribute to the ${\bmr}\cdot{\bmE}$ interactions
of ultrasoft gluons with the $\QQbar$ pair in pNRQCD.
}
\end{figure}

The final theory, pNRQCD, is formulated in terms of a quark-antiquark
pair and a single gluon field that describes ultrasoft fluctuations.
The Lagrangian in configuration space
reads\,\cite{Pineda1,Brambilla1}
\begin{eqnarray}
\label{pNRQCDLagrangian}                         
& & {\cal L}_{\rm pNRQCD} =
{\rm Tr} \,\Biggl\{ {\rm S}^\dagger \left( i\partial_0 
- c_k{{\bf p}^2\over m} +c_4{{\bf p}^4\over 4m^3}
- V^{(0)}_s(r) - {V_s^{(1)} \over m}- {V_s^{(2)} \over m^2}+ \dots  \right) {\rm S}
\nonumber \\
&& \nonumber 
\qquad \qquad + {\rm O}^\dagger \left( iD_0 - c_k{{\bf p}^2\over m}
- V^{(0)}_o(r) 
+\dots  \right) {\rm O} \Biggr\}
\nonumber\\
& &\qquad + g V_A {\rm Tr} \left\{  {\rm O}^\dagger {\bf r} \cdot {\bf E} \,{\rm S}
+ {\rm S}^\dagger {\bf r} \cdot {\bf E} \,{\rm O} \right\} 
+ g {V_B \over 2} {\rm Tr} \left\{  {\rm O}^\dagger {\bf r} \cdot {\bf E} \, {\rm O} 
+ {\rm O}^\dagger {\rm O} {\bf r} \cdot {\bf E}  \right\}  
\nonumber\\
& &\qquad- {1\over 4} G_{\mu \nu} G^{\mu \nu}
\,,
\end{eqnarray}
where ``$S$'' describes a $\QQbar$ color singlet and ``$O$'' a $\QQbar$
color octet, and where the $c_i$ and $V_i$ are Wilson coefficients.
The terms displayed in Eq.~(\ref{pNRQCDLagrangian}) are sufficient
for a  next-to-next-to-leading logarithmic (NNLL) 
description of the $\QQbar$ singlet dynamics.
The operators involving two $\QQbar$ fields and the ultrasoft gluon 
${\bf E}$-field describe the leading order interaction of an ultrasoft
gluon with a $\QQbar$ pair. In the matching procedure 
this interaction is the gauge-invariant combination of NRQCD diagrams
with (ultrasoft) gluon interactions with each of the quarks and the
(potential) gluon that is exchanged between the quark pair (see 
Fig.~\ref{figpnrqcdradiation}).
The bilinear $\QQbar$ operators involving the $V$'s describe
interactions between 
$\QQbar$ pairs, the potentials. This is the origin of the ``p'' in
pNRQCD. 
Because pNRQCD distinguishes between ultrasoft and
potential energy and momenta, the multipole expansion is carried
out, i.e.\ the dependence of the gluon field on the relative distance
between the quarks is expanded out and 
all gluon fields in Eq.~(\ref{pNRQCDLagrangian}) are only
functions of the center-of-mass coordinate and time.
The explicit $r$-dependence of the operators describing ultrasoft
gluon radiation arises from the multipole expansion.
Due to the multipole expansion the potentials only depend on the
relative distance between the quarks and not on time, i.e.\ they are
instantaneous, but non-local in space.
Up to order $1/m^2$ the singlet potentials read  ($C_F=4/3$, $C_A=3$)
\begin{eqnarray}  
V^{(0)}_{s} & = &
  - C_F {\alpha_{V_{s}} \over r}
\,,
\qquad\qquad
V^{(0)}_{o} \, = \,
\left({C_A \over 2}-C_F \right)
{\alpha_{V_{o}} \over r}
\,,
\nonumber\\
{V^{(1)}_s \over m} & = &
-{C_FC_A D^{(1)}_s \over 2mr^2}
\,,
\label{pnrqcdpotentialslist}
\\
{V^{(2)}_s \over m^2} & = & 
- { C_F D^{(2)}_{1,s} \over 2 m^2} \left\{ {1 \over r},{\bf p}^2 \right\}
+ { C_F D^{(2)}_{2,s} \over 2 m^2}{1 \over r^3}{\bf L}^2
+ {\pi C_f D^{(2)}_{d,s} \over m^2}\delta^{(3)}({\bf r})
\nonumber\\
& & + {4\pi C_F D^{(2)}_{S^2,s} \over 3m^2}{\bf S}^2 \delta^{(3)}({\bf r})
+ { 3 C_F D^{(2)}_{LS,s} \over 2 m^2}{1 \over r^3}{\bf L} \cdot {\bf S}
+ { C_F D^{(2)}_{S_{12},s} \over 4 m^2}{1 \over r^3}S_{12}({{\bmr}})
\,,
\nonumber
\end{eqnarray}
where $S_{12}({\bmr}) \equiv 3 {\bmr}\cdot\bfsigma_1 \, 
{\bmr}\cdot\bfsigma_2/r^2 
- \bfsigma_1\cdot\bfsigma_2$,
${\bf S} = \bfsigma_1/2 + \bfsigma_2/2$, and
the $D$'s are Wilson coefficients.
The term $V^{(0)}_{s/o}$ is the static singlet/octet
potential. 
The pNRQCD Lagrangian can in principle also be written in momentum
space and/or in terms of quark and antiquark fields. However, such a
formulation has not yet been given explicitly. 
Nevertheless, in cases where the transfer to
the momentum picture is straightforward,
I will use the momentum space representation at
some instances in the following presentation.

Because NRQCD is treated in a $1/m$ expansion also the pNRQCD
amplitudes are power-counted in $1/m$ during the matching
computations.\cite{Pineda1,Brambilla1} This is necessary because the
kinetic energy 
term provides an IR cut-off for the Coulomb singularity (in the
non-relativistic power counting), and because IR regularization in full
and effective theory has to be equivalent to obtain the correct
matching conditions. Thus NRQCD and pNRQCD amplitudes are matched
order by order in $1/m$ and in the multipole expansion.
In Ref.~\citebk{Brambilla1} 
Brambilla et al. carried out the matching calculation
with Wilson loops that describe a static quark-antiquark pair with
fixed relative distance $r$ interacting for a very large time
$T\gg\frac{1}{r}$. In terms of quarks and gluons and in momentum space
this matching procedure corresponds 
to comparing $\QQbar$ scattering amplitudes, where the
incoming and outgoing quarks have off-shell non-relativistic momenta
$(0,\pm\bmp)$ and $(0,\pm\bmp^\prime)$, respectively. In this
kinematic situation all pNRQCD loop diagrams (in the $1/m$ expansion)
are scaleless integrals because the quark propagator behaves like a
static propagator, 
\begin{eqnarray}
\frac{i}{k^0-\frac{(\bmk+\bmp)^2}{2m}+i\delta} 
& \stackrel{1/m\,\, \mbox{\scriptsize expansion}}{=} & 
\frac{i}{k^0+i\delta} 
+ \frac{i}{(k^0+i\delta)^2}\frac{(\bmk+\bmp)^2}{2m}
+ \ldots
\,,
\end{eqnarray}
and because radiation of gluons as well as potential interactions depend
polynomially on the external momenta due to the multipole expansion. 
Thus the NRQCD scattering amplitudes give directly the pNRQCD matching
coefficients. 
As an example, consider the matching calculation for the singlet
static potential in the limit $m=\infty$. The pNRQCD scattering
amplitude to all orders in $\alpha_s$ at the scale $\mu=\mu_s$ reads
\begin{eqnarray}
\frac{4 \pi i C_F \alpha_{V_s}(\mu_s)}{(\bmp-\bmp^\prime)^2}
\,.
\label{pNRQCDstaticpotential}
\end{eqnarray}
The corresponding NRQCD amplitudes are built from static quark
propagators, the quark gluon coupling in the $D_0$ covariant
derivative and the gluon self-interactions. Quark pinch-singularities
are removed by the exponentiation of the static
potential.\cite{exponentiation}
The calculation was carried out up to two loops 
and the result reads ($\bmk=(\bmp-\bmp^\prime)$)
\begin{eqnarray}
&& 
\frac{4 \pi i C_F \alpha_s(\mu_s)}{\bmk^2}\,
\bigg\{\, 1 +
\Big(\frac{\alpha_s(\mu_s)}{4 \pi}\Big)\,\Big[\,
-\beta_0\,\ln\Big(\frac{\bmk^2}{\mu_s^2}\Big) + a_1
\,\Big]
\nonumber\\[2mm] & & \quad
 + \Big(\frac{\alpha_s(\mu_s)}{4\,\pi}\Big)^2\Big[
\beta_0^2\,\ln^2\Big(\frac{\bmk^2}{\mu_s^2}\Big)  
- \Big(2\,\beta_0\,a_1 +
\beta_1\Big)\,\ln\Big(\frac{\bmk^2}{\mu_s^2}\Big) 
+ a_2
\Big]
\bigg\}
\,,\quad
\label{staticpotentialschroeder}
\end{eqnarray}
where $\beta_0=11-\frac{2}{3}\, n_\ell$ and $\beta_1=102-\frac{38}{3}\, n_\ell$
are the one- and two-loop beta functions, and
$a_1=\frac{31}{3}-\frac{10}{9}\, n_\ell$ (see Refs.~\citebk{Fischler1}), 
$a_2=456.749 - 66.354 \,n_\ell + 1.235\, n_\ell^2$
(see Refs.~\citebk{Schroeder1}). 
The constant $n_\ell$ denotes the number of massless 
quark species.
Finite light quark mass effects at two loops are also known,\cite{Melles1}
but are not discussed here. A numerical estimate of the three-loop
corrections for massless light quarks based on Pad\'e approximation
techniques has been carried out in Ref.~\citebk{Chishtie1}. 
The two-loop matching condition for $\alpha_{V_s}$ is obtained by
demanding equality of Eqs.~(\ref{pNRQCDstaticpotential}) and
(\ref{staticpotentialschroeder}). The matching conditions for the
other pNRQCD coefficients shown in Eqs.~(\ref{pNRQCDLagrangian}) 
and (\ref{pnrqcdpotentialslist}), which are 
needed for a NNLL description of the $\QQbar$ dynamics,
have been given in Ref.~\citebk{Pineda3}. Because the matching
procedure uses off-shell scattering amplitudes, the pNRQCD potential
operators basis is different from vNRQCD, where on-shell
matching is carried out.

\begin{figure}[ht] 
\begin{center}
\leavevmode
\epsfxsize=4.3cm
\leavevmode
\epsffile[200 390 400 480]{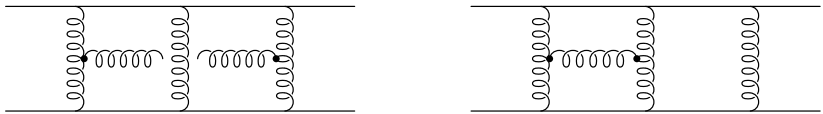}
 \end{center}
%
%
 \caption{\label{figADMIRdivergence}   
Graphs contributing to the three-loop IR divergence of the QCD static
potential in perturbation theory.
}
\end{figure}

\begin{figure}[ht] 
\begin{center}
\leavevmode
\epsfxsize=4.4cm
\leavevmode
\epsffile[120 360 320 430]{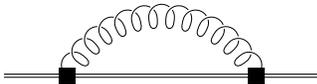}
 \end{center}
%
%
 \caption{\label{figpnrqcdselfenergy}   
Selfenergy of a $\QQbar$ pair due to ultrasoft gluon radiation
in pNRQCD. 
}
\end{figure}

Appelquist, Dine and Muzinich (ADM)\,\cite{Appelquist2} pointed out
that at three 
loops the potential of a static $\QQbar$ pair with relative distance
$r$ has an IR divergence from graphs of the
form in Fig.\ \ref{figADMIRdivergence}. 
ADM showed that this IR divergence could be
avoided by summing diagrams with an arbitrary number of gluon rungs.
This summation gives a factor $\exp([V_s(r)-V_o(r)]T)$
for the configuration space propagation of the intermediate
color-octet $\QQbar$ pair, which regulates the IR divergence. 
In the pNRQCD framework the diagrams considered by ADM are obtained in
the intermediate theory NRQCD, where the heavy quarks are treated in
the static approximation.
In Ref.~\citebk{Brambilla2} Brambilla et al. made the important
observation that, by construction, 
the ADM IR divergence is equivalent to IR divergences of (unresummed)
pNRQCD graphs that describe the selfenergy of a quark-antiquark pair
due to an 
ultrasoft gluon (see Fig.\ \ref{figpnrqcdselfenergy}). 
Thus the static potential defined as the pNRQCD
matching coefficient $\alpha_{V_s}$ is an infrared-safe quantity.
As pointed out in Ref.~\citebk{Brambilla1}, this also means that the static
potential defined via the Wilson loop
\begin{eqnarray} 
 V_{\rm stat}(r) = \lim_{T\to\infty} {1\over T} 
 \ln \vev{ {\rm Tr}\, P\exp \Big[- 
  i g \oint_C A_\mu dx^\mu\Big]}
\,, 
\label{wilson}
\end{eqnarray}
where $P$ refers to pathordering, 
is not equivalent to $V_s^{(0)}(r)$ at short distances, because the
former contains the full 
ultrasoft selfenergy contributions just mentioned before.

The anomalous dimensions of the pNRQCD coefficients are determined 
from UV divergences in $\QQbar$ scattering diagrams, where again the
$1/m$ expansion with off-shell static $\QQbar$ pairs at fixed 
distance $r$ (or with external momenta $(0,\pm\bmp)$ and
$(0,\pm\bmp^\prime)$ for the incoming and outgoing quark) is
used.\cite{Brambilla1,Pineda4} 
At this stage, loop diagrams with UV divergences that occur in potential
loop integration, i.e.\ when potential interactions are iterated, are
neglected and only UV divergences from ultrasoft gluon exchange are
considered.\cite{Pineda3} In particular, the iteration of static
potential exchanges between the quarks is treated effectively in $D=4$
dimensions by using the resummed singlet/octet static $\QQbar$
propagator $\sim (-i\partial_0 + V_{s,o}^{(0)}(r))^{-1}$, see
Ref.~\citebk{Brambilla1}.  

\begin{figure}[ht] 
\begin{center}
\leavevmode
\epsfxsize=3.8cm
\leavevmode
\epsffile[80 470 490 700]{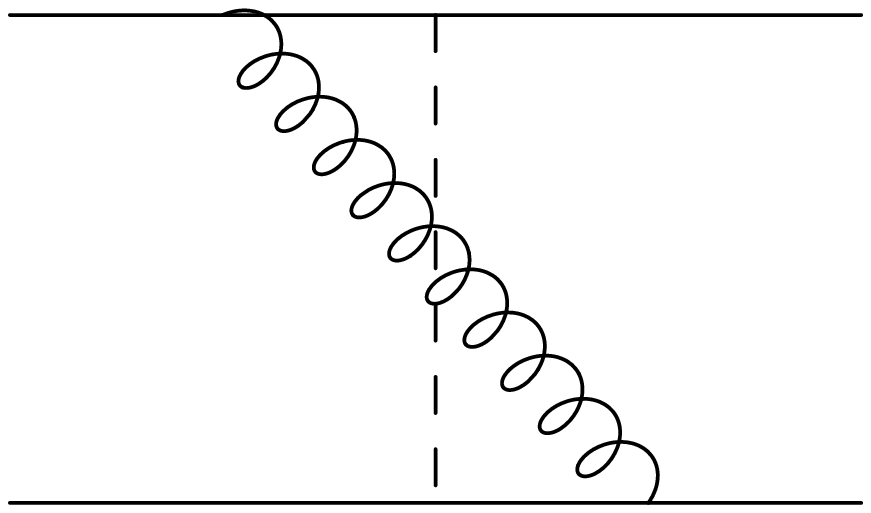}
\leavevmode
\epsfxsize=3.8cm
\leavevmode
\epsffile[80 470 490 700]{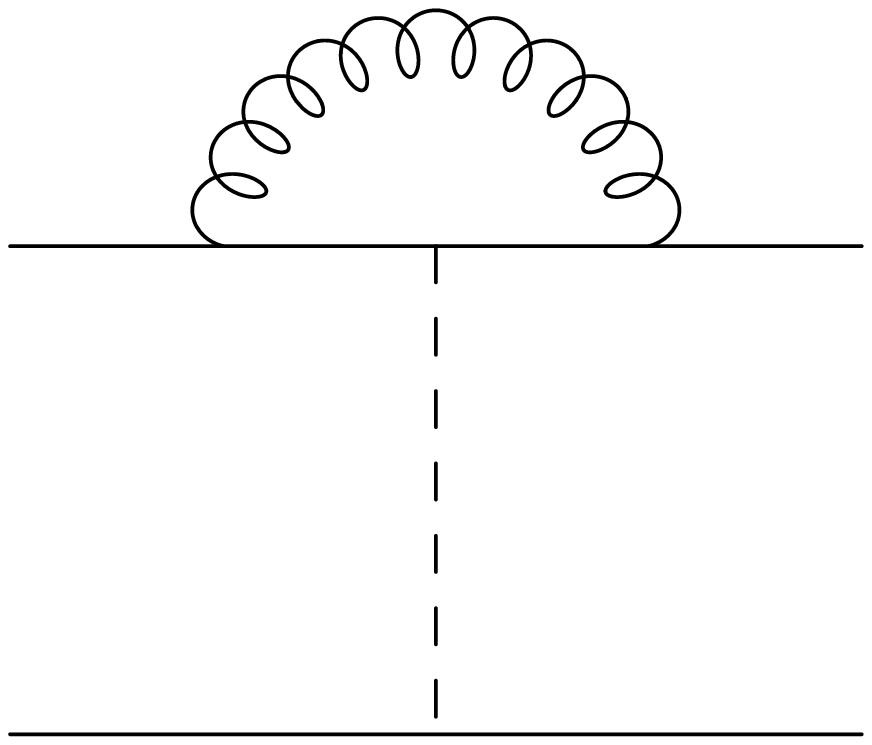}
 \end{center}
%
%
%
\vskip  0cm
 \caption{\label{figusoft1um2pnrqcd}   
Diagrams in pNRQCD that contribute to the anomalous dimension of 
the spin-independent $1/m^2$ potentials. The solid lines are static
quark propagators.
The dashed line represents
the static potential.
}
\end{figure}

At leading order in the multipole expansion
the entire ultrasoft renormalization of the potential coefficients is
obtained from the effective one-loop selfenergy shown in 
Fig.\ \ref{figpnrqcdselfenergy},
where the double line represents the static $\QQbar$ propagator.
For example, the leading logarithmic (LL) 
running of the potentials $\propto 1/m^2$ is
induced by the ``one-loop'' diagrams where the double line contains only
one static potential exchange, see Fig.~\ref{figusoft1um2pnrqcd} for 
typical graphs. 
At this level the LL expression for the coupling of the static
potential is sufficient. Because the coupling of the static potential
to the quarks does not run at LL order in pNRQCD, it is just
$\alpha_s(\mu_s)$. Due to gauge cancellations
only the ${\bmp}.{\bmA}/m$ coupling of the ultrasoft gluons to quarks
(which only runs trivially with $\alpha_s(\mu)$ at LL order)
contributes and generically leads to the following LL anomalous
dimension of the spin-independent $1/m^2$ coefficients,
\begin{eqnarray}
\frac{d}{d\ln\mu}\Big\{\,
D^{(2)}_{1,s}(\mu,\mu_s), D^{(2)}_{d,s}(\mu,\mu_s)
\,\Big\} 
& \propto &
\alpha_s(\mu)\,\alpha_s(\mu_s)
\,.
\end{eqnarray}
At this order, the spin-dependent potentials do not run in pNRQCD.

\begin{figure}[t] 
\begin{center}
\leavevmode
\epsfxsize=5.5cm
\leavevmode
\epsffile[90 530 480 650]{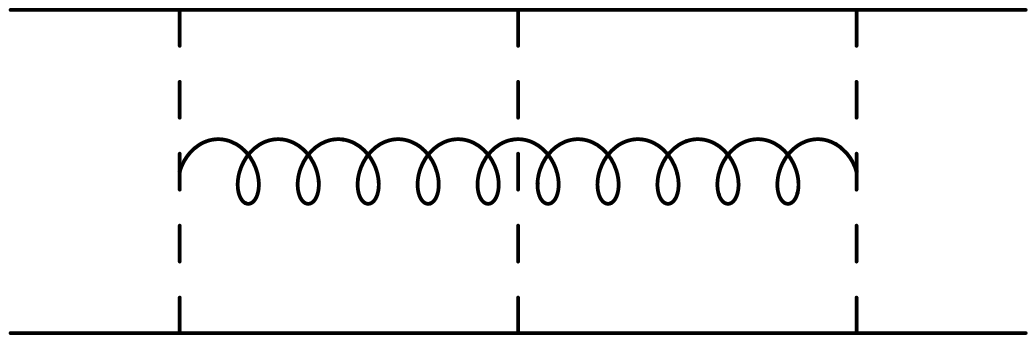}
\leavevmode
\epsfxsize=5.5cm
\leavevmode
\epsffile[90 530 480 650]{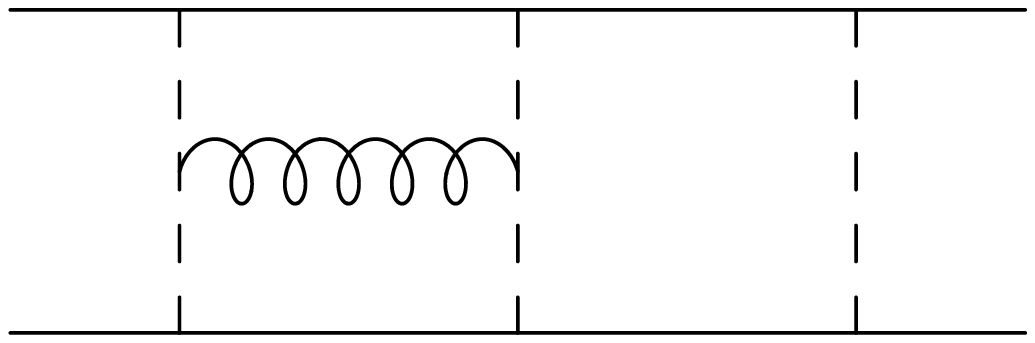}
 \end{center}
%
%
%
\vskip  0cm
 \caption{\label{figusoftstaticpnrqcd}   
Diagrams in pNRQCD that contribute to the three-loop 
anomalous dimension of the static potential. The solid lines
are static quark propagators.
The static potential
does not run at the one- and two-loop level.
}
\end{figure}

A more involved example is the NNLL running of the static singlet potential
$\alpha_{V_s}$, which was considered in
Refs.~\citebk{Brambilla2,Pineda4}. An ultrasoft 
anomalous dimension of $\alpha_{V_s}$ is generated for the first time
by UV divergences in 
the ``three-loop'' diagrams in Fig.~\ref{figusoftstaticpnrqcd}, which are
also contained in 
the one-loop selfenergy of Fig.~\ref{figpnrqcdselfenergy}. 
The ultrasoft loop with the
gluon propagator is UV-divergent. The two potential insertions are
obtained by expanding the static $\QQbar$ propagator in terms of
the static potential. At this order its coupling can again be
approximated by $\alpha_s(\mu_s)$.
In momentum space the computation
consists of an ultrasoft gluon loop in $D=4-2\epsilon$ dimensions,
where the coupling of ultrasoft gluons to the static potential is
$\alpha_s(\mu)$, and  
static quark loops in $D=4$ dimensions, where the couplings of the
potentials to the quarks are $\alpha_s(\mu_s)$. The incoming and outgoing
quark momenta are $(0,\pm\bmp)$ and $(0,\pm\bmp^\prime)$, respectively.
The ultrasoft loop is UV-divergent, and the static quark loops lead to 
the overall factor $1/(\bmp-\bmp^\prime)^2$.
For a color singlet $\QQbar$ pair the amplitude reads
\begin{eqnarray}
\mbox{Figs.\ref{figusoftstaticpnrqcd}} & = &
-i\frac{C_F\,C_A^3}{6}\,
\frac{[\alpha_s(\mu_s)]^3\,[\alpha_s(\mu)\,\mu^{2\epsilon}]}
{(\bmp-\bmp^\prime)^2}
\bigg[\,\frac{1}{\epsilon_{\rm UV}}-\frac{1}{\epsilon_{\rm IR}}\,\bigg]
\,.
\label{coulombusoftdiv}
\end{eqnarray}
The IR divergence $\propto 1/\epsilon_{\rm IR}$ is equivalent to the
ADM infrared divergence mentioned above. 
The anomalous dimension for the static singlet potential obtained from
the UV divergence reads
\begin{eqnarray}
\frac{d}{d\ln\mu} \alpha_{V_s}(\mu,\mu_s) 
& = &
\frac{C_A^3}{12\pi}\,
[\alpha_{V_s}(\mu_s)]^3\,\alpha_s(\mu)
\,.
\label{coulombusoftrge}
\end{eqnarray}
Using the initial condition of Eq.~(\ref{staticpotentialschroeder})
the NNLL result for the static singlet potential at the scale
$\mu=\mu_u$ reads\,\cite{Pineda4}
\begin{eqnarray}
\alpha_{V_s}(\mu_u,\mu_s) & = &
\Big[\alpha_{V_s}(\mu_s)\Big]_{\mbox{\scriptsize two-loop}} 
+  \bigg[\, \frac{C_A^3}
 {6\beta_0}\alpha_s^3(\mu_s) 
\ln\bigg(\frac{\alpha_s(\mu_s)}{\alpha_s(\mu_u)}\bigg) 
 \,\bigg]  \,,
\label{staticpotentialusoft}
\end{eqnarray}
where $[\alpha_{V_s}(\mu_s)]_{\mbox{\scriptsize two-loop}}$
is the two-loop matching result obtained from 
Eqs.\ (\ref{pNRQCDstaticpotential}) and
(\ref{staticpotentialschroeder}).

\subsection{Potential Loops in pNRQCD}

\begin{figure}[t] 
\begin{center}
\begin{picture}(100,30)(1,1)
\put(-140,-225) { \epsfxsize=8.5cm 
               \hbox{\epsfbox{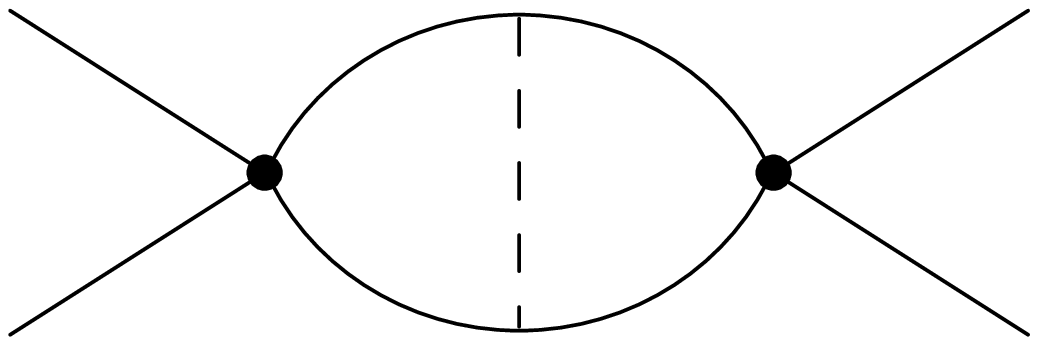}} }
\put(0,-225) { \epsfxsize=8.5cm 
               \hbox{\epsfbox{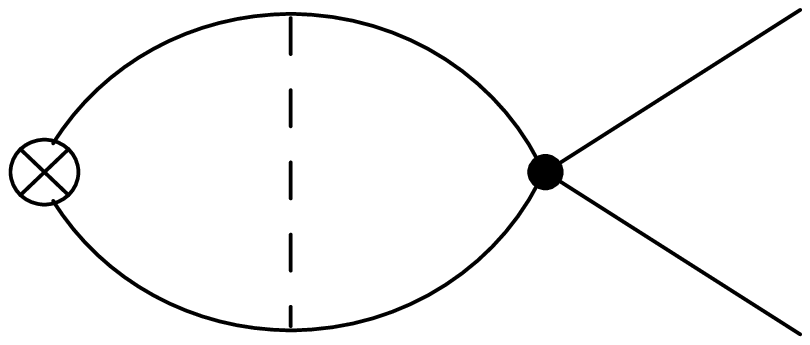}} }
\put(-95,25){$a)$}
\put(-17,5){$V_s^{(0)}$}
\put(-62,22){$D_{d,s}^{(2)}$}
\put(3,22){$D_{d,s}^{(2)}$}
\put(70,25){$b)$}
\put(123,5){$V_s^{(0)}$}
\put(143,22){$D_{d,s}^{(2)}$}
\end{picture}
 \end{center}
\vskip  0.4cm
 \caption{\label{figpotentialUVpNRQCD}   
Diagrams in pNRQCD that contribute to the two-loop anomalous dimension
of the $D_{d,s}^{(2)}$ potential (a) and the $\QQbar$ production
current (b).
}
\end{figure}

If pNRQCD is applied to dynamical quarks, 
the counting $D_0\sim k^0 \sim \frac{\bmat{D}^2}{m}\sim
\frac{\bmag{k}^2}{2m}$ is reinstated and the potential region is
reintroduced into the theory. As mentioned before, it is claimed in
Refs.~\citebk{Pineda1,Pineda2,Brambilla1} that the potential region
is always contained in the pNRQCD framework, but hidden in a subtle
way as long as the $1/m$ power counting is imposed. 
Potential loops, which are generated by iterations of potentials,
can lead to UV divergences. Consider for example the $\QQbar$ scattering
diagram in Fig.\ \ref{figpotentialUVpNRQCD}a 
with two $D_{d,s}^{(2)}$ potentials
and one static potential. The $D_{d,s}^{(2)}$ potential is a
$\delta$-function in configuration space, and 
the graph is UV divergent inducing a
two-loop next-to-leading logarithmic (NLL) 
anomalous dimension to $D_{d,s}^{(2)}$. The diagram
in Fig.\ \ref{figpotentialUVpNRQCD}b 
describes $\QQbar$ production and induces a two-loop
NLL anomalous dimension to the coefficients of the $\QQbar$
production currents such as the 
${}^{2s+1}L_j={}^3S_1$ current
$O_{\psi\chi}=\psi^\dagger\bsigma(i\sigma_2)\chi$. 
Thus the LL running of 
$D_{d,s}^{(2)}$ and $\alpha_{V_s}$ obtained in NRQCD and from ultrasoft
gluon radiation in pNRQCD mixes into $D_{d,s}^{(2)}$ and
$c_{\psi\chi}$ at NLL order through UV 
divergences in potential loop diagrams. In order to quantify this
mixing a correlation between the soft scale $\mu_s$, the ultrasoft
scale $\mu_u$ and the scale for potential loops, $\mu_p$, needs to be
imposed. Such a correlation does, by itself, not exist in the pNRQCD 
framework, since the quarks are treated as static. So the correlation
needs to be imposed by hand, once potential loops and the 
counting $D_0\sim k^0 \sim \frac{\bmat{D}^2}{m}\sim
\frac{\bmag{k}^2}{2m}$ are reintroduced into the theory.

From the physical point of view this correlation should
reflect the energy-momentum relation of the heavy quark:
$\mu_p\sim\mu_s\sim\sqrt{\mu_u m}$. This concept has been first
realized in the framework of vNRQCD,\,\cite{Luke1} where the correlation of
scales arises naturally from the theory
itself. In Ref.~\citebk{Pineda5} Pineda proposed the following prescription
for pNRQCD: after NRQCD is evolved to $\mu=\mu_s$ and the ultrasoft
pNRQCD evolution is carried out down to the scale $\mu=\mu_u$ as
described before, the relations
\begin{equation}
\mu_s \, = \, \mu_p\,,\quad
\mu_u \, = \, \frac{\mu_p^2}{m}
\label{pNRQCDscalecorrelation}
\end{equation}
are imposed. This scale correlation has to be used for the
determination and the integration of the RGE's obtained from potential
UV divergences as well as for the determination of matrix elements,
i.e.\ in particular in the solution of the Schr\"odinger equation.
For the determination of all matrix elements the $1/m$ expansion is
abandoned and the non-relativistic power counting 
from the regions in Eq.\ (\ref{momentumregions}) is
reestablished.  
For the integration of the renormalization group equations
$\mu_p$ has to be run from the hard
scale $m$ down to the soft scale of order $m v$. 
This procedure means in particular that, for the description of 
dynamical quarks, the end points of the running in NRQCD ($\mu_s$) 
and the ultrasoft running in pNRQCD ($\mu_u$) are generally not the 
soft and the ultrasoft scales of order $m v$ and $m v^2$, respectively.
It is claimed in Ref.~\citebk{Pineda5} that this prescription sums
properly all large logarithmic terms.

The NLL anomalous dimensions induced by Figs.~\ref{figpotentialUVpNRQCD}a
and b have the generic form
\begin{eqnarray}
\frac{d}{\ln\mu_p} D_{d,s}^{(2)}\Big(\frac{\mu_p^2}{m},\mu_p\Big)
& \sim & 
\bigg[D_{d,s}^{(2)}\Big(\frac{\mu_p^2}{m},\mu_p\Big)\bigg]^2\,
\alpha_{V_s}\Big(\frac{\mu_p^2}{m},\mu_p\Big)
\,,
\nonumber
\\[2mm]
\frac{d}{\ln\mu_p} c_{\psi\chi}(\mu_p)
& \sim &  
c_{\psi\chi}(\mu_p)\,
D_{d,s}^{(2)}\Big(\frac{\mu_p^2}{m},\mu_p\Big)\,
\alpha_{V_s}\Big(\frac{\mu_p^2}{m},\mu_p\Big)
\,.
\label{potentialandim}
\end{eqnarray}
At this order, only the LL results need to be known on the RHS.
The full NLL running of $D_{d,s}^{(2)}$ has actually not been
calculated so far; it would be relevant for the description of the
$\QQbar$ dynamics only at N$^3$LL order. The NLL runnning of
$c_{\psi\chi}$ in the pNRQCD framework 
has been determined in Ref.~\citebk{Pineda5}. 

\subsection{Open Issues}
\label{subsectionopenissues}

There are issues in the pNRQCD framework that might require some
investigations. 
One obvious point is that the matching procedures in the
scheme of Eq.~(\ref{pNRQCDscheme}) appear to strongly rely on specific
choices for the kinematic situation where the matching conditions can be
calculated. It would be desirable that the matching
procedure could be carried out for any 
kinematic situation, for which the effective theory is applicable. For
example,  
the NRQCD matching is carried out for on-shell scattering amplitudes with
external quarks exactly at the threshold point, where the quark
velocity is zero. If on-shell scattering amplitudes with some small but
non-zero quark velocity are employed the diagrams in full QCD also
contain contributions from, let's say, the potential momentum region,
which cannot be reproduced in the $1/m$ expansion employed for the
NRQCD computations. 

Another issue concerns the prescription of Ref.~\citebk{Pineda5} for the
treatment of UV 
divergences from potential loops after the endpoints of the NRQCD and
ultrasoft pNRQCD running, $\mu_s$ and $\mu_u$, are correlated to the
potential scale, $\mu_p=\mu_s=\sqrt{\mu_u m}$. The potentials that arise
order-by-order from matching pNRQCD to NRQCD are independent of the
matching scale $\mu_s$ up to higher order terms, which are small
corrections in the matching conditions for $\mu_s\sim
(\bmp-\bmp^\prime)$, see for example Eq.~(\ref{staticpotentialschroeder})
for the static potential. On the other hand, in the anomalous
dimensions for $D_{d,s}^{(2)}$ and $c_{\psi\chi}$ in
Eqs.~(\ref{potentialandim})  
the residual dependence of for example $\alpha_{V_s}$ on $\mu_s$ leads
to the summation of unphysical logarithms unless the corresponding
logarithmic terms in the matching computation for the potential are
summed too. This means in principle that the higher order matching
corrections to the potentials cannot be considered as small when
correlated running according to Ref.~\citebk{Pineda5} is employed.
 
It is a prediction of the pNRQCD framework that the Coulomb
potential in the Schr\"odinger equation 
that describes dynamical $\QQbar$ pairs is equal to the static
potential determined in the $1/m$ expansion in
pNRQCD.\cite{Brambilla1} The 
ultrasoft anomalous dimension of the static potential is determined
from Eq.~(\ref{coulombusoftdiv}), which does not contain any scale
dependence from the 
static ladder diagrams that iterate the potentials. 
On the other
hand, after the scale 
correlation $\mu_p=\mu_s=\sqrt{\mu_u m}$ and the non-relativistic
power counting are imposed,
the corresponding UV divergent contribution from the diagram in
the N$^3$LL matrix element describing let's say on-shell
$\QQbar$ scattering in analogy to Fig.~\ref{figusoftstaticpnrqcd}
reads 
\begin{eqnarray}
i\frac{C_F\,C_A^3}{6}\,
\frac{[\alpha_s(\mu_s)\,\mu_s^{2\epsilon}]^3\,
[\alpha_s(\mu_u)\,\mu_u^{2\epsilon}]}
{(\bmp-\bmp^\prime)^2}
\bigg[\,\frac{1}{\epsilon}\,\bigg]
\,,
\label{staticdynamicpnrqcd}
\end{eqnarray}
where the dependence on $\mu_s$ and $\mu_u$ has been made manifest.
Here, the scale dependence of the potential loops is included,
because the quarks are not static. 
The $1/\epsilon$ term is canceled by the counterterm of the
Coulomb potential determined from Eq.~(\ref{coulombusoftdiv}).
However, only a part of the $\ln\mu_p$ terms 
is canceled by the order $\alpha_s^4$
ultrasoft logarithm summed into $\alpha_{V_s}$ by
Eq.~(\ref{coulombusoftrge}).
The remaining $\ln\mu_p$ terms seem to remain uncanceled in the final 
N$^3$LL result for $\QQbar$ scattering.

\subsection{$\lqcd \sim mv^2$ or larger}
\label{subsectionlqcdlarger}

So far pNRQCD has been discussed for the case $m\gg m v\gg m
v^2\gg\lqcd$, where the counting
$\alpha_s(m)\sim\alpha_s(mv)\sim\alpha_s(mv^2)\ll 1$ was used. In this
situation the description of the $\QQbar$ dynamics is predominantly
perturbative and the effective theories serve mainly as a tool to
handle the non-trivial perturbative calculations and summations in a
manageable and systematic fashion. Non-perturbative effects appear as
local gluon and light 
quark condensates in an expansion in $\lqcd/mv^2$, see 
Refs.~\citebk{Shifman1,Voloshin1,Leutwyler1}.
One motivation in the construction of pNRQCD was that its multi-layer
structure should make it also applicable to systems where $\lqcd$ is
of order $mv^2$ or larger. In this case also the
non-relativistic dynamics of the charmonium or of higher bottomonium
excitations could 
be studied in a systematic and quantitative manner. Qualitative
discussions of the application of pNRQCD in cases where $\lqcd$ is of
order $mv^2$ or larger have been given in Ref.~\citebk{Brambilla1}.
For $mv > \lqcd \gsim mv^2$ pNRQCD can still be matched perturbatively
to NRQCD at the soft scale $\mu_s$, but $\lqcd$ screens the ultrasoft
scale (or any other lower scale) and $\alpha_s(mv^2)$ is of order
$1$. However, the coupling of ultrasoft ($\sim\lqcd$) gluons to the
quark-antiquark pair $\propto {\bmr}.{\bmE}$ is still small because
of the multipole 
expansion. The effects of ultrasoft gluons enter as 
non-perturbative non-local condensates, that are small
corrections. For $\lqcd\gsim mv$, the matching of NRQCD to QCD at the
hard scale can still be carried out perturbatively, but the full
non-relativistic $\QQbar$ dynamics is screened by $\lqcd$. In this
case NRQCD (Sec.\ \ref{sectionNRQCD}) should describe $\QQbar$
production and decay, but all non-relativistic $\QQbar$ dynamics is
non-perturbative.

\vspace{1cm}

\section{Effective Theories III: Velocity NRQCD}
\label{sectionvNRQCD}

The effective theory vNRQCD was first proposed by Luke, Manohar and
Rothstein.\cite{Luke1} It is devised for systems where the relation
$m\gg mv\gg mv^2\gg \lqcd$ is valid. Its main fields of application
are top pair production close to threshold, sum rules for $\Upsilon$
mesons and maybe low radial excitations of bottomonia.
The theory vNRQCD provides the separation of all different degrees of 
freedom in a single step at the hard scale and is defined 
in dimensional regularization.

\subsection{Basic Idea}

The basic idea behind the construction of vNRQCD is that all resonant
degrees of freedom (dof's) 
(i.e.\ dof's that can develop a particle pole in the momentum regions
of Eq.\ (\ref{momentumregions})) are separated and
that all off-shell fluctuations (i.e.\ fluctuations that cannot come
close to their mass-shell in the regions of  Eq.\
(\ref{momentumregions})) are integrated out at the hard scale. This
means that the theory contains different gluon field operators for the
gluonic fluctuations with soft and ultrasoft energies and momenta. The
theory vNRQCD is matched to QCD at the scale $m$ and the
renormalization group equations are constructed such that large logarithms
of the scale ratios $m/\langle p\rangle$, $m/\langle E\rangle$ and 
$\langle p\rangle/\langle E\rangle$ are summed simultaneously into
the coefficients of the operators when the theory is scaled down. This
is achieved by having different renormalization scales for
potential/soft and ultrasoft loop integrals, which are correlated in
accordance with the quark equations of motion. The
scale correlation is fixed by the theory itself and not imposed by
hand. The correlation of the scales and the renormalization group
evolution are expressed in terms of the 
``velocity'' scaling parameter $\nu$. The hard scale corresponds to
$\nu=1$ and all large logarithmic terms are summed for $\nu=v$.

\subsection{Effective Lagrangian}

The effective Lagrangian is formulated in terms of quark and gluonic fields
describing modes that can become on-shell in the momentum regions of 
Eq.~(\ref{momentumregions}). For convenience, massless
quarks will be referred to as gluons 
in the following presentation. Heavy quarks are simply referred to as
quarks. On-shell fluctuations for quarks exist only for the
potential regime where $(k_0,{\bmk})\sim(mv^2,mv)$ and on-shell gluonic
fluctuations exist in the soft and the ultrasoft regime where
$(k_0,{\bmk})\sim(mv,mv)$ and  $(k_0,{\bmk})\sim(mv^2,mv^2)$,
respectively. These three kinds of modes are
referred to as ``potential quarks'', ``soft gluons'' and ``ultrasoft
gluons''. Although soft gluons cannot be produced in a
non-relativistic $\QQbar$ system with total energy of the order
$mv^2$, they are kept as relevant dof's in vNRQCD. This is necessary
because interactions with soft gluons are involved in the
renormalization of vNRQCD operators. In fact, one can consider vNRQCD
as a theory that can describe interactions of a quark with soft gluons
as well, such as in Compton scattering. 

\begin{figure}[ht] 
\begin{center}
\leavevmode
\epsfxsize=2.2cm
\leavevmode
\epsffile[150 60 450 720]{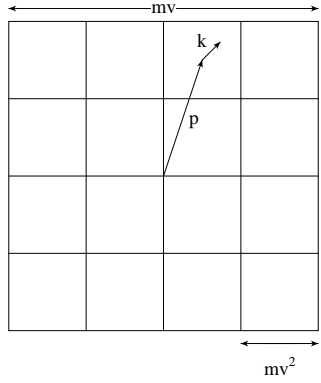}
 \end{center}
%
%
 \caption{\label{figcubes} 
Momentum space of size $mv$ is divided into boxes of size $mv^2$. A
  point in momentum space is labeled by $\mathbf p$ and $\mathbf k$.
}
\end{figure}

In the effective Lagrangian
only ultrasoft energies and momenta are treated as continuous
variables. Soft energies and momenta are treated as discrete indices
for potential quarks and soft gluons. For a heavy quark this is
achieved by writing its energy $E$ and three-momentum $\bmP$ as
\begin{equation}
E \, = \, k^0\,,\qquad
\bmP \, = \, \bmp + \bmk
\end{equation}
where the three-vector $\bmp$ is of order of the soft scale $mv$ and
the four-vector $k=(k^0,\bmk)$ is of order of the ultrasoft scale
$mv^2$. This can be understood as subdividing the full
non-relativistic
momentum space for the quark three-momentum
(with side length of order $mv$) 
into small cubes with ultrasoft side length of order $mv^2$, such
that the number of ultrasoft cubes is $(\frac{mv}{mv^2})^3$. The
center location of each ultrasoft cube is given by the index $\bmp$
and a point within an ultrasoft cube is labeled by $\bmk$, 
see Fig.\ \ref{figcubes}.
Only the ultrasoft four-momenta $k$ are treated as continuous
variables, the soft three-momentum $\bmp$ is a discrete index.
The number soft three-momenta is equal to the number of ultrasoft
cubes. The quark and antiquark fields in vNRQCD are
written as two-component spinor fields
\begin{equation}
   \psi_{\bmp} (x)\,,
   \quad \chi_{\bmp} (x)
   \,,
\label{quarkfieldsvnrqcd}
\end{equation}
where $\bmp$ is the soft index and $x$ the continuous variable that
describes ultrasoft fluctuations. This procedure is similar to
HQET,\,\cite{HQETpapers} where the four-momentum $q^\mu$ of a heavy quark is
split into $mv^\mu$ of order $m$ and a residual four-momentum $k$ of
order $\lqcd$, $q^\mu=mv^\mu+k^\mu$. The index $v^\mu$ is a discrete
variable one has to sum, whereas $k^\mu$ is a continuous variable one
has to integrate.
In vNRQCD one has to sum over $\bmp$ and integrate over $x$ (or $k$ in
momentum space representation).

The energy $E$ and three-momentum $\bmP$ of a soft gluon is written in
an analogous way as
\begin{equation}
E \, = \,p^0 +  k^0\,,\qquad
\bmP \, = \, \bmp + \bmk
\,,
\end{equation}
where $p=(p_0,\bmp)$ is a four-vector of the order of the soft scale
$mv$ and $k=(k_0,\bmk)$ is a four-vector of the order of the ultrasoft
scale $mv^2$. 
The corresponding field in vNRQCD that annihilates and creates a soft
gluon with index $p$ in vNRQCD is $A_p^\mu(x)$.
One has to sum over $p$ and integrate over $x$ (or $k$).
The energy $E$ and the three-momentum $P$ of an ultrasoft gluon is
just 
\begin{equation}
E \, = k^0\,,\qquad
\bmP \, = \bmk
\,,
\end{equation}
where $k=(k_0,\bmk)$ is a four-vector of the order of the ultrasoft
scale $mv^2$.  
The corresponding field in vNRQCD that annihilates and creates an
ultrasoft gluon vNRQCD is $A^\mu(x)$.

The decomposition into a soft and an ultrasoft momentum component of
let's say the quark is not unique. Taking an ultrasoft three-momentum
$\bmq$ one can redefine 
\begin{equation}
\bmk \to \bmk+\bmq\,,\qquad \bmp \to \bmp-\bmq
\,.
\label{reparametrizationtrafo}
\end{equation}
This reparametrization invariance is similar to HQET,\cite{HQETpapers}
but it 
does not affect the spin axis of the quarks, because the latter is
fixed by the choice of the center of mass frame, which is not affected
by the transformation in Eq.\ (\ref{reparametrizationtrafo}). So the
consequences of 
reparametrization invariance in vNRQCD are the same as for HQET for
spin-zero particles. The basic outcome of reparametrization invariance
in vNRQCD is that derivatives of $\psi_{\bmp}(x)$ or $\chi_{\bmp}(x)$
are always of the form $(i\bmp+\bmnabla)$.

The effective Lagrangian is written in terms of quark fields,
$\psi_{\bmp}$, antiquark fields, $\chi_{\bmp}$, soft gluon fields,
$A_{\bmq}^\mu$, and an ultrasoft gluon field, $A^\mu$. The covariant
derivative is $D^\mu = (D^0,{\bmD})$ with $D^0=\partial^0+i g A^0$,
${\bmD}={\bmnabla}-i g {\bmA}$, and it involves only the ultrasoft gluon
field. The effective Lagrangian is manifestly gauge-invariant with
respect to ultrasoft gauge transformations, i.e.\ with slowly varying
gauge phases involving frequencies of order $mv^2$. It is argued
in Ref.~\citebk{Luke1} that the full gauge
invariance including also higher frequencies of order $mv$ is
recovered by combination of reparametrization invariance and ultrasoft
gauge invariance. A mathematical proof of this conjecture does,
however, not yet exist.

\begin{figure}[ht] 
\begin{center}
\leavevmode
\epsfxsize=3.cm
\leavevmode
\epsffile[180 380 390 540]{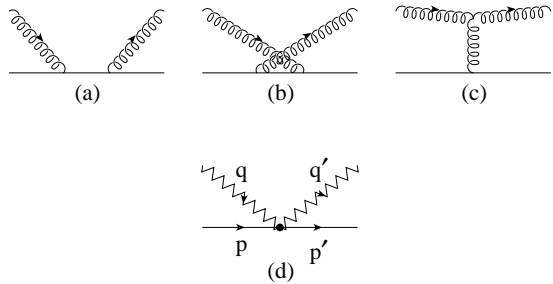}
\end{center}
%
%
\vskip  1.3cm
 \caption{\label{figsoftvnrqcd}   
The Compton scattering diagrams (a,b,c) in full QCD generate the soft
gluon coupling (d) in vNRQCD.
Soft gluon modes are denoted by a zigzag line
}
\end{figure}

\begin{figure}[ht] 
\begin{center}
\leavevmode
\epsfxsize=2.5cm
\leavevmode
\epsffile[180 430 390 460]{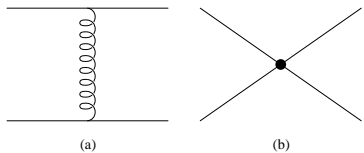}
\end{center}
%
%
\vskip  1.3cm
 \caption{\label{figpotentialvnrqcd}   
The scattering Born diagram (a) generates the potential four-quark
interaction (b) in vNRQCD.
}
\end{figure}

In the center of mass frame the Lagrangian has the
form\,\cite{Luke1,Manohar2,Manohar3} 
\begin{eqnarray}
\lefteqn{
{\cal L}_{\rm vNRQCD} \, = \,
 \sum_{\mathbf p}\,\bigg\{
   \psi_{\bmp}^\dagger   \bigg[ i D^0 - {\left({\bf p}-i{\bf D}\right)^2
   \over 2 m} +\frac{{\mathbf p}^4}{8m^3} + \ldots \bigg] \psi_{\bmp}
 + (\psi \to \chi)\,\bigg\}\quad\mbox{}
}
\nonumber \\[2mm] && 
 - g^2 \sum_{{\bmp},{\bmp^\prime},q,q^\prime,\sigma} \bigg\{ 
  \frac{1}{2}\, \psi_{\bmp^\prime}^\dagger
 [A^\mu_{q^\prime},A^\nu_{q}] U_{\mu\nu}^{(\sigma)} \psi_{\bmp}
 + (\psi \to \chi) + \ldots\, \bigg\}
\nonumber \\[2mm] &&  
+ \frac{2ig^2}{(\bmp^\prime-\bmp)^4} f^{ABC}
  {(\bmp-\bmp^\prime)}.(g{\bmA}^C) [\psi_{\bmp^\prime}^\dagger
  T^A \psi_{\bmp} \chi_{-\bmp^\prime}^\dagger \bar T^B \chi_{-\bmp} ] 
+\ldots \qquad\quad\mbox{}
\nonumber \\[3mm] && 
-\, V({\bmp,\bmp^\prime})\,\psi_{\bmp^\prime}^\dagger \psi_{\bmp}
   \chi_{-\bmp^\prime}^\dagger \chi_{-\bmp}
\nonumber \\[3mm] &&  
 -{1\over 4}G^{\mu\nu}G_{\mu \nu} + \sum_{p} \abs{p^\mu A^\nu_p -
 p^\nu A^\mu_p}^2 + \ldots
\,,
\label{vNRQCDLagrangian}
\end{eqnarray} 
where ($\bmk=(\bmp-\bmp^\prime)$)
\begin{eqnarray}
\lefteqn{
 V({\bmp},{\bmp^\prime}) \, = \,  (T^A \otimes \bar T^A) \bigg[
 \frac{{\cal V}_c^{(T)}}{\bmk^2}
 + \frac{{\cal V}_k^{(T)}\pi^2}{m|{\bmk}|}
 + \frac{{\cal V}_r^{(T)}({\bmp^2 + \bmp^{\prime 2}})}{2 m^2 \bmk^2}
 + \frac{{\cal V}_2^{(T)}}{m^2}
 + \frac{{\cal V}_s^{(T)}}{m^2}{\bmS^2}
}
\nonumber \\ && 
 +\, \frac{{\cal V}_\Lambda^{(T)}}{m^2}\Lambda({\bmp^\prime ,\bmp}) 
 + \frac{{\cal V}_t^{(T)}}{m^2}\,
 T({\bmk})\bigg] 
 + (1\otimes 1)\bigg[
  \frac{{\cal V}_k^{(1)}\pi^2}{m|{\bmk}|}
 + \frac{{\cal V}_2^{(1)}}{m^2}
 + \frac{{\cal V}_s^{(1)}}{m^2}{\bmS^2}
\bigg]
\,,\qquad\quad
\nonumber \\[2mm]
\lefteqn{
{\bmS} \, = \, { {{\bmsigma}_1 + {\bmsigma}_2} \over 2}\,,
 \qquad 
\Lambda({\bmp^\prime},{\bmp}) \, = \, -i {{\bmS} \cdot ( {\bmp^\prime}
 \times {\bmp}) \over  {\bmk}^2 }\,,
}
\nonumber \\[2mm]
\lefteqn{
 T({\bmk}) \, = \, {{\bmsigma}_1 \cdot {\bmsigma}_2} - {3\, {{\bmk}
 \cdot {\bmsigma}_1}\,  {{\bmk} \cdot {\bmsigma}_2} \over {\bmk}^2}
  \,,
}
\label{vNRQCDpotential}
\end{eqnarray}
and
\begin{eqnarray}
 U^{(0)}_{00} &=&  \frac{1}{q^0}\,,\quad
 U^{(0)}_{0i}  = -\frac{{({2\bmp^\prime}-2 {\bmp}-{\bmq})}^i}{({\bmp}-{\bmp}^\prime)^2}\,, 
\nonumber \\  
 U^{(0)}_{i0} &=&
 -\frac{({\bmp}-{\bmp}^\prime-{\bmq})^i}{({\bmp}-{\bmp}^\prime)^2}
\,, \quad
 U^{(0)}_{ij}  = \frac{-2 q^0 \delta^{ij} }{({\bmp}-{\bmp}^\prime)^2}
\,.
\label{vNRQCDsoft}
\end{eqnarray}
The matrices $T^A$ and $\bar T^A$ are the color matrices for the 
$\bf 3$ and $\bf \bar 3$ representations,
$i g G^{\mu\nu}=\left[D^\mu,D^\nu\right]$ and $m$ is the heavy quark
pole mass. 
Interactions involving ghosts and light quarks are not displayed.
The effective Lagrangian (second line) contains interactions between
quarks and soft gluons.  
Due to energy conservation at least two soft
gluons participate in interactions with quarks. Graphically the
interaction of a quark with two soft gluons is represented by Fig.\
\ref{figsoftvnrqcd}d.
The explicit form of the leading order soft interactions is displayed
in Eq.~(\ref{vNRQCDsoft}). The $1/q^0$ term that is visible for example in
$U_{00}^{(0)}$ arises from the propagation of an off-shell ``soft
quark'', see Fig.~\ref{figsoftvnrqcd}. The term does not have an
$i\delta$ prescription and does not lead to poles or
pinch-singularities when the loop integral over $q_0$ is
carried out by contours. Terms 
in the expansion of full QCD such as
$$\frac{1}{k_0+i\delta} -\frac{1}{k_0-i\delta}$$ lead to 
additional four-quark interactions with soft gluons in the matching
procedure.\cite{Luke1,Manohar2} 
There are terms in the Lagrangian where ultrasoft gluons couple to
potential operators. The term in the third line of 
Eq.~(\ref{vNRQCDLagrangian}) 
for example is generated by a QCD diagram
similar to the third diagram in Fig.\ \ref{figpnrqcdradiation}.
There are potential-type four-quark interactions (fourth line) that
depend non-locally on the soft indices, but are local interactions
with respect to the dynamical ultrasoft fluctuations. They are
graphically represented by Fig.\ \ref{figpotentialvnrqcd}b.
In Eq.\ (\ref{vNRQCDpotential}) the leading order Coulomb potential
interaction and the $v$- and $v^2$-suppressed potentials are
displayed. The basis in 
terms of $(1\otimes 1)$ and $(T^A \otimes \bar T^A)$ can be converted
to the color singlet and octet basis by the transformation
\begin{eqnarray}
 \left[\begin{array}{c} V_{\rm singlet} \cr V_{\rm octet} \end{array}\right]
 =\left[\begin{array}{ccc} 1 &  & -C_F \cr
    1 &  & {1\over 2} C_A - C_F \cr
 \end{array}\right]
 \left[\begin{array}{c} V_{1\otimes 1} \cr V_{T\otimes T} 
 \end{array}\right]\,.
\label{convertpotentials}
\end{eqnarray}
The multipole expansion has to be applied strictly for all terms in
the effective Lagrangian in order to achieve separation of the modes
in the different regions. For example the term $({\bmp}-i{\bmD})^2$ in
the bilinear quark sector has to be considered as an expansion in
$\bmD$ because $i\bmD\ll\bmp$. 
Interactions between soft and ultrasoft gluons without (heavy) quarks
do not exist because ultrasoft modes cannot resolve the ``high
frequency'' soft gauge transformations (that are realized through
reparametrization invariance). On the other hand, the soft gauge
fields feel ultrasoft gauge transformations just as a global phase
transformation.\cite{Bauer2}

The velocity power counting of the fields can be derived from
demanding that the action for the kinetic terms is of order
$v^0$ and reads\\[2mm]
\begin{equation}
\mbox{
\mbox{}\hfill
\begin{tabular}{|c|c|c|c|c|c|}
\hline
$\bmp$ & $\psi_{\bmp},\ \chi_{\bmp}$ & $A_{p}^\mu$ & $D^0$ & $\bmD$ & $A^\mu$\\[1mm]
\hline
$v$   & $v^{3/2}$   & $v$ &    $v^2$     &  $v^2$   &   $v^2$\\
\hline  
\end{tabular}.
\hfill\mbox{}
}
\label{vnrqcdfieldscounting}
\end{equation}
Soft and ultrasoft massless quark fields count as $v^{3/2}$ and $v^3$,
respectively.

External currents describing quark production, annihilation or decay
are also built from the quark fields in Eq.\
(\ref{quarkfieldsvnrqcd}). For example, consider $\QQbar$ production
at NNLL order in $e^+e^-$ annihilation.
One needs the ${}^3S_1$
(vector) current with dimension three and five,
\begin{eqnarray}
 \O{p}{1} & = & {\psip{p}}^\dagger\, \bsigma(i\sigma_2)\, {\chip{-p}^*} \,, 
 \nonumber  \\[2mm]
 \O{p}{2} & = & \frac{1}{m^2}\, {\psip{p}}^\dagger\, 
    {\bmp}^2\bsigma (i\sigma_2)\, {\chip{-p}^*}\,,
\label{vectorcurrents}
\end{eqnarray}
and the ${}^3P_1$ (axial-vector) current with dimension four
\begin{eqnarray}
 \O{p}{3} & = & \frac{-i}{2m}\, {\psip{p}}^\dagger\, 
      [\,\bsigma,\bsigma\cdot{\bmp}\,]\,(i\sigma_2)\,
   {\chip{-p}^*} \,. 
\label{axialvectorcurrents}
\end{eqnarray}
The currents describe production of a $\QQbar$ pair with momentum
$\pm\bmp$. The corresponding annihilation currents
$\O{p}{1-3}^\dagger$ are obtained from complex conjugation.

\subsection{Loops and Correlation of Scales}

\begin{figure}[ht] 
\begin{center}
\mbox{}\hspace{3mm}
\leavevmode
\epsfxsize=2.5cm
\leavevmode
\epsffile[0 200 630 570]{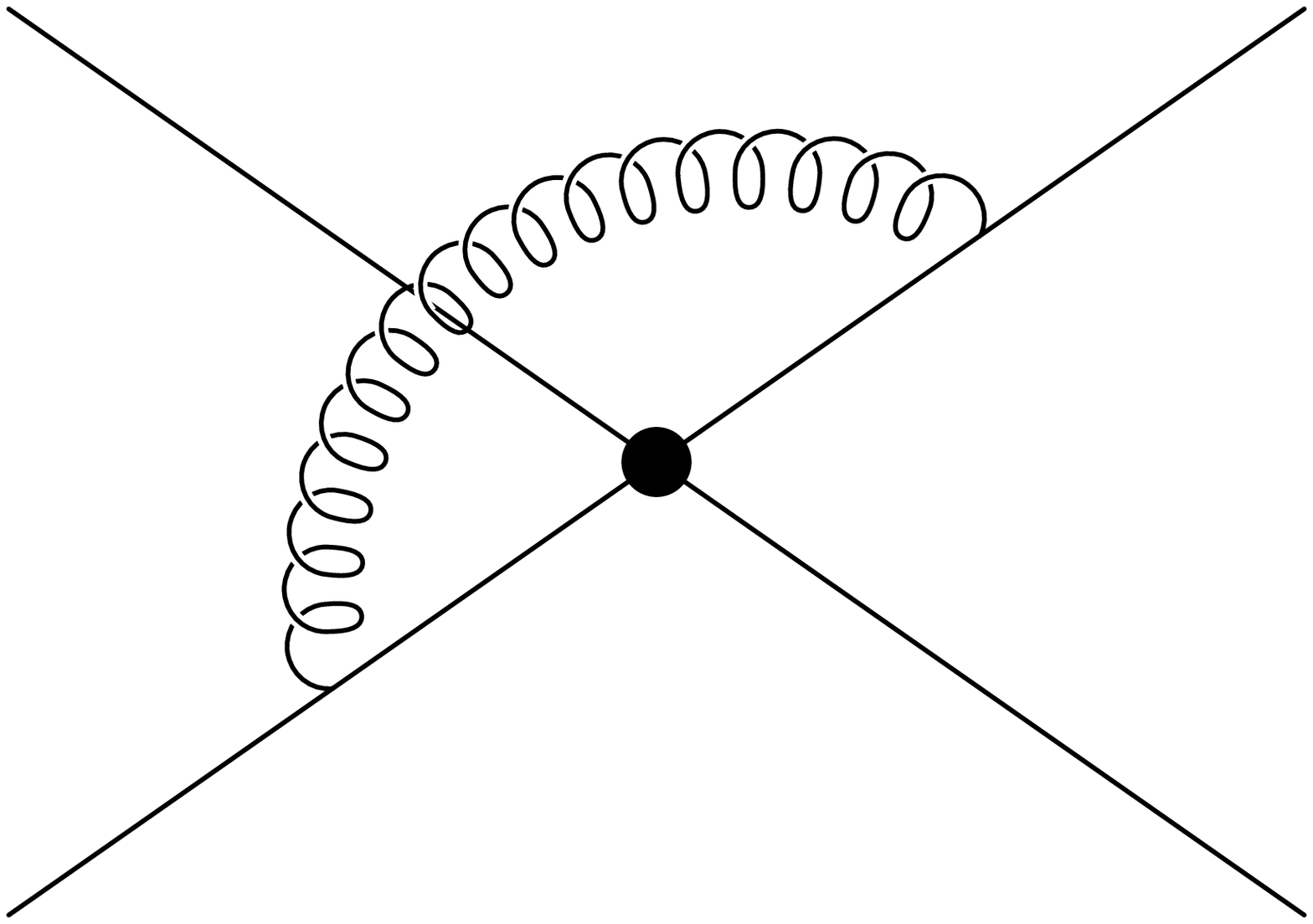}
\hspace{13mm}
\leavevmode
\epsfxsize=2.5cm
\leavevmode
\epsffile[80 420 230 520]{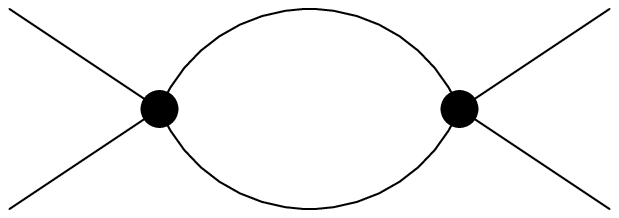}
\hspace{7mm}
\leavevmode
\epsfxsize=2.5cm
\leavevmode
\epsffile[0 200 630 570]{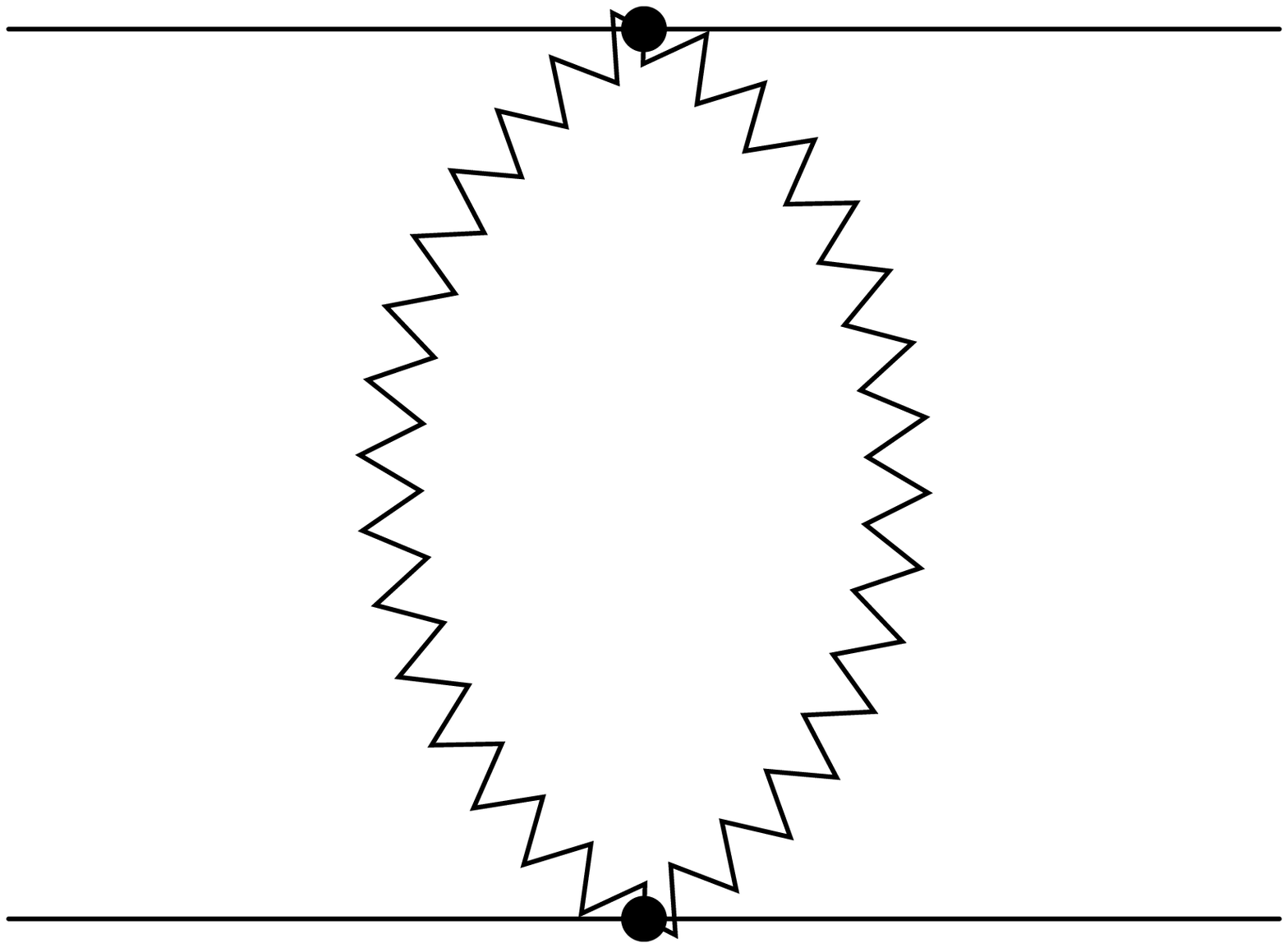}
\begin{picture}(0,0)(1,1)
\put(-295,35){$a)$}
\put(-205,35){$b)$}
\put(-83,35){$c)$}
\end{picture}
 \end{center}
%
%
\vskip  -0.2cm
 \caption{\label{figvnrqcdloops}   
Typical one-loop diagrams in vNRQCD containing ultrasoft (a),
potential (b) and soft (c) loop integrals.
}
\end{figure}

A general multi-loop graph in vNRQCD is divided into soft, potential and
ultrasoft loop subgraphs. The diagram in Fig.~\ref{figvnrqcdloops}a
shows a typical ultrasoft loop. 
Because an ultrasoft gluon cannot change the soft
index of a quark the loop does not involve any sum over soft indices
of the quark. In dimensional regularization the integration measure
of the ultrasoft loop reads ($D=4-2\epsilon$)
\begin{equation}
\mu_U^{4-D}\int\frac{d^D k}{(2\pi)^D}
\,,
\end{equation}
where $\mu_U$ is the common renormalization scale for the ultrasoft
dynamical fluctuations introduced to keep
couplings dimension-less. In vNRQCD $\mu_U$ is rewritten as
\begin{equation}
\mu_U \, \equiv \, m \nu^2\,,
\end{equation}
where $\nu$ is called the ``velocity scaling parameter'' and
$\nu=1$ corresponds to $\mu_U=m$. 
The reason for introducing $\nu$ will become clear below.
The natural choice for the velocity
scaling parameter in matrix elements is $\nu\approx v$. The integral for the
ultrasoft loop is carried out over the full
$D$-dimensional non-relativistic space. The information that
the loop is dominated by the ultrasoft region is implemented by
the pole structure of the propagators and the multipole expansion.

In Fig.\ \ref{figvnrqcdloops}b a typical potential loop diagram is
displayed. Its integration measure reads
\begin{equation}
\mu_U^{4-D}\sum\limits_{\bmp}\int\frac{d^D k}{(2\pi)^D}
\label{potentialloopsuma}
\end{equation}
and involves an integral over an ultrasoft momentum and a sum over all
possible soft indices of intermediate quark-antiquark pairs. In
practical analytical calculations it is 
convenient to rewrite the combination of the sum and the ultrasoft
integration as an integral over the full $D$-dimensional
non-relativistic momentum space. The number of terms in 
the sum is $(\frac{\mu_S}{\mu_U})^3$, where
\begin{equation}
\mu_S \, \equiv \, m \nu
\,.
\end{equation}
Thus Eq.\ (\ref{potentialloopsuma}) can be written as
\begin{equation}
\bigg[\,
\mu_U^{4-D}
\Big(\frac{\mu_S}{\mu_U}\Big)^{3}
\Big(\frac{\mu_S}{\mu_U}\Big)^{1-D}\,\bigg]\,
\sum\limits_{\bmp}^{D\mbox{\scriptsize-dim.}}\int\frac{d^D k}{(2\pi)^D}
\, = \,
\mu_S^{4-D}\int\frac{d^D p}{(2\pi)^D}
\,,
\label{potentialloopsumb}
\end{equation}
and we find that vNRQCD itself ``recognizes'' that the proper renormalization
scale for a potential loop is the soft scale $\mu_S$. The pole
structure of the propagators and the multipole 
expansion determine that the loop is dominated by the potential region.
For a typical
soft loop such as in Fig.\ \ref{figvnrqcdloops}c the same
renormalization scale $\mu_S$ arises, when the combination of sum and
ultrasoft integration is written as an integral over the full
$D$-dimensional non-relativistic momentum space,
\begin{equation}
\mu_U^{4-D}\sum\limits_{q}\int\frac{d^D k}{(2\pi)^D}
\, \to \,
\mu_S^{4-D}\int\frac{d^D q}{(2\pi)^D}
\,.
\label{softloopsum}
\end{equation}
The pole structure of the propagators and the multipole 
expansion determine that the loop is dominated by the soft region.
For more complicated multi-loop diagrams the same rules apply. Soft
and potential loops have the renormalization scale $\mu_S$ and purely
ultrasoft loops the renormalization scale $\mu_U$. Whether a loop is
either soft, potential or ultrasoft is determined by the pole
structure of the propagators and the multipole expansion.

In vNRQCD there are two renormalization scales, which are correlated
such that a ``subtraction velocity'' $\nu$ can be defined. The parameter
$\nu$ serves as the natural scaling parameter for all types of loop
integrals in vNRQCD. The scale correlation is not imposed by hand, but
an intrinsic property  of the theory itself. The
physical origin of the correlation $\mu_U=\frac{\mu_S^2}{m}$ is the
equation of motion of the quark, which relates the soft and ultrasoft
energy scales. The renormalization scales can be naturally
incorporated in the effective Lagrangian in $D=4-2\epsilon$ dimensions
to keep the couplings dimensionless.\cite{Manohar4}
In general, a vertex with $2+i$
quarks, $j$ soft gluons and $k$ ultrasoft gluons receives a factor 
$[\mu_S^{(i+j)\epsilon}\,\mu_U^{k\epsilon}]$.

\begin{figure}[ht] 
\begin{picture}(280,60)(0,-10)
  \put(50,-3){$c_1$} \put(87,-3){${\cal V}_c$} \put(127,-3){${\cal V}_c$}
  \put(195,-3){$c_1$} \put(228,-3){$\delta{\cal V}_{2,r}$} 
       \put(275,-3){${\cal V}_c$}
\qquad\,\,   \epsfxsize=4.3cm \raise4pt \hbox{\epsfbox{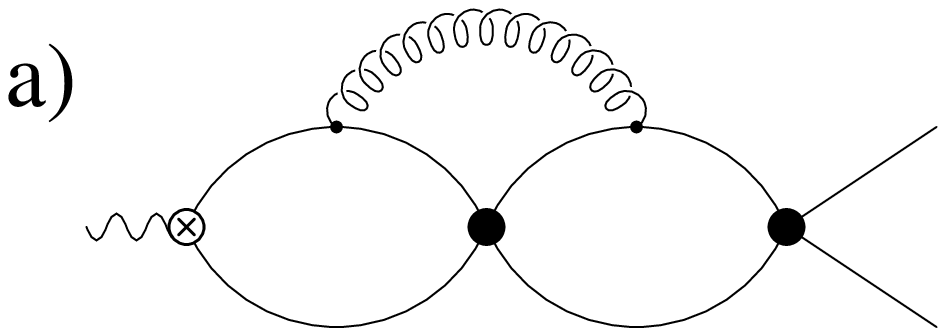}} \quad\quad
   \epsfxsize=4.3cm \raise4pt \hbox{\epsfbox{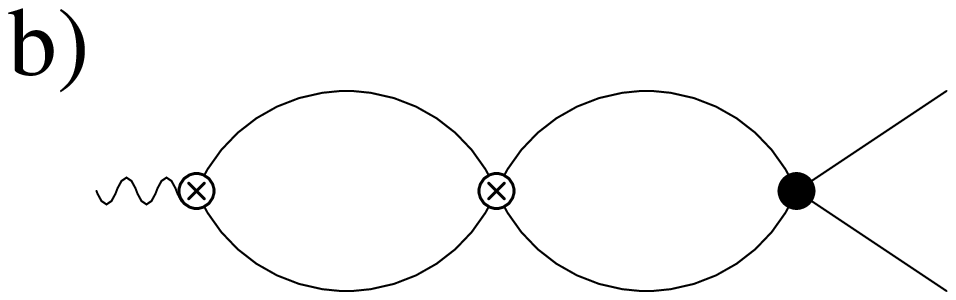}} 
\end{picture}
\vskip -0.3cm 
\caption{   
Examples of graphs that contribute to the three-loop
anomalous dimension for $c_1$. $\delta {\cal V}_{2,r}$ are one loop
counterterms. The coupling of the ultrasoft gluon to the quarks
is $g {\bmp}.{\bmA}/m$, where the scale of g is $m\nu^2$.
\label{fig3loopvnrqcd}}
\end{figure}

Without the correlation of scales vNRQCD becomes inconsistent.
Consider, for example, the three-loop diagram in 
Fig.~\ref{fig3loopvnrqcd}a, which contributes to the renormalization of the
dimension-three ${}^3S_1$ $\QQbar$ current 
${\cal
O}_{\bmp,1}=\psi_{\bmp}^\dagger\bsigma(i\sigma_2)\chi_{-\bmp}^*$.
The operator ${\cal O}_{\bmp,1}$ 
describes for example the leading order production of
$\QQbar$ pairs in $e^+e^-$ annihilation. After subtraction of all UV
subdivergences for $\epsilon\to 0$ through the diagrams in Fig.\
\ref{fig3loopvnrqcd}b the remaining overall UV divergence leads to the
following counter term contribution for the coefficient $c_1$ of the
operator ${\cal O}_{\bmp,1}$,
\begin{equation}
\delta c_1 \, \sim \, \alpha_s(m\nu^2)\,
[{\cal V}_c^{(T)}(\nu)]^2
\bigg[\,
\frac{\#}{\epsilon^2} + \frac{\#}{\epsilon}\,
\Big(\,
\ln\mu_U - 2\ln\mu_S + \#
\,\Big)
\,\bigg]
\,,
\label{c1runningcorrelation}
\end{equation}
where $\#$ symbolizes real numbers.
The anomalous dimension for $c_1$
can be determined unambiguously only if the correlation between
$\mu_U$ and $\mu_S$ is accounted for.\cite{Hoang3}

The computation of matching conditions and anomalous dimensions
proceeds in the canonical way taking the velocity scaling parameter
$\nu$ as the fundamental renormalization scaling variable in
dimensional regularization. This is the origin of the ``v'' in vNRQCD.

\subsection{Power Counting}

The velocity power counting can be carried out using the velocity
scaling rules of energies and momenta in the soft, potential and
ultrasoft regions of Eq.~(\ref{momentumregions}). At this point the
power counting 
is similar to the one in the threshold expansion and relies on
assigning powers of $v$ to all loop measures, propagators and vertices
in vNRQCD diagrams. An equivalent but more convenient power counting
prescription has been derived in Ref.~\citebk{Luke1}. It relies only on
$v$-counting for the vertices of the effective Lagrangian and the
topology of the soft momentum component in loop diagrams.
Let $V_k^{(S)}$, $V_k^{(U)}$ and $V_k^{(P)}$ denote the number of
soft, ultrasoft and potential vertices of order $v^k$ in a given
graph, where the velocity counting of the fields in
Eq.~(\ref{vnrqcdfieldscounting}) is included. Here,  
ultrasoft vertices involve only ultrasoft fields, potential vertices
involve at least one quark and no soft fields, and soft vertices at
least one soft field. For example, the vertices 
$$\psi_{\bmp}^\dagger\frac{{\bmp}.{\bmD}}{m}\psi_{\bmp}\, ,\quad
\psi_{\bmp^\prime}^\dagger\frac{1}{q^0}[A_{q^\prime},A_q]\psi_{\bmp}\, ,\quad
{\rm and}\quad
\psi_{\bmp^\prime}^\dagger\psi_{\bmp}\frac{1}{(\bmp-\bmp^\prime)^2}
\chi_{-\bmp^\prime}^\dagger\chi_{-\bmp}$$ have 
$V_6^{(P)}$, $V_4^{(S)}$ and $V_4^{(P)}$, respectively.
The ultrasoft gauge kinetic term $G^{\mu\nu}G_{\mu\nu}$ has
$V_8^{(U)}$.
For soft and potential vertices vertices $k\ge 4$ and for ultrasoft
vertices $k\ge 8$. 
A given diagram is then of order $v^\delta$, where\,\cite{Luke1}
\begin{equation}
\delta \, = \, 
5 + \sum_k \left[(k-8) V^{(U)}_k+(k-5) V^{(P)}_k +(k-4) V^{(S)}_k\right] 
 -N_S 
\,,
\label{vnrqcdpowercounting}
\end{equation}
$N_S$ being the number of connected soft components of the diagram.
Ultrasoft and soft vertices can only give positive contributions to
the sum. Potential vertices give positive contributions except for the
Coulomb potential interaction, which has $V_4^{(P)}$. An insertion of
$n$ Coulomb potentials lowers $\delta$ by $n$, but leaves the
$v$-counting unchanged if one also counts the Coulomb coupling as
${\cal V}_c\sim\alpha_s\sim v$. Thus each additional insertion of a
Coulomb potential gives a factor $\alpha_s v^{-1}$.
It is therefore convenient to count the Coulomb potential
$\propto\alpha_s/{\bmk}^2$ simply as
of order  $\alpha_s v^{-1}$, the $1/m|{\bmk}|$ potentials as of order
$\alpha_s v^{0}$ and the  $1/m^2$ potentials as of order
$\alpha_s v^{1}$, etc.. 

\begin{figure}[ht] 
\begin{picture}(280,60)(-70,-3)
 \put(70,10){${\cal V}_c$} \put(117,10){${\cal V}_c$}
\qquad\,\,   \epsfxsize=4.cm \raise4pt \hbox{\epsfbox{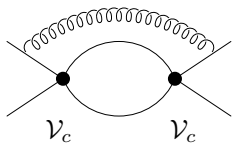}}
\end{picture}
\vskip -0.5cm 
\caption{   Examples of graph or order $\alpha_s^3\,v^5$.
The coupling of the ultrasoft gluon to the quarks
is $g {\bmp}.{\bmA}/m$.
\label{figvnrqcdpowercounting}}
\end{figure}

For illustration let us consider the graph displayed in
Fig.~\ref{figvnrqcdpowercounting}, which has
$V_6^{(P)}=2$ and $V_4^{(P)}=2$.
Thus one obtains $\delta = 5 + (6-5)\times 2 + (4-5)\times 2 = 5$, so
the graph is of order $\alpha_s^3\,v^5$.
Dividing this result by the velocity counting of the four external
quark fields ($v^6$), we find that the amputated graph is of order
 $\alpha_s^3\,v^{-1}$. The same result is found from the 
$v$-counting of the loop measures, propagators and vertices  
based on the velocity scaling of the ultrasoft and potential regions
in Eq.~(\ref{momentumregions}).
The diagram in Fig.~\ref{figvnrqcdpowercounting} is
suppressed by $\alpha_s\,v^2$ with respect to the same graph
without the ultrasoft gluon, which is of order $\alpha_s^2 v^3$.
Each additional ultrasoft gluon gives an additional factor $\alpha_s
v^2$. If we apply the same counting rules to QED we find that
the Lamb-shift, which is caused by ultrasoft photons, is suppressed by
a factor $\alpha v^2\sim \alpha^3$ with respect to the leading Coulomb
interaction.

\subsection{Matching}

The determination of the matching conditions of the coefficients in
the effective Lagrangian at the hard scale, $\nu=1$, is carried
out with amplitudes for on-shell quarks and gluons. This avoids
operators that vanish by the equation of motion and gauge-dependent
matching conditions. Any kinematic situation described by the
effective theory can be used for the matching calculation. For
example, at the Born level the matching conditions for the potentials
obtained from the on-shell graphs in Fig.\
\ref{figpotentialvnrqcd} are\,\cite{Manohar2}
\begin{eqnarray}
 {\cal V}_c^{(T)}(1) &=& 4 \pi \alpha_s(m)\,, \qquad
 {\cal V}_r^{(T)}(1) = 4 \pi \alpha_s(m)\,, \qquad 
 {\cal V}_s^{(T)}(1) = -\frac{4 \pi \alpha_s(m)}{3}\,,
\nonumber\\ 
 {\cal V}_\Lambda^{(T)}(1) &=& -6 \pi \alpha_s(m) \,,\qquad
 {\cal V}_t^{(T)}(1) = -\frac{\pi \alpha_s(m)}{3} \,. \qquad
\label{vnrqcdbornmatching}
\end{eqnarray} 
Matching conditions that vanish and annihilation contributions are not
shown in Eq.\ (\ref{vnrqcdbornmatching}).
The on-shell matching conditions for the potentials at the one-loop
level were computed in Ref.~\citebk{Manohar4}.  
For the order $1/(m\bmk)$ 
potentials the results read
\begin{eqnarray}
  {\cal V}_k^{(T)}(1) &=& \alpha_s^2(m) \Big( \frac{7 C_A}{8}-\frac{C_d}{8}
  \Big) \,,\qquad
   {\cal V}_k^{(1)}(1) = \alpha_s^2(m) \frac{C_1}{2} \,,
\label{Vkresult}
\end{eqnarray}
where $C_d=8C_F-3C_A$ and $C_1=C_F(\frac{1}{2}C_A-C_F)$.
For SU$(N)$ we have $$C_F=\frac{N^2-1}{2N}\,,\quad  C_A=N\,,\quad
C_d=\frac{N^2-4}{N}\,,\quad {\rm and} \quad C_1=\frac{N^2-1}{4N^2}\,.$$
For the description of the $\QQbar$ dynamics at NNLL order the results
in Eqs.\ (\ref{vnrqcdbornmatching}) and (\ref{Vkresult}) are
sufficient.
The coefficients of the Coulomb potential do not receive any higher
order matching corrections up to order $\alpha_s^3$, see
Ref.~\citebk{Hoang4}. 
In general, a non-zero matching corrections appears, when there is a
an off-shell region such as the hard one that contributes in the
matching condition.

The Coulomb potential $\frac{{\cal V}_c}{\bmk^2}$ is equivalent to the
potential $\propto\frac{1}{\bmk^2}$ that describes the $\QQbar$
binding effects in the Schr\"odinger equation only at the leading
logarithmic level. 
The known one- and two-loop corrections obtained in
Refs.~\citebk{Fischler1,Schroeder1}
are not contained in  $\frac{{\cal V}_c}{\bmk^2}$. In vNRQCD these
corrections arise in form of time-ordered products of the soft
vertices in Eq.\ (\ref{vNRQCDsoft}), such as the one-loop diagram in
Fig.\ \ref{figvnrqcdloops}c, i.e.\ they are contributions from matrix
elements and not part of the matching coefficient
of the Coulomb potential.

The matching conditions for the two-quark soft vertices are determined
by Compton scattering diagrams in full QCD. At Born level the diagrams
in Fig.\ \ref{figsoftvnrqcd}a,b,c are matched onto the local soft
operators in the second line of Eq.\ (\ref{vNRQCDLagrangian}). At
leading order in $v$ ($U_{\mu\nu}^{(\sigma=0)}$) the results are
displayed in Eq.~(\ref{vNRQCDsoft}). 
The results up to next-to-next-to-leading
order in $v$ ($U_{\mu\nu}^{(\sigma=1,2)}$) have been
determined in Ref.~\citebk{Manohar2}. Interestingly, the higher order
corrections to the two-quark soft vertices can be determined by
computing the corrections in the HQET Lagrangian (see e.g.\
Ref.~\citebk{Bauer1}), and then 
matching their time-ordered products to the two-quark soft
vertices.\cite{Manohar2} This works 
because only the soft momentum region is relevant for the
renormalization of the two-quark
soft vertices. 
One outcome is that there are no higher order perturbative corrections
to the leading order ($\sigma=0$) soft vertices in 
Eq.~(\ref{vNRQCDsoft}).

\begin{figure}[t!]
\begin{picture}(180,60)(10,1)
  \put(190,46){$c_1$} \put(215,46){${\cal V}_k$}
  \put(263,46){$c_1$} \put(290,46){${\cal V}_c$} \put(317,46){${\cal V}_c$}
   $\begin{array}{c}
  \centerline{
   \raisebox{15pt}{$\left( \begin{array}{c} \\  \\[2mm] \end{array}\right.
     \!\!\!\!\!\!$}
   \epsfxsize=1.7cm \raise2pt \hbox{\epsfbox{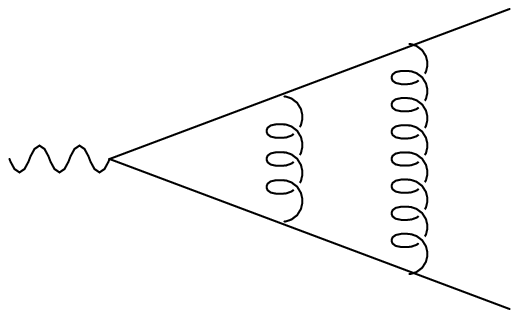}} 
   \raisebox{15pt}{$ + $}
   \epsfxsize=1.7cm \raise2pt \hbox{\epsfbox{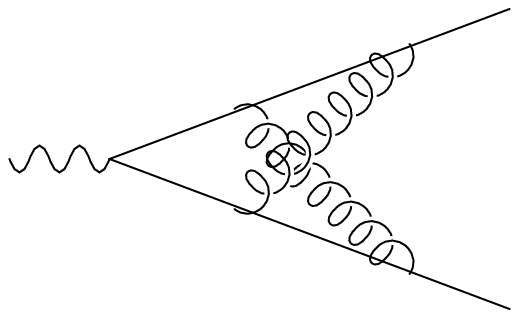}}
   \raisebox{15pt}{$ +\ \ldots $}
   \raisebox{15pt}{$\!\!\!\! \left. \begin{array}{c} \\  \\[2mm] \end{array}
     \right)$}
   \raisebox{15pt}{$-$} 
   \raisebox{15pt}{$\left( \begin{array}{c} \\ \\[2mm] \end{array}\right.
      \!\!\!\!$}
   \epsfxsize=1.7cm \raise8pt \hbox{\epsfbox{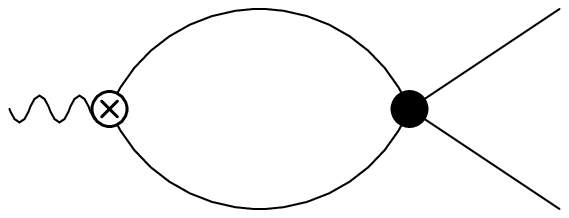}} \quad
   \raisebox{15pt}{$ + $}
   \epsfxsize=2.7cm \raise8pt \hbox{\epsfbox{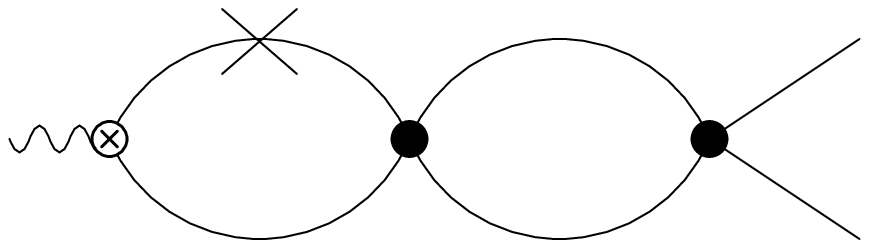}} 
   } \\[5mm]
   \put(28,33){$c_1$} \put(50,33){${\cal V}_{2,s,r}$} \put(82,33){${\cal V}_c$}
  \put(130,33){$c_1$} \put(155,33){${\cal V}_c$} \put(180,33){${\cal V}_{2,s,r}$}
  \put(230,33){$c_2$} \put(257,33){${\cal V}_c$} \put(284,33){${\cal V}_c$}
   \centerline{
   \raisebox{18pt}{$ + $}
   \epsfxsize=2.7cm \raise11pt \hbox{\epsfbox{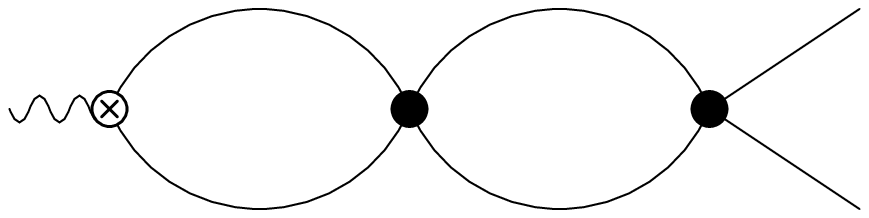}} \quad
   \raisebox{18pt}{$ + $}
   \epsfxsize=2.7cm \raise11pt \hbox{\epsfbox{c11b.eps}} \quad
   \raisebox{18pt}{$ + $}
   \epsfxsize=2.7cm \raise11pt \hbox{\epsfbox{c11b.eps}} \quad
   \raisebox{18pt}{$\left. \begin{array}{c} \\ \\[2mm] \end{array}\right)$}
   } 
   \end{array}$
\end{picture}
\vskip 1.cm 
\caption{   Difference of full QCD
and order $\alpha_s^2 v^0$ vNRQCD
graphs which gives the two loop matching for $c_1(1)$. The $\times$ denotes an
insertion of the ${\bmp}^4/(8m^3)$ operator, and graphs with this insertion on
a different propagator are understood.\label{figmatchingc1}}
\end{figure}

The matching conditions for external currents are determined in the
same way as the matching conditions for the potentials. For example,
for the description of $\QQbar$ production at NNLL order in $e^+e^-$
annihilation one needs the
matching conditions for the ${}^3S_1$ vector current up to dimension
five,  
${\bf J}^v_{\bmp}= c_1 \O{p}{1} + c_2 \O{p}{2}$
and the ${}^3P_1$ axial-vector current up to dimension
four,
${\bf J}^a_{\bmp}= c_3 \O{p}{3}$, see Eqs.~(\ref{vectorcurrents}) and
(\ref{axialvectorcurrents}).
The matching condition for $c_1$ needs to be known at order
$\alpha_s^2$. For $c_{2,3}$ Born matching is sufficient and is
obtained by expanding the full QCD currents $Q\gamma^\mu Q$ and 
$Q\gamma^\mu\gamma_5 Q$ at tree level,
\begin{eqnarray}
  c_2(\nu=1) = -1/6 \,, \qquad\qquad c_3(\nu=1)=1 \,.
\label{matchingLLc23}
\end{eqnarray}
The one-loop matching condition for $c_1$ is well known and
scheme-independent for mass-independent regularization
schemes. The two-loop computation involves computing the
difference between graphs in full QCD and in the effective theory as
shown in Fig.\ \ref{figmatchingc1}.
%
%
The two-loop vertex corrections in the full theory were computed in
Refs.~\citebk{Czarnecki3,Hoang5}. The matching can be carried out at the level of
amplitudes as shown in Fig.\ \ref{figmatchingc1}, which involves a
cancellation of IR divergent Coulomb phases, or at the level 
of the total cross section, where the Coulomb phases are absent.
The latter type of matching is called ``direct matching''.\cite{Hoang1}
The result reads\,\cite{Hoang3}
\begin{equation}
c_1(1) =  1- \frac{2 C_F}{\pi}\: {\alpha_s(m)} +   
  \alpha_s^2(m) \bigg[C_F^2\Big(\frac{\ln 2}{12}-\frac{25}{24}
  -\frac{2}{\pi^2}\Big) + C_A C_F\Big(\ln 2-1\Big) + \frac{\kappa}{2}
  \bigg]\,, 
\label{c1matchingnrqcd}
\end{equation}
where\,\cite{Hoang13}
$
\kappa =
  -2.556 + 0.08256\, n_\ell
$,
and $n_\ell$ is the number of light quark species.
At order $\alpha_s^2$ the matching condition is dependent on the
subtraction scheme and on the definition of the operators in the
effective theory. The result in Eq.\ (\ref{c1matchingnrqcd}) is
obtained in the \ms scheme and with the form of all operators 
given in $D=4$ dimensions.  
The result is different from
Ref.~\citebk{Beneke2}, 
where the potentials were defined with an explicit dependence on
$D=4-2\epsilon$, such that the matching conditions agrees with the
contribution of the hard region in Eq.\ (\ref{momentumregions}) in the
threshold expansion.

\subsection{Soft IR Divergences}

One of the most important (and interesting) conceptual aspects of
vNRQCD is the existence of the ultrasoft gluons at the hard scale
through the relation of scales $\mu_U=\frac{\mu_S^2}{m}=m\nu^2$. This
relation is an intrinsic property of the effective theory. One
consequence is for example that ultrasoft gluons
contribute in the determination of the matching conditions at $\nu=1$,
or the running of the coefficients for any value for $\nu$ smaller
than $1$. The
necessity for having the ultrasoft degrees of freedom separated at the
hard scale can be seen explicitly in multi-loop diagrams, where
potential (or soft) and ultrasoft loops occur at the same time. An
unambiguous determination of the anomalous dimensions is only
possible, if the scale correlation is taken into account, see
e.g.\ Eq.~(\ref{c1runningcorrelation}).

This seems to be in contradiction to the fact that the ultrasoft
gluons clearly arise from splitting the original gluon field into soft
($\sim mv$) gluons and ultrasoft ($\sim mv^2$) gluons. The latter
clearly describe fluctuations at scales much smaller than $mv$. For the
threshold expansion the presence of the ultrasoft momentum regions and
the necessity for a proper expansion (separation) of the ultrasoft
momentum region for any choice of renormalization scale is
unproblematic conceptually -- the threshold expansion is a technical
prescription to obtain an expansion of Feynman diagrams and not an
effective theory, where renormalization issues are relevant.

In vNRQCD the problem is resolved by the property of IR divergences in
soft loop integrations. The mechanism to account for the scale
correlation $\mu_U=\frac{\mu_S^2}{m}=m\nu^2$ and the fact that soft
and ultrasoft degrees of freedom fluctuate at different length scales
is to treat the IR divergences in soft loops as UV
divergences.\cite{Manohar3,Hoang4} This means that soft IR and soft
UV divergences contribute to the anomalous dimensions of the operators
in the effective Lagrangian. 
For illustration let us discuss the following simplified situation
where only diagrams with single $\frac{1}{\epsilon}$ poles and either
purely soft or ultrasoft gluons are discussed and real
Coulombic IR divergences are dropped.\cite{Hoang4} A general diagram with a
divergent soft loop then gives an amplitude with the divergence
structure 
\begin{eqnarray}
 i{\cal A}^S &=& {A\over \epsilon_{\rm UV}} + {B \over \epsilon_{\rm IR}} +
   C \left( {1\over \epsilon_{\rm UV}}-{1\over \epsilon_{\rm IR}} \right)
  = {A+B \over \epsilon_{UV}} +
   (C-B) \left( {1\over \epsilon_{\rm UV}}-{1\over \epsilon_{\rm IR}}
\right)\,.
\nonumber\\\mbox{}
\label{Asoft}
\end{eqnarray}
Pure dimensional regularization does not distinguish between UV and IR
divergences, but an identification of let's say UV divergences can be
made by performing the calculation with an additional IR
regulator. The term $C$ represents diagrams involving scaleless
integrals such as tadpole graphs, and $A$ and $B$ represent all other
graphs. For example, for the hydrogen Lamb shift (see Sec.\
\ref{subsectionlambshift}) only $C$ is non-zero, while for
positronium or quarkonium also $A$ and $B$ are non-zero. At the same
order in the power counting as Eq.\ (\ref{Asoft}) there is an
amplitude with a divergent ultrasoft gluon loop, which has the form
\begin{eqnarray} 
 i{\cal A}^U  &=&  { (C-B) \over \epsilon_{\rm UV}} + {D \over
 \epsilon_{\rm IR}}\,. 
\label{Ausoft}
\end{eqnarray}
In general, $D$ is independent of $(C-B)$ since ultrasoft loops are in
general not proportional to $(\frac{1}{\epsilon_{\rm UV}}
-\frac{1}{\epsilon_{\rm IR}})$. Examples are the two- and three-loop
diagrams contributing to the 
renormalization of ${\cal V}_k$ (Ref.~\citebk{Manohar3})
and ${\cal V}_c$ (Ref.~\citebk{Hoang4}), respectively.
The IR divergence matches with an IR divergence in full QCD. The
non-trivial issue is that the coefficient of the UV divergence in Eq.\
(\ref{Ausoft}) matches with the IR divergence in the soft amplitude in
Eq.\ (\ref{Asoft}). This statement can be understood intuitively from
the fact that the original gluon field is split into soft and
ultrasoft modes in the effective theory. Thus any soft IR divergence
cannot be an IR divergence of the full theory, because it is not
associated with scales below $mv^2$. In other words, in Eq.\
(\ref{Asoft}) the $\epsilon_{\rm UV}$'s correspond to the scale $m$
and the $\epsilon_{\rm IR}$'s correspond to the scale $mv$, while in
Eq.\ (\ref{Ausoft}) the  $\epsilon_{\rm UV}$'s correspond to the scale $mv$
and the $\epsilon_{\rm IR}$'s correspond to the scale $mv^2$. 
The validity of this argument for complicated multiloop diagrams,
where soft, potential or ultrasoft loops exist simultaneously, is
conjectured, and not yet proven mathematically. However, examples such as
in Eq.~(\ref{c1runningcorrelation}), seem to support its validity.

The running of the operators is determined by UV divergences. However,
examining  $i{\cal A}^S+ i{\cal A}^U$ one finds that the term $(C-B)$
in the soft amplitude acts like a tadpole contribution that pulls
the $\frac{1}{\epsilon_{\rm UV}}$ in the ultrasoft amplitude up to the
hard scale. It is argued in Ref.~\citebk{Hoang4} that the
scale-dependence of the coefficient $(C-B)$ in Eq.\ (\ref{Ausoft})
does not affect this argument since the scale-dependence of the
$(C-B)$ term is Eq.~(\ref{Asoft}) can be chosen arbitrary. The final
outcome is that  
interpreting all $\frac{1}{\epsilon}$ terms in the soft amplitude as
UV divergences, running the ultrasoft modes from $m$ to $mv^2$ with
an anomalous dimension $\propto(C-B)$ and running the soft
modes from $m$ to $m v$ with an anomalous dimension $\propto(A+B)$
correctly performs the running between the scales. This is the basis
of computations of the evolution of the couplings and the summation
of large logarithmic terms in vNRQCD. 

The correspondence in Eqs.\ (\ref{Asoft}) and (\ref{Ausoft}) provides
a useful tool. If the UV divergences $(A+C)$ in the soft diagrams are
known, and the combination $(B-C)$ is determined from the UV
divergences in the ultrasoft diagrams, one arrives at
$(A+C)+(B-C)=A+B$, which is the combination needed to determine the
soft anomalous dimension. This means that one can recycle results for
the running of coefficients in HQET for the renormalization of
two-quark soft vertices in vNRQCD.\cite{Manohar2}
An explicit demonstration for the mechanism described above is given
in Sec.~\ref{sectionvNRQCDvspNRQCD}.

\subsection{Renormalization Group Equations and Running}
\label{subsectionvNRQCDrunning}

The renormalization procedure is carried out in the canonical way. 
For soft loops UV and IR divergences are not distinguished
and both contribute to the anomalous dimensions. The renormalization
group equations
are formulated in terms of the velocity scaling parameter $\nu$.
For illustration let us consider the NNLL order running of the color
singlet Coulomb potential $\frac{{\cal V}_c^{(s)}}{{\bmk}^2}$ in some 
detail.\cite{Hoang4} As mentioned previously, the hard matching
conditions only contain an order $\alpha_s$ Born contribution, 
${\cal V}_c{(T)}=4\pi\alpha_s(m)$, ${\cal V}_c{(1)}=0$. 
In the following discussion it is understood that $\alpha_s$ runs with the
three-loop $\beta$-function.
As for the matching calculation, the running is determined using
on-shell four-quark amplitudes. 

At LL order the only effective theory graphs
$\propto\frac{\alpha_s^2}{{\bmk}^2}$ are the soft diagrams in Fig.\
\ref{figvnrqcdloops}c, where the soft gluon vertices are given in Eq.\ 
(\ref{vNRQCDsoft}). The result reads
\begin{eqnarray}  \label{eftvc1}
\begin{picture}(60,30)(1,1)
 \epsfxsize=2.0cm \lower16pt \hbox{\epsfbox{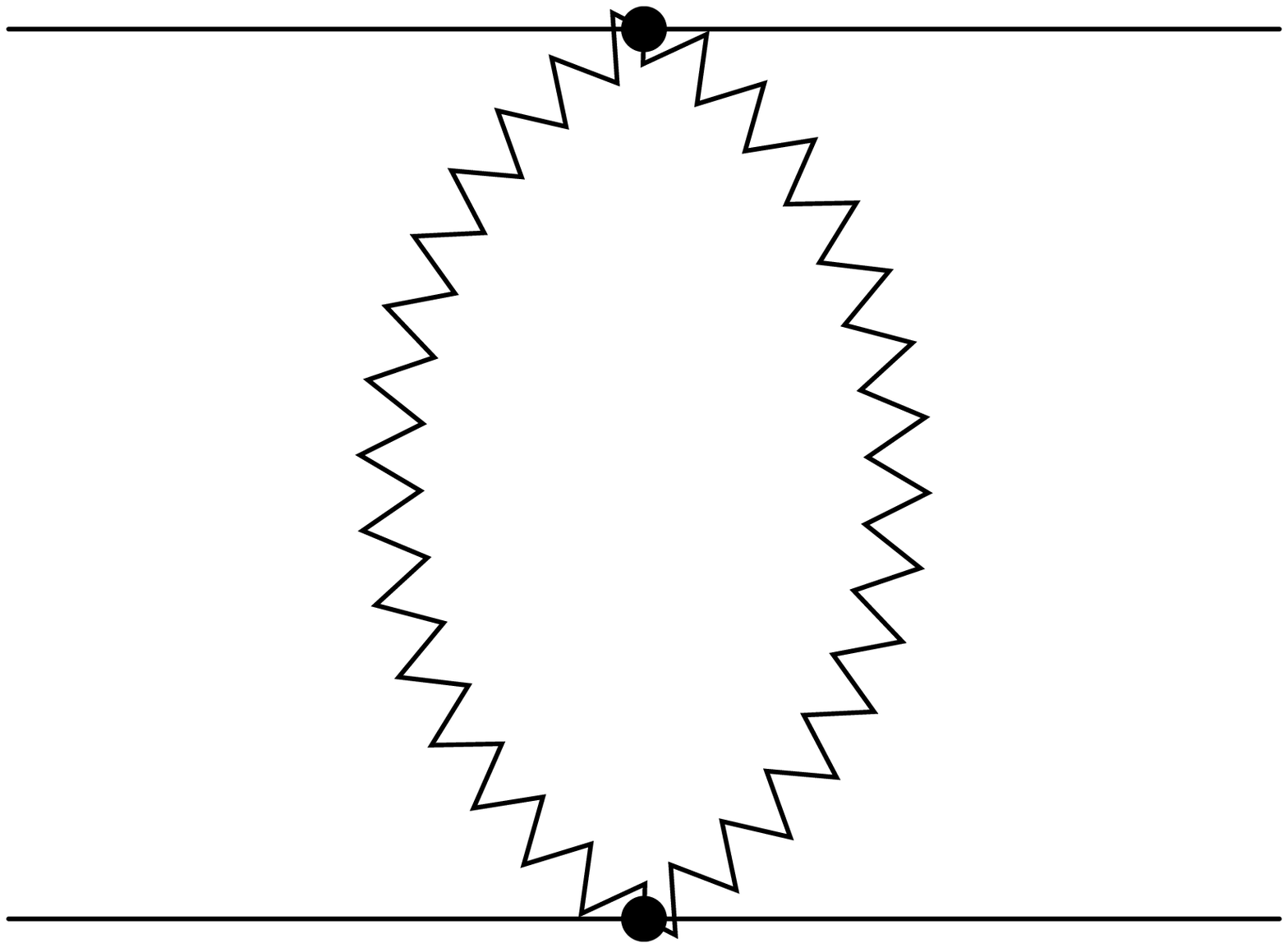}}
\end{picture}
  &=& \frac{-i \mu_S^{2\epsilon}\alpS^2(\nu)}{\bf k^2} (T^A  \otimes \bar T^A) 
  \bigg[ \frac{\beta_0}{\epsilon} + \beta_0
   \ln\Big( \frac{\mu_S^2}{\bf k^2}\Big)  + a_1 \bigg] \,, \\[-1pt]\nn
\end{eqnarray}
where $\beta_0=11-\frac{2}{3}\,n_\ell$ and 
$a_1=\frac{31}{3}-\frac{10}{9}\, n_\ell$ in
the $\msb$ scheme, and $n_\ell$ is the number of massless quarks.
The divergence is canceled by a counterterm for the operator
$\frac{{\cal V}_c^{(T)}}{{\bmk}^2}$, which causes the coefficient
${\cal V}_c^{(T)}$ to run with the anomalous dimension
$-2\beta_0\alpha_s(m\nu)$, i.e.\ ${\cal V}_c^{(T)}=4\pi\alpha_s(m\nu)$ 
at LL order. The remaining terms in the soft graphs
are identical to the one-loop static potential calculation of Fischler
and Billoire.\cite{Fischler1} This can be understood because the $1/m$
expansion of the static calculation in Ref.~\citebk{Fischler1} just
separates out the soft region. 

\begin{figure}[t]
 \centerline{\hbox{\epsfxsize=2.3cm \epsfbox{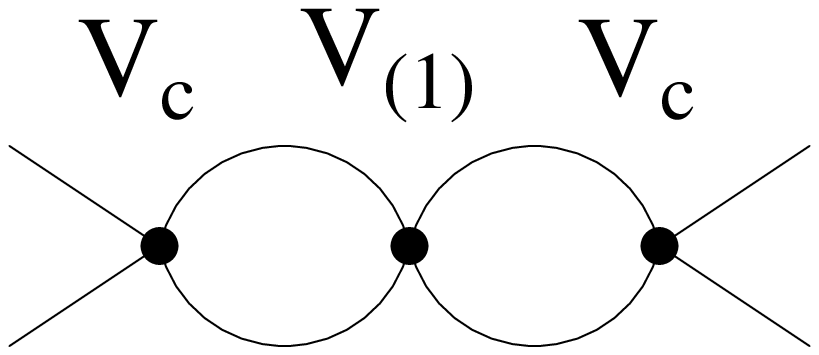}} \qquad 
             \hbox{\epsfxsize=2.3cm \epsfbox{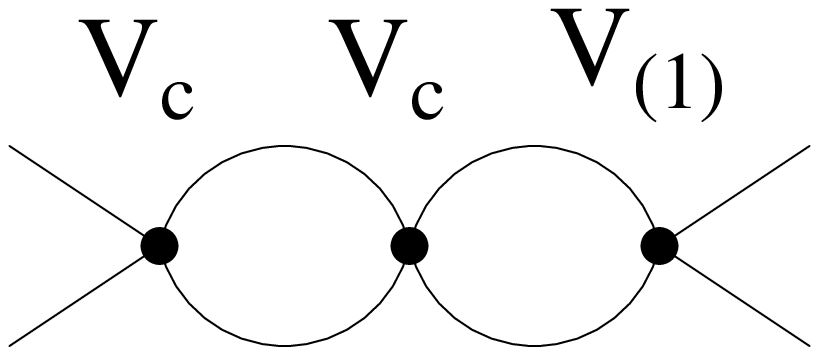}} \qquad
             \hbox{\epsfxsize=2.3cm \epsfbox{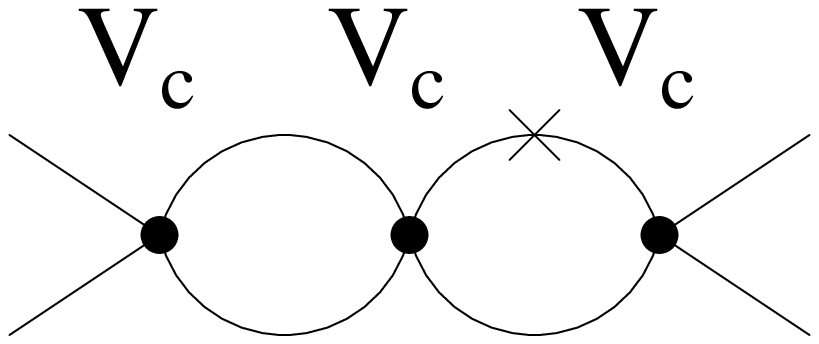}} \qquad
 \lower1pt
             \hbox{\epsfxsize=1.8cm \epsfbox{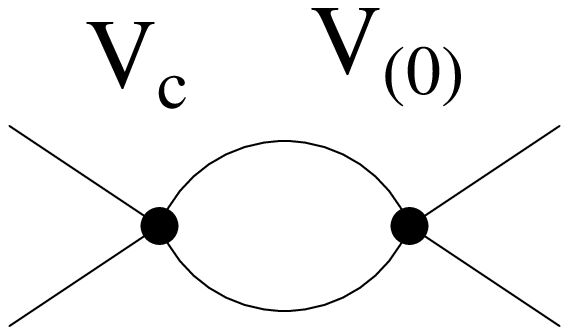}} }
 \caption{\label{figtwoloopCoulombpotential}   
 Order $\alpha_s^3/v$ diagrams with potential 
 iterations. The $\times$ denotes an insertion of the ${\bf p^4}/8m^3$ 
 relativistic correction to the kinetic term, ${\cal V}_{(0)}$ stands
 for a ${\cal V}_k$ potential and  ${\cal V}_{(1)}$ for a $1/m^2$
 potential.
} 
\end{figure}

\begin{figure}
 \centerline{\hbox{\epsfxsize=1.5cm \epsfbox{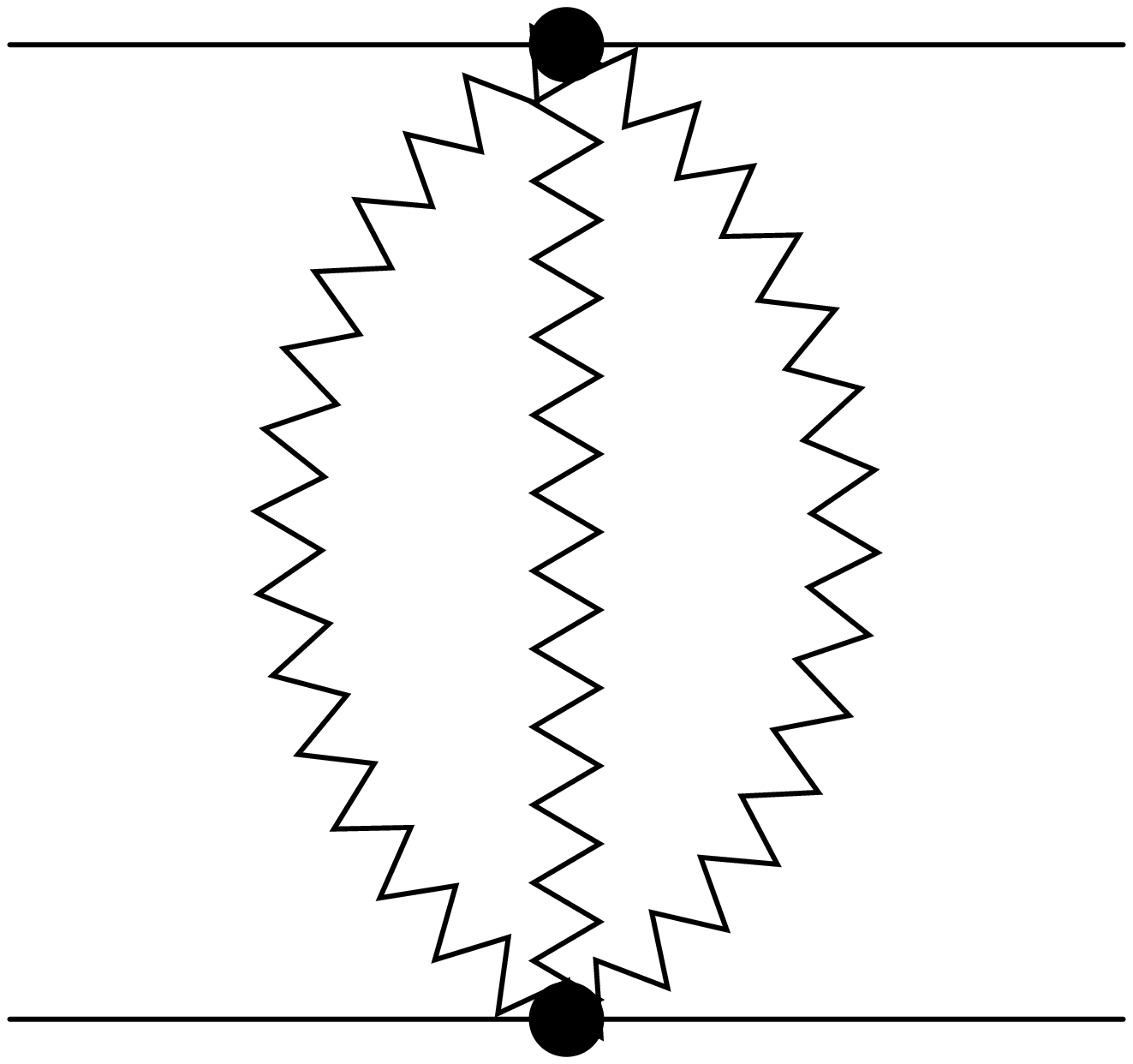}} \quad
             \hbox{\epsfxsize=1.5cm \epsfbox{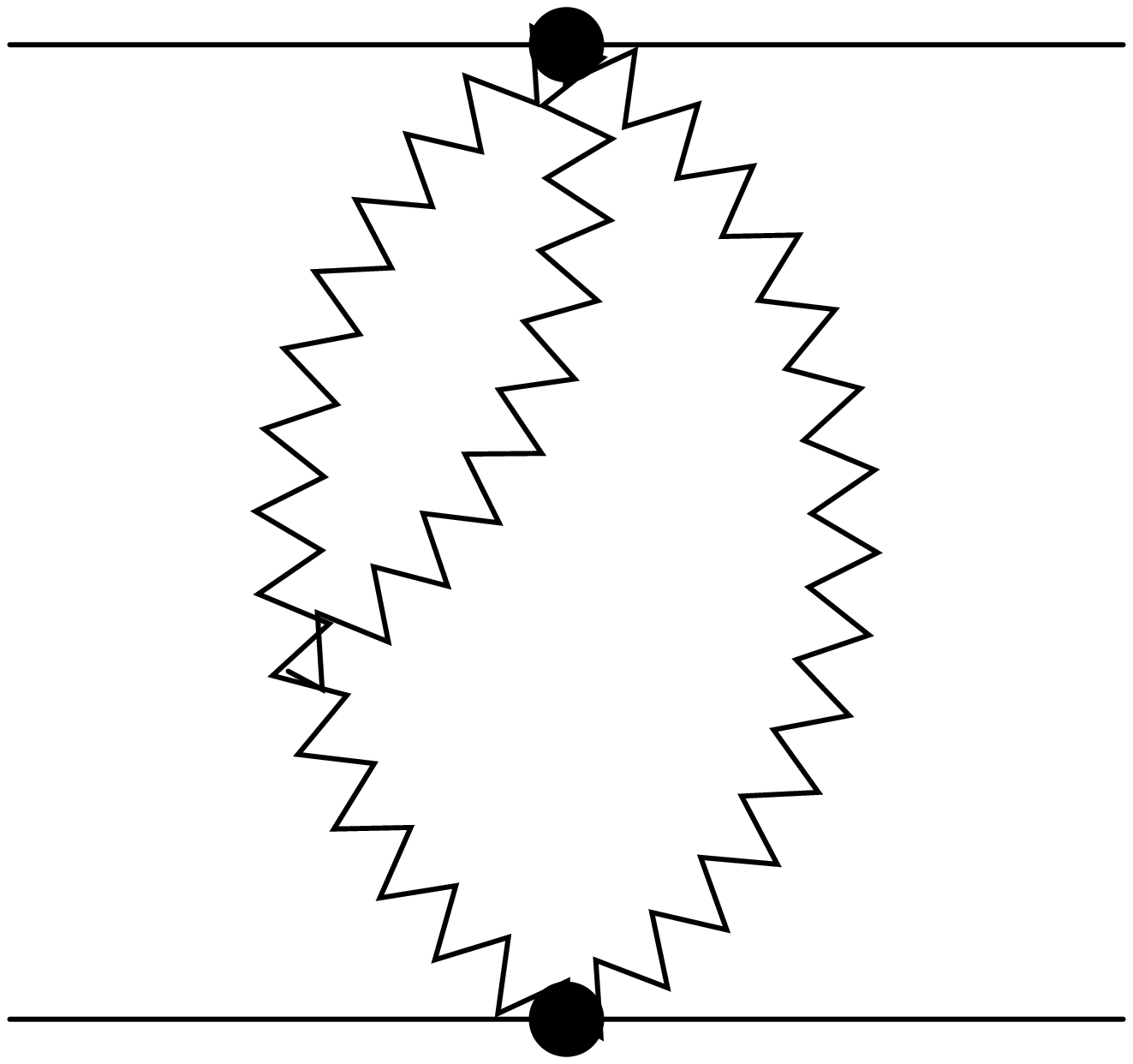}} \quad
             \hbox{\epsfxsize=1.5cm \epsfbox{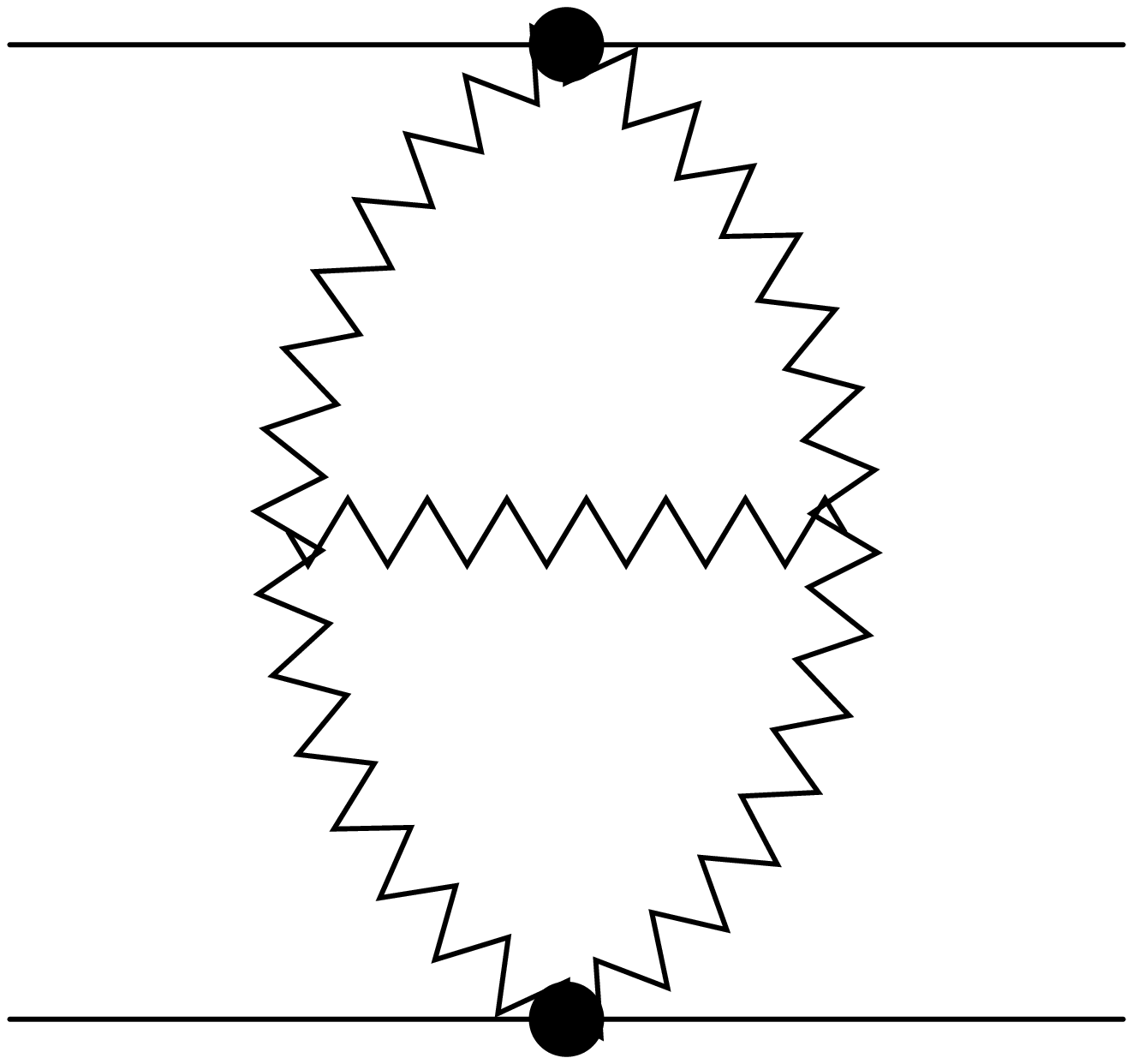}} \quad
           \lower2pt  \hbox{\epsfxsize=1.77cm \epsfbox{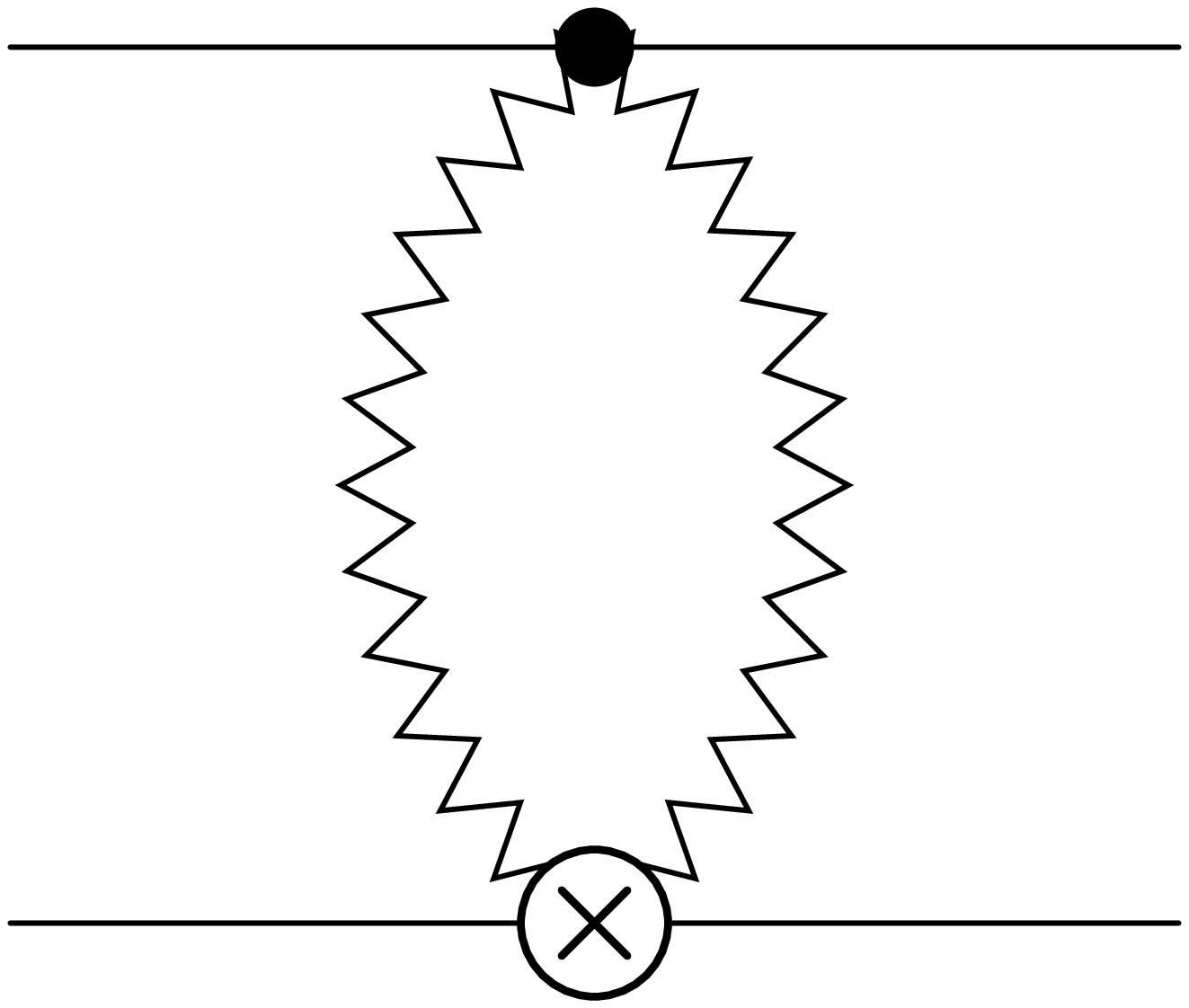}} 
 \raise20pt \hbox{\Large $\ \ \ldots$} } \medskip
 \caption{\label{figtwoloopCoulombsoft}   
 Examples of order $\alpha_s^3/v$ diagrams with soft vertices.
 The vertex with a cross denotes an insertion of a one-loop counterterm. }
\end{figure}

\begin{figure}
 \centerline{
 \hbox{ \epsfxsize=3.3cm\epsfbox{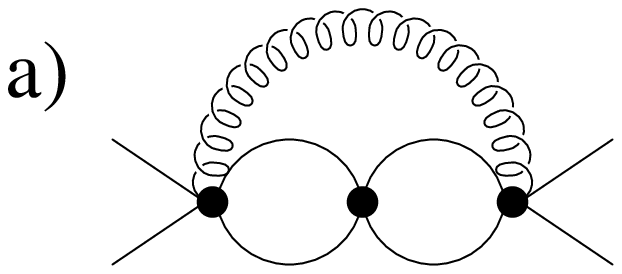}} \qquad 
 \hbox{\epsfxsize=3.3cm\epsfbox{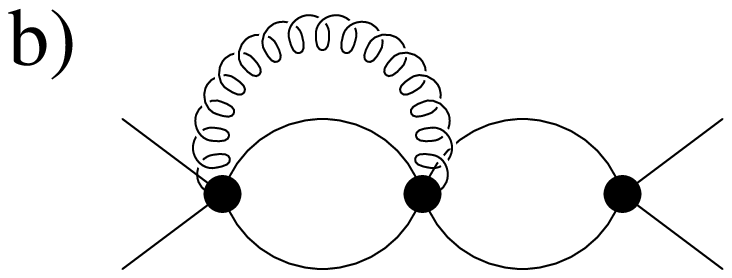}} \qquad
 \hbox{\epsfxsize=3.cm\epsfbox{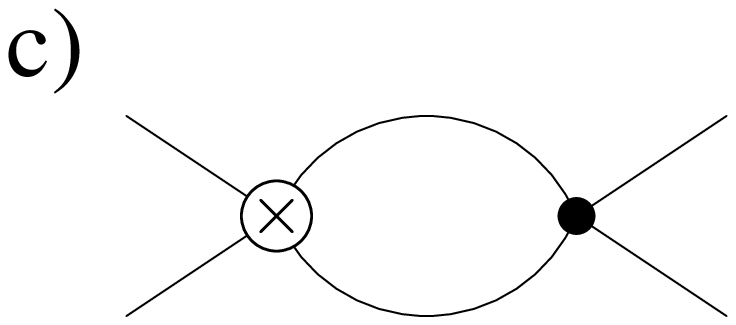}} 
 }
 \caption{\label{figCoulombusoft}   
 Graphs with an ultrasoft gluon, which contribute to the 
 three-loop running of the Coulomb potential.}
\end{figure}

At NLL order the effective theory diagrams
$\propto\frac{\alpha_s^3}{{\bmk}^2}$ have iterations of potentials as
shown in Fig.\ \ref{figtwoloopCoulombpotential}
%
%
and purely soft diagrams such as shown in Fig.\ \ref{figtwoloopCoulombsoft}.
%
%
The potential diagrams are UV-finite and reproduce the Coulomb
singularities in full QCD. The soft diagrams are identical to
the static QCD calculation of Peter and Schr\"oder\,\cite{Schroeder1}
up to the pinch-singularities caused by the $i\delta$ prescription
of the static quark propagators. 
In the static QCD calculation the
pinch-singularities are removed by the exponentiation of the static
potential\,\cite{exponentiation} (Sec.\ \ref{sectionpNRQCD}).
The effective theory computation does not have pinch-singularities
because there is no $i\delta$
prescription in the soft Feynman rules. After subtraction of
subdivergences the remaining divergence is canceled by a two-loop
counter term for $\frac{{\cal V}_c^{(T)}}{{\bmk}^2}$. Up to order
$\alpha_s^2$ the counterm reads
\begin{eqnarray} 
 Z_c = 1- \frac{\alpha_s(m\nu)\beta_0}{4\pi\epsilon}  + 
 \frac{\alpha_s^2(m\nu)}{(4\pi)^2} \bigg[ 
 \frac{\beta_0^2}{\epsilon^2} - \frac{\beta_1}{\epsilon} \bigg]
  \,,
\end{eqnarray}
where $\beta_1=102-\frac{38}{3}\,n_\ell$ is the two-loop $\beta$-function,
so that ${\cal V}_c^{(T)}=4\pi\alpha_s(m\nu)$ also at NLL order.
At NNLL order, diagrams with ultrasoft gluons have to be considered for
the first time. In Coulomb gauge we need to consider graphs with
${\bmp}.{\bmA}/m$ vertices as well as the coupling of ultrasoft gluons to
the Coulomb potential. After subtraction of subdivergences through
counterterms of the potentials $\propto 1/(m|{\bmk}|)$ and $1/m^2$, the
diagrams with ultrasoft gluons that have UV divergences not completely
canceled are shown in Fig.\ \ref{figCoulombusoft}a and b.
Figure~\ref{figCoulombusoft}c shows a counterterm diagram needed to
subtract a UV subdivergence associated with the $1/(m|{\bmk}|)$ potential.
%
%
Because the energy in the potential loop is of order $mv^2$, the
potential and the ultrasoft loops are not separable. 
The sum of the diagrams in the singlet channel reads
\begin{eqnarray}
\lefteqn{
\mbox{Fig.\ \ref{figCoulombusoft}} \, = \,
-\frac{4i}{3}\frac{C_FC_A^3}{8}\,
\frac{[{\cal V}_c^{(T)} \mu_S^{2\epsilon}]^3\,
      [\alpha_s(m\nu^2) \mu_U^{2\epsilon}]}{(4\pi)^3\,{\bmk}^2}\,
\bigg[\,\frac{1}{\epsilon} + \ldots\,\bigg]
}
\nonumber \\ &&= 
-\frac{4i}{3}\frac{C_FC_A^3}{8}\,
\frac{[\alpha_s(m\nu)]^3\,\alpha_s(m\nu^2)\,\mu_S^{2\epsilon}}
{{\bmk}^2}\,
\bigg[\,\frac{1}{\epsilon}+ 
\ln\Big(\frac{\mu_U^2}{E^2}\Big) +
 2\ln\Big(\frac{\mu_S^2}{\bf k^2}\Big)  + \ldots\,\bigg]
.\quad\quad
\label{vnrqcdcoulombultrasoft}
\end{eqnarray}
The total anomalous dimension in the singlet channel
from the ultrasoft diagrams with respect
to the velocity scaling parameter $\nu$ reads
\begin{eqnarray} 
  \gamma_U &=& -\,\frac{4 C_FC_A^3}{3}\,[\alpha_s(m\nu)]^3\,\alpha_s(m\nu^2)
\,.
\label{Coulombusoftanomdimvnrqcd}
\end{eqnarray}
After subtraction of subdivergences
the three-loop soft graphs contain a UV divergence associated with the
three-loop \ms $\beta$-function and a UV divergence that is induced by
the UV divergence in the ultrasoft graphs discussed above.
So the sum of all soft graphs gives the following contribution to the
anomalous dimension
\begin{eqnarray} 
  \gamma_S &=& C_F C_A^3\,[\alpha_s(m\nu)]^4 
+ 2C_F\beta_2\frac{[\alpha_s(m\nu)]^4}{(4\pi)^2}
\,,
\label{Coulombsoftanomdimvnrqcd}
\end{eqnarray}
where $\beta_2=\frac{2857}{2}-\frac{5033}{18}\, n_\ell +
\frac{325}{54}\, n_\ell^2\,.$
The result in Eq.\ (\ref{Coulombsoftanomdimvnrqcd}) has been determined
in Ref.~\citebk{Hoang4} without an explicit 
calculation. The first term on the RHS of 
Eq.~(\ref{Coulombsoftanomdimvnrqcd})
can be obtained directly from Eq.~(\ref{vnrqcdcoulombultrasoft}), 
because the overall dependence on the
scales and on $\alpha_s$ of
the corresponding divergence in the sum of the soft diagrams
is proportional to $[\alpha_s(m\nu)\mu_S^{2\epsilon}]^4/\epsilon$.
Taking into account the matching condition for the singlet coefficient
at the hard scale, ${\cal V}_c^{(s)}(1)=-4\pi C_F\alpha_s(m)$, the
full NNLL result reads\,\cite{Hoang4}
\begin{eqnarray} \label{Vcfin}
  {\cal V}_c^{(s)}(\nu) &=& -4\pi C_F \alpha_s(m\nu)
\nonumber\\[2mm] &+ &
 \frac{8\pi C_F C_A^3}{3\beta_0}\alpha_s^3(m)
 \bigg[ \frac{11}{4}- 2z- \frac{z^2}{2} - \frac{z^3}{4} + 4\ln(w)
\bigg] \,, 
\label{Coulombvnrqcdfinal}
\end{eqnarray}
where $z=\alpha_s(m\nu)/\alpha_s(m)$ and 
$w=\alpha_s(m\nu^2)/\alpha_s(m)$.
This result is needed for a NNLL order description of the
non-relativistic $\QQbar$ dynamics. At the same order the coefficients
of the $1/m^2$ potentials are needed at LL order, and the coefficients of
the $1/(m|{\bmk}|)$ potentials are needed at NLL. The corresponding
computations were carried out in Refs.~\citebk{Manohar2} and
\citebk{Manohar3}, 
respectively.

\begin{figure}[t!] 
\begin{center}
 \leavevmode
 \epsfxsize=2.3cm
 \leavevmode
 \epsffile[63 432 213 532]{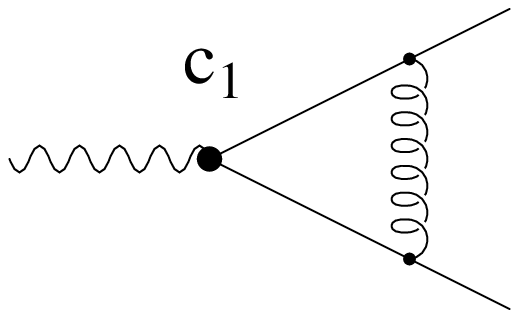}
 \hskip 1cm
 \epsfxsize=2.3cm
 \epsffile[63 432 213 532]{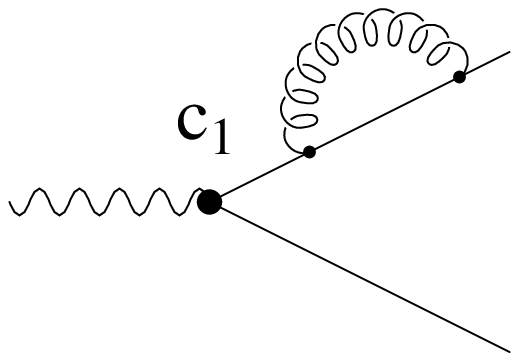}
 \hskip 1cm
 \epsfxsize=2.3cm
 \epsffile[63 432 213 532]{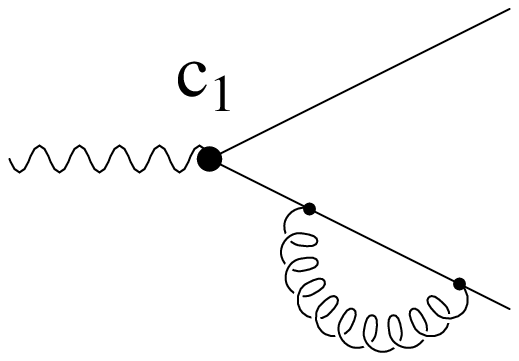} 
 \vskip  0.2cm
 \caption{   Graphs with an ultrasoft gluon with ${\bmp}.{\bmA}$ couplings 
 which contribute to the running of $c_2(\nu)$.
 \label{figusoftc2} }
\end{center}
\end{figure}

To describe the production of $\QQbar$ pairs at NNLL order in $e^+e^-$
annihilation one needs the LL evolution of the coefficients
$c_2$ and $c_3$ (see Eq.\ (\ref{matchingLLc23})) 
and the NNLL order evolution of $c_1$. 
The complete NNLL running of $c_1$ is not yet known. The NLL running was
determined in Refs.~\citebk{Manohar3,Hoang6} and arises from UV divergent one- and
two-loop graphs with insertions of the $1/(m|{\bmk}|)$ and $1/m^2$
potentials. At LL order $c_1(\nu)=1$, i.e.\ it does not run.
The LL evolution of the 
axial-vector coefficient $c_3$ would be determined by UV divergent
diagrams of order $\alpha_s v$, but such diagrams do not exist and
$c_3(\nu)=c_3(1)=1$. The LL running of $c_2$ was determined in
Ref.~\citebk{Hoang3} and arises from UV divergences in the ultrasoft 
graphs of order
$\alpha_s v^2$ shown in Fig.\ \ref{figusoftc2}.
%
%
The pull-up to the hard scale is provided by tadpole-like
soft one-loop diagrams that vanish in dimensional regularization.
The soft graphs arise from those in Fig.\
\ref{figusoftc2} by replacing the ultrasoft gluons by soft gluons and
contracting the internal quark line to a point. 
The resulting anomalous dimension reads
\begin{eqnarray}
 \nu{\partial\over\partial\nu} c_2(\nu) &=& \frac{8C_F}{3\pi} 
   \alpha_s(m\nu^2)\, c_1(\nu) \,,
\end{eqnarray}
with the solution
\begin{eqnarray}
 c_2(\nu)&=& -\frac{1}{6}-
 \frac{8 C_F}{3\beta_0} \ln\bigg[ \frac{\alpha_s(m\nu^2)}
 {\alpha_s(m)} \bigg] \,,
\label{c2ll}
\end{eqnarray}
where Eq.~(\ref{matchingLLc23}) is used as the initial condition.

\vspace{1cm}

\section{vNRQCD versus pNRQCD}
\label{sectionvNRQCDvspNRQCD}

The effective theories vNRQCD and pNRQCD {\it are not
equivalent}. From the phenomenological point of view pNRQCD is the by
fare more ambitious theory, because it was devised with the 
aim to treat systems with the hierarchy $m\gg mv\gg m v^2\gg\lqcd$ as
well as systems, where $\lqcd$ is larger. In addition, it is supposed
to treat, at the same time, systems with dynamical quark pairs and
systems where the quarks are static.\cite{Brambilla1}
It is fair to say that some
conceptual and technical aspects are not yet fully worked out, as the
authors admit at some occasions, but it is argued that this
does not affect the concepts of pNRQCD.\cite{Pineda5}
On the other hand, vNRQCD has been designed only for the case $m\gg
mv\gg m v^2\gg\lqcd$, with the primary aim to provide a theory that
is fully worked out technically and conceptually for this
special case. The treatment of systems where $\lqcd$ is larger is by
construction difficult in vNRQCD.
The theories pNRQCD and vNRQCD are not equivalent, because
they lead to different predictions in the case $m\gg
mv\gg m v^2\gg\lqcd$. 

It is very instructive to have a closer look at the
mechanisms working in vNRQCD and pNRQCD by means of two examples. One,
the order $m_e\alpha^5\ln\alpha$ Lamb shift in hydrogen in
QED, where both effective theories agree, and the other, the Coulomb
potential in the QCD 
Schr\"odinger equation for a color singlet $\QQbar$ pair, where both
theories disagree. For simplicity, I will only concentrate on the
mechanism of renormalization and the structure of logarithmic terms,
which is sufficient to pinpoint the conceptual and technical
differences.
We note that both theories also disagree on their results for the
$1/m$ potentials and the $\QQbar$ production currents. Those cases
will not be discussed here, but the origin of the disagreement is the same
as for the Schr\"odinger Coulomb potential. 

\subsection{Hydrogen Lamb Shift}
\label{subsectionlambshift}

The Lamb shift is the $2S_{1/2}$--$2P_{1/2}$ level
splitting in the hydrogen spectrum. For infinite proton mass the order
$m_e\alpha^5\ln\alpha$ term reads 
\begin{eqnarray}
\Delta E & = &
-\frac{m_e\,\alpha^5}{3\,\pi}\,\ln\alpha
\,,
\label{lambshift}
\end{eqnarray}
where $\alpha=1/137$ is the fine structure constant and $m_e$ the
electron mass. The result in Eq.~(\ref{lambshift}) can be derived in
vNRQED from the LL running of the ${\cal V}_2$  
potential\,\cite{Manohar5} (Eq.~(\ref{vNRQCDpotential})) 
and in the framework of pNRQED from the  LL  
running of the $D_{d,s}^{(2)}$ potential 
(Eq.~(\ref{pnrqcdpotentialslist})) 
(see Ref.~\citebk{PinedaLamb} for a fixed order calculation).

\begin{figure}[t] 
\begin{center}
\leavevmode
\epsfxsize=2.5cm
\leavevmode
\epsffile[0 200 630 570]{9923_fd8.eps}
\leavevmode
\epsfxsize=2.5cm
\leavevmode
\epsffile[0 200 630 570]{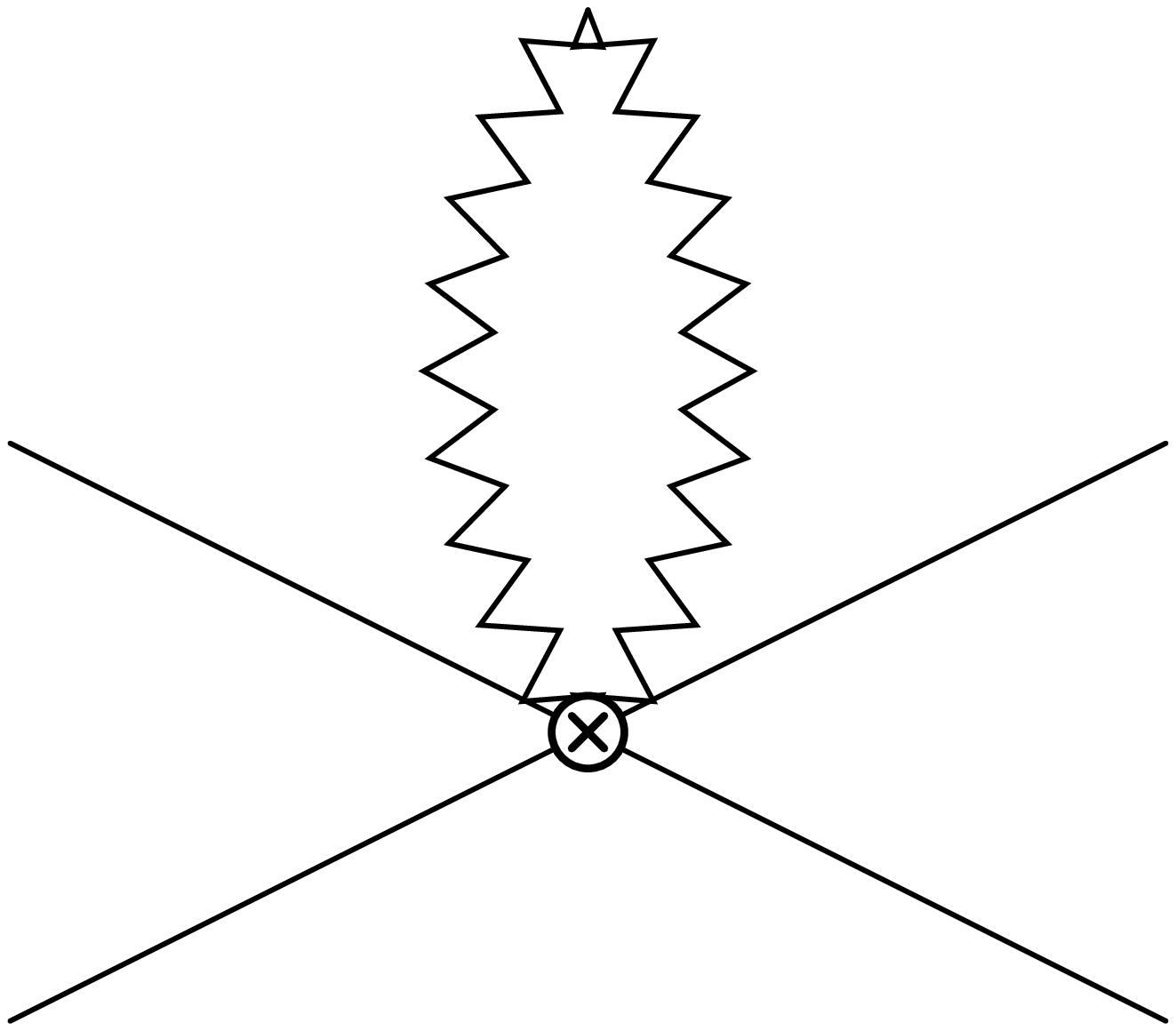}
\leavevmode
\epsfxsize=2.5cm
\leavevmode
\epsffile[0 200 630 570]{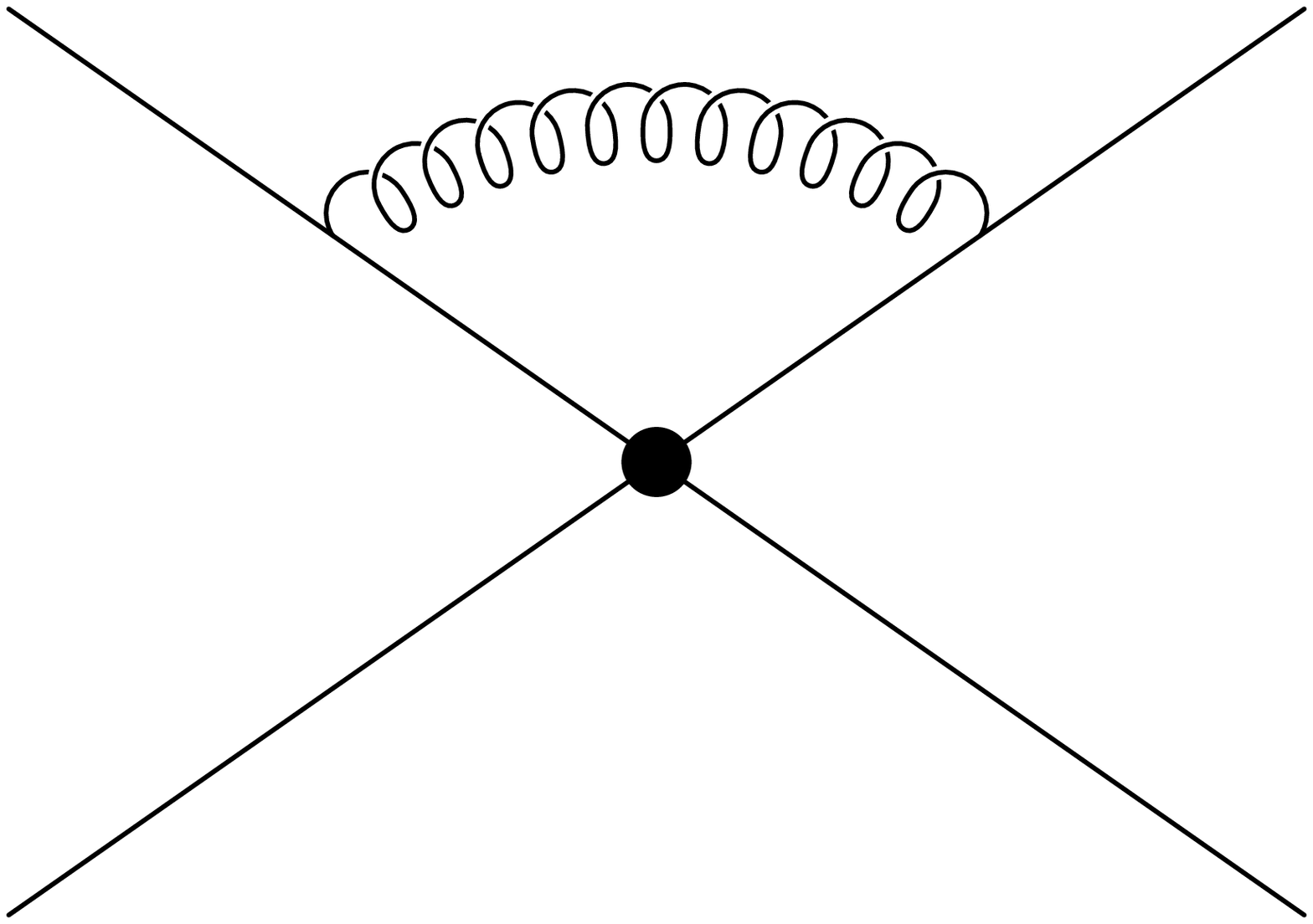}
\leavevmode
\epsfxsize=2.5cm
\leavevmode
\epsffile[0 200 630 570]{9923_fd3.eps}
\begin{picture}(0,0)(1,1)
\put(-303,35){$a)$}
\put(-220,35){$b)$}
\put(-152,35){$c)$}
\put(-77,35){$d)$}
\end{picture}
 \end{center}
%
%
\vskip  0cm
 \caption{\label{figvnrqedV2anomdim}   
One-loop vNRQCD diagrams contributing to the LL anomalous dimension of
${\cal V}_2$.
}
\end{figure}
In vNRQED the LL anomalous dimension of ${\cal V}_2$ comes from soft
scattering loop diagrams, such as in Fig.\ \ref{figvnrqedV2anomdim}a
and b and from
ultrasoft graphs, such as in Fig.\ \ref{figvnrqedV2anomdim}c and d. In
QED the leading 
soft two-fermion vertex is of order $1/m$ and there is no
contribution from the soft diagram in Fig.\ \ref{figvnrqedV2anomdim}a, 
if the proton
mass is infinite. The soft diagram in 
Fig.\ \ref{figvnrqedV2anomdim}b gives a massless
tapole and vanishes in dimension regularization. Distinguishing
between soft UV and IR divergences the result reads
\begin{eqnarray}
\mbox{Fig.\ \ref{figvnrqedV2anomdim}b} & = &
-i\,\frac{2}{3m_e^2}\,
[\alpha\,\mu_S^{2\epsilon}]^2\,
\Big[\,\frac{1}{\epsilon_{\rm UV}}-\frac{1}{\epsilon_{\rm IR}}\,
\Big]
\,.
\label{lambvnrqedsoft}
\end{eqnarray} 
In Coulomb gauge the ultrasoft contributions are generated by
ultrasoft photons with ${\bmp}.{\bmA}/m$ couplings
dressing the Coulomb potential. Typical diagrams are displayed in
Figs.\ \ref{figvnrqedV2anomdim}c and d. 
For infinite proton mass only the diagram in Fig.\
\ref{figvnrqedV2anomdim}c contributes. For on-shell external electrons
($E=\frac{\bmp^2}{2m}$) and including wave function renormalization
constants the result reads
\begin{eqnarray}
\mbox{Fig.\ \ref{figvnrqedV2anomdim}c} & = &
-i\,\frac{2}{3m_e^2}\,
[\alpha\,\mu_U^{2\epsilon}]\,[\alpha\,\mu_S^{2\epsilon}]\,
\Big[\,\frac{1}{\epsilon_{\rm UV}}-\frac{1}{\epsilon_{\rm IR}}\,
\Big]
\,.
\label{lambvnrqedusoft}
\end{eqnarray}
The ultrasoft IR divergence in Eq.\ (\ref{lambvnrqedusoft}) is present
also in full QED and not relevant for the renormalization
procedure. This can be seen from the fact that the IR divergence is
regularized for an off-shell electron with $E\neq\frac{\bmp^2}{2m}$.
The ultrasoft UV divergence is canceled by a counter term in the
unrenormalized ${\cal V}_2$ potential. The soft IR divergence is
unphysical and does not exist in full QED. It is associated with the
soft scale $m_e v$ and matches with the ultrasoft UV divergence,
because in vNRQED the photon from full QED has been split into soft photon
fields and an ultrasoft photon field. The full soft tadpole
contribution pulls the ultrasoft running up to the hard scale. This
feature exists for any ultrasoft running in vNRQCD.
The final result for the LL anomalous dimension of ${\cal V}_2$ is
determined from the ultrasoft UV divergence in Eq.\
(\ref{lambvnrqedusoft}) and reads~\footnote{
In Ref.~\citebkcap{Manohar5} the coefficient ${\cal V}_2$ is
actually called $U_2$. The coefficient is defined
with an additional minus sign with respect to the convention for
${\cal V}_2^{(T)}$ in vNRQCD of Eq.~(\ref{vNRQCDpotential}).
}
\begin{eqnarray}
\nu\frac{\partial}{\partial\nu}{\cal V}_2(\nu) & = &
\frac{8}{3}\,\alpha^2
\,.
\label{V2anomdimvnrqed}
\end{eqnarray}
In practical calculations the UV and IR divergences in soft loops do
not need to be distinguished, because the pull-up is an automatic
mechanism. This means that in the case of the Lamb shift the soft
diagram would not have to be calculated at all. However, the IR divergences in
ultrasoft loops have to be identified, because they are IR divergences
of full QED and do not take part in the renormalization of vNRQED. The
anomalous dimension in Eq.\ (\ref{V2anomdimvnrqed}) can be trivially
integrated giving
\begin{eqnarray}
{\cal V}_2(\nu) & = &
-\frac{\pi\,\alpha}{2} + \frac{8}{3}\,\alpha^2\,\ln\nu
\,,
\label{V2anomdimvnrqedsolved}
\end{eqnarray}
where ${\cal V}_2(1)=-\frac{\pi\alpha}{2}$ is obtained from tree level
matching. Interestingly, Eqs.~(\ref{V2anomdimvnrqed}) and
(\ref{V2anomdimvnrqedsolved}) show that there is no infinite series of
logarithmic terms $\propto\alpha(\alpha \ln\nu)^n$. The series
terminates because $\alpha$ does
not run and because there is no non-trivial mixing from other running
operators.\cite{Manohar5} 
Since the ${\cal V}_2$ potential is the delta function in 
configuration space the contribution to the $2S_{1/2}$--$2P_{1/2}$ level
splitting can be readily calculated,
\begin{eqnarray}
\label{Lambshiftformula}
\left\langle\, 2S\,\left|-\frac{{\cal V}_2(\nu=\alpha)}{m_e^2}
\,\right|\,2S\,\right\rangle
& = & -\,
\left|\Psi_{n=2}(0)\right|^2\,\frac{{\cal V}_2(\alpha)}{m_e^2}
\,,
\end{eqnarray}
where $|\Psi_{n}(0)|^2=m_e^3\alpha^3/(\pi n^3)$, 
which gives the known order $m_e\alpha^5\ln\alpha$ term in Eq.\
(\ref{lambshift}). A complete calculation of
the order $m_e\alpha^5\ln\alpha$ Lamb shift for arbitrary masses in
vNRQED can be found in Ref.~\citebk{Manohar5}. 
(The same work also contains the 
determination of the order $m_e\alpha^8\ln^3\alpha$ Lamb shift, the
order  $m_e\alpha^7\ln^2\alpha$ hyperfine splitting for arbitrary
masses, and the order $\alpha^2\ln\alpha$ and $\alpha^3\ln^2\alpha$
corrections to ortho- and para-positronium decay in vNRQED
using the renormalization group equations.) 

\begin{figure}[ht] 
\begin{center}
\leavevmode
\epsfxsize=4.4cm
\leavevmode
\epsffile[150 530 480 680]{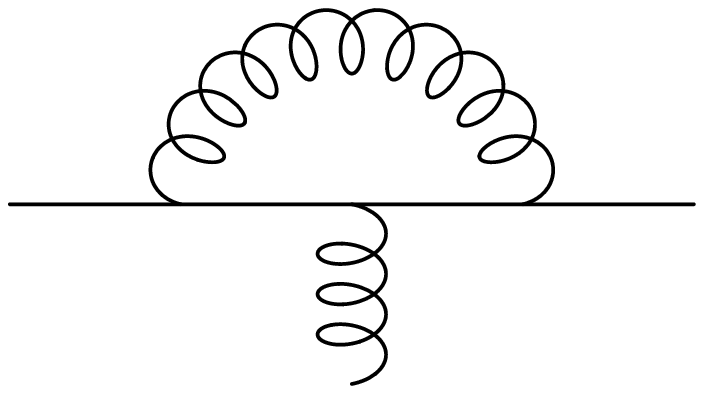}
\leavevmode
\epsfxsize=4.4cm
\leavevmode
\epsffile[150 530 480 680]{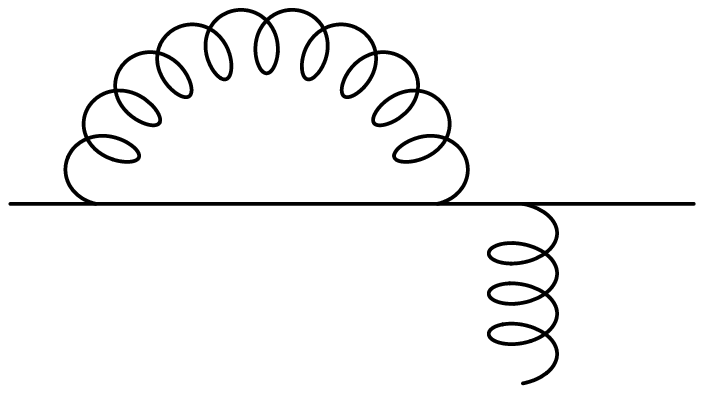}
 \end{center}
%
%
\vskip  0cm
 \caption{\label{figcdnrqcd}   
One-loop HQET vertex and wave function diagrams that
contribute to the anomalous dimension of $c_D$. The photon in the loop 
couples with ${\bmp}.{\bmA}/m_e$ and the external photon with $A^0$.
}
\end{figure}

\begin{figure}[t] 
\begin{center}
\begin{picture}(100,30)(1,1)
\put(-35,-170) { \epsfxsize=6.cm 
               \hbox{\epsfbox{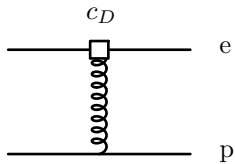}} }
\put(47.5,25){$c_D$}
\put(98,12){\mbox{e}}
\put(98,-27){\mbox{p}}
\end{picture}
 \end{center}
\vskip  0.8cm
 \caption{\label{figpnrqcdmatching}   
NRQCD diagram contributing to the Born matching condition
of the $D_{d,s}^{(2)}$ potential in pNRQCD.
}
\end{figure}

In pNRQED the determination of $D_{d,s}^{(2)}$ is divided into two
steps. In NRQED (supplemented by the $1/m$ expansion) the running of
the Darwin coefficient $c_D$ is carried out from $\mu=m$ to
$\mu=\mu_s$. The anomalous dimension of $c_D$ is obtained from
one-loop vertex corrections and wave function diagrams such as
displayed in Fig.\ \ref{figcdnrqcd}
and has been
carried out before in the framework of HQET.\cite{Bauer1}
In these diagrams the photon in the loop couples with ${\bmp}.{\bmA}/m_e$
to the electron. For on-shell electrons with momenta $\bmp$ and
$\bmp^\prime$ the sum of diagrams
relevant for the renormalization of $c_D$ reads 
\begin{eqnarray}
\mbox{Figs.\ \ref{figcdnrqcd}} & = &
-\frac{i}{6\pi}\,\frac{(\bmp-\bmp^\prime)^2}{m_e^2}\,
[g\,\mu^\epsilon]\,[\alpha\,\mu^{2\epsilon}]\,
\Big[\,\frac{1}{\epsilon_{\rm UV}}-\frac{1}{\epsilon_{\rm IR}}\,
\Big]
\,,
\label{lambpnrqedsoft}
\end{eqnarray}
when IR and UV divergences are distinguished.
Interestingly, the result seems to be equivalent to the soft vNRQCD
amplitude in Eq.\ (\ref{lambvnrqedusoft}), when the photon exchange to
the proton is included. However, here the IR divergence is the IR
divergence from full QED, because NRQED is supposed to describe the
whole momentum space below the scale $m_e$. At this point this
observation seems to be only a curious fact, but it is at the
heart of the conceptual differences between vNRQCD and pNRQCD. 
The UV divergence in Eq.~(\ref{lambpnrqedsoft}) is canceled by a
counter term in the 
unrenormalized $c_D$ interaction and the anomalous dimension of $c_D$
reads
\begin{eqnarray}
\mu \frac{\partial}{\partial \mu}c_D(\mu) & = &
-\frac{8\,\alpha}{3\,\pi}
\label{Darwinanomdimnrqed}
\end{eqnarray}
with the solution
\begin{eqnarray}
c_D(\mu) & = &
1 - \frac{8\,\alpha}{3\,\pi}\,\ln\Big(\frac{\mu}{m_e}\Big)
\,,
\end{eqnarray}
where $c_D(m_e)=1$ is obtained from tree level matching.
The second step is carried out in pNRQED. The matching condition for
the $D_{d,s}^{(2)}$ potential at $\mu=\mu_s$ is obtained
by calculating the NRQED Born scattering diagram in 
Fig.\ \ref{figpnrqcdmatching}
for off-shell electron momenta $(0,\bmp)$ and $(0,\bmp^\prime)$ for
the incoming 
and outgoing electron, respectively. The little square indicates the 
$c_D$-coupling. This gives
\begin{eqnarray}
D_{d,s}^{(2)}(\mu=\mu_s) & = &
\frac{\alpha}{2}\,c_D(\mu_s)
\, = \,
\frac{\alpha}{2} - 
\frac{4\,\alpha^2}{3\,\pi}\ln\Big(\frac{\mu_s}{m_e}\Big)
\,.
\label{Dsd2pnrqedmatching}
\end{eqnarray}
The running of $D_{d,s}^{(2)}$ in pNRQCD is determined as described in
Sec.\ \ref{sectionpNRQCD} from the ultrasoft one-loop diagrams in 
Fig.\ \ref{figusoft1um2pnrqcd}b
in the $1/m$ expansion. For electrons with external momenta $(0,\bmp)$
and $(0,\bmp^\prime)$ the result reads
\begin{eqnarray}
\mbox{Fig.\ \ref{figusoft1um2pnrqcd}b} & = &
-i\,\frac{2\,\alpha}{3m_e^2}\,
[\alpha\,\mu^{2\epsilon}]\,
\Big[\,\frac{1}{\epsilon_{\rm UV}}-\frac{1}{\epsilon_{\rm IR}}\,
\Big]
\,.
\label{lambpnrqedusoft}
\end{eqnarray}
The IR divergence is again the full QED IR divergence. The UV
divergence is canceled by a counterterm in the unrenormalized
$D_{d,s}^{(2)}$ potential and leads to the following LL anomalous
dimension for $D_{d,s}^{(2)}$,
\begin{eqnarray}
\frac{d}{d\ln\mu} D_{d,s}^{(2)}(\mu)
& = &
-\frac{4\,\alpha^2}{3\,\pi}
\,.
\end{eqnarray}
This is again integrated trivially giving
\begin{eqnarray}
D_{d,s}^{(2)}(\mu_u) & = & 
D_{d,s}^{(2)}(\mu_s) 
-\frac{4\,\alpha^2}{3\,\pi}\,\ln\Big(\frac{\mu_u}{\mu_s}\Big)
\, = \,
\frac{\alpha}{2} - 
\frac{4\,\alpha^2}{3\,\pi}\ln\Big(\frac{\mu_u}{m_e}\Big)
\nonumber\\
& = &
\frac{\alpha}{2}\,c_D(\mu_u)
\end{eqnarray}
at the scale $\mu=\mu_u$. For $\mu_u=m_e\alpha^2$ this agrees 
with ${\cal V}_2$ up to a convention-dependent factor and also 
leads to Eq.\ (\ref{lambshift}) for the order $m_e\alpha^5\ln\alpha$
Lamb shift. Comparing the computations we see that the difference 
between vNRQED and pNRQED appears to be simply that the running is
distributed in a different way. In vNRQED running happens in one step and
in pNRQED it happens in two steps.
Agreement is in fact found for the LL running of all $1/m^2$
potentials in vNRQCD (Refs.~\citebk{Manohar2,Hoang6}) and
pNRQCD (Ref.~\citebk{Pineda3}). 

One of the main 
differences between QED and QCD is that the coupling constant runs
itself. If we consider the Lamb shift of muonic hydrogen and treat
the electron as massless we can ``simulate'', in a simplified way, 
one difference of the calculation presented above to the
QCD case. 
If massless electrons are included, the fine structure constant 
becomes scale-dependent using let's say the \ms renormalization
scheme. 
This makes the couplings in all potentials scale-dependent and 
makes the $2S_{1/2}$--$2P_{1/2}$ level splitting more complicated.
As shown in Sec.\ \ref{subsectionschroedingercoulomb}, 
the $\alpha$ in the Coulomb-potential has to be set at the scale 
$\mu_s$ or $m\nu$ and, as the consequence, also the $\alpha$ in the 
wavefunction at the origin in Eq.~(\ref{Lambshiftformula}). In addition,  
also the $\alpha$ in the Spin-Orbit potential (the $D^{(2)}_{LS,s}$ term 
in Eq. (\ref{pnrqcdpotentialslist}) and the ${\cal V}_\Lambda$ terms in 
Eq. (\ref{vNRQCDLagrangian})), has to be set at the scale $\mu_s$ or 
$m\nu$ and now contributes logarithmic terms to the 
$2S_{1/2}$--$2P_{1/2}$ level splitting. 
For the purpose of this discussion we only consider the coefficients 
$D_{d,s}^{(2)}$ and ${\cal V}_2$ in some detail, because they contain a
non-trivial ultrasoft running involving the scales $\mu_u$ or $m\nu^2$.
In the pNRQCD computation for $c_D$ the coupling in the
anomalous dimension in Eq.\
(\ref{Darwinanomdimnrqed}) is $\alpha(\mu)$ and the result for
$c_D$ is
\begin{eqnarray}
c_D(\mu) & = & 1+\frac{16}{3\,\bar\beta_0}\,
\ln\bigg(\frac{\alpha(\mu)}{\alpha(m_\mu)}\bigg)
\,,
\end{eqnarray}
where $\bar\beta_0 = -\frac{4}{3}n_\ell$, $n_\ell=1$.
The $\alpha$ in the matching condition of Eq.\
(\ref{Dsd2pnrqedmatching}) is evaluated 
at $\mu=\mu_s$ and the LL anomalous dimension for $D_{d,s}^{(2)}$
reads
\begin{eqnarray}
\frac{d}{d\ln\mu} D_{d,s}^{(2)}(\mu) & = &
-\frac{4\,\alpha(\mu_s)}{3\,\pi}\,\alpha(\mu)
\,.
\end{eqnarray}
The coefficient of the LL static potential
is set at the scale $\mu_s$ and does not run below $\mu_s$.
The solution just reads
$$D_{d,s}^{(2)}(\mu_u) = \frac{\alpha(\mu_s)}{2}c_D(\mu_u)\,.$$
In vNRQED with massless electrons the
anomalous dimension for ${\cal V}_2$ also gets a contribution from
loops in Fig.\ \ref{figvnrqedV2anomdim}a and b where the soft photons
are replaced by soft (massless) electrons. 
The modified version of Eq.\ (\ref{V2anomdimvnrqed}) then
reads
\begin{eqnarray}
\frac{d}{d\ln\nu}{\cal V}_2(\nu) & = &
\frac{8}{3}\,\alpha(m_\mu\nu)\,\alpha(m_\mu\nu^2)
+ \frac{\bar\beta_0}{4}\,[\alpha(m_\mu\nu)]^2\,c_{4\mu 2e}(\nu)
\,,
\nonumber\\[2mm]
c_{4\mu 2e}(\nu) & = & c_D(m_\mu \nu^2)  
\,,
\label{V2anomdimvnrqedrunning}
\end{eqnarray}
where $c_{4\mu 2e}$ is
the coefficient of the 6-fermion operator contained in Fig.\
\ref{figvnrqedV2anomdim}b. (This type of operators, involving
four heavy quarks and two soft gluons or massless quarks,
has been missed in Ref.~\citebk{Manohar2}. Their numerical 
contribution is small~\cite{Hoang6} and does not affect 
conclusions drawn based on the results of Ref.~\citebk{Manohar2}.)

It turns out that this coefficient is equal
to the $c_D$ obtained 
in NRQED at the scale $\mu=m_\mu \nu^2$.
Integrating Eq.\ (\ref{V2anomdimvnrqedrunning}) with the initial
condition ${\cal V}_2(1)=-\pi\alpha(m)/2$ gives
\begin{eqnarray}
{\cal V}_2(\nu) & = &
-\frac{\pi\,\alpha(m_\mu \nu)}{2}\,c_D(m_\mu \nu^2)
\,,
\end{eqnarray}
which agrees again with the pNRQED result up to a convention-dependent
factor. 
The vNRQCD and pNRQCD computations agree, because only one-loop
diagrams contribute to the LL anomalous dimensions of the $1/m^2$
potentials. So at this level it seems to be merely a
choice of convention whether one uses vNRQCD or pNRQCD, because the
running is just distributed in a different way.
The actual reason for agreement, however, it that at LL order 
the (one-loop) diagrams cannot contain non-separable
ultrasoft and potential loops at the same time, which connects in
a subtle way the evolution with respect to soft and ultrasoft scales.

\subsection{Schr\"odinger Coulomb Potential at NNLL Order}
\label{subsectionschroedingercoulomb}

The subtle difference in the renormalization of vNRQCD and pNRQCD
becomes apparent beyond the LL approximation, when non-trivial multi-loop
diagrams with UV divergences contain loops dominated by the ultrasoft
and the potential region. Examples are the Coulomb potential $\propto
1/{\bmk}^2$ in the
NNLL order Schr\"odinger equation for a color singlet $\QQbar$ pair
(see Ref.~\citebk{Hoang4} for vNRQCD and Ref.~\citebk{Pineda4} for
pNRQCD) or the 
$1/(m|{\bmk}|)$
potential at NLL order (see Ref.~\citebk{Manohar3} for vNRQCD and
Refs.~\citebk{Pineda3,Brambilla3} 
for pNRQCD). Because the $1/(m|{\bmk}|)$ potentials contribute to the NLL
running of the $\QQbar$ production current, vNRQCD and pNRQCD
predictions also differ for the latter case (see
Refs.~\citebk{Manohar3,Hoang6} for vNRQCD 
and Ref.~\citebk{Pineda5} for pNRQCD).
Let us discuss the NNLL order Coulomb potential that appears in the
Schr\"odinger equation. Its determination in vNRQCD and pNRQCD has
already been described in previous sections and will not be repeated
in detail.

In vNRQCD the NNLL $1/(\bmp-\bmp^\prime)^2$ potential that appears in
the Schr\"odin\-ger equation, called $V_c^{v}$ in the following,
consists of the sum of the Coulomb 
potential ${\cal V}_c{(s)}/(\bmp-\bmp^\prime)^2$ at NNLL order and the
time-ordered products of soft vertices proportional to
$1/(\bmp-\bmp^\prime)^2$ at one (Figs.\ \ref{figvnrqcdloops}c and 
Eq.~(\ref{eftvc1})) 
and two loops (see Figs.\ \ref{figtwoloopCoulombsoft}) that
determine the running of ${\cal V}_c$ at LL and NLL order.
Im momentum space for $D=4$ the sum is ($\bmk=(\bmp-\bmp^\prime)$)
\begin{eqnarray}
\lefteqn{
\tilde V_c^{v}({\bmp},{\bmq}) 
  = 
 \frac{{\cal{V}}_c^{(s)}(\nu)}{{\bmk}^2}\,
 -\,\frac{4\pi C_F\, \alpha_s(m\nu)}{\bmk^2}\bigg\{
 \frac{\alpha_s(m\nu)}{4\pi}\,\bigg[
 -\beta_0\,\ln\Big(\frac{\bmk^2}{m^2\nu^2}\Big) + a_1
 \bigg]
} \nonumber\\[2mm] 
& & + \bigg(\frac{\alpha_s(\mu_s)}{4\pi}\bigg)^2 \bigg[
 \beta_0^2\,\ln^2\Big(\frac{\bmk^2}{m^2\nu^2}\Big)  
 - \Big(2\,\beta_0\,a_1 +
 \beta_1\Big)\,\ln\Big(\frac{\bmk^2}{m^2\nu^2}\Big) + a_2 \bigg]
 \,\bigg\} \qquad
 \nonumber\\[4mm]
& = &
\Big[\mbox{Eq.\ (\ref{staticpotentialschroeder})} \Big] + 
 \frac{8\pi C_F C_A^3}{3\beta_0\,{\bmk}^2} \,\alpha_s^3(m)\, 
 \bigg[ \frac{11}{4}- 2z- \frac{z^2}{2} - \frac{z^3}{4} + 4\ln(w) \bigg] \,, 
 \label{VCoulombvnrqcd}
\end{eqnarray}
where $z=\alpha_s(m\nu)/\alpha_s(m)$ and 
$w=\alpha_s(m\nu^2)/\alpha_s(m\nu)$. 
In principle it is inappropriate to consider the soft time-ordered
products together with the ${\cal V}_c^{(s)}$ term as a single
quantity for $D=4$ because they are matrix elements depending
non-trivially on $D=4-2\epsilon$ dimensions. Therefore, they should
not be considered for $\epsilon\to 0$ independently of the other matrix
elements relevant for a given process. However, as long as insertions 
of the soft time-ordered products do not generate UV divergences
relevant for the 
process, Eq.~(\ref{VCoulombvnrqcd}) can be used. This is also the case
for the NNLL order description of $\QQbar$ pairs, when the quarks are
stable.  

In the pNRQCD framework the NNLL $1/(\bmp-\bmp^\prime)^2$ potential
that appears in the Schr\"odinger equation, 
called $V_c^{p}$ in the following, is obtained directly from
the two-step matching and running through NRQCD and pNRQCD. The NNLL
result reads 
\begin{eqnarray}
\tilde V_c^{p}(\bmp,\bmq) & = &
-\frac{4 \pi C_F \alpha_{V_s}(\mu_u,\mu_s)}{\bmk^2}
\nonumber\\[2mm]
 & = &
\Big[\mbox{Eq.\ (\ref{staticpotentialschroeder})} \Big]
-  \frac{2 \pi C_F C_A^3}
 {3\beta_0\,{\bmk}^2}\alpha_s^3(\mu_s) 
\ln\bigg(\frac{\alpha_s(\mu_s)}{\alpha_s(\mu_u)}\bigg) 
\,.
\label{VCoulombpnrqcd}
\end{eqnarray}
The first term is the two-loop pNRQCD matching condition and the
second term arises from the pNRQCD running from $\mu_s$ to
$\mu_u$. All computations are carried out in the static limit
abandoning the non-relativistic power counting for $\QQbar$ pairs
because otherwise the NRQCD computations cannot be properly defined.
It is one of the most important assumptions in the pNRQCD framework
that the static calculation of the running coefficients can be applied
also for moving dynamical quark pairs. 

Comparing Eqs.\ (\ref{VCoulombvnrqcd}) and (\ref{VCoulombpnrqcd}) one
finds disagreement in the contributions associated with the ultrasoft
gluons. Expanding the second term in Eq.\ (\ref{VCoulombvnrqcd})
(times a factor ${\bmk}^2$) in
$\alpha_s(m)$, we find
\begin{eqnarray} \label{expVc}
  -\frac{1}{3}\, {C_F C_A^3}\,  \alpha_s^4(m) \ln(\nu) 
  +\frac{2\beta_0}{3\pi}\, {C_F C_A^3}\, \alpha_s^5(m) \ln^2(\nu) 
  +  \ldots \,.
\end{eqnarray} 
The analogous result for the static calculation in Eq.\
(\ref{VCoulombpnrqcd}) for $\mu_s=m\nu$ and  $\mu_u=m\nu^2$ reads
\begin{eqnarray} \label{expVstat}
 -\frac{1}{3}\, {C_F C_A^3}\,  \alpha_s^4(m) \ln(\nu) 
  +\frac{3\beta_0}{4\pi}\, {C_F C_A^3}\, \alpha_s^5(m) \ln^2(\nu) 
  +  \ldots \,.
\end{eqnarray}
The single $\ln\nu$ terms agree, but the higher order logarithms
differ.
From the phenomenological point of view the difference is quite small
for most applications, but both theories claim to sum the correct
higher order logarithms. 

In vNRQCD the result arises from the UV divergences in the three-loop
ultrasoft-potential diagrams in Fig.~\ref{figCoulombusoft} and the
associated UV 
divergences in three-loop soft diagrams. The individual quark
propagators $i/(k_0-{\bmk}^2/m+i\delta)$ in the ultrasoft-potential
diagrams 
describe dynamical quarks. The resulting anomalous dimension in Eq.\
(\ref{Coulombusoftanomdimvnrqcd}) describes running from $\nu=1$ to
$\nu=v$ and includes 
the scale-dependence from all three loops, i.e.\ the factors
$\mu_S^{4\epsilon}\,\mu_U^{2\epsilon}$ for the ultrasoft-potential
diagrams and $\mu_S^{6\epsilon}$ for the soft 
as well as the scale-dependence
of all the coupling constants that occur in the diagrams.
In pNRQCD the result arises from the UV divergence of the one-loop
diagram Fig.\ \ref{figpnrqcdselfenergy} with one ultrasoft gluon and a
static $\QQbar$ propagator 
$i/(i\partial_0+(\frac{1}{2}C_A-C_F)\alpha_{V_o}/r)$. This propagator
sums an infinite number of static potential rungs in $D=4$ dimensions.
The resulting anomalous dimension in Eq.\ (\ref{coulombusoftrge})
describes running from $\mu=\mu_s\sim mv$ to $\mu=\mu_u\sim mv^2$ and
includes only the scale dependence of the ultrasoft loop, i.e.\ the
factor $\mu^{2\epsilon}$ and the ultrasoft coupling $\alpha_s(\mu)$.
The quarks in the loops with the Coulomb rungs are not dynamical.
Clearly, if the calculation would involve only true one-loop diagrams,
as was the case for the Lamb shift discussed before, we
would find again agreement, because loops involving multiple iterations of
potentials wouldn't exist.

At the conceptual level the difference can be understood as the
difference between a calculation for static and dynamical quarks. In
the dynamical vNRQCD calculation the correlation between soft and
ultrasoft scales through the quark equation of motion is maintained at
all times. In the static pNRQCD calculation the scale correlation is
broken at intermediate steps through the $1/m$ expansion, i.e.\ energy
and three-momentum are treated independently. The discussion above
shows that dynamical and static $\QQbar$ systems are different.
In particular, it can be problematic to use results obtained
in static quark systems for problems involving dynamical quarks.

At this point it is, once more, instructive to see with a very simple
example\,\cite{Hoang4} how the $1/m$ expansion breaks the non-relativistic
power counting. In the static calculation the time-ordered product of
a $1/(m|\bmk|)$ and a $1/\bmk^2$ potential is
\begin{eqnarray}
\int \frac{d^4k}{(2\pi)^4}\,\frac{1}{m|\bmk|}
\,\frac{1}{\left(k_0+i\delta\right)}
\,\frac{1}{\left(k_0-i\delta\right)}
\,\frac{1}{\bmk^2} & \,\sim\, &
\frac{1}{m}
\end{eqnarray}
and hence suppressed by $1/m$. 
\begin{figure}[t] 
\begin{center}
\leavevmode
\epsfxsize=2.5cm
\leavevmode
\epsffile[70 430 200 520]{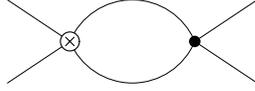}
 \end{center}
\vskip  0cm
 \caption{\label{figpotentialTproduct}   
Loop graph with an insertion of the $1/(m|{\bf k})$ and 
 $1/{\bf k^2}$ potentials.
}
\end{figure}
In the dynamical calculation (see Fig.\
\ref{figpotentialTproduct}) the time-ordered product of
a $1/(m|\bmk|)$ and a $1/\bmk^2$ potential is
\begin{eqnarray}
\lefteqn{
\int \frac{d^4k}{(2\pi)^4}\,\frac{1}{m|\bmk|}
\,\frac{1}{\left(k_0-\frac{\bmk^2}{2m}+i\delta\right)}
\,\frac{1}{\left(k_0+\frac{\bmk^2}{2m}-i\delta\right)}
\,\frac{1}{\bmk^2}
}
\nonumber\\  & \sim &
\int \frac{d^3k}{(2\pi)^3}\,\frac{1}{m|\bmk|}
\,\frac{m}{\left(\bmk^2-i\delta\right)}
\,\frac{1}{\bmk^2}
\,\sim\, 1
\end{eqnarray}
and of order $1$.

\vspace{1cm}

\section{Heavy Quark Mass and Renormalons}
\label{sectionquarkmass}

In the discussions of the previous sections on non-relativistic
effective theories we have assumed the pole mass definition $m$ for
the heavy quarks. From the effective Lagrangians one can derive the
one-particle-irreducible two-point function for a quark with energy
$E=\sqrt{s}-2m^{}_{\rm pole}$, $\sqrt{s}$ being the total
center-of-mass energy, and three-momentum ${\bmp}$,
\begin{equation}
E\,-\,\frac{{\bmp}^2}{2m^{}_{\rm pole}} \,+\,
\frac{{\bmp}^4}{8m^{3}_{\rm pole}}\,+\,\ldots
\,, 
\label{OPI2point1}
\end{equation}
which vanishes for an on-shell quark in the center-of-mass frame
($E={\bmp}=0$). The on-shell condition is useful for the determination
of matching conditions and anomalous dimensions in the effective
theories. So at this stage the use of the pole mass definition is
appropriate and useful.
For a general mass definition $M=m_{\rm pole}+\delta
M$ the effective Lagrangians contain additional bilinear quark
operators $\delta M\psi^\dagger\psi$, $\delta
M\psi^\dagger\frac{{\bmp}^2}{M^2}\psi$, etc., and the 
one-particle-irreducible two-point function for a quark reads
\begin{equation}
E_M^{}\,+\,2\delta M
\,-\,\frac{{\bmp}^2}{2M}\bigg(1+\frac{\delta M}{M}+\ldots\bigg) 
\,+\,\frac{{\bmp}^4}{8M^3}\,+\,\ldots
\,,
\label{OPI2point2}
\end{equation}
where $E_M^{}=\sqrt{s}-2M$.

\subsection{Pole Mass and Static Potential}
\label{subsectionpolemassVstat}

Since quark masses, just like coupling constants such as
$\alpha_s$, are parameters of the Lagrangian and not physical 
observables, in
principle any quark mass definition can be employed as long as its
perturbative definition is known and well defined. A suitable mass
definition should also have reasonable properties such as being
gauge-invariant and IR-safe. The pole mass definition certainly belongs
to this class.\cite{Kronfeld1} However, there are good reasons not to
use the 
pole mass definition for the analysis of experimental data, because it
induces artificially large perturbative corrections. In the framework
of the theory of asymptotic series these perturbative
corrections can, with certain assumptions on their large order behavior, 
be interpreted as an intrinsic ambiguity of the pole mass definition
of the order $\lqcd\approx 300$~MeV caused by a strong linear
sensitivity to small momenta of order $\lqcd$. (The exact amount of
the uncertainty is 
difficult to estimate, and I have just given a personal (and probably
biased) number to be definite. It is also not intended to rely on a
specific scheme when the notion of ``$\lqcd$'' is used.)
This unpleasant property is, at least from the
perturbative point of view, not characteristic for all quark mass
definitions. For the pole mass this property was first discussed in
the analysis of B meson decays in the framework of
HQET.\cite{Bigi1,BenekeBraun}   
In the context of non-relativistic $\QQbar$ systems the same problem
was first realized by Aglietti and Ligeti\,\cite{Ligeti1} in an
analysis of the large order behavior of the perturbative static
$\QQbar$ potential. Initially, however, the connection to the
properties of the pole mass parameters was not noticed. The problems
associated with the pole mass definition are not academic. For heavy
quarkonium systems it is important to consider the issue seriously,
since the parametric perturbative uncertainty of present
next-to-next-to-leading order computations is already at a level where
an ambiguity of a mass definition of order $\lqcd$ is relevant.

Aglietti  and Ligeti\,\cite{Ligeti1} considered the color singlet
static potential in the ``large-$\beta_0$'' approximation,
\begin{eqnarray}
\tilde V_{\rm static}^{\beta_0}({\bmk}) & = &
-\frac{4\pi C_F\alpha_s(\mu)}{{\bmk}^2}\,
\sum\limits_{n=0}^\infty
\bigg[-\frac{\alpha_s(\mu)}{4\pi}\,\beta_0\,
\ln\Big(\frac{{\bmk}^2 e^c}{\mu^2}\Big)\,\bigg]^n 
\,,
\label{Vstatlargeorder}
\end{eqnarray}
where $c=-5/3$ in the \ms scheme.
The large-$\beta_0$ approximation assumes that the highest power of
$\beta_0$ in each order of perturbation theory dominates numerically
the rest of the contributions.\cite{Beneke3} At this point the
distinction between 
the static pNRQCD potential and the Schr\"odinger Coulomb potential in
vNRQCD is irrelevant, because the large-$\beta_0$ approximation is the
same in both cases. The large order behavior can then be analyzed through
the Borel transform of $\tilde V_{\rm static}^{\beta_0}$ with respect
to $\alpha_s\beta_0/(4\pi)$,
\begin{eqnarray}
B[\tilde V_{\rm static}^{\beta_0}](u) & = &
-\frac{16 \pi^2 C_F}{\beta_0\,{\bmk}^2}\,
\sum\limits_{n=0}^\infty
\frac{1}{n!}\,
\bigg[-u\,\beta_0\,
\ln\Big(\frac{{\bmk}^2 e^c}{\mu^2}\Big)\,\bigg]^n 
\nonumber\\
& = &
-\frac{16 \pi^2 C_F}{\beta_0\,{\bmk}^2}\,
\frac{\mu^{2u}e^{-cu}}{({\bmk}^{2})^{1+u}}
\,.
\end{eqnarray}
The inverse Borel transform reads 
$$\tilde V_{\rm stat}^{\beta_0}({\bmk})=
\int_0^\infty du\,
B[\tilde V_{\rm stat}^{\beta_0}](u)
\, \exp\Big(-\frac{4\pi u}{\beta_0\alpha_s(\mu)}\Big)\,.$$ 
The impact of the large
order behavior of Eq.\ (\ref{Vstatlargeorder}) in bound state
calculations can be visualized by considering the static potential in
configuration space, describing the static energy of a $\QQbar$ pair
with distance $r\ll\frac{1}{\lqcd}$ in the pole mass scheme,
\begin{eqnarray}
B[V_{\rm static}^{\beta_0}](u) & = &
\int\frac{d^3{\bmk}}{(2\pi)^3}\,e^{i{\bmk}.{\bmr}}\,
B[\tilde V_{\rm static}^{\beta_0}](u)
\nonumber\\ & = &
-\frac{4 C_F e^{-c u}}{\beta_0\,r}\,(\mu r)^{2u}\,
\frac{\Gamma(\mbox{$\frac{1}{2}+u$})\Gamma(\mbox{$\frac{1}{2}-u$})}
{\Gamma(2u+1)}
\,.
\label{BorelstaticVr}
\end{eqnarray}
The RHS of Eq.\ (\ref{BorelstaticVr}) has poles on the positive real
axis at $u=(2k+1)/2$, 
$k=0,1,2,3,\ldots$, with residues $\propto (\mu r)^{2k+1}/r$.
These poles cause ambiguities  in $V_{\rm stat}^{\beta_0}(r)$
proportional to 
$$\frac{(\mu r)^{2k+1}}{r}\exp\Big(-\frac{4\pi}{\beta_0\alpha_s(\mu)}
\frac{2k+1}{2}\Big)
\sim \frac{(\lqcd \, r)^{2k+1}}{r}\,,$$ 
since there is no {\em a priori}
prescription how the poles on the positive real axis should be treated
in the inverse Borel transformation. The pole closest to the origin at
$u=1/2$ is dominant and causes an ambiguity of order
$\lqcd$. It corresponds to an asymptotic large order behavior of
$V_{\rm stat}^{\beta_0}(r)$ proportional to
$$\alpha_s(\mu)\,\mu\,\left(\frac{\alpha_s(\mu)\beta_0}{2\pi}\right)^n n!\,\,.$$ 
One might wonder
how such an $r$-independent behavior can arise in the static
potential, which has an overall factor $1/r$. It was shown in
Ref.~\citebk{Hoang7} that it arises from an exponential structure of the 
$\ln(\mu r)$ terms in the order $\alpha_s^{n+1}$ term in
$V_{\rm stat}^{\beta_0}(r)$ of the form 
\begin{eqnarray}
\lefteqn{
\sim
\frac{\alpha_s}{r} \left(\frac{\alpha_s\beta_0}{2\pi}\right)^n\,n!\,
\bigg[\,\frac{1}{(n-1)!}\ln^{n-1}(\mu r)+ \ldots 
+\frac{1}{2}\ln^2(\mu r)
+\ln(\mu r)+1
\,\bigg] 
}
\nonumber\\[2mm]
& \approx &
\frac{\alpha_s}{r} \left(\frac{\alpha_s\beta_0}{2\pi}\right)^n\,
n!\,\,e^{\ln(\mu r)}\, = \,
\alpha_s\,\mu\,\left(\frac{\alpha_s\beta_0}{2\pi}\right)^n\, n!
\,.
\hspace{4cm}
\label{largeorderbehaviorVstat}
\end{eqnarray}
It should be noted that the large-$\beta_0$ approximation is just a
heuristic concept. However,
explicit multi-loop calculations for example for the perturbative
relation between \ms and pole
mass\,\cite{Broadhurst1,Melnikov1,Chetyrkin1} have 
shown a remarkable 
consistency with the large-$\beta_0$ approximation. On the other
hand, it has also been reported\,\cite{Kiyo1,Hoang8} that the
large-$\beta_0$ 
approximation gives only order of magnitude estimates for higher order 
corrections in mass relations that do not involve the pole mass
definition. Numerical
evidence for the bad behavior of the static potential in configuration
space was in fact noticed quite early, see
e.g.\ Refs.~\citebk{Jezabek1}, but 
the observations were not interpreted as a consequence of the pole
mass definition. 

If the previously described large order behavior would be
real, the determination of perturbative quark mass parameters would be
forever limited to an uncertainty of order $\lqcd$.
However, it
was found by Hoang et al.\,\cite{Hoang9} and Beneke\,\cite{Beneke4} (see
also Refs.~\citebk{Uraltsev1})
that the ambiguity of order $\lqcd$ is unphysical and merely an artifact 
of using the pole mass definition. The pole mass counterterm
subtracts, apart from the usual UV divergent term, also a finite piece
that correspond to the static quark selfenergy,
\begin{equation}
\label{staticselfenergy}
\frac{1}{2}\int\frac{d^3{\bmk}^3}{(2\pi)^3}\tilde V_{\rm
stat}({\bmk})\,.
\end{equation}
This expression is just the leading term in the 
non-relativistic expansion of the finite piece in the pole mass
counterterm. Its asymptotic large order behavior is exactly 
$1/2$ times the asymptotic large order behavior of the static
potential displayed in Eq.\ (\ref{largeorderbehaviorVstat}).
If a mass definition is chosen that does not subtract
the static selfenergy, the order $\lqcd$ ambiguity and the
corresponding bad large order behavior of the perturbative series
described above do not arise in the first place.
Hoang et al.\,\cite{Hoang9} and Beneke\,\cite{Beneke4} demonstrated that
the total static energy,
\begin{eqnarray}
E_{\rm stat} & = & 2m_{\rm pole} \, + \, V_{\rm stat}(r)
\,,
\label{totalstatic}
\end{eqnarray}
is free of the ambiguity of order $\lqcd$, at least from the point of
view of perturbation theory. 

It was shown by Smith and Willenbrock\,\cite{Smith1} that even if the
quark is unstable and decays with a finite lifetime smaller than
$1/\lqcd$, the  static quark selfenergy in Eq.~(\ref{staticselfenergy})
%
produces the same large corrections described above.
Thus the problematic behavior of the pole mass definition remains also
for short-lived quarks such as the top quark with $\Gamma_t
\approx 1.5$~GeV in the Standard Model.  

The previous discussion disqualifies the pole mass a priori for the
use in analyses of experimental data, since it induces artificially
large corrections that are compensated e.g.\ by shifts in the pole mass
value itself, when higher order corrections are included. In
principle, the pole mass can still be employed as an order- and
scale-dependent correlated quantity, similar to the value of a matrix 
element.\,\cite{Manohar6} In such an approach the
large perturbative corrections associated with the pole mass are
contained in its numerical value. Since these corrections depend on
the order and on the choice for renormalization scales and couplings,
the numerical value of the pole mass needs to be treated as a function of
these parameters in order to achieve the proper cancellation of the
large corrections, if the pole mass value is used in computations for 
other mass-dependent quantities. It should be noted, however, that
this strategy can become increasingly unreliable for orders $n$ 
where the corrections induced by the pole mass are large enough that
their numerical cancellation is incomplete.  
Results for analyses in the pole mass scheme for the $t\bar t$ production
close to threshold in $e^+e^-$ annihilation and for bottom mass
extraction from  $b\bar b$ sum rules  are discussed in
Secs.~\ref{sectionttbar} and \ref{sectionsumrules}.

The subdominant ambiguity in the static potential is order $\lqcd^2
r\sim\lqcd^2/(m\alpha_s)$ and quite small parametrically. The
associated large order behavior of the perturbative series 
was analyzed by Brambilla et al. in
Ref.~\citebk{Brambilla1}, and it was found 
that it cancels with the large order behavior of the leading
ultrasoft corrections shown in Fig.~\ref{figpnrqcdselfenergy} coming
from two insertions of the ${\bmr}.{\bmE}$ interaction. The
cancellation is independent of the choice for the quark mass
definition.
The ambiguity of order  $\lqcd^2 r$ can be ignored for any practical
analysis of the non-relativistic $\QQbar$ dynamics.

\subsection{\ms Mass and Threshold Masses}
\label{subsectionthresholdmasses}

To avoid the problems of the pole mass definition it is advantageous
to use quark mass definitions that are less sensitive to low momenta
and do not contain the static quark selfenergy in the mass counter
term. Such quark mass definitions are called ``short-distance'' masses
and have a parametric ambiguity or order $\lqcd^2/M$ or smaller. In
principle, the most natural choice of a short-distance mass would be
the \ms (or MS) mass, because its mass counter term subtracts
only the UV poles $\propto 1/\epsilon^n$ and does not
contain any infrared-sensitive subtraction. This means that the \ms
mass is only sensitive to scales of order or larger than $m$ and that
it is from the conceptual point of view not allowed to lower the scale
of the \ms mass below $m$. The relation
between the pole mass and the \ms mass $\overline m(\overline m)$ 
reads ($\bar\alpha_s=\alpha_s^{(n_\ell+1)}(\overline m(\overline m))$)
\begin{eqnarray}
r_m 
& \equiv & 
\frac{m_{\rm pole}}{\overline m(\overline m)} \, = \,
1
+\frac{4\bar\alpha_s}{3\pi} +
\Big(\frac{\bar\alpha_s}{\pi}\Big)^2\,\Big( 13.4434 - 1.0414 n_\ell\Big)
\nonumber\\ & & \qquad\quad
+ \Big(\frac{\bar\alpha_s}{\pi}\Big)^3\,
\Big( 190.595 - 26.655 n_\ell+0.6527
n_\ell^2\Big)
+\ldots
\label{polemsbar}
\end{eqnarray}
for $n_\ell$ massless light quark species. 
The two-loop correction was determined in Ref.~\citebk{Broadhurst1}
and the three-loop term in Ref.~\citebk{Melnikov1} 
(see Ref.~\citebk{Chetyrkin1} for a numerical computation).
For b quarks with $\bar\alpha_s=0.22$ and $n_\ell=4$ the series on the 
RHS of Eq.~(\ref{polemsbar}) reads $1 -0.093 + 0.045+ 0.032 $. Its
rather poor 
convergence properties at already low orders illustrate the 
infrared sensitivity of the pole mass. The use of the \ms mass for
non-relativistic $\QQbar$ systems would in principle be welcome,
because it is also the preferred mass definition for high energy
processes or electroweak precision observables. Unfortunately, the \ms
mass breaks the non-relativistic power counting. The parametric size of
the terms of the one-particle-irreducible two-point function of Eq.\
(\ref{OPI2point1}) is of order $mv^2\sim 
m\alpha_s^2$. In the \ms scheme, however, $\delta m_{\overline{\rm
MS}}$ is of order $m\alpha_s$ and exceeds all the other terms in Eq.\
(\ref{OPI2point1}), which would in principle require that $\delta
m_{\overline{\rm MS}}$ is not treated as a correction. This leads
essentially back to the pole mass definition.\cite{Beneke4} Explicit 
examples for the use of the \ms mass for $t\bar t$ production close to
threshold in $e^+e^-$ annihilation and for bottom mass extractions
from $b\bar b$ sum rules have been given in Ref.~\citebk{Hoang10}. 
From the physical point of view, the incompatibility of the \ms mass
with non-relativistic $\QQbar$ systems is a consequence of the fact
that it is a mass definition adapted to the situation where the heavy
quark has a virtuality $|q^2-m^2|$ of order or larger than $m^2$, $q$
being the heavy quark four-momentum. Thus the \ms mass is the
preferred mass definition to describe virtual effects of a heavy
particle in low energy processes with characteristic energies
$E\ll m$ or mass effects in high energy processes with characteristic
energies  $E\gg m$. 

An appropriate mass definition to use for non-relativistic $\QQbar$
systems is a short-distance mass that also complies with the
non-relativistic power counting, i.e.\ $\delta M$ is parametrically of
order $Mv^2\sim M\alpha_s^2$. Such masses are called ``threshold
masses'', since they are adapted to the situation where the
quark virtuality is small.\cite{Hoang11} Obviously an arbitrary number
of such threshold 
masses can be invented, because the number of subtractions possible to  
achieve $\delta M\sim M\alpha^2$ is infinite. Meanwhile, a
considerable number of threshold mass definitions can be found in the
literature. In the following I briefly comment on the threshold
masses in the current literature in historic order.\\

\noindent
{\it Kinetic Mass}\\[1mm]
The kinetic mass has been proposed by Bigi et
al. in Ref.~\citebk{Bigi2}. Originally 
it was devised as a mass definition for the description of B
mesons. It is defined as 
\begin{eqnarray}
M_{\rm kin}^{}(\mu_{\rm kin}^{}) 
& = &
m_{\rm pole} - \left[\bar\Lambda(\mu_{\rm kin}^{})\right]_{\rm pert} 
- \left[\frac{\mu^2_\pi(\mu_{\rm kin}^{})}{2 m_{\rm pole}}\right]_{\rm pert}
+\ldots
\nonumber \\[2mm] & = &
m_{\rm pole} - \frac{16}{9}\,\frac{\alpha_s}{\pi}\,\mu_{\rm kin}^{}
\, + \, \ldots
\,,
\label{kindef}
\end{eqnarray}
where $\left[\bar\Lambda(\mu^{}_{\rm kin})\right]_{\rm pert}$ and
$\left[\mu_\pi^2(\mu^{}_{\rm kin})\right]_{\rm pert}$ are perturbative
evaluations of HQET matrix elements that describe the difference
between the pole and the B meson mass. The two-loop contributions to
the kinetic mass have been calculated in Ref.~\citebk{Czarnecki5}.
The three-loop contributions, which correspond to NNLO in the
non-relativistic power counting, are known in the large-$\beta_0$
approximation. In the first line of Eq.~(\ref{kindef}) the ellipses
indicate matrix elements of operators with higher dimension, which
have not been included in any analysis so far.
The kinetic mass is dependent on the scale $\mu_{\rm kin}^{}$, which
is the momentum cutoff for the evaluation of the matrix elements. 
The cutoff has to be chosen of  order $m\alpha_s$ to
comply with the non-relativistic power counting. Thus the order
$\mu_{\rm kin}^{}\alpha_s$ term is LO in the non-relativistic 
power counting,  
the order $\mu_{\rm kin}^{}\alpha_s^2$ term is NLO, etc..
If the kinetic mass is used in the perturbative series for quantities
that do not have non-relativistic power counting, the explicit
counting in powers of $\alpha_s$ is used, i.e.\ $\mu^{}_{\rm kin}$ is
formally counted of order $m$. This is the case for example in the
relation between the \ms and the kinetic mass or when the kinetic mass
is used in computations for B mesons. \\

\noindent
{\it Potential Subtracted Mass}\\[1mm]
The potential subtracted (PS) mass $m_{\rm PS}$ has been proposed by
Beneke\,\cite{Beneke4} and is defined by
\begin{eqnarray}
M_{\rm PS}(\mu_{\rm PS})
 & = &
m_{\rm pole} + \frac{1}{2}\,
\int\limits^{|{\scriptsize\mathbf q}|<\mu_{\tiny\rm PS}^{}}
\frac{d^3{\bmk}}{(2\pi)^3}\,
\tilde V_{\rm stat}({\bmk})
\nonumber \\[2mm] & = &
m_{\rm pole} - \frac{4}{3}\,\frac{\alpha_s}{\pi}\,\mu_{\rm PS}^{}
\, + \, \ldots
\,,
\label{PSdef}
\end{eqnarray}
where $\tilde V_{\rm stat}$ is the singlet static potential in momentum
space representation (see Eqs.~(\ref{staticpotentialschroeder}) and 
(\ref{VCoulombpnrqcd})). 
In the last line of Eq.\ (\ref{PSdef}) the first term in an
expansion in $\alpha_s$ is displayed.
The additional subtraction of the PS mass is
equal to the non-relativistic selfenergy mentioned before and can be
regarded as a kind of minimal way to cancel the large higher order
corrections in the static potential in Eq.\ (\ref{totalstatic}).
This subtraction has in fact been proposed before by Bigi et
al. in Ref.~\citebk{Bigi1}, but no explicit mass definition resulted
from the subtraction prescription.
The PS mass is dependent on the scale $\mu_{\rm PS}^{}$, which is the
cutoff for the selfenergy integration. The cutoff has to be chosen of
order $m\alpha_s$ to comply with the non-relativistic power counting. 
Thus the order
$\mu_{\rm PS}^{}\alpha_s$ term is LO in the non-relativistic 
power counting, the order $\mu_{\rm PS}^{}\alpha_s^2$ term is NLO,
etc.. At NNLL order in perturbation
theory the PS mass receives an additional dependence on the pNRQCD
renormalization scale $\mu_u$ due to the ultrasoft anomalous dimension
of the static potential caused by the selfenergy diagram shown in
Fig.\ \ref{figpnrqcdselfenergy}. This scale is correlated to the
renormalization scale $\mu_p$ used in the computation of the pNRQCD
matrix elements by Eq.\ (\ref{pNRQCDscalecorrelation}).
If the PS mass is used in perturbative series for quantities that do
not have non-relativistic power counting, the explicit counting in
powers of $\alpha_s$ is used, i.e.\ $\mu_{\rm PS}^{}$ is formally
counted of order of $m$.  
A modified version of the PS mass, called the ${\overline{\rm PS}}$
mass was proposed by Yakovlev and Groote.\cite{Yakovlev1} In the 
${\overline{\rm PS}}$ scheme additional $1/m$-suppressed corrections
from the non-relativistic expansion of the quark selfenergy are included
in Eq.~(\ref{PSdef}).\\

\noindent
{\it 1S Mass}\\[1mm]
The 1S mass has been proposed by Hoang et
al.\,\cite{Hoang7,Hoang12} and is 
defined as half of the perturbative series for the mass of the $n=1$,
${}^{2s+1}L_j={}^3S_1$ quarkonium bound state,
\begin{eqnarray}
M_{\rm 1S}^{} & = & \frac{1}{2}\,
\left[\,M_{\Upsilon_{\QQbar}(1\,{}^3S_1)}
\,\right]_{\rm pert}
\nonumber \\[2mm] & = &
m_{\rm pole} - \frac{2}{9}\,\alpha_s^2\,m_{\rm pole} 
\, + \, \ldots
\,.
\label{1Sdef}
\end{eqnarray}
In the last line of Eq.\ (\ref{1Sdef}) the LO term is displayed.
The 1S mass is scale-independent, and it is determined with the same 
methods that are used for the computations of the non-relativistic
quantities for which it is used as the quark mass definition.
By construction, the value of the 1S mass very weakly correlated to other
parameters such as $\alpha_s$ or the renormalization scale in
extractions from the $\QQbar$ spectrum. As such it will in general have
small perturbative uncertainties. 
If the 1S mass is used in perturbative series for quantities that do
not have non-relativistic power counting, the order $\alpha_s^2$ term
has to be counted as order $\alpha_s$, the order  $\alpha_s^3$ term
has to be counted as order $\alpha_s^2$, etc.. This prescription 
is called the ``upsilon expansion''.\cite{Hoang7}
In the case $m\gg mv\gg mv^2\gg\lqcd$ the
physical $1^3S_1$ quarkonium mass is equal to $2M_{\rm 1S}$ up to
non-perturbative corrections, which are parametrically of order
$m\lqcd^4/(m\alpha)^4$ and which can be related to local condensates
of the operators product expansion of Shifman et al.\,\cite{Shifman1}
(Sec.~\ref{sectionspectrum}).\\

\noindent
{\it Renormalon Subtracted Mass}\\[1mm]
The renormalon subtracted mass has been proposed by
Pineda\,\cite{Pineda6} and is defined as the perturbative series that
results from subtracting all non-analytic pole terms from the Borel
transform of the pole-\ms mass relation at $u=1/2$ with a fixed choice
for the renormalization scale $\mu=\mu_{\rm RS}^{}$. The scale
$\mu_{\rm RS}^{}$ is then kept independent from the renormalization scale 
used for the computation of the non-relativistic quantities of
interest. The terms in the relation between the pole mass and the RS
mass are formally known to all orders, but the numerical value of the
individual coefficients of the series are known only approximately due
to yet undetermined subleading contributions to the residue at
$u=1/2$. To order $\alpha_s$ the relation between RS mass and pole
mass reads, \begin{eqnarray}
M_{\rm RS}^{}(\mu_{\rm RS}^{}) 
& = &
m_{\rm pole} - c\,\alpha_s\,\mu_{\rm RS}^{}
\, + \, \ldots
\,,
\label{RSdef}
\end{eqnarray}
where the constant $c$ depends on the number of light quark species
and has an uncertainty because the residue at $u=1/2$ in  the Borel
transform of the pole-\ms mass relation is known only approximately.
The scale $\mu_{\rm RS}^{}$ has to be chosen of  order $m\alpha_s$ to  
comply with the non-relativistic power counting. If the RS mass is
used in perturbative series for quantities that do 
not have non-relativistic power counting, $\mu_{\rm RS}^{}$ is formally
counted of order of $m$. Pineda also proposed a slightly modified
version of the RS scheme, which he called RS$^\prime$ scheme.
\\

At this point I refrain from a general numerical comparison of the
various threshold masses. Comparisons for some cases can be
found in the references given above (see also Ref.~\citebk{Beneke5}).  
To allow for a comparison between the results that have been obtained
for the various threshold masses in the literature it has become
practice to determine in a second step the \ms mass 
$\overline m(\overline m)$. The perturbative uncertainty
in the \ms mass that is obtained from the conversion 
can be larger than the perturbative uncertainty in the threshold
masses, because the large order $\alpha_s$ term in the pole-\ms
mass relation (Eq.~(\ref{polemsbar})) is quite sensitive to the
uncertainty of $\alpha_s$.

\subsection{Threshold Masses and Heavy Quark Decay}
\label{subsectionquarkdecay}

The application of threshold masses is not restricted to
non-relativistic $\QQbar$ systems. Threshold masses are in general
useful for systems, where the virtuality of the heavy quarks is 
small, i.e.\ if the heavy quarks are close to their mass-shell. For
non-relativistic $\QQbar$ systems the virtuality $q^2-m^2$ is of
order $m^2 v^2\ll m^2$. A similar situation arises in  B mesons, where
the heavy quark virtuality is of order $m\lqcd\ll m^2$.

As an example consider the inclusive semileptonic $B\to X_u e \nu$
decay rate, which is relevant for the determination of the CKM matrix
element $|V_{ub}|$. 
The inclusive semileptonic partial rate into light hadrons
has been analyzed in the 1S\,\cite{Hoang7} 
and the kinetic mass scheme\,\cite{Uraltsev2} (see also
Refs.~\citebk{Melnikov1,Beneke5}).
The semileptonic $B\to X_u e \nu$
partial rate has a particularly strong dependence on
the bottom quark mass $\propto G_F^2 (m^b_{})^5$ and its
perturbative series illustrates the properties of the various mass
definitions. In the following I show the behavior of the perturbative
series of the semileptonic $B\to X_u e \nu$
partial rate in the pole, \mbox{\ms\hspace{-1mm},} 1S and kinetic mass schemes.
For simplicity, $\alpha_s=0.22$ at the scale of the respective mass
definition has been used as the expansion parameter.
For a realistic phenomenological application of the results in the 1S
and kinetic mass schemes including the effects of non-perturbative
contributions and I refer to Ref.~\citebk{LEPBdecays}.

In the pole mass scheme the perturbative series for the inclusive
decay rate reads 
\begin{eqnarray}
\Gamma\left(\bar B\to X_u e\bar\nu\right) & = & 
\frac{G_F^3|V_{ub}|^2}{192\pi^3}\,\Big(m_{\rm pole}^b\Big)^5\,
\Big[\,
1\,-\,0.17\,-\,0.10\,-\ldots
\,\Big]
\,,
\label{decayratepole}
\end{eqnarray}
where $G_F$ is the Fermi constant, and the second and third term in
the brackets correspond to the one- 
and two-loop\,\cite{vanRitbergen1} corrections. 
The contributions of
non-perturbative matrix elements are not displayed.
In the large-$\beta_0$ approximation the two-loop term is estimated
as $-0.13$, which agrees reasonably well with the true result $-0.10$;
the three-loop term is unknown and estimated as $-0.12$ in the
large-$\beta_0$ approximation. The convergence of the series is
not very good. In particular, the estimate for
the three-loop correction indicates that two- and three-loop
corrections are of the same size. This is an artifact of the pole mass
definition. If the \ms scheme is employed the series reads~\footnote{ 
The order $\alpha_s^2$ contributions in Eqs.~(\ref{decayratemsbar}) and
(\ref{decayrate1S}) have been taken from Ref.~\citebkcap{Ligeti2},
where the new results from Ref.~\citebkcap{vanRitbergen1} were
included. 
}  
\begin{eqnarray}
\Gamma\left(\bar B\to X_u e\bar\nu\right) & = & 
\frac{G_F^3|V_{ub}|^2}{192\pi^3}\,
\Big(\overline m^b(\overline m^b)\Big)^5\,
\Big[\,
1\,+\,0.30\,+\,0.13\,+\ldots
\,\Big]
\,.
\label{decayratemsbar}
\end{eqnarray}
In the large-$\beta_0$ approximation the two-loop term is estimated
as $+0.19$ and the three-loop term is estimated as $+0.05$.
While the use of the \ms mass leads to a cancellation of the
bad pole mass behavior  
at high orders, the corrections are still large at low
orders. 
On the other hand, in the 1S mass scheme the series for the inclusive
semileptonic $B\to X_u e \nu$ decay rate reads
\begin{eqnarray}
\Gamma\left(\bar B\to X_u e\bar\nu\right) & = & 
\frac{G_F^3|V_{ub}|^2}{192\pi^3}\,
\Big(M^b_{\rm 1S}\Big)^5\,
\Big[\,
1\,-\,0.115\,-\,0.031\,-\ldots
\,\Big]
\,,
\label{decayrate1S}
\end{eqnarray}
where the upsilon expansion\,\cite{Hoang7} has to be used to obtain
the terms in the series.
In the large-$\beta_0$ approximation the two-loop term is estimated
as $-0.035$, which agrees well with the exact result $-0.031$;
the three-loop term is estimated as $+0.005$ in the
large-$\beta_0$ approximation. So the series in the 1S scheme has
considerably better convergence properties than in the pole and the
\ms scheme. 
In the kinetic mass scheme for $\mu_{\rm kin}^{}=1$~GeV the decay rate 
reads\,\cite{Uraltsev2} 
\begin{eqnarray}
\Gamma\left(\bar B\to X_u e\bar\nu\right) & = & 
\frac{G_F^3|V_{ub}|^2}{192\pi^3}\,
\Big(M^b_{\rm kin}\Big)^5\,
\Big[\,
1\,-\,0.022\,-\,0.006\,-\ldots
\,\Big]
\,.
\label{decayratekinetic}
\end{eqnarray}
The series converges even better than in the 1S scheme, which is a
consequence of the specific choice $\mu_{\rm kin}^{}=1$~GeV.
In Ref.~\citebk{Hoang7} all inclusive semileptonic B decay modes were studied
in the 1S mass scheme including an analysis of the semileptonic B
decay form factors. Other applications of threshold mass schemes in 
B decays can be found e.g.\ in Refs.~\citebk{Bdecaysthresholdmass}.

\vspace{1cm}

\section{Heavy Quark Pair Production at Threshold}
\label{sectionQQbarproduction}

The production of heavy quark-antiquark pairs in the kinematic regime
close to threshold $\sqrt{s}\approx 2m$ is currently the most
important application of non-relativistic effective theories for
quark-antiquark systems in QCD. Moments of the total cross section
$\sigma(e^+e^-\to b\bar b+X)$ in the threshold region are an important
tool for precise determinations of the bottom quark mass parameter,
and $t\bar t$ production close to threshold is a major part of the top
physics program of the future Linear Collider with the aim to provide 
high precision measurements of the top mass and other parameters. In
this section I will review the calculational steps necessary
for a computation of the total production cross section of a genuine
color singlet $\QQbar$ pair close to threshold assuming the
hierarchy $m\gg mv\gg mv^2\gg\lqcd$ and concentrating mainly on
$e^+e^-$ annihilation. For simplicity the quarks are assumed stable
and electroweak effects are neglected except for the production
mechanism. Non-perturbative effects are also neglected. 
Applications to $t\bar t$ and $b\bar b$ production including
discussions on non-perturbative effects (to the extent they are known)
are given in Secs.~\ref{sectionttbar} and \ref{sectionsumrules}. For
an application on $\tau$ pair 
production close to threshold in QED see Ref.~\citebk{Ruiz-Femenia1}.
I will first review a recent computation of the total cross section at
NNLL order in the framework of vNRQCD and then comment on 
computations in NRQCD at
NNLO in the fixed order approach, i.e.\ without summation of QCD
logarithms of $v$.

\subsection{$\QQbar$ Production At NNLL Order}
\label{subsectioncrosssectionvnrqcd}

Generically, the normalized total heavy $\QQbar$ cross section  
in $e^+e^-$ annihilation
at NNLL order in the non-relativistic expansion takes the form
\begin{eqnarray}
 R & = & \frac{\sigma(e^+e^-\to\QQbar)}{\sigma_{\mu^+\mu^-}}
 \, = \,
 v\,\sum\limits_k \left(\frac{\alpha_s}{v}\right)^k
 \sum\limits_i \left(\alpha_s\ln v \right)^i
\nonumber\\ & & \hspace{1cm}
 \times
 \bigg\{1\,\mbox{(LL)}; \alpha_s, v\,\mbox{(NLL)}; 
 \alpha_s^2, \alpha_s v, v^2\,\mbox{(NNLL)}\bigg\}
 \,,
 \label{RNNLLorders}
\end{eqnarray}
where the indicated terms are LL, NLL and NNLL order.  The
summation of logarithms can be performed using the renormalization
group equations in the framework of the non-relativistic effective
theories discussed in Secs.\ \ref{sectionpNRQCD} and
\ref{sectionvNRQCD}.
In the following I will review the computation of 
$\sigma(e^+e^-\to\QQbar)$ in the framework of vNRQCD using dimension
regularization in the \ms scheme.\cite{Hoang3} In the framework of
pNRQCD such a computation has not yet been carried out, 
but follows essentially the same lines. For $\QQbar$ production with a
different initial state ($\gamma\gamma$, etc.) or production of other 
colored massive particles such as squarks or for QED systems the
same method can be applied with straightforward modifications.\\

\noindent
{\it Schr\"odinger Equation and Potentials}\\[1mm]
At NNLL order ultrasoft corrections enter only through mixing into the
coefficients of potentials and currents (Sec.\
\ref{sectionvNRQCD}). Therefore the NNLL order non-relativistic
$\QQbar$ dynamics is described by a common time-independent two-body
Schr\"odinger equation. In momentum space and with a threshold 
mass definition $M$ (Sec.\ \ref{sectionquarkmass}) the NNLL order
Schr\"odinger equation takes the form  
\begin{eqnarray} 
\lefteqn{
 \bigg[\, \frac{\bmp^2}{M}\bigg(1+\frac{\delta M}{M}\bigg) -
 \frac{\bmp^4}{4M^3} - (E +2\, \delta M) \, \bigg]
 \, \tilde G({\bmp},{\bmp}^\prime,E)
}
\nonumber\\[2mm] & &\quad
 + \int D^n{\bmk}\,\tilde V({\bmp},{\bmk})\, 
 \tilde G({\bmk},{\bmp^\prime},E)
 = \, (2\pi)^n\,\delta^{(n)}({\bmp}-{\bmp}^\prime) \,,
\label{NNLLSchroedinger}
\end{eqnarray}
where $$E=\sqrt{s}-2M\equiv Mv^2\,,\quad n=3-2\epsilon\,,\quad {\rm and}
\quad   
D^n{\bmk}\equiv
e^{\epsilon(\gamma_E-\ln 4\pi)}\mu_S^{2\epsilon}\,d^n{\bmk}/(2\pi)^n\,.$$
The term $$\delta M = M-m_{\rm pole}^{}=
\delta M^{}_{\rm LL}+\delta M^{}_{\rm NLL}+
\delta M^{}_{\rm NNLL}$$ is the difference between pole and threshold
mass. For the term ${\bmp}^2 \delta M/M^2$ only the LL
expression needs to be taken into account.
The Schr\"odinger equation in Eq.\ (\ref{NNLLSchroedinger}) is
obtained 
from the NNLL order equation of motion of a $\QQbar\QQbar$ four point
function with center-of-mass momenta 
$(\frac{E}{2}\pm p_0,\pm{\bmp})$ and 
$(\frac{E}{2}\pm p_0^\prime,\pm{\bmp}^\prime)$  
for the incoming and outgoing quarks, respectively, supplemented by a
contour integration over $p_0$ and $p_0^\prime$ (see
e.g.\ Ref.~\citebk{Hoang12}). The Green function 
$\tilde G({\bmp},{\bmp^\prime},E)$ describes production and
annihilation of an off-shell $\QQbar$ pair with relative momenta
$2{\bmp}$ and $2{\bmp}^\prime$, respectively, at total center-of-mass
energy $\sqrt{s}=2M+E$. The potential $\tilde V$ arises from the
potential-type operators up to order $1/m^2$ 
(see Eq.~(\ref{vNRQCDpotential}) and Fig.~\ref{figpotentialvnrqcd}b)
and from  time-ordered products of the two-quark soft interactions
(see Eq.~(\ref{vNRQCDsoft}) and Figs.~\ref{figsoftvnrqcd},
\ref{figvnrqcdloops}c). The full potential at NNLL order for a
$\QQbar$ pair in a S-wave state reads
\begin{eqnarray}
 \tilde V({\bmp},{\bmp}^\prime) \, = \, 
 \tilde V_c({\bmp},{\bmp}^\prime) + 
 \tilde V_\delta({\bmp},{\bmp}^\prime) + 
 \tilde V_r({\bmp},{\bmp}^\prime) +
 \tilde V_k({\bmp},{\bmp}^\prime)  
 \,,
\label{Vsdetail}
\end{eqnarray} 
where $\tilde V_c$ has been given before in Eq.\
(\ref{VCoulombvnrqcd}) and (${\bmq}={\bmp}-{\bmp}^\prime$)
\begin{eqnarray}  
 \tilde V_k({\bmp},{\bmp}^\prime) & = & \frac{\pi^2}{M |{\bmq}|}\, 
    {\cal{V}}_k^{(s)}(\nu) \,,
\qquad\quad
 \tilde V_\delta({\bmp},{\bmp}^\prime) \, = \,
\frac{{\cal{V}}_2^{(s)}(\nu) 
    +{\mathbf S}^2 {\cal{V}}_s^{(s)}(\nu)}{M^2} \,, 
\nonumber \\[2mm]
 \tilde V_r({\bmp},{\bmp}^\prime) & = & 
\frac{({\bmp}^2+{\bmp}^{\prime\,2})}{2 M^2 {\bmq}^2}\, 
    {\cal{V}}_r^{(s)}(\nu) \,. 
\end{eqnarray}
The color singlet components of the potential are obtained from Eq.\
(\ref{convertpotentials}). 
The running of the potential coefficients was determined in
Refs.~\citebk{Manohar2,Manohar3,Hoang4,Hoang6}. For $\nu\sim v$ all
QCD logarithms of $v$ that arise 
from the non-relativistic $\QQbar$ dynamics at NNLL order are summed
into the coefficients of the potentials. 
In $e^+e^-$ annihilation the $\QQbar$ pair
is predominantly produced in a ${}^3S_1$ state where the spin
functions in Eq.\ (\ref{vNRQCDpotential}) are ${\mathbf S}^2=2$ and
$\Lambda=T=0$. For a ${}^1S_0$ state (e.g.\ $\gamma\gamma\to\QQbar$)
the spin functions read ${\mathbf S}^2=\Lambda=T=0$.
The spin function are evaluated in the convention that the traces of
the Pauli-matrices are carried out in three dimensions. The difference
between using three and $n$ dimensional Pauli matrices is a change in 
the renormalization scheme.\cite{Hoang3} A similar scheme dependence
arises in chiral perturbation theory.\cite{Savage1}\\

\noindent
{\it Currents}\\[1mm]
To describe $\QQbar$ production
at NNLL order in $e^+e^-$ annihilation
one needs the ${}^3S_1$
vector current with operators of dimension three and five and
the  ${}^3P_1$ axial-vector current with dimension four,
\begin{eqnarray}
{\bf J}^v_{\bmp} & = & c_1(\nu) \O{p}{1}(\nu) + c_2(\nu) \O{p}{2}(\nu)
\,,
\nonumber\\
 {\bf J}^a_{\bmp} & = & c_3(\nu) \O{p}{3}(\nu)
\,,
\label{vnrqcdcurrents}
\end{eqnarray}
where the currents $\O{p}{1-3}$ are given in Eqs.\
(\ref{vectorcurrents}) and (\ref{axialvectorcurrents}).
In this operators basis there is an additional dimension five vector
current, $\O{p}{4}=\frac{1}{m^2}\, {\psip{p}}^\dagger\, 
   ({\bmp} ({\bsigma}.{\bmp})-\bsigma \frac{\bmp^2}{3})
   \,(i\sigma_2)\,{\chip{-p}^*}$.
However it produces a $D$-wave quark-antiquark pair and therefore does not
contribute at NNLL order. The current $\O{p}{1}$ is dominant and
contributes at LL order. The currents  $\O{p}{2-3}$ are suppressed and
only contribute at NNLL order. Thus $c_1$ has to be known at NNLL
order, i.e.\ one needs to know the two-loop matching condition and the
three-loop anomalous dimension. The two-loop matching condition has
been determined in Ref.~\citebk{Czarnecki3,Hoang3} and is displayed in
Eq.\ (\ref{c1matchingnrqcd}). The anomalous dimension is currently
only known at NLL order.\cite{Manohar3,Hoang6} The coefficients $c_2$
and $c_3$ are needed at LL order and have been determined in 
Ref.~\citebk{Hoang3}, see Eq.\ (\ref{c2ll}).\\

\noindent
{\it Total Cross Section}\\[1mm]
In full QCD the expression for the total cross
section $\sigma_{\rm tot}^{\gamma,Z}(e^+e^-\to \gamma^*,Z^*\to
\QQbar)$ for quarks at center of mass energy $\sqrt{s}$ is
\begin{eqnarray}
  \sigma_{\rm tot}^{\gamma,Z}(s) = \sigma_{\rm pt} 
  \Big[\, F^v(s)\,R^v(s) +  F^a(s) R^a(s) \Big] \,,
\label{totalcross}
\end{eqnarray}
where $\sigma_{\rm pt}=4\pi\alpha^2/(3 s)$. The vector and axial-vector
$R$-ratios are
\begin{eqnarray} \label{fullR}
 R^v(s) \, =  \,\frac{4 \pi }{s}\,\mbox{Im}\,\left[-i\int d^4x\: e^{i q\cdot x}
  \left\langle\,0\,\left|\, T\, j^v_{\mu}(x) \,
  {j^v}^{\mu} (0)\, \right|\,0\,\right\rangle\,\right] \,, \nn\\[2pt]
 R^a(s) \, =  \,\frac{4 \pi }{s}\,\mbox{Im}\,\left[-i\int d^4x\: e^{i q\cdot x}
  \left\langle\,0\,\left|\, T\, j^a_{\mu}(x) \,
  {j^a}^{\mu} (0)\, \right|\,0\,\right\rangle\,\right] \,, 
\end{eqnarray}
where $q=(\sqrt{s},0)$ and $j^{v}_\mu$ ($j^{a}_\mu$) is the vector
(axial-vector) current that produces a quark-antiquark pair. With both $\gamma$
and $Z$ exchange the prefactors in Eq.~(\ref{totalcross}) are
\begin{eqnarray}
  F^v(s) &=& \bigg[\, Q_q^2 - 
   \frac{2 s\, v_e v_q Q_q}{s-m_Z^2} + 
   \frac{s^2 (v_e^2+a_e^2)v_q^2}{(s-m_Z^2)^2}\, \bigg]\,,
\quad 
  F^a(s) \, = \, \frac{s^2\, (v_e^2+a_e^2)a_q^2}{ (s-m_Z^2)^2 } \,,
\nonumber \\[0mm]
  v_f & = & \frac{T_3^f-2 Q_f \sin^2\theta_W}{2\sin\theta_W \cos\theta_W}\,,
  \qquad\qquad
  a_f \, = \, \frac{T_3^f}{2\sin\theta_W \cos\theta_W} \,.
\end{eqnarray}
Here $Q_f$ is the charge for fermion $f$, $T_3^f$ is the third component of weak
isospin, and $\theta_W$ is the weak mixing angle.

In vNRQCD at NNLL order the current correlators are replaced by the correlators
of the non-relativistic currents $\O{p}{i}$, so that
\begin{eqnarray} \label{Rveft}
 R^v(s) & = & \frac{4\pi}{s}\,
 \mbox{Im}\Big[\,
 c_1^2(\nu)\,{\cal A}_1(v,M,\nu) + 
 2\,c_1(\nu)\,c_2(\nu)\,{\cal A}_2(v,M,\nu) \,\Big] \,,
\\[0mm] \label{Raeft}
 R^a(s) & = &  \frac{4\pi}{s}\,
 \mbox{Im}\Big[\,c_3^2(\nu)\,{\cal A}_3(v,M,\nu)\,\Big] \,,
\end{eqnarray}
with  
\begin{eqnarray}
 {\cal A}_1 & = & i\,
 \sum\limits_{\mbox{\scriptsize\boldmath $p$},\mbox{\scriptsize\boldmath $p'$}}
 \int\! d^4x\: e^{i \hat{q} \cdot x}\:
 \Big\langle\,0\,\Big|\, T\, \O{p}{1}(x){\Od{p'}{1}}(0)
 \Big|\,0\,\Big\rangle \,, \nn
\\[0mm]
 {\cal A}_2 & = &
 \frac{i}{2}\, 
 \sum\limits_{\mbox{\scriptsize\boldmath $p$},\mbox{\scriptsize\boldmath $p'$}}
 \int\! d^4x\: e^{i \hat{q}\cdot x}\:
 \Big\langle\,0\,\Big|\,
 T\,\Big[ \O{p}{1}(x){\Od{p'}{2}}(0)+\O{p}{2}(x){\Od{p'}{1}}(0)
 \Big] \Big|\,0\,\Big\rangle \,, \nn
\\[0mm]
 {\cal A}_3 & = & i\, 
 \sum\limits_{\mbox{\scriptsize\boldmath $p$},\mbox{\scriptsize\boldmath $p'$}}
 \int\! d^4x\: e^{i \hat{q}\cdot x}\:
 \Big\langle\,0\,\Big|\, T\,
 \O{p}{3}(x){\Od{p'}{3}}(0)\Big|\,0\,\Big\rangle \,.
\label{correlatordef}
\end{eqnarray}
Here $\hat{q}\equiv(\sqrt{s}-2M,0)$.
The correlators ${\cal A}_i$ are functions of the quark mass $M$, the
velocity scaling parameter $\nu$, and the velocity
\begin{eqnarray} \label{vdefwidth}
  v & = &
 \left(\frac{\sqrt{s}-2M+2\delta M^{}_{LL}}{M}\right)^{\frac{1}{2}}
 \,,
\end{eqnarray}
where $\delta M^{}_{LL}$ is the LL contribution in the relation
between pole mass and the threshold mass that is used for the
computation.
The correlator ${\cal A}_2$ can
be related to ${\cal A}_1$ using the quark equation of motion (see
also Ref.~\citebk{Gremm1}) giving
\begin{eqnarray}
  {\cal A}_2(v,M,\nu) & = & {v^2}\,{\cal A}_1(v,M,\nu) \,.
\end{eqnarray}
The correlators can be computed from the Green function of the
Schr\"odinger equation\ (\ref{NNLLSchroedinger}),
\begin{eqnarray}
{\cal A}_1(v,M,\nu) 
 \, &=& \, 6\,N_c\, \int D^n{\bmp}\,D^n{\bmp}^\prime\, 
      \tilde G({\bmp},{\bmp}^\prime)  \,, \nn\\[2mm]
{\cal A}_3(v,M,\nu) 
 \, &=& \, \frac{12\,N_c}{m^2\:n}\, \int D^n{\bmp}\,
      D^n{\bmp}^\prime\, ({\bmp}.{\bmp}^\prime)\,
      \tilde G({\bmp},{\bmp}^\prime)  \,,
\label{Acorrelators}
\end{eqnarray}
where $N_c=3$ is the number of colors. 

The computation can be divided into two parts. The higher order
contributions coming from the Coulomb potential $\tilde V_c$ do not
lead to any divergences in the absorptive part of ${\cal A}_1$, so
they can be determined in $n=3$ dimensions. An extensive technology
has been developed to carry out the corresponding computations. 
For example, in
Refs.~\citebk{Beneke2,Penin1,Melnikov2} 
various analytic methods were used within time-ordered  
perturbation theory,
\begin{eqnarray}
\delta \tilde G({\bmp},{\bmp}^\prime) & = &
-\int D^n{\bmk}_1 D^n{\bmk}_2
   \tilde G_c({\bmp},{\bmk}_1)
   \delta\tilde V({\bmk}_1,{\bmk}_2)
   \tilde G_c({\bmk}_2,{\bmp}^\prime) +
\ldots
\,,\quad
\end{eqnarray}
where $\tilde G_c$ is the Coulomb Green function that solves the LL
Schr\"odinger equation. In a different approach the Coulombic corrections
were determined by solving the corresponding Schr\"odinger equation
exactly. In Ref.~\citebk{Melnikov3,Yakovlev2,Nagano1} the computations
were carried out in  configuration space, and in 
Refs.~\citebk{Hoang13,Hoang12,Hoang3} they were done in momentum
space based on numerical routines developed in Ref.~\citebk{Jezabek2}. 
The corrections to ${\cal A}_1$ from the $1/(m|{\bmk}|)$ and the
$1/m^2$ potentials, on the other hand, lead to UV divergences, which
are associated with the NLL anomalous dimension of the leading order
current $\O{p}{1}$ and the evolution of $c_1$. 
Here, the computations need to be carried out in $n=3-2\epsilon$
dimensions in the \ms scheme. In general, the scheme has to be the
same as for the determination of the anomalous dimensions and the
matching conditions. The results for ${\cal A}_3$
and for the UV-divergent corrections to  ${\cal A}_1$ in
$n=3-2\epsilon$ dimensions in the \ms scheme have been determined in
Ref.~\citebk{Hoang3}.
It is also necessary to treat the different orders in $\delta M$ 
in the same way as the corrections to the potentials to achieve a
complete cancellation of the large unphysical corrections associated
with the pole mass scheme.

\subsection{$\QQbar$ Production In Fixed Order Perturbation Theory}
\label{subsectionQQbarfixedorder}

In renormalization-group-improved perturbation theory
the quantity $\alpha_s\ln v$ is considered as a
quantity of order $1$. It is summed to all orders in $\alpha_s$ as
indicated in Eq.\ (\ref{RNNLLorders}). After scaling all
coefficients and couplings of the effective theory down from $\nu=1$
to $\nu=v$ in vNRQCD (or from $\mu_p=m$ to $\mu_p=m v$ in pNRQCD) 
QCD logarithms of $v$ are summed into the coefficients
and couplings, and matrix elements of the operators are free of QCD
logarithms of $v$. This is the approach that I have discussed up to this
point in this review and that is used in most modern QCD
computations involving scale hierarchies. Experience has shown that
summing potentially large logarithms can lead to better controlled QCD
predictions. 

It is also possible to consider $\ln v$ not as large. In this
case the term  $\alpha_s\ln v$ is counted of order $\alpha_s$ and the
normalized cross section has the generic form
\begin{eqnarray}
 R & = & \frac{\sigma(e^+e^-\to\QQbar)}{\sigma_{\mu^+\mu^-}}
 \, = \,
 v\,\sum\limits_k \left(\frac{\alpha_s}{v}\right)^k
\nonumber\\ & & \hspace{1cm}
 \times
 \bigg\{1\,\mbox{(LO)}; \alpha_s, v\,\mbox{(NLO)}; 
 \alpha_s^2, \alpha_s v, v^2\,\mbox{(NNLO)}\bigg\}
 \,.
 \label{RNNLorders}
\end{eqnarray}
I will call this counting scheme ``fixed order perturbation
theory''. The indicated terms are leading order (LO), next-to-leading
order (NLO) and next-to-next-to-leading order (NNLO) in fixed order
perturbation theory and each are a subset of the LL order, NLL order
and NNLL order corrections in renormalization-group-improved
perturbation theory.
Of course, the counting $\alpha_s/v\sim 1$ is maintained,
since the summation of the Coulomb singuarity is at the heart of the
non-relativistic expansion and is what makes up the Schr\"odinger
equation.

At present,
most computations of the total $\QQbar$ cross section have in fact
been carried out in fixed order perturbation theory.
In Refs.~\citebk{Hoang1,Melnikov3,Hoang13} cutoff schemes were
employed using NRQCD as the conceptual basis for the computation.
In Ref.~\citebk{Beneke2} dimensional regularization and the threshold
expansion were used.
Since the cutoff schemes still employed the running QCD coupling
$\alpha_s$ from full QCD, the computations that used a cutoff were
unavoidably dependent on several scales, which are a priori
independent.
I will report in Secs.~\ref{subsectionfixedorderttbar} and 
\ref{sectionsumrules} on the fixed order results for
$t\bar t$ production close to threshold and non-relativistic $b\bar b$
sum rules. 
In the fixed order computations only a partial summation of QCD
logarithms is carried out, for example by setting the scale of 
$\alpha_s$ to a non-relativistic scale. From the point of view of the
effective theories pNRQCD and vNRQCD these summations are inconsistent.
For the rest of this section I review the basic technical steps in the
computation of the total $\QQbar$ production cross section at NNLO in
fixed order perturbation theory and discuss the main differences to
the renormalization-group-improved calculation.\\

\begin{figure}[t] 
\begin{center}
\hspace{7mm}
\leavevmode
\epsfxsize=3cm
\leavevmode
\epsffile[250 470 450 720]{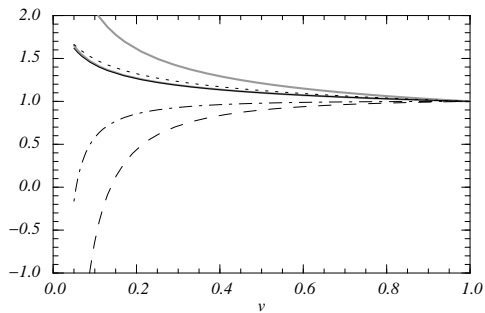}
\begin{picture}(0,0)(1,1)
\end{picture}
 \end{center}
%
%
\vskip  -0.cm
 \caption{\label{figrunningpotential}   
Evolution of renormalization-group-improved potential coefficients in
vNRQCD (black lines) and fixed order coefficients (gray
lines) for $\nu<1$ in the $t\bar t$ system.
}
\end{figure}

\noindent
{\it Schr\"odinger Equation and Potentials}\\[1mm]
At NNLO in fixed order perturbation theory ultrasoft corrections do
not enter anywhere in the computations. In addition, the coefficients
of the operators do not run apart from the evolution of $\alpha_s$ that is
already know from full QCD. The NNLO non-relativistic
$\QQbar$ dynamics is also described by a common
time-independent Schr\"odinger equation, which is similar to Eq.\
(\ref{NNLLSchroedinger}). Here, 
$\delta M=\delta M_{\rm LO}^{}+\delta M_{\rm NLO}^{}+
\delta M_{\rm NNLO}^{}$. The potential $\tilde V$ is constructed
from the NRQCD interactions using Labelle's multipole
expansion\,\cite{Labelle1} (Sec.\ \ref{subsectionnrqcdcutoff}).
The full fixed order potential at NNLO for a $\QQbar$ pair in a
${}^3S_1$ state reads
\begin{eqnarray}
 \tilde V^f({\bmp},{\bmp}^\prime) \, = \, 
 \tilde V_c^f({\bmp},{\bmp}^\prime) + 
 \tilde V_\delta^f({\bmp},{\bmp}^\prime) + 
 \tilde V_r^f({\bmp},{\bmp}^\prime) +
 \tilde V_k^f({\bmp},{\bmp}^\prime)  
 \,,
\label{Vsdetailfixed}
\end{eqnarray} 
where ($a_s=\alpha_s(\mu)$)
\begin{eqnarray}  
 \tilde V_c^f({\bmp},{\bmp}^\prime) & = &
-\,\frac{4\,\pi\,C_F\,a_s}{{\bmq}^2}\,
\bigg\{\, 1 +
\Big(\frac{a_s}{4\,\pi}\Big)\,\Big[\,
-\beta_0\,\ln\Big(\frac{{\bmq}^2}{\mu^2}\Big) + a_1
\,\Big]
\nonumber\\[2mm] & & \quad
 + \Big(\frac{a_s}{4\,\pi}\Big)^2\,\Big[\,
\beta_0^2\,\ln^2\Big(\frac{{\bmq}^2}{\mu^2}\Big)  
- \Big(2\,\beta_0\,a_1 +
\beta_1\Big)\,\ln\Big(\frac{{\bmq}^2}{\mu^2}\Big) 
+ a_2
\,\Big]
\,\bigg\}
\,,
\nonumber \\[2mm]
\tilde V_k^f({\bmp},{\bmp}^\prime) & = & 
 -\frac{\pi^2 C_F(2C_A-C_F)}{2 M |{\bmq}|}\, a_s^2 \,,
\qquad\quad
 \tilde V_\delta^f({\bmp},{\bmp}^\prime) \, = \,
\frac{8 \pi C_F}{3 M^2} a_s \,, 
\nonumber \\[2mm]
 \tilde V_r^f({\bmp},{\bmp}^\prime) & = & 
-\frac{2\pi C_F ({\bmp}^2+{\bmp}^{\prime\,2})}{ M^2 {\bmq}^2}\,a_s \,. 
\end{eqnarray}
In some publications (see e.g.\ Refs.~\citebk{Hoang1,Melnikov3,Hoang13})
slightly different conventions, employing off-shell potentials $\propto
({\bmp}^2-{\bmp}^{\prime 2})$ were used. This
does, however, not affect the final results.
The coefficients of the potentials in the fixed order approach do not
run except for the evolution of $\alpha_s$. To get an impression of
the impact 
of the summation of QCD logarithms, it is instructive to consider
the numerical difference between the fixed order coefficients and the
coefficients in renormalization-group-improved perturbation theory for
scales below $M$. In Fig.\ \ref{figrunningpotential} the ratios 
${\cal V}_c^{(s)}(\nu)/{\cal V}_c^{(s)}(1)$ (solid line),
${\cal V}_k^{(s)}(\nu)/{\cal V}_k^{(s)}(1)$ (dashed line),
$({\cal V}_2^{(s)}(\nu)+2{\cal V}_s^{(s)}(\nu))/
({\cal V}_2^{(s)}(1)+2{\cal V}_s^{(s)}(1))$ (dotted line) and
${\cal V}_r^{(s)}(\nu)/{\cal V}_r^{(s)}(1)$ (dash-dotted line) are
displayed for $\nu<1$ for $M=175$~GeV and $\alpha_s(M)=0.1074$
using four-loop running of the strong coupling. 
The results show the evolution of the potential
coefficients in vNRQCD for
the $t\bar t$ system. For the $b\bar b$ system the results are similar.
The results in pNRQCD agree for the $1/m^2$
potentials, but disagree for the Coulomb and the
$1/(m|{\bmk}|)$ potential. On the horizontal scale of Fig.\
\ref{figrunningpotential} there is no notable numerical difference
between the Coulomb coefficient in vNRQCD (${\cal V}_c^{(s)}$) and the
corresponding expression in pNRQCD, see Eqs.\ (\ref{VCoulombvnrqcd})
and (\ref{VCoulombpnrqcd}).
In the fixed order approach the ratios simply reduce to 
$\alpha_s^2(M v)/\alpha_s^2(M)$ for the $\tilde V_k$ potential 
(upper gray line) and to 
$\alpha_s(M v)/\alpha_s(M)$ for all the other potentials (lower
gray line). 
The curve for $\alpha_s(M v)/\alpha_s(M)$ is on top of the
solid line that displays the running of 
${\cal V}_c^{(s)}(\nu)/{\cal V}_c^{(s)}(1)$. This shows that the
contributions in the three-loop anomalous dimension of 
${\cal V}_c^{(s)}(\nu)$ that are induced by the ultrasoft corrections
are quite small for the $t\bar t$ system. 

We find that the evolution in the fixed order and the
renormalization-group-improved approach roughly agrees for the Coulomb
coefficient and for the $\tilde V_\delta$ potential. Up to small
corrections, both evolve basically with the QCD $\beta$-function. For
the $\tilde V_k$ and the $\tilde V_r$ potentials, however, the
evolution of the fixed order and the renormalization-group-improved
coefficients is profoundly different. Whereas the fixed order
coefficients grow with $\alpha_s$ for smaller scales, the
renormalization-group-improved coefficients become smaller. 
Expanding for example the vNRQCD coefficient ${\cal V}_k^{(s)}(\nu)$ 
in terms of $\alpha_s(m)$, we find\,\cite{Manohar3}
\begin{eqnarray} 
 {\cal V}_k^{(s)}(\nu) &=& 
  \Big( \frac{C_F^2}{2}-C_FC_A \Big) \alpha_s^2(m)  
  -\frac{\beta_0}{\pi}\Big( \frac{C_F^2}{2}-C_FC_A \Big)
  \alpha_s^3(m)\ln(\nu)
\nonumber\\ [1mm]
 &&  -\,\frac{4 C_A C_F (C_A+2 C_F)}{3\pi}\: {\alpha_s^3(m)}\ln(\nu) +\ldots 
  \,. 
\end{eqnarray}
In the expansion of the corresponding fixed order coefficient the
order $\alpha_s^3\beta_0$ term agrees, but the third term is missing,
because the anomalous dimension of the $\tilde V_k$ potential is not
accounted for properly.
Numerically the coefficient of the third term is larger than the
coefficient of the second term and has a
different sign. This explains the different running for scales below
$M$, and leads to a profoundly different behavior of  higher order 
corrections in both approaches.
\\

\noindent
{\it Currents and Total Cross Section}\\[1mm]
In the fixed order approach the $\QQbar$ production currents can be
defined in analogy to Eq.\ (\ref{vnrqcdcurrents}), see
e.g.\ Refs.~\citebk{Hoang12}, and the computation of the correlators 
${\cal A}_{1-3}$ follows the lines of Sec.\
\ref{subsectioncrosssectionvnrqcd}.  
Since a consistent summation of higher order QCD logarithms is not
intended, the basic requirement is that the regularization scheme used
in the solution of the Schr\"odinger equation in Eq.\
(\ref{NNLLSchroedinger}) is also employed in the matching procedure.
For the currents $\O{p}{2-3}$ Born matching is
sufficient.
For the dominant current $\O{p}{1}$ two-loop matching has to be
carried out. 

For example, in Ref.~\citebk{Hoang12} a cutoff scheme was used,
where the Schr\"odinger equation in Eq.\ (\ref{NNLLSchroedinger}) was
solved in a three-dimensional momentum space sphere with radius
$\Lambda\sim M$. The coefficient $c_1$ was then computed by demanding
that the total cross section in NRQCD equals the total cross section
in full QCD at order $\alpha_s^2$ in the limit $\alpha_s\ll v\ll 1$.
The resulting expression for $c_1$ is a complicated function of 
$M$ and $\Lambda$ and also contains the power counting breaking
effects discussed at the end of Sec.\ \ref{subsectionnrqcdcutoff}.
This matching procedure at the level of the total cross section
is called ``direct matching''\,\cite{Hoang1} and also works for ad-hoc 
regularization schemes that do not have a straightforward definition
in mathematical terms. For example
in Refs.~\citebk{Hoang1} a cutoff $\Lambda$ was suggested, where the
solution of Eq.\ (\ref{NNLLSchroedinger}) was expanded for
$\Lambda\to\infty$. Logarithmic divergent terms were kept and linear,
quadratic, etc. divergences dropped by hand. This scheme was later
adopted by all fixed order computations at NNLO except for
Refs.~\citebk{Beneke2,Beneke6}. In this scheme also the 
power counting breaking effects just mentioned are removed by hand. 
The resulting expression for $c_1$ is simpler than in
Ref.~\citebk{Hoang12} and has the same order
$\alpha_s$ contribution as Eq.\ (\ref{c1matchingnrqcd}).
The scale of $\alpha_s$ in $c_1$ obtained in this scheme is  
independent of the scale of $\alpha_s$ in the correlators.

\vspace{1cm}

\section{Top Quark Pair Production at Threshold}
\label{sectionttbar}

Top--antitop quark pair production close to the threshold
will provide an integral part of the top quark physics program at the
Linear Collider, which is supposed to be the next major accelerator
project after the LHC. It is fair to say that top threshold physics
was the main motivation for the conceptual and technical progress in
heavy quarkonium physics in the recent years.
The theoretical interest in the top--antitop
quark threshold arises from the fact that the large top quark mass 
and width ($\Gamma_t\approx 1.5$~GeV) lead to a suppression of
non-perturbative effects.\cite{topthresholdfoundations,Fadin1}
Effectively, the top quark velocity is 
\begin{eqnarray}
v_{\rm eff}^{}=\left(\frac{\sqrt{s}-2M+2\delta M+i\Gamma_t}{M}
\right)^{1/2}
\,,
\label{veffdef}
\end{eqnarray}
so the hierarchy 
$M_t\gg M_t|v_{\rm eff}^{}|\gg M_t |v_{\rm eff}^{}|^2\gg \lqcd$
is satisfied for any energy in the threshold region.
This makes top threshold physics an ideal application of the
non-relativistic effective theories reviewed in the previous
sections. In particular, perturbative methods are a reliable tool to
describe the physics of non-relativistic $t\bar t$ pairs and allow
for measurements of top quark properties directly at the parton
level. 

Due to the large top width the total $t\bar t$ production cross
section line shape is a smooth function of the energy, which rises
rapidly at the point where the remnant of a toponium 1S resonance can
be formed. From the energy where this increase occurs, the top quark
mass can be determined, whereas shape and height of the cross section
near  threshold  can be used to determine $\Gamma_t$, the coupling
strength of top quarks to gluons and, if the Higgs boson is not heavy,
the top Yukawa coupling.\cite{Strassler1,Harlander1}
For the total top quark width only very few other methods to determine
it are known. From differential
quantities, such as the top momentum
distribution,\cite{Jezabek2,Suminoold} the forward--backward
asymmetry or certain leptonic
distributions,\cite{otherNLO,NLOnonfactordist} one can obtain
measurements of $\Gamma_t$, the top quark spin and possible anomalous 
couplings.\cite{Jezabek3}

\begin{figure}[t] 
\begin{center}
\epsfig{file=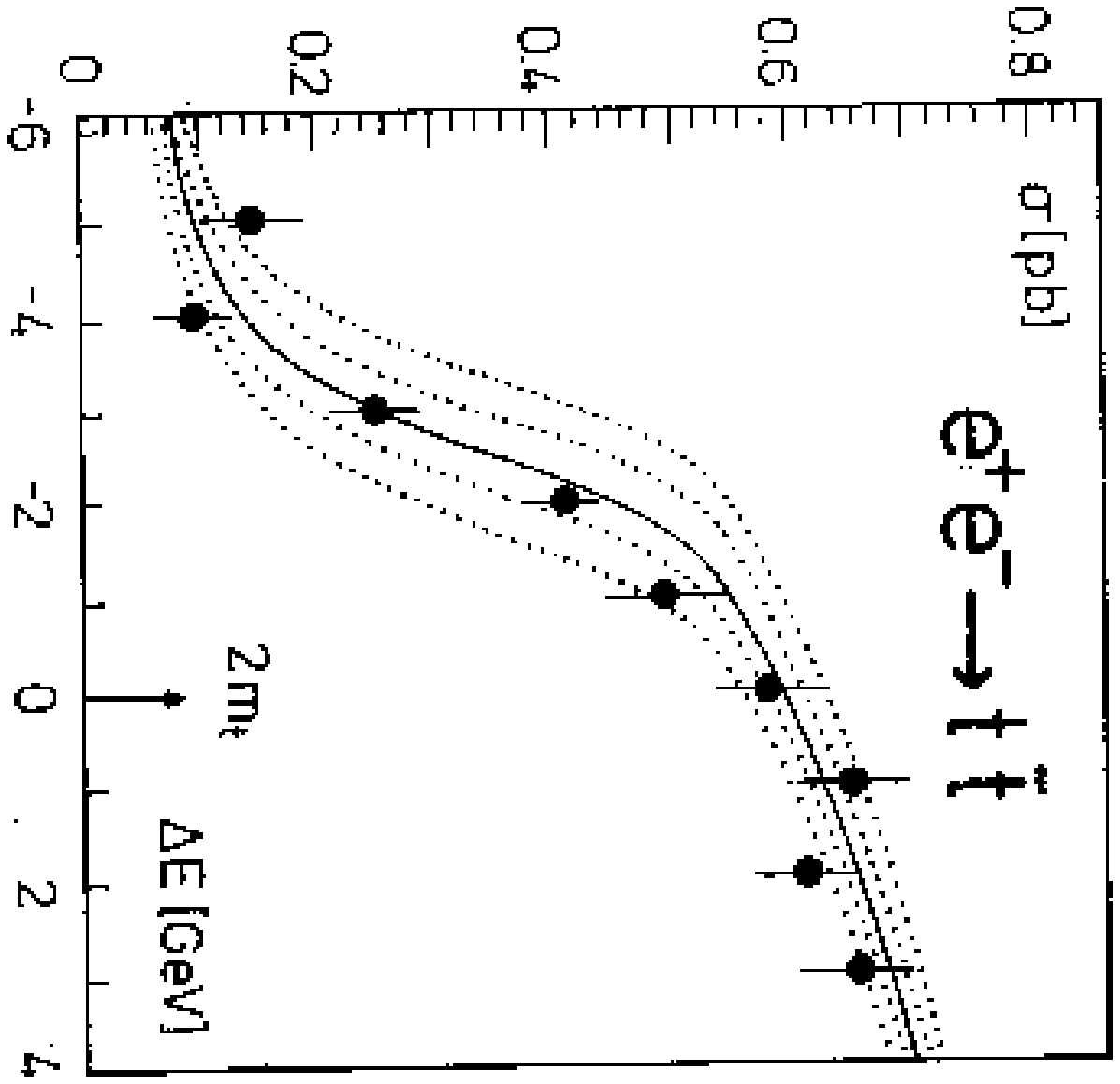,width=3.5cm,angle=89}
\hspace{10mm}
\leavevmode
\epsfxsize=2.6cm
\epsffile[20 20 370 120]{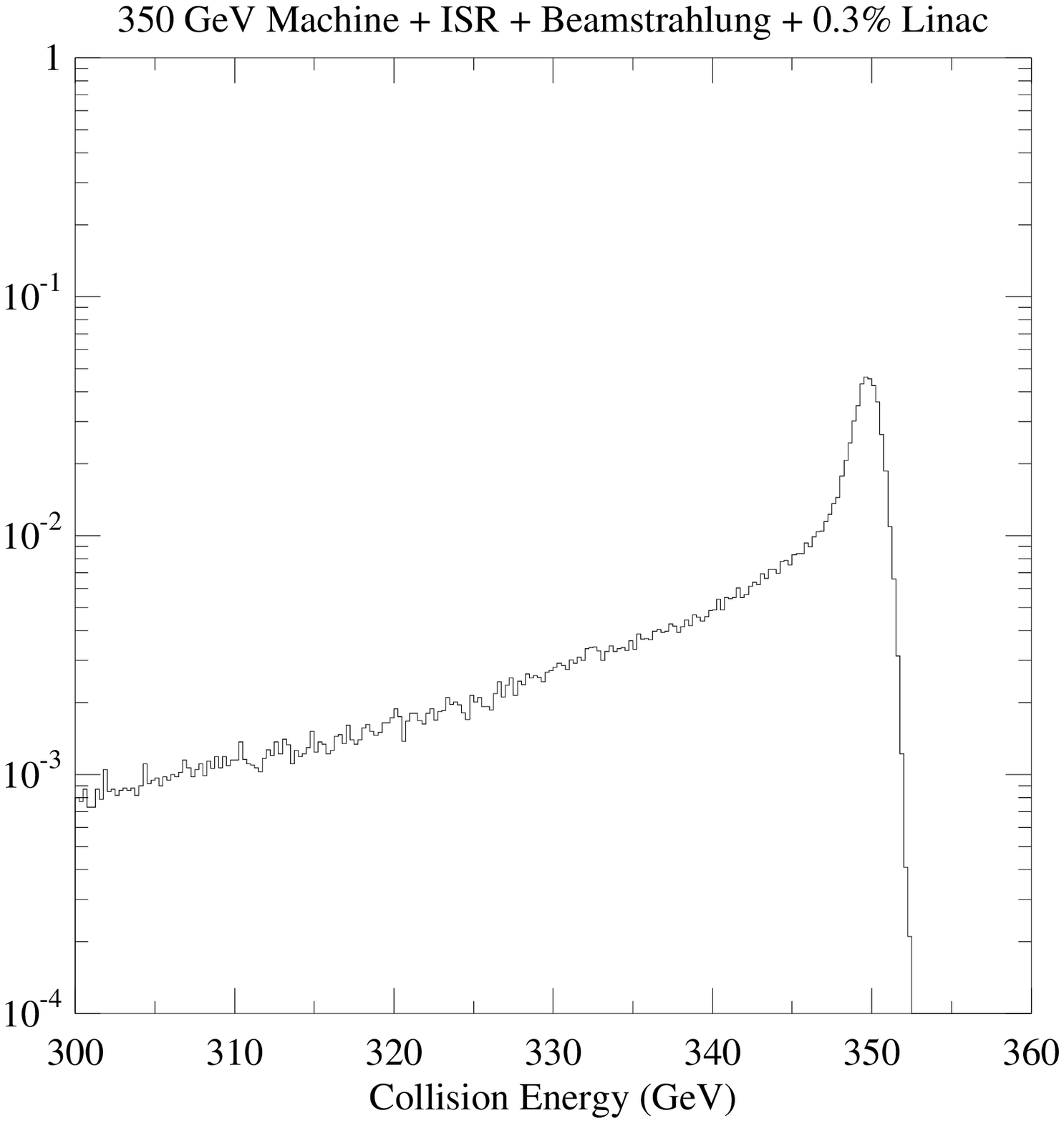}
\hspace{10mm}\mbox{}
\begin{picture}(0,0)(1,1)
\put(-260,90){$a)$}
\put(-115,90){$b)$}
\end{picture}
 \end{center}
%
%
\vskip  -0.cm
 \caption{
(a) Sensitivity of the $t\bar t$ line shape, convoluted by the TESLA
$e^+e^-$ beam luminosity spectrum, to the top quark mass. The solid
line corresponds to the nominal mass, while the dotted lines to values
which differ from the nominal value by $\pm 200$ and $\pm 400$~MeV. The
diagram is taken from an early DESY simulation 
study in Ref.~\protect\citebkcap{Bagliesi1}, where the
``experimental'' points and error bars each correspond to total
luminosity of $1$~fb$^{-1}$. Theoretical uncertainties are not taken
into account. 
(b) Typical form of the luminosity spectrum at the Linear Collider at nominal
beam energy of $350$~GeV due to initial state radiation, beamstrahlung
and beam energy spread from Ref.~\protect\citebkcap{Cinabro1}. 
\label{figtopthreshexp}
}
\end{figure}
A considerable number of experimental 
studies were carried out in the past
to assess the feasibility of such
measurements.\cite{Bagliesi1,ttbarexperimenta,Cinabro1,Peralta1,Martinez1}
The measurement of the top quark
mass from a threshold line shape scan is
particularly interesting, see Fig.~\ref{figtopthreshexp}a for an early
simulation study.\cite{Bagliesi1} In contrast to the standard top mass
determination method, which relies on the reconstruction of the
invariant mass of jets originating from the decay of
a single top quark, the line shape measurement  
has the advantage that only color-singlet $t\bar t$ events have to be
counted. Therefore, the effects of final state interactions, 
uncertainties from hadronization modeling, etc. are suppressed, and
systematic uncertainties in the top mass determination are small. 
Recent simulation studies have shown that, for a
total luminosity of $100$~fb$^{-1}$, statistical and
systematical experimental uncertainties in the top mass determination 
are well below 50~MeV.\cite{Peralta1,Martinez1}
Such a precision would be relevant for example in supersymmetric
scenarios, where the lightest CP-even Higgs mass can have a very
strong dependence on the top quark mass due to top-stop
loops.\cite{tophiggsmssm}

\subsection{Theoretical Achievements}

With the excellent experimental prospect in view it is obvious that a
careful analysis and 
assessment of theoretical uncertainties in the
prediction of the total cross section and various distributions 
is mandatory, in order to
determine whether the theoretical precision can meet the experimental
one. Initially, a number of leading order\,\cite{LOttbar} and
next-order
computations,\, \cite{Strassler1,Jezabek2,Suminoold,otherNLO,NLOnonfactordist,NLOttbar}  
were carried out. The latter relied basically on QCD-inspired
potential models that used phenomenological input from
$\Upsilon$ and 
charmonium data. As such they did not represent true first-principles
QCD calculations, and there was no systematic way how the computations
could be consistently improved to include higher order radiative or
relativistic corrections.

After adoption of the concepts of effective theories in the framework
of NRQCD (Sec.\ \ref{sectionNRQCD}), consistent fixed order NNLO computations 
appeared in 
Refs.~\citebk{Hoang1,Hoang13,Melnikov3,Yakovlev2,Beneke2,Nagano1,Hoang12,Penin2,Yakovlev1}.
The results were not just some new higher order corrections, but 
led to a number of surprising and important insights.
The NNLO corrections to the location where the cross section rises
and the height of the cross section were found to be much larger than
expected from the results at NLO. The large corrections to  
the location of the rise were found to be an artifact of the on-shell
pole mass definition and it was realized that  
the top pole mass cannot be extracted with an uncertainty
smaller than ${\cal{O}}(\Lambda_{\rm QCD})$ from non-relativistic
heavy quark--antiquark systems 
(Sec.\ \ref{sectionquarkmass}).
Subsequently, carefully designed threshold mass definitions were
suggested to allow for a stable extraction of the top quark
mass parameter (Sec.\ \ref{sectionquarkmass}). 
The remaining uncertainties in the normalization of the 
cross section were estimated\,\cite{Hoang11} to about $20$\% and seemed to
jeopardize precise measurements of the top width, 
the top quark coupling to gluons and the Higgs boson. 
Some N$^3$LO logarithmic corrections to the cross section were found to
be large too\,\cite{Kniehl2,Kniehl3} and at present the only way to
possibly improve 
the situation in the fixed order approach seems to be the
determination of N$^3$LO corrections in order to learn more
about the structure of the perturbative series.      

An alternative way is to reorganize the perturbative series by using
re\-nor\-ma\-li\-zation-group-improved perturbation theory 
(Sec.\ \ref{subsectioncrosssectionvnrqcd}), which counts 
$\alpha_s\ln v$ of order one and sums all QCD logarithms of $v$
at a certain order 
into the coefficients of the effective theory. Such a treatment has
become standard in many other areas of QCD where physics is governed
by widely separated scales.
The theoretical
framework that allows to carry out such a summation consistently is
more sophisticated than NRQCD (Sec.\ \ref{sectionNRQCD}). Currently,
there are two alternative  
formulations in the literature, called pNRQCD (Sec.\
\ref{sectionpNRQCD}) and vNRQCD (Sec.\
\ref{sectionvNRQCD}). The two effective theories are not equivalent
(Sec.\ \ref{sectionvNRQCDvspNRQCD}). 
Recently, a renormalization-group-improved 
calculation in vNRQCD of the total cross section at NLL order and
including most NNLL order corrections has been provided in
Refs.~\citebk{Hoang3}. It was found that the normalization uncertainties are
an order of magnitude smaller and the remaining relative uncertainty
in the normalization of the cross section was
estimated at the level of $3$\%.

In the following sections I discuss how the top quark width 
and non-perturbative effects are
implemented into computations of the non-relativistic $t\bar t$
dynamics, and I give a more detailed
discussion of results for the total cross section in fixed order
perturbation theory and the recent results obtained in vNRQCD.

\subsection[Top Width and Non-Perturbative Effects]
{Top Width and Non-Perturbative Effects in the Total Cross Section}

For computations of the total cross section
the top width can be understood as a modification of the coefficients  
of the effective theory caused by electroweak corrections, since one
is not interested in the dynamics of the top quark decay. (In general,
the top decay would be associated with external currents similarly to
the current that produces the $t\bar t$ pair.)
Because the particles involved in these corrections are lighter
than the top quark, they can lead to non-zero imaginary parts in
the coefficients of the effective theory. This is a
known concept in quantum mechanics of inelastic processes.
In the context of top quark threshold production the effects of the
top quark decay are the most important part of a whole array of
electroweak corrections. 

The dominant effect of the top quark width is obtained by including
the electroweak top quark selfenergy in the matching conditions for
the coefficients of the bilinear top operators. Since the matching is
determined for on-shell top quarks, the dominant effect of the top
decay is that of an imaginary mass term proportional to the top quark
width being added to the effective Lagrangian.  
For example, in vNRQCD  the operators
\begin{eqnarray} \label{Gtop}
 \delta {\cal L} = \sum_{{\bmp}} \psip{p}^\dagger \: \frac{i}{2} \Gamma_t\: 
  \psip{p} \ +\  
  \sum_{{\bmp}}   \chip{p}^\dagger\: \frac{i}{2} \Gamma_t\: \chip{p} \,.
\end{eqnarray}
have to be added to the Lagrangian, $\Gamma_t$ being the on-shell top
decay width.
In the Standard Model the dominant decay channel is $t\to bW^+$ and gives a
width of $\Gamma_t=1.43\,{\rm GeV}$ for $m^t_{\rm pole}=175$~GeV
including one-loop electroweak\,\cite{Beenakker1} 
(see also Ref.~\citebk{JezabektopWwidth})
and two-loop QCD\,\cite{Czarnecki6} 
effects. 
In the effective theories, $\Gamma_t$ is treated as
a fixed parameters that can be determined from experiment. 
Since the typical energy of the top quark is $E\sim M|v_{\rm eff}^{}|^2\sim 
4\,{\rm GeV}$, it is natural in the Standard Model to count $\Gamma_t$
of order $Mv^2$. Thus, the propagator of a top quark
with momentum $(p^0,{\bmp})$ is 
\begin{eqnarray}
  \frac{i}{p^0 - \frac{{\bmp}^2}{2M}+\delta M + \frac{i}{2}\Gamma_t +
i\delta} 
\,.
\end{eqnarray}
In extensions of the Standard Model, where the total top width is
different, the $v$-counting might have to be modified.
The use of this propagator is equivalent to the replacement\,\cite{Fadin1} $E\to
E+i\Gamma_t$ in $t\bar t$ Green functions obtained from
the Schr\"odinger equation for stable quarks
in Eq.\ (\ref{NNLLSchroedinger}), i.e.\ it is
sufficient to replace $v$ by the effective velocity $v_{\rm eff}^{}$
defined in Eq.\ (\ref{veffdef}) in the Schr\"odinger equation of Eq.\
(\ref{NNLLSchroedinger}). 
For the total $t \bar t$ cross section the
operators in Eq.~(\ref{Gtop}) have been shown to be sufficient to
account for all electroweak effects at LO\,\cite{Fadin1} and 
NLO\,\cite{nonfactorizable} in the fixed order approach. At NNLO the
treatment of electroweak effects is more involved, since also first
order electroweak corrections to top quark production and decay and
background processes consistent with the top decay final state need to
be taken into account.  
The complete set of electroweak effects at NNLO order is currently
unknown. However, some one-loop electroweak corrections to the 
$t\bar t$ production current\,\cite{Guth1} and some non-resonant
background contributions\,\cite{Biernacik1} have been computed.
Also, the structure of some higher order effects
associated with the electroweak top quark selfenergy is understood.
The next order corrections from the selfenergy leads
to a top wave function correction in the effective theory, 
which is up to a different sign equal to the corresponding
vertex correction to the gluon-quark coupling contained in the Coulomb
potential. In computations of energy levels both contributions cancel
due to gauge-invariance. The remaining subdominant width effect
arises from the non-relativistic expansion of the top quark spinors
of full QCD. In vNRQCD this leads to the following additional operators 
\begin{eqnarray} \label{Gtop2}
 \delta {\cal L} = - \sum_{{\bmp}} \psip{p}^\dagger \: \frac{i}{2} \Gamma_t\: 
  \frac{{\bmp}^2}{2 M^2}
  \psip{p} \ -\  
  \sum_{{\bmp}}   \chip{p}^\dagger\: \frac{i}{2} \Gamma_t\: 
  \frac{{\bmp}^2}{2 M^2} \chip{p} \,.
\end{eqnarray}
Since the effective Lagrangian describes the non-relativistic $t\bar
t$ dynamics in the center-of-mass frame, these operators 
can be interpreted as a time-dilatation effect 
$\sim \Gamma_t(1-\langle\frac{{\bmp}^2}{2M^2}\rangle+\ldots)$
that arises from
the Fermi motion of the individual top quarks in the $t\bar t$ system
at rest.
The cancellation between the electroweak wave function corrections and
the electroweak vertex corrections to the Coulomb potential was 
verified at the one-loop level by explicit
calculation in Ref.~\citebk{Modritsch1}
(see also Ref.~\citebk{Jezabekwidth}). For muonic atoms 
it was noted before in Ref.~\citebk{Uberall1} and later again in 
Ref.~\citebk{Czarnecki7}.
An analogous discussion for the $B_c$ lifetime can be found in
Ref.~\citebk{Beneke7}. 

\begin{figure}[th] 
\begin{center}
\leavevmode
\epsfxsize=5cm
\leavevmode
\epsffile[190 590 420 660]{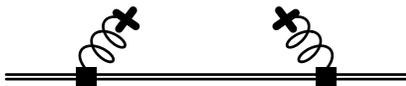}
 \end{center}
%
%
 \caption{\label{fignonpertG2}   
Leading non-perturbative correction in the operator product expansion
due to the vacuum average of two chromoelectric fields.
}
\end{figure}

The large top quark width plays a crucial role in suppressing the size of
non-perturbative hadronic contributions governed by the scale $\Lambda_{\rm
QCD}$. Even arbitrarily close to the  threshold point,  
the $t\bar t$ system has an effective energy
set by the perturbative scale $\Gamma_t$, see Eq.\
(\ref{veffdef}). For stable quarks, an operator product 
expansion\,\cite{Voloshin1,Leutwyler1} in powers of 
$\lqcd/E$ can be
used to incorporate non-perturbative contributions.  
The first non-perturbative correction arises from the radiation and
absorption of a gluon with an energy of order $\lqcd$. 
In the framework of pNRQCD and in configuration space
representation the correction can be understood from the insertion of
two ${\bmr}.{\bmE}$ interactions, where the gluon field carries no
energy and momentum as shown in Fig.\ \ref{fignonpertG2}.
Due to Lorentz and gauge invariance the one-particle-irreducible
two-point function of the two chromoelectric fields is related to the
vacuum average (or vacuum condensate) of two gluon field strength
tensors, $$\langle 0|E_i^A E^{B}_{k}|0\rangle=
-\frac{1}{96}\delta_{ik}\delta^{AB}\langle 0|G_{\mu\nu}^C 
G^{\mu\nu\,C}_{}|0\rangle\,.$$ The resulting leading correction to the
color singlet Green function of the Schr\"odinger equation in
configuration space representation reads\,\cite{Voloshin1}
\begin{eqnarray}
\delta G_s({\bmr},{\bmr}^\prime) & = &
-\frac{\pi}{18}\langle 0|\alpha_s\,G_{\mu\nu} G^{\mu\nu}_{}|0\rangle
\,
\int\! d^3{\bmx}\,\int\! d^3{\bmy}\,{\bmx}.{\bmy}\,
 G_s({\bmr},{\bmx}) G_o({\bmx},{\bmy}) G_s({\bmy},{\bmr}^\prime)
\,,
\nonumber\\
\label{condensate1}
\end{eqnarray}
where $G_o$ stands for the Green function of the Schr\"odinger
equation with the octet Coulomb potential.
For $t\bar t$ production the relative size of the gluon condensate
correction is of order $[\Lambda_{\rm QCD}/(E+i\Gamma_t)]^4\times
v_{\rm eff}^{2}\sim 10^{-4}$. The first factor arises for dimensional
reasons and the second term from the ${\bmp}.{\bmA}/M$
interaction that comes from the multipole expansion.
The radiation of the gluon off the potential 
(Fig.\ \ref{figpnrqcdradiation}) gives the same estimate. 
An explicit expression for the gluon condensate corrections to the
correlator ${\cal A}_1$ was derived in Ref.~\citebk{FadinYakovlev1}, 
\begin{eqnarray}
\lefteqn{
\delta {\cal A}_1(v_{\rm eff}^{},M,\nu) \, = \,
\frac{9\,i\,C_F^2\,\alpha_s^2}{8\,M^2\,v_{\rm eff}^{7}}\,\,
\langle 0|\alpha_s\,G_{\mu\nu} G^{\mu\nu}_{}|0\rangle\,
}
\nonumber\\ & & \qquad \times \,
\sum_{n=0}^\infty\,
\frac{\Gamma(n+4)\Gamma^2(n-\lambda)}
{(n+2+\frac{\lambda}{8})\Gamma(n+1)\Gamma^2(n+5-\lambda)}
\hspace{1cm}
\end{eqnarray}
where $\lambda=-C_F\alpha_s/(2v_{\rm eff}^{})$ and 
$\Gamma$ is the gamma function.   
The size of this correction for the (conservative) literature value
$\langle 0|\alpha_s\,G_{\mu\nu} G^{\mu\nu}_{}|0\rangle=0.05\pm
0.03\,\mbox{GeV}^4$ 
is consistent with the previous parametric
estimate and can be safely neglected in view of the current perturbative
uncertainties. It is a quite interesting question, whether this
estimate of the size of non-perturbative corrections is
indeed correct. 

\subsection{Cross Section at NNLO in the Fixed Order Approach}
\label{subsectionfixedorderttbar}

In the fixed order approach NNLO computations of the $t\bar t$ cross
section close to threshold in $e^+e^-$ annihilation have been carried
out by Hoang et al.,\,\cite{Hoang13,Hoang12} 
Melnikov et al.,\,\cite{Melnikov3}  Yakovlev,\,\cite{Yakovlev2}
Beneke et al.,\,\cite{Beneke2}
Nagano et al.,\,\cite{Nagano1} and
Penin et al.\,\cite{Penin2} Penin et al.\,\cite{Penin2} have also
provided a NNLO computation of the $t\bar t$ cross section in
$\gamma\gamma$ collisions. Since the fixed order computations
basically use NRQCD as the  
conceptual framework (Sec.\ \ref{sectionNRQCD}), which does not fix
a preferred convention for the regularization of UV divergences,
the results obtained by the various groups are, partly, quite
different.  The differences of the results in the various conventions
and regularization schemes
can be interpreted as differences in the treatment of corrections
from beyond NNLO and serve as an important tool to 
estimate the inherent uncertainties in the fixed order approach.

In Ref.~\citebk{Hoang13} Hoang et al. solved the momentum space
Schr\"odinger equation in Eq.\ (\ref{NNLLSchroedinger}) exactly for
the Coulomb potential based on the numerical routines of
Ref.~\citebk{Jezabek2}; the other $1/m$ suppressed contributions where
calculated with 
time-ordered perturbation theory using a momentum cutoff keeping only 
terms that depend logarithmically on the cutoff and discarding
terms that depend linearly, quadratically, etc. on the cutoff.
In Ref.~\citebk{Hoang12} Hoang et al. solved the complete NNLO
Schr\"odiner equation exactly in a momentum sphere with radius
$\Lambda\sim M$. In Ref.~\citebk{Melnikov3} Melnikov et al. solved the
NNLO Schr\"odinger equation exactly in configuration space
representation for the Green function $G(0,r_0)$, where the distance 
$r_0$ served as the cutoff parameter. They also kept only terms that a
logarithmically dependent on $r_0$ and discarded power-like divergences.
The method was also employed by Yakovlev in Ref.~\citebk{Yakovlev2}
and Nagano et al. in Ref.~\citebk{Nagano1}. A similar regularization
method was employed by 
Penin et al in Refs.~\citebk{Penin2}, but the entire Schr\"odinger
equation was solved with 
time-ordered perturbation theory supplemented by partial resummations
to avoid multiple energy poles. All methods used the direct matching
procedure\,\cite{Hoang1} to determine the coefficient $c_1$ of the
dominant $t\bar t$ production current. The results obtained in these
cutoff schemes had a 
residual dependence on three different scales, the renormalization
scale $\mu$  of $\alpha_s$ in the potentials, the cutoff
and the renormalization scale of $\alpha_s$ in the coefficient $c_1$. 
A different strategy was used by Beneke et al. in 
Refs.~\citebk{Beneke2}. They used dimensional regularization identifying
the contributions from the various momentum regions in Eq.\
(\ref{momentumregions}), i.e.\ potentials, coefficients, etc., 
using the threshold expansion and solving the Schr\"odinger equation
in time-ordered perturbation theory. In this approach the potentials
and coefficients were determined without an explicit matching 
calculation. The result had  a
residual dependence on two different scales, the renormalization scale
$\mu$ of $\alpha_s$ in the potentials, and the factorization scale.
In addition, a summation of some NLL order logarithms was 
included in $c_1$. The prescription used for this summation, however,
is not consistent with the known NLL order anomalous dimensions in
pNRQCD or vNRQCD. 
In addition, $c_1$ was not included as a global factor of the
correlator ${\cal A}_1$ but expanded out.  

\begin{figure}[t!] 
\begin{center}
\leavevmode
\epsfxsize=3.cm
\leavevmode
\epsffile[220 580 420 710]{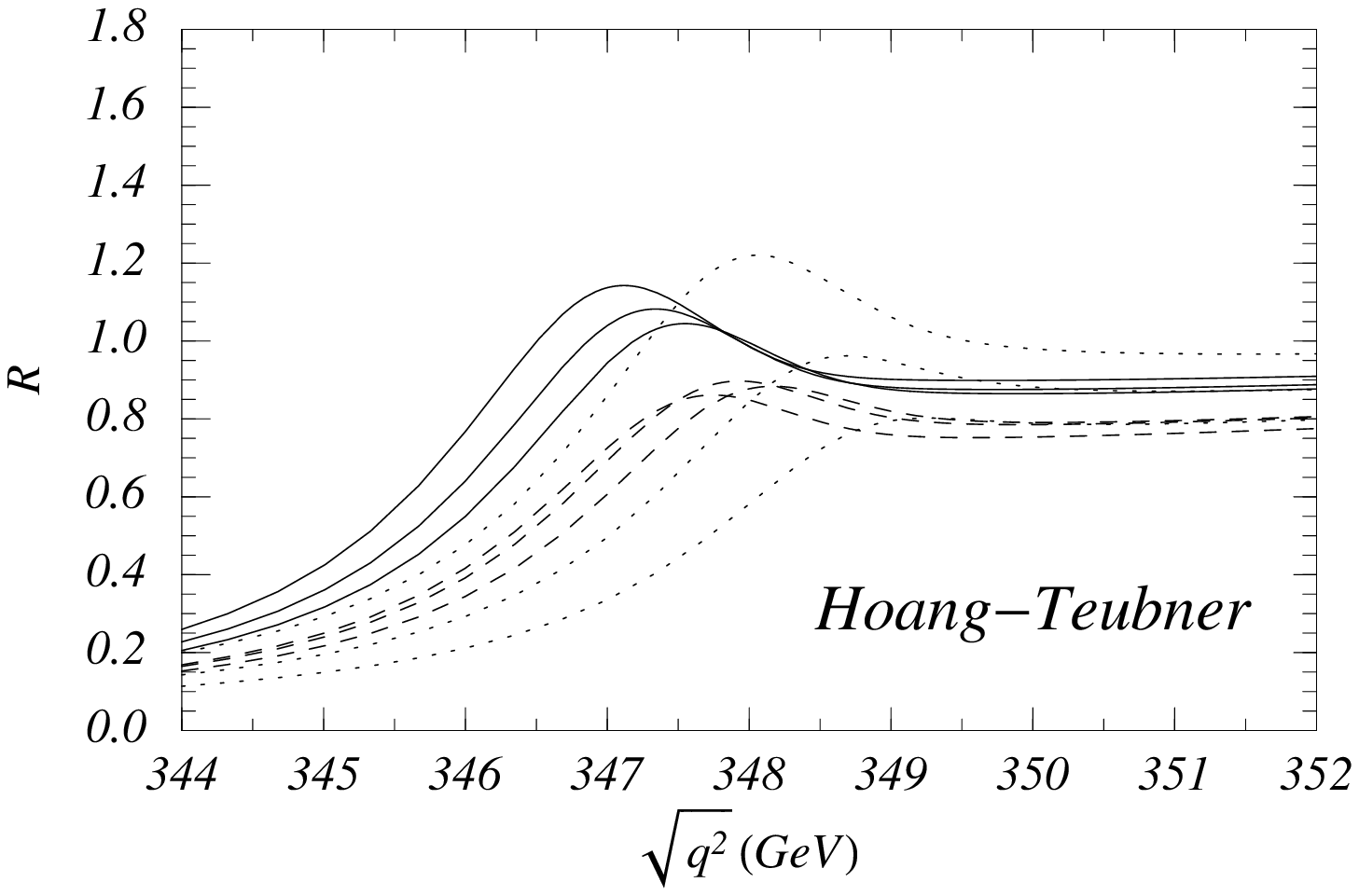}
\hspace{3.1cm}
\epsfxsize=3.cm
\leavevmode
\epsffile[220 580 420 710]{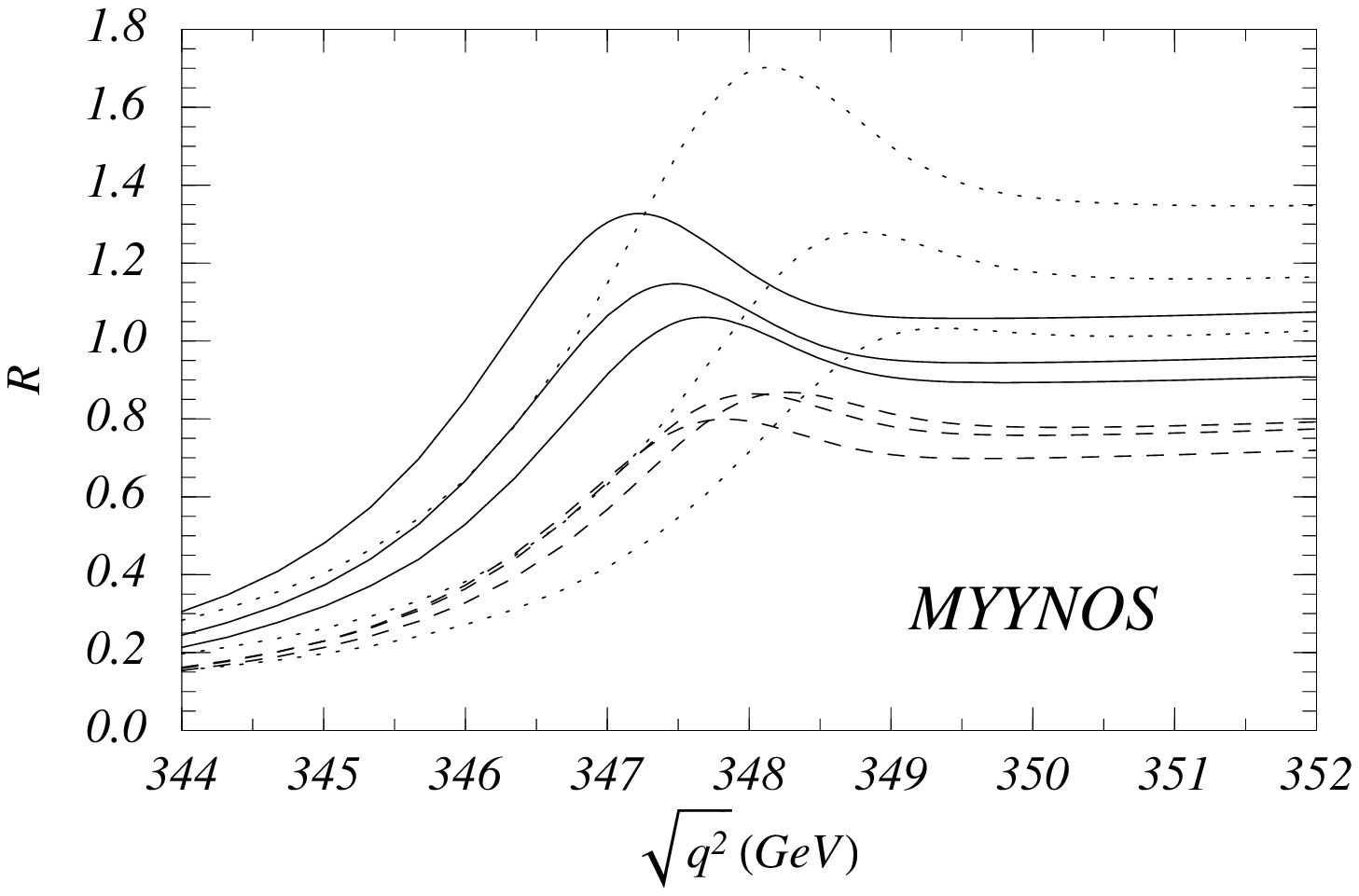}
\vskip 2.3cm
\leavevmode
\epsfxsize=3.cm
\leavevmode
\epsffile[220 580 420 710]{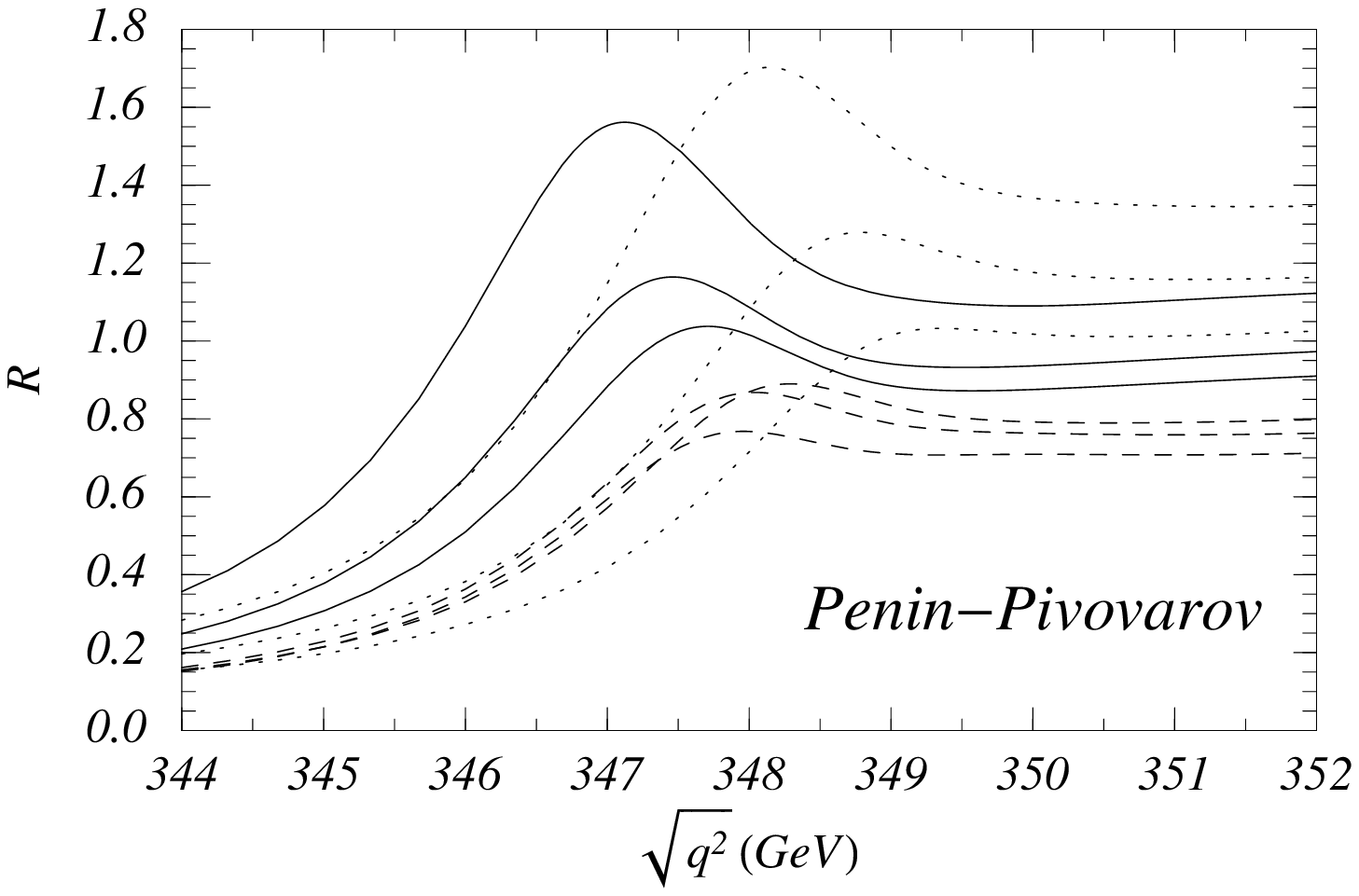}
\hspace{3.1cm}
\epsfxsize=3.cm
\leavevmode
\epsffile[220 580 420 710]{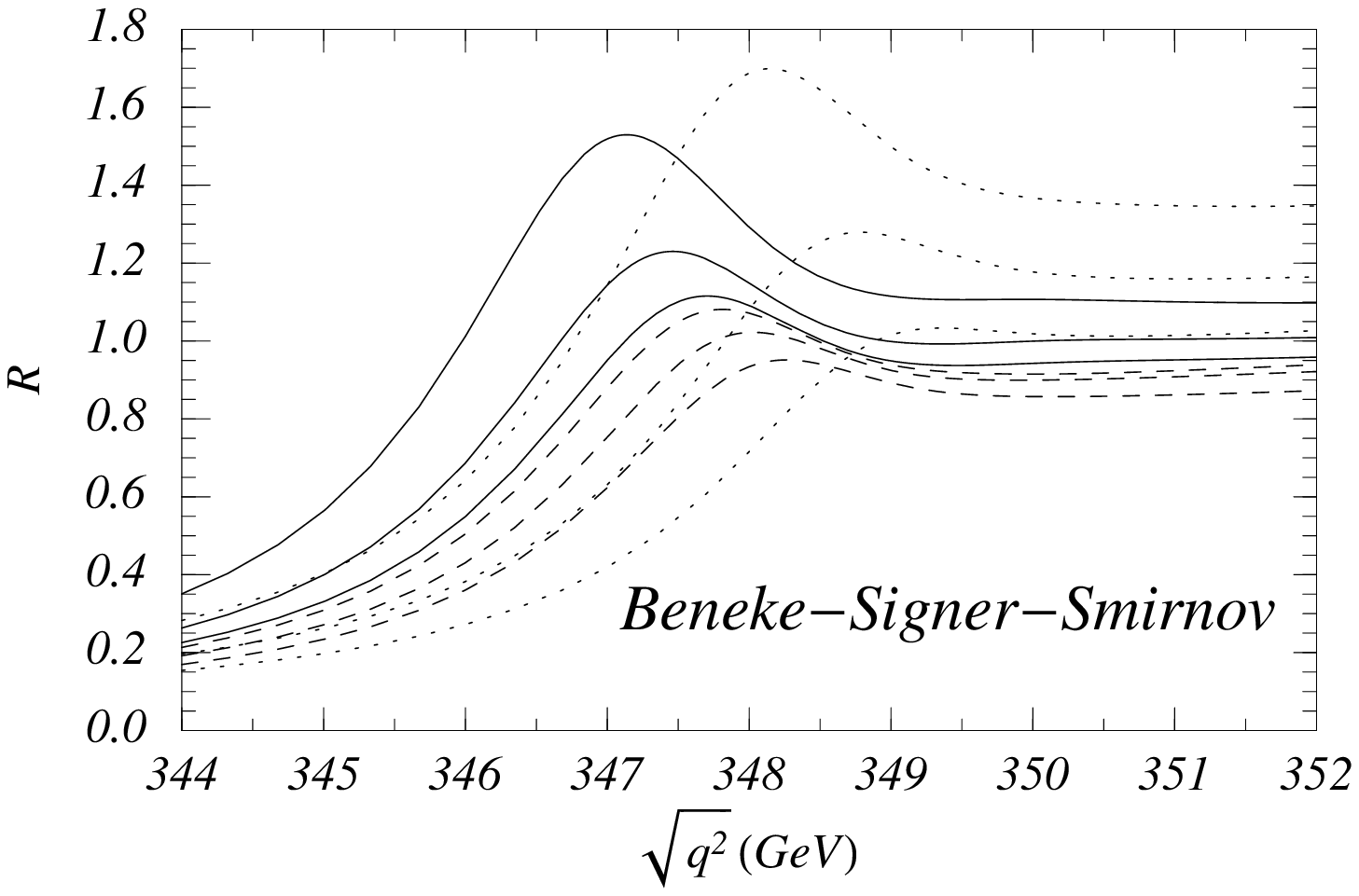}
\vskip  1.8cm
 \caption{\label{figpolettbarfixed}   
The total normalized photon-induced $t\bar t$ cross section  $R$ at
the Linear Collider versus the c.m. energy in the threshold regime at LO (dotted
curves), NLO (dashed) and NNLO (solid) in the pole mass scheme for
$m_t^{\rm pole}=175.05$~GeV,   
$\alpha_s(M_Z)=0.119$ , $\Gamma_t=1.43$~GeV and $\mu=15$,
$30$, $60$~GeV.  
The plots were given in Ref.~\protect\citebkcap{Hoang11} from results provided by
Hoang-Teubner (HT), 
Melnikov-Yelkhovsky-Yakovlev-Nagano-Ota-Sumino
(MYYNOS), Penin-Pivovarov (PP) and Beneke-Signer-Smirnov (BSS).
}
 \end{center}
\end{figure}

In Ref.~\citebk{Hoang11} the results for the normalized total photon-induced
cross section $Q_t^2 R^v=\sigma(e^+e^-\to\gamma\to t\bar t)/\sigma_{\rm pt}$
of all groups were compiled  
and compared numerically in detail using an equivalent set of
parameters to analyze the scheme-dependence of the fixed order
approach. Since the axial-vector current contributions are only a small
correction at the few percent level it is justified to consider
only the photon-induced cross section.
Figure\ \ref{figpolettbarfixed} shows the results obtained in 
Ref.~\citebk{Hoang11} in the pole mass scheme for $m^{}_{\rm pole}=175.05$~GeV,
$\alpha_s(M_Z)=0.119$, $\Gamma_t=1.43$~GeV and $\mu=15,30$ and
$60$~GeV. The other scales where fixed at  
$M_{\rm pole}$.
The effects of the luminosity spectrum of
the $e^+e^-$ beams, which leads to a smearing of the line shape, 
were not taken into account.
Since the method used by Melnikov et al., Yakovlev et al. and Nagano
et al. were equivalent, a single figure is displayed for their
results. The result of Hoang and Teubner was based on the momentum
sphere method of Ref.~\citebk{Hoang12} and has the smallest variations.
In particular, the LO result and the NLO and NNLO corrections are
smaller than for the other groups.  
The results in Fig.\ \ref{figpolettbarfixed} demonstrate the
instability of the ``peak position'' in the pole mass scheme and the
necessity of using threshold masses. It was concluded in Ref.~\citebk{Hoang11}
that the perturbative uncertainty in an extraction of the pole mass parameter
from the peak position (i.e.\ when the beam smearing effects are neglected)
is around $300$~MeV. The results in Fig.\
\ref{figpolettbarfixed} also show a large uncertainty in the
normalization of the cross section. This uncertainty is particularly
puzzling, because there is no obvious physical reason for its
existence. At present it does not seem to be related to
renormalon-type higher order corrections, although it has also been
speculated that the large size of the corrections could have some
infrared origin.\cite{Kiyo2} On the other hand, it has been shown  
in Ref.~\citebk{Kniehl3} that the dominant N$^3$LO logarithmic corrections to
the cross section of (relative) order $\alpha_s^3 \ln^2v$
are at the level of $10$\% and not small.
It was estimated in Ref.~\citebk{Hoang11} that the present theoretical
relative uncertainty in the normalization of the cross section in the
fixed order approach is around $20$\%. 

\begin{figure}[t!] 
\begin{center}
\leavevmode
\epsfxsize=3.cm
\leavevmode
\epsffile[220 580 420 710]{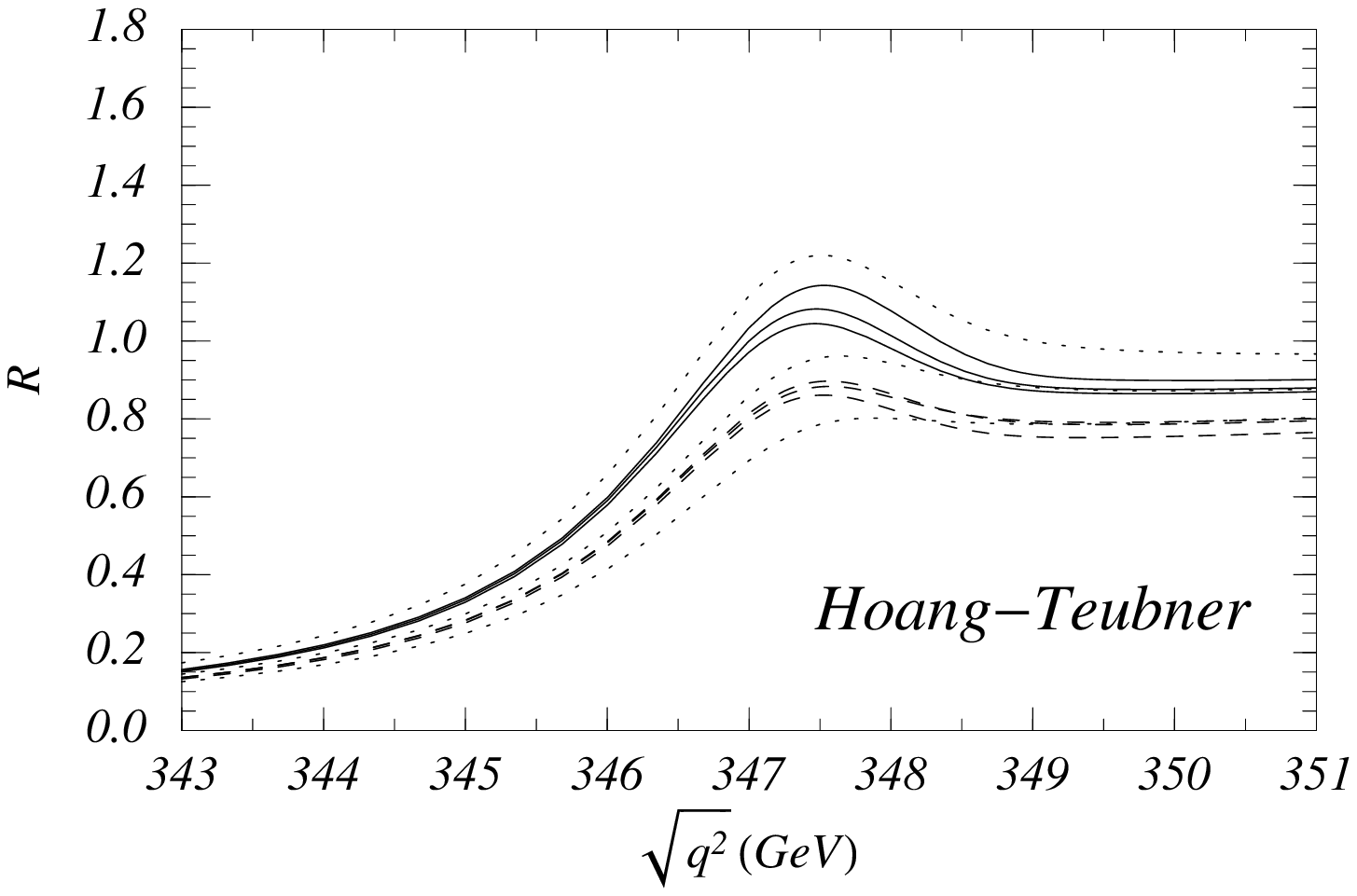}
\hspace{3.1cm}
\epsfxsize=3.cm
\leavevmode
\epsffile[220 580 420 710]{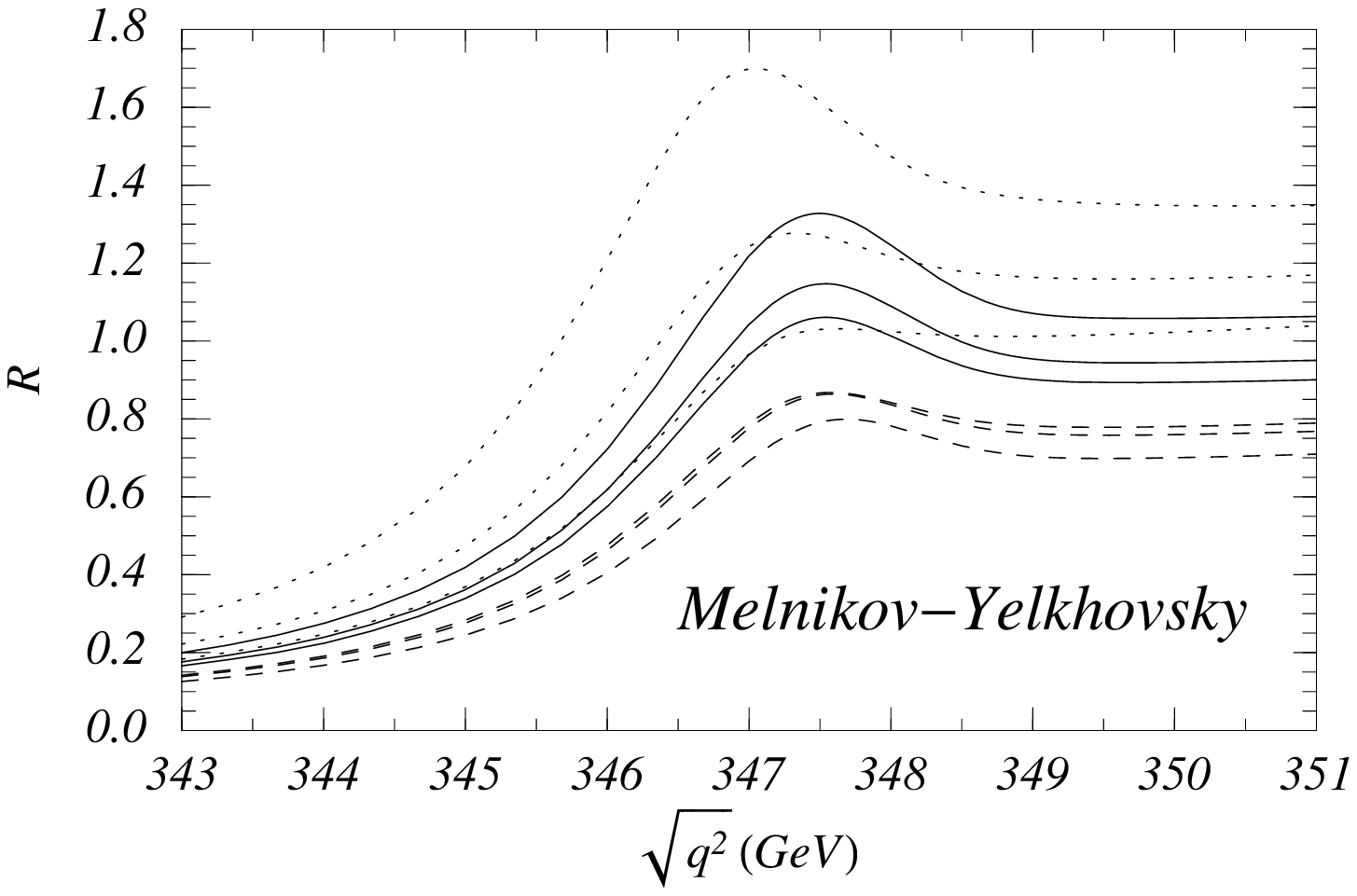}
\vskip 2.3cm
\leavevmode
\epsfxsize=3.cm
\leavevmode
\epsffile[220 580 420 710]{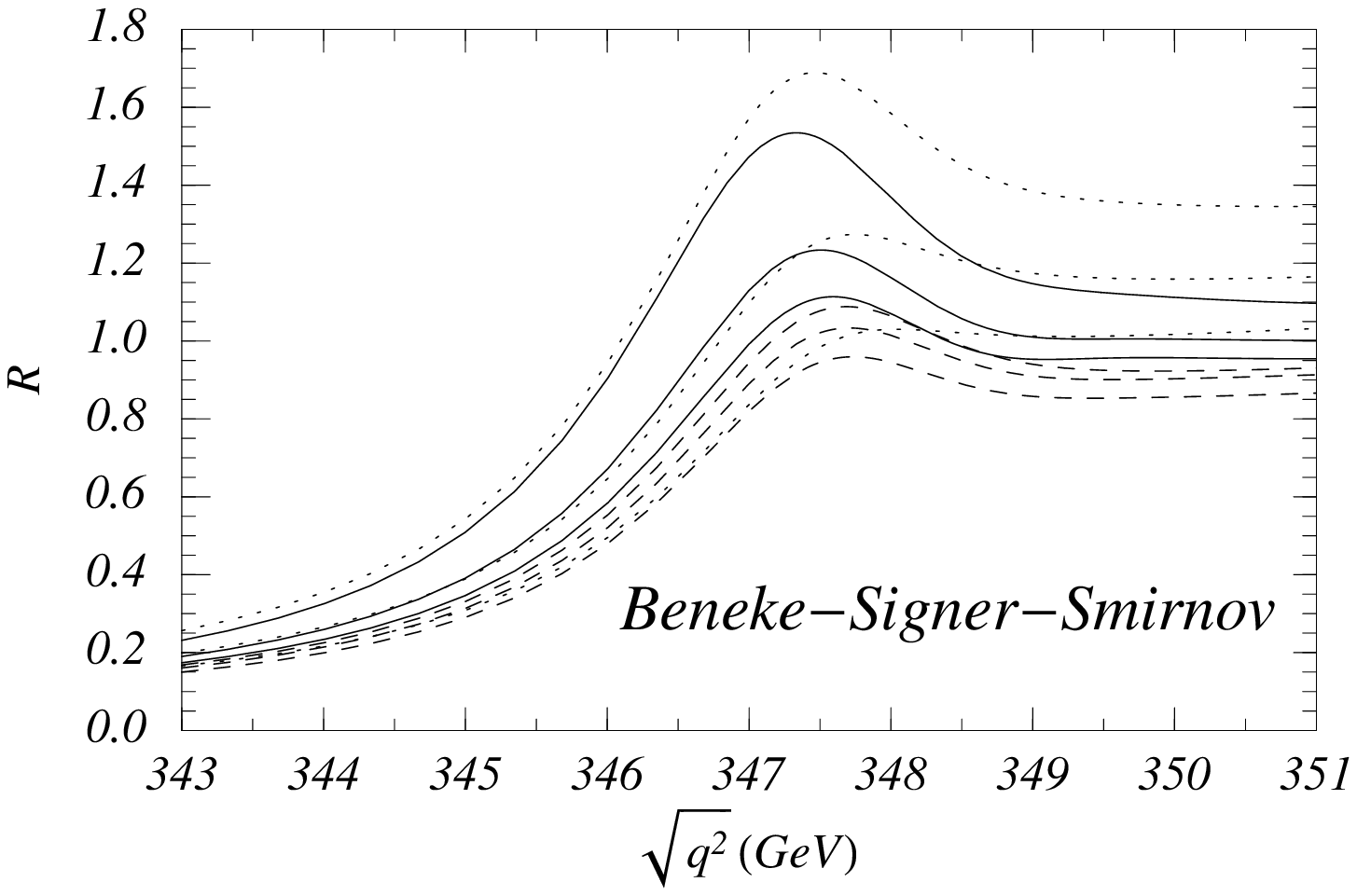}
\hspace{3.1cm}
\epsfxsize=3.cm
\leavevmode
\epsffile[220 580 420 710]{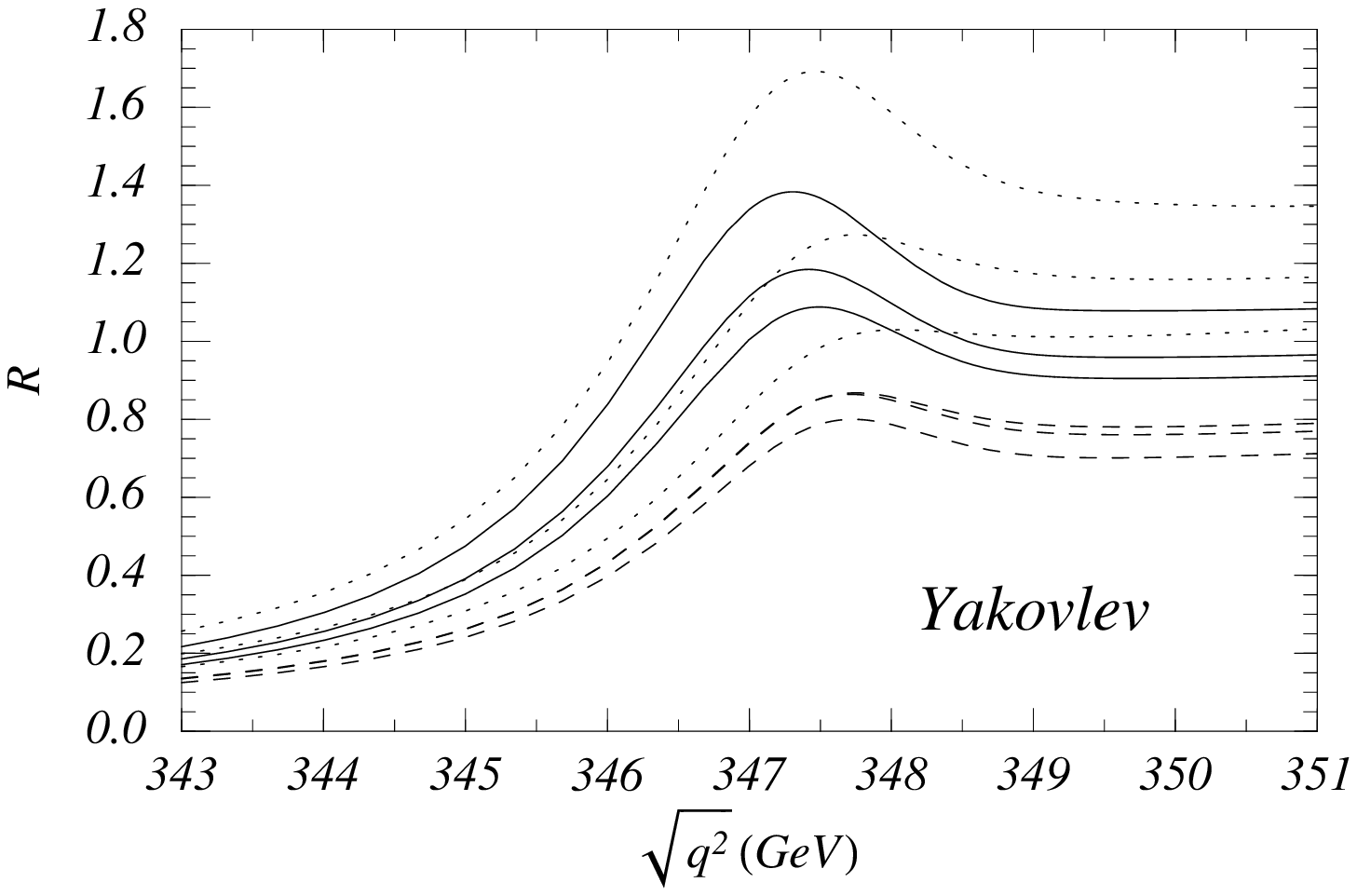}
\vskip  1.8cm
 \caption{\label{figthresholdmasses}   
The total normalized photon-induced $t\bar t$ cross section  $R$ at
the Linear Collider versus the c.m. energy in the threshold regime at
LO (dotted 
curves), NLO (dashed) and NNLO (solid) for
$\alpha_s(M_Z)=0.119$ , $\Gamma_t=1.43$~GeV and $\mu=15$,
$30$, $60$~GeV. Hoang--Teubner used the 1S mass
$M_{1S}^{}=173.68$~GeV, Melnikov--Yelkhovsky the kinetic
$M_{\rm kin}(15~\mbox{GeV})=173.10$~GeV, and 
Beneke--Signer--Smirnov and Yakovlev the PS mass 
$M_{\rm PS}(20~\mbox{GeV})=173.30$~GeV. 
The plots were given in Ref.~\protect\citebkcap{Hoang11} from results
provided by  Hoang--Teubner (HT), Melnikov--Yelkhovsky (MY),
Beneke--Signer--Smirnov (BSS) and Yakovlev. 
}
 \end{center}
\end{figure}

In Ref.~\citebk{Hoang11} it was also studied how well the peak position 
can be stabilized, when threshold mass schemes are employed (Sec.\
\ref{subsectionthresholdmasses}).
Figure\ \ref{figthresholdmasses} shows the results for the photon
induced total cross section obtained in  
Ref.~\citebk{Hoang11}
where the respective values of the threshold masses were obtained from
the top \ms mass value $\overline m(\overline m)=165$~GeV as a
reference point. Numerically, the series (in an expansion in powers of
$\alpha_s$ for $\mu=\overline m$) that relate the pole mass and
the threshold masses used in Fig.\ \ref{figthresholdmasses} to the \ms
mass are as follows:\,\cite{Hoang11} 
\begin{eqnarray}
m_{\rm pole}^{} & = & 
[\,165+7.64+1.64+0.52+0.25^{(*)}\,]~\mbox{GeV}
\nonumber\\
M_{\rm 1S} ^{} & = & 
[\,165+7.21+1.24+0.23+0.05^{(*)}\,]~\mbox{GeV}
\nonumber\\
M_{\rm PS}^{}(20~\mbox{GeV}) & = & 
[\,165+6.72+1.21+0.29+0.08^{(*)}\,]~\mbox{GeV}
\nonumber\\
m_{\rm kin}^{}(15~\mbox{GeV}) & = & 
[\,165+6.68+1.15+0.27^{(*)}\,]~\mbox{GeV}
\,.
\label{msbarothermasses}
\end{eqnarray}
The numbers indicated by the superscript $(*)$ are estimated in
the large-$\beta_0$ approximation and $\alpha_s(M_Z)=0.119$ (which
corresponds to $\alpha_s(165~\mbox{GeV})=0.1091$ using four-loop
running) has been used. 
The order $\alpha_s^3$ terms for the threshold masses are about a
factor two smaller than for the pole mass.
The results shown in Fig.~\ref{figthresholdmasses} demonstrate that
the peak position is significantly stabilized when threshold masses
are employed. It was concluded in Ref.~\citebk{Hoang11} that the
perturbative 
uncertainty in the determination of threshold masses from the peak
position (i.e.\ when beam smearing effects are neglected), is between
$50$ and $80$~MeV. It was also pointed out that the \ms mass
can be determined with a comparable perturbative precision, only if
$\alpha_s(M_Z)$ is known with an uncertainty of around $0.001$. This
restriction arises from the relatively large order $\alpha_s$ correction
in the relation between threshold masses and \ms mass.
For example, for a given measurement of the 1S mass, let's say
$M_{\rm 1S}^{}=175$~GeV$\,\pm\,\delta M_{\rm 1S}^{}$, and
$\alpha_s(M_Z)=0.118\pm x\,0.001$ the result for $\overline
m(\overline m)$ reads\,\cite{Hoang12}
\begin{eqnarray}
\overline m(\overline m)  & = & 
\bigg[\,
175 - 7.58\,  - \, 
0.96\,  - \, 
0.23\,
\, \pm \, \delta M_{1S}^{}
\, \pm \,  x\,0.07 
\,\bigg]~\mbox{GeV}
\,\qquad
\label{MSmassestimate}
\end{eqnarray} 
where the first four numbers represent the perturbative series.

The results in Fig.~\ref{figthresholdmasses} also show that the
threshold masses do not affect the normalization of the cross section;
the uncertainty of around $20$\% remains in threshold mass schemes.
This can be understood from the fact that the transfer from the pole
mass to threshold masses corresponds predominantly to a shift in the
threshold point. The large normalization uncertainties could have 
severe consequences for experimental measurements.
For the measurement of the top quark mass it is conceivable that,
after the luminosity spectrum of a specific collider design is folded
into the calculations for the cross section,
the resulting ``peak-less'' line-shape (see
Fig.\ \ref{figtopthreshexp}a) could lead to a transfer of
normalization uncertainties into the top mass measurement.
In general, such beam effects are smaller for $\mu^+\mu^-$ colliders, 
because beamstrahlung is suppressed by the muon mass, see e.g.\
Ref.~\citebk{Berger1}.    
On the other hand, the large normalization uncertainties would make
precise measurements of $\alpha_s$, $\Gamma_t$  or the top quark
Yukawa coupling to the Higgs, $y_t$ impossible. In the fixed order
approach, effects of a Standard Model Higgs can be implemented by a
Higgs potential $\tilde V_{\rm tth}^{}=-y_t^2/[2({\bmq}^2+m_h^2)]$ 
and a correction to the coefficient
$c_1$.\cite{Strassler1,Harlander1} For a light Higgs 
they are typically at the level of several percent.
Figure\ \ref{fig:outlook} shows the variation of the
theoretical total cross section $\sigma(e^+e^-\to\gamma,Z\to t\bar t$)
for changes in the input parameters $\alpha_s$, $\Gamma_t$ and $y_t$.

The question how well the top mass and the strong coupling can be
determined from a threshold scan in $e^+e^-$ collisions with the
TESLA-LC design\,\cite{Aguilar1} was addressed by Peralta et
al.\,\cite{Peralta1} in a simulation study based on the theoretical
NNLO calculations for the cross section from Hoang et
al. in Ref.~\citebk{Hoang12}. The  
theoretical computations from Ref.~\citebk{Hoang12} have the smallest
theoretical variations of all NNLO computations, so the estimates for
the theoretical uncertainties obtained in the experimental study of
Ref.~\citebk{Peralta1} are likely to be larger, if 
the NNLO computation of the other groups are used in a similar
analysis. However, the results of Ref.~\citebk{Peralta1} illustrate
very well the  dependence of the measurements on theoretical
uncertainties. In the simulation Peralta et al. included the TESLA
luminosity sprectrum and realistic assumptions on
efficiencies, background and other systematics. In the study nine scan
points were placed onto the $t\bar t$ line shape similar to Fig.\
\ref{figtopthreshexp}a and a tenth scan point was placed well below
the threshold to evaluate the background. At each point an integrated
luminosity of $10$~fb$^{-1}$ was spent.
\begin{figure}[t] 
\begin{center}
\leavevmode
\epsfxsize=1.8cm
\leavevmode
\epsffile[200 580 400 670]{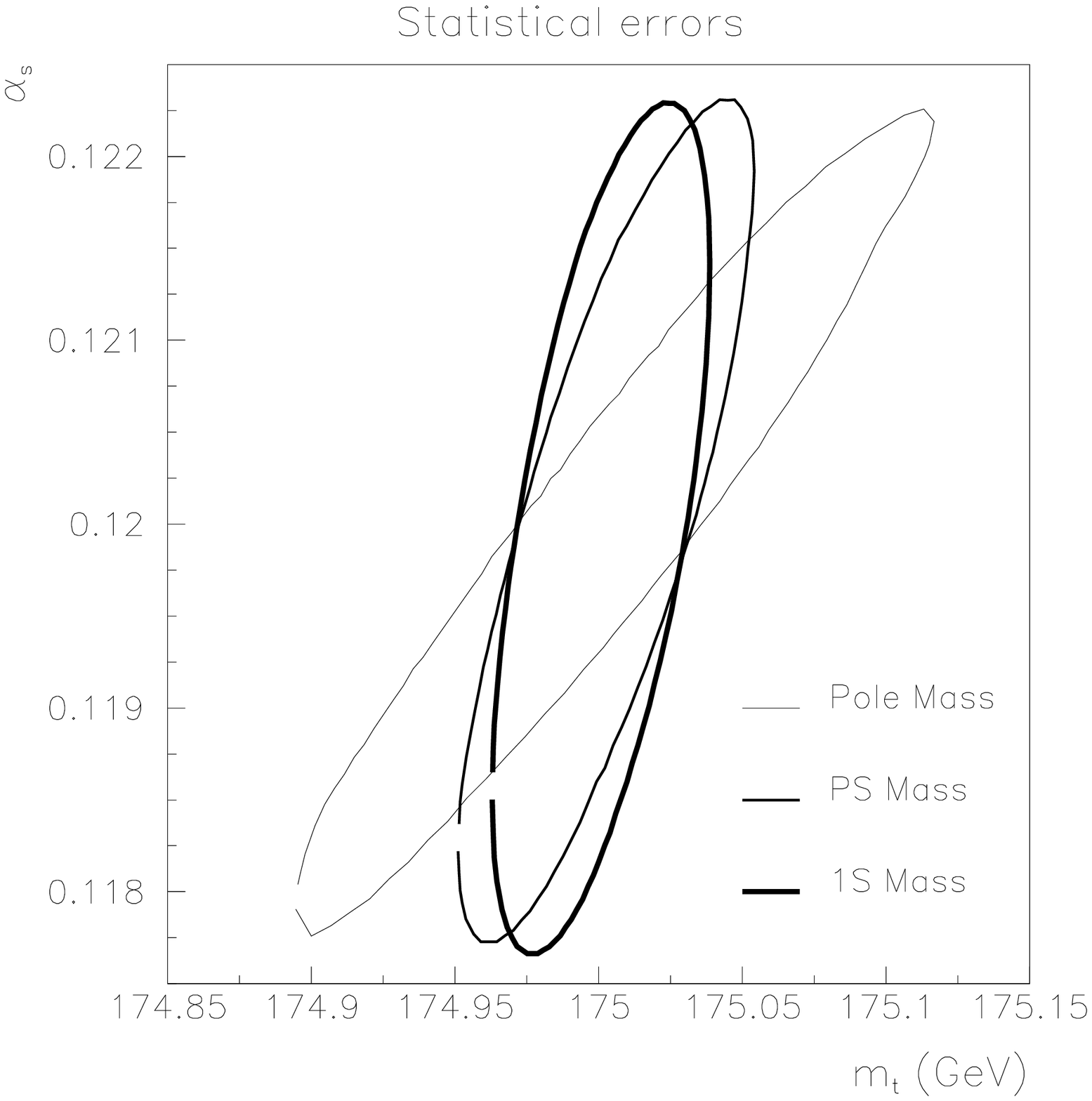}
\mbox{\hspace{3.8cm}}
\leavevmode
\epsfxsize=1.8cm
\leavevmode
\epsffile[200 580 400 670]{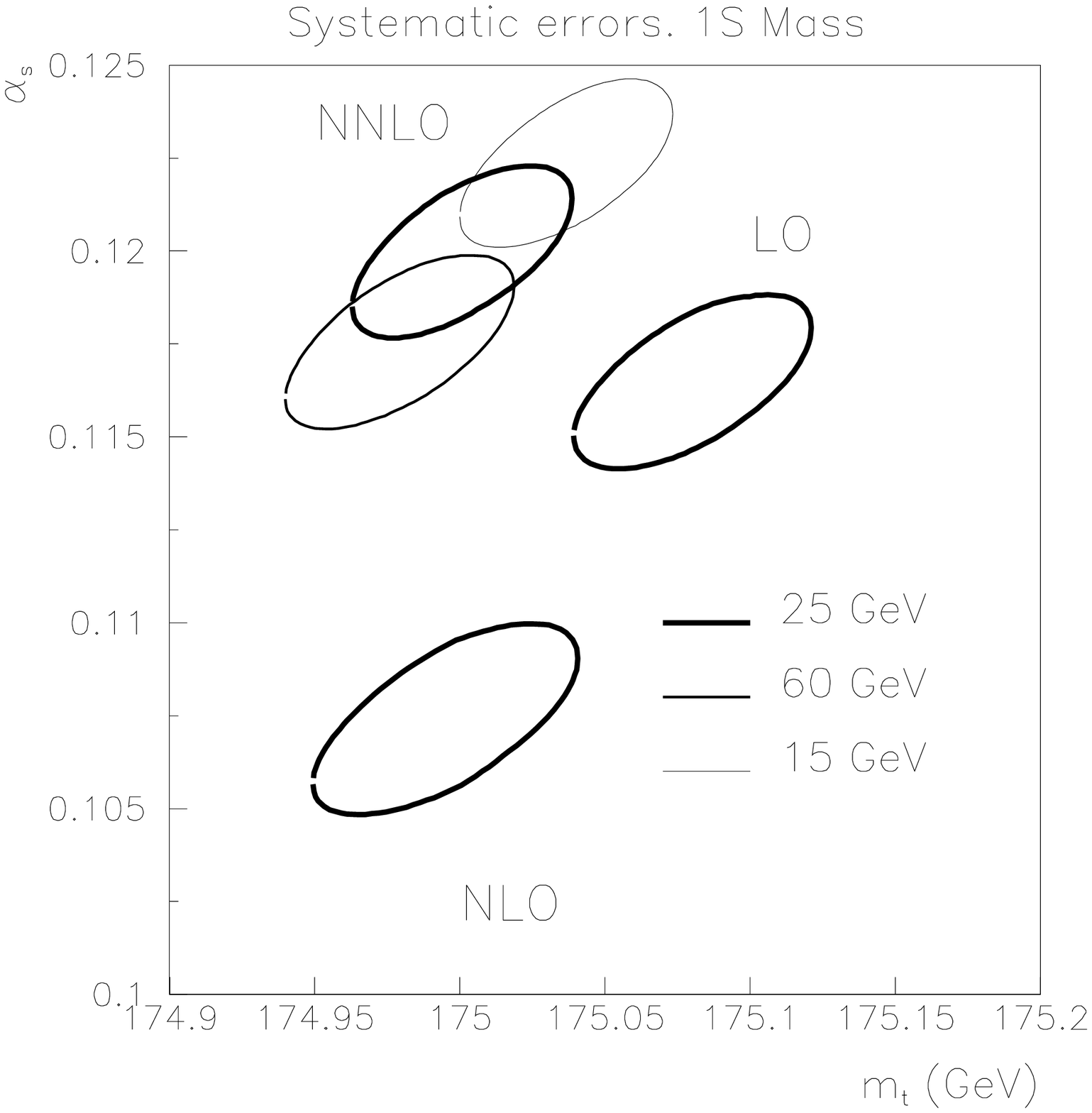}
\begin{picture}(0,0)(1,1)
\put(-278,10){$a)$}
\put(-110,10){$b)$}
\end{picture}
\vskip 3.6cm
 \caption{   
(a) Results of a two-parameter ($M$--$\alpha_s(M_Z)$) $\chi^2$ fit
   ($\Delta\chi^2=1$) for the pole, 1S and PS 
   ($\mu_{\rm PS}^{}=20$~GeV) masses using the same 
   NNLO theoretical cross section ($\mu=25$~GeV) for ``experiment'' 
   and theory. The ellipses show the experimental errors.
(b) Result for a two-parameter $M_{\rm 1S}^{}$--$\alpha_s(M_Z)$
   $\chi^2$ fit ($\Delta\chi^2=1$) using different theoretical cross
   sections for 
   the ``experiment'' while keeping the theory fixed to NNLO with
   $\mu=25$~GeV. The shifts in the locations of the ellipses show the
   theoretical errors. The analysis was carried out in 
   Ref.~\protect\citebkcap{Peralta1} based on the
   computations in 
   Ref.~\protect\citebkcap{Hoang12}.
  \label{figperalta} }
\end{center}
\end{figure}
For a two-parameter fit of top mass and $\alpha_s(M_Z)$ Peralta el al.
found the correlations shown in Fig.\ \ref{figperalta}a. The error
ellipses correspond to $\Delta\chi^2=1$. Because the corrections to the
${}^3S_1$ toponium ground state mass are zero in the 1S scheme, the
correlation is smallest for the 1S mass leading to the smallest
experimental uncertainties. The experimental uncertainty is $40$~MeV
($50$~MeV) for  $M_{\rm 1S}^{}$ ($M_{\rm PS}^{}(20~\mbox{GeV})$) and
$120$~MeV for the pole mass. The correlation is larger in the pole
mass scheme. In order to test the impact of the theoretical
uncertainties in the normalization of the cross section the fits were
repeated for ``experimental'' data generated from the LO, NLO and NNLO
theoretical predictions and for three different values of the
renormalization scale, $\mu=15,25,60$~GeV. The difference in the
location of the ellipses are a measure for the theoretical
uncertainties; see Fig.\ \ref{figperalta}b
for the result in the 1S scheme ($\Delta\chi^2=1$). For 
$M_{\rm 1S}^{}$ and $M_{\rm PS}^{}(20~\mbox{GeV})$ the theoretical
uncertainties were found to be at the level of $10$--$40$~MeV, whereas
the error in $\alpha_s(M_Z)$ was 0.013, which is several times larger
then the current world average. For the pole mass the theoretical
uncertainty was found to be about $100$~MeV, and the error in
$\alpha_s(M_Z)$ was 0.012. This result shows that in a two-parameter
fit the large
normalization uncertainties of the total cross section are fed almost
entirely into $\alpha_s$ and essentially leave the measurement of
threshold mass parameters unaffected. 

Peralta et al.\,\cite{Peralta1} also carried out a constraint fit taking
$\alpha_s(M_Z)=0.120\pm 0.001$ as an input. Here it was found that the
experimental uncertainties for $M_{\rm 1S}^{}$,
$M_{\rm PS}^{}(20~\mbox{GeV})$ and the pole mass are slightly smaller
than in the two-parameter fit, but that the theoretical uncertainties
for the masses are about $3$--$4$ time larger. This behavior can be
understood from the correlation between masses and $\alpha_s$ shown in
Fig.\ \ref{figperalta}a and from the fact that the large normalization
uncertainties of the total cross section are partly absorbed into the
mass for a constraint fit.

In a more recent simualtion study by Martinez\,\cite{Martinez1}
the size of the experimental uncertainties for simultaneous
measurements of the $1S$ mass, $\alpha_s(M_Z)$, $\Gamma_t$ and $y_t$ in a
top threshold run at an $e^+e^-$ Linear Collider was reexamined taking
into account ``data'' on total cross section, top three-momentum
distribution and forward-backward asymmetry for a 9+1 point threshold
scan using a total integrated luminosity of
$300$~fb$^{-1}$. The intention was to determine which experimental
uncertainties could be eventually achieved, and not to estimate 
theoretical uncertainties.
Fixing $y_t$ to the Standard Model (SM) value  Martinez et al. obtained 
$\Delta M_{\rm 1S}=18$~MeV, $\Delta\alpha_s(M_Z)=0.0012$ and
$\Delta\Gamma_t=30$~MeV from a three-parameter fit.
In a one parameter fit, fixing all other parameters, they obtained
$\delta y_t/y_t$ between $12$ and $24$\%. In a fit where $\Gamma_t$ is
fixed to the SM value and $\alpha_s$ is constrained to
$\alpha_s(M_Z)=0.120\pm 0.001$ they obtained $\delta M_{\rm
  1S}^{}=25$~MeV and $\delta y_t/y_t=(+0.31, -0.49)$.  
In a fit where only $\alpha_s$ is constrained to
$\alpha_s(M_Z)=0.120\pm 0.001$ they obtained $\delta M_{\rm
  1S}^{}=30$~MeV, $\delta\Gamma_t=33$~MeV
and $\delta y_t/y_t=(+0.33,-0.57)$ for a Higgs mass of $m_h=120$~GeV.
For simulation studies for $\mu^+\mu^-$ colliders see for example
Ref.~\citebk{Berger1}.

At present it appears questionable whether the experimental precision
for measurements of $\alpha_s$, $\Gamma_t$ and $y_t$ can be matched by
the precision of the theoretical computations in the fixed order approach.

\subsection{Cross Section at NNLL Order in vNRQCD}
\label{subsectionttbarcrossvNRQCD}

At present there exists only one computation of the total $t\bar t$
production cross section close to threshold in the
renormalization-group-improved approach by Hoang et al.\,\cite{Hoang3} in
the framework of vNRQCD. Since $c_1$ does not run and the Coulomb
potential evolves trivially with $\alpha_s$ at LL order, the LL cross
section in the renormalization-group-improved approach and the LO
fixed order cross section are equivalent. The NLL cross section in
vNRQCD and the NLO fixed order cross section differ because $c_1$ runs
non-trivially at NLL order. Thus the NLL order cross section and the
fixed order NLO cross section differ by a normalization factor.
At NNLL order all relevant coefficients for
potentials and currents run non-trivially, and the difference to the
fixed order NNLO results can be significant with respect to the
normalization as well as to the shape of the cross section.  
Hoang et al.\,\cite{Hoang3} determined the cross section at LL, NLL and NNLL
order. The NNLL order result is incomplete, since it does not include
the three-loop evolution of $c_1$, which is presently unknown.

In Refs.~\citebk{Hoang3} the calculations were carried out in $D=4-2\epsilon$
dimensions in the \ms scheme. The Schr\"odinger equation in Eq.\
(\ref{NNLLSchroedinger}) has been solved exactly with numerical
methods for all Coulombic contributions, and the corrections from the
kinetic energy term ${\bmp}^4/(4M^3)$ and the $1/m$-suppressed
potentials were determined in time-independent perturbation theory.
\begin{figure}[t] 
\begin{center}
 \leavevmode
\epsfxsize=3cm
\leavevmode
\epsffile[205 580 415 700]{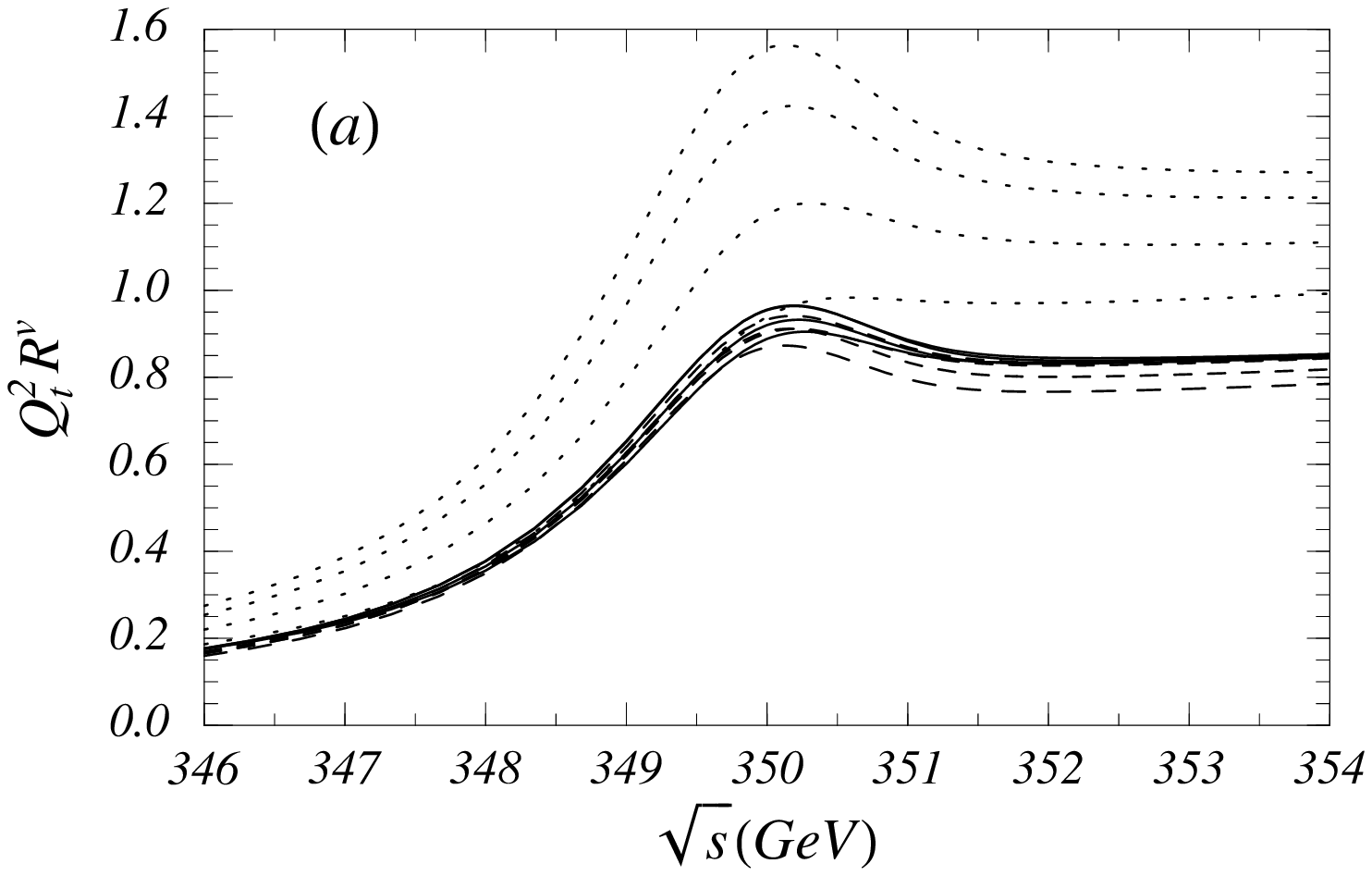}
\hspace{2.9cm}
\leavevmode
\epsfxsize=3cm
\leavevmode
\epsffile[205 580 415 700]{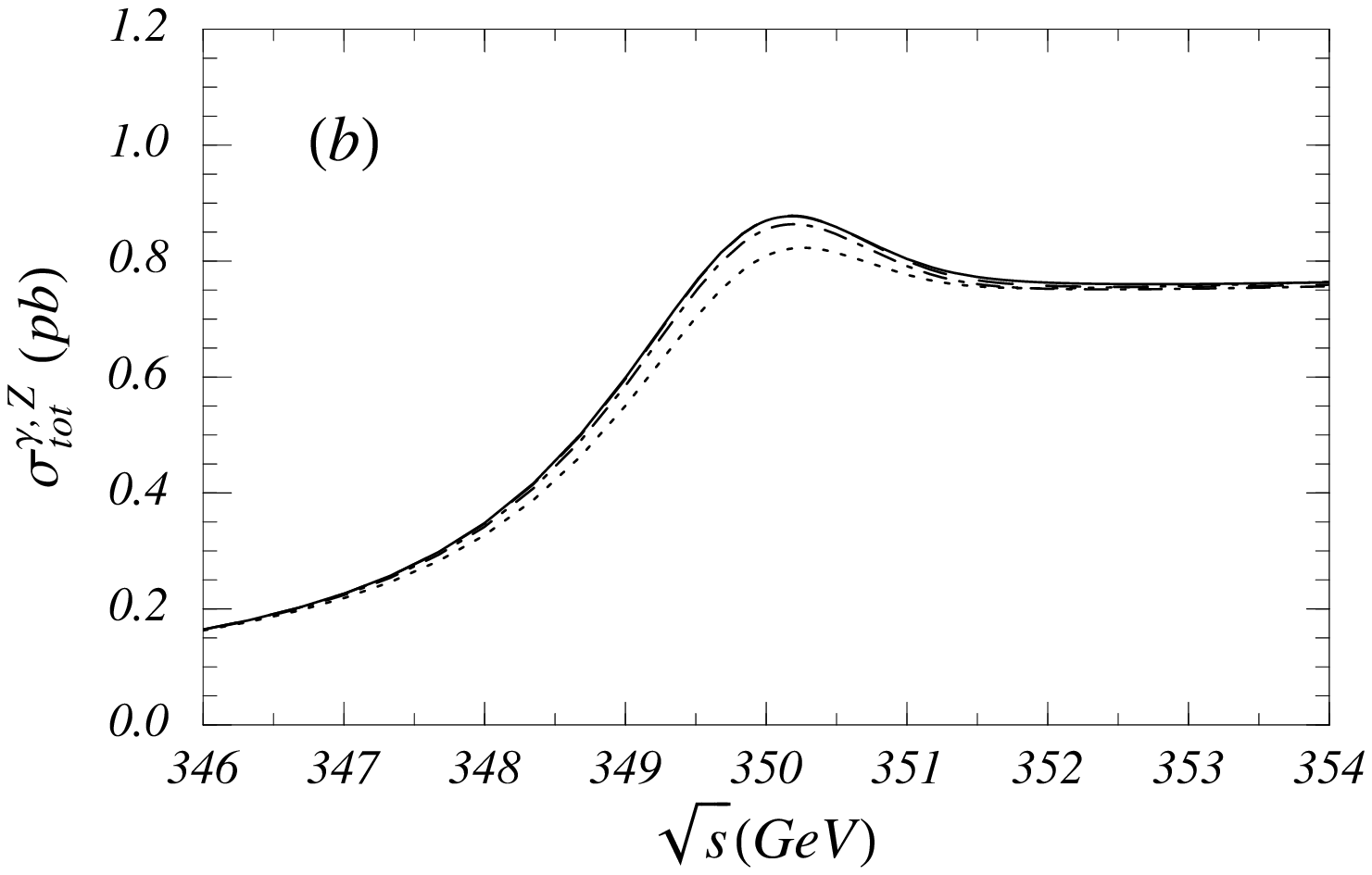}
\vskip 1.8cm
\caption{ \label{figrgecross}    
 (a) Normalized photon-induced cross section $Q_t^2 R^v$
  at LL (dotted lines), NLL (dashed lines) and NNLL (solid lines)
  order for $M_{\rm 1S}^{}=175$~GeV, $\alpha_s(M_Z)=0.118$,
  $\Gamma_t=1.43$~GeV and $\nu=0.1,0.125,0.2,0.4$ at each order.
 (b) Total NNLL order cross section in units of pb for the same parameters.
  The dotted, dashed, dot-dashed, and solid curves correspond to 
  $\nu=0.1,0.15,0.2,0.4$. The figures were obtained with
 renormalization-group-improved perturbation theory in
 Ref.~\protect\citebkcap{Hoang3} using vNRQCD.
 } \end{center}
\end{figure}
In Fig.\ \ref{figrgecross}a the normalized photon-induced cross
section obtained in Ref.~\citebk{Hoang3} is displayed 
at LL (dotted lines), NLL (dashed lines) and NNLL
(solid lines) order in the 1S mass scheme for 
$M_{\rm 1S}^{}=175$~GeV and four different choices of the velocity
scaling parameter, $\nu=0.1,0.125,0.2,0.4$.
The peak position is
stable due to the use of a threshold mass scheme. In contrast to the
fixed order computations discussed in the previous subsection
an excellent convergence of the
normalization and a decreasing sensitivity to changes in $\nu$ at
higher orders is observed. It was shown that this
behavior is a consequence of the evolution of the vNRQCD coefficients,
which sum QCD logarithms of $v$ if $\nu$ is chosen of order
$v\sim\alpha_s$. In particular, the non-trivial evolution of the
potential coefficients (Fig.\ \ref{figrunningpotential})
leads to a partial cancellation of NNLL order corrections because 
some coefficients develop a relative sign difference when $\nu$ is
lowered. Such a behavior does not exist at NNLO in the fixed order
approach, where all coefficients scale trivially with $\alpha_s$ and
grow in the same relative direction when the scale $\mu$ is lowered.
In Fig.\ \ref{figrgecross}b the total NNLL order $t\bar t$ cross
section including $\gamma$ and $Z$ exchange in units of pb determined
in Refs.~\citebk{Hoang3} is displayed for $\nu=0.1,0.15,0.2,0.4$.

From the size of the NNLL order corrections,the $\nu$-variation of
the the cross section and an examination of some known higher order
corrections, Hoang et al. estimated the remaining perturbative
uncertainty of the normalization of the cross section as $3$\%, which
is a reduction by almost an order of magnitude with respect to
estimates in the fixed order approach. If the uncertainty of the
renormalization-group-improved approach is established, the prospects
for the size of theoretical uncertainties in measurements of top mass,
$\alpha_s$, $\Gamma_t$ and the Yukawa coupling $y_t$ from a top
threshold scan are significantly  better than in the fixed order
approach. In particular, for fits where the top couplings are fixed,
the danger that the perturbative normalization uncertainties are fed
into the top mass measurement is avoided.
On the other hand, measurements of $\alpha_s$, 
$\Gamma_t$ and $y_t$ with acceptable theoretical uncertainties appear
only to be possible in the renormalization-group-improved approach.  
\begin{figure}[t] 
\begin{center}
\leavevmode
\epsfxsize=2.7cm
\leavevmode
\epsffile[220 580 420 710]{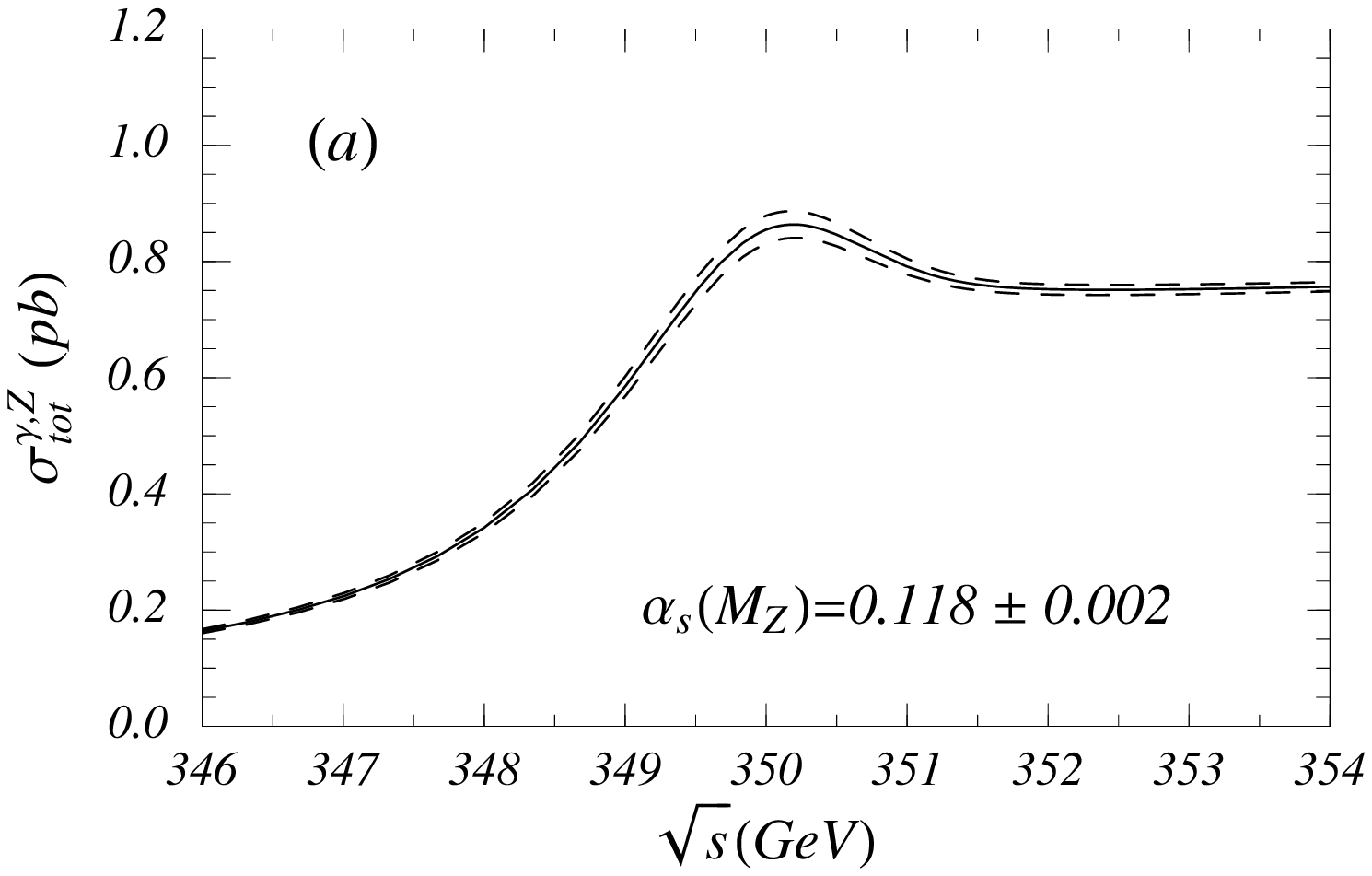}
\mbox{\hspace{3.cm}}
\leavevmode
\epsfxsize=2.7cm
\leavevmode
\epsffile[220 580 420 710]{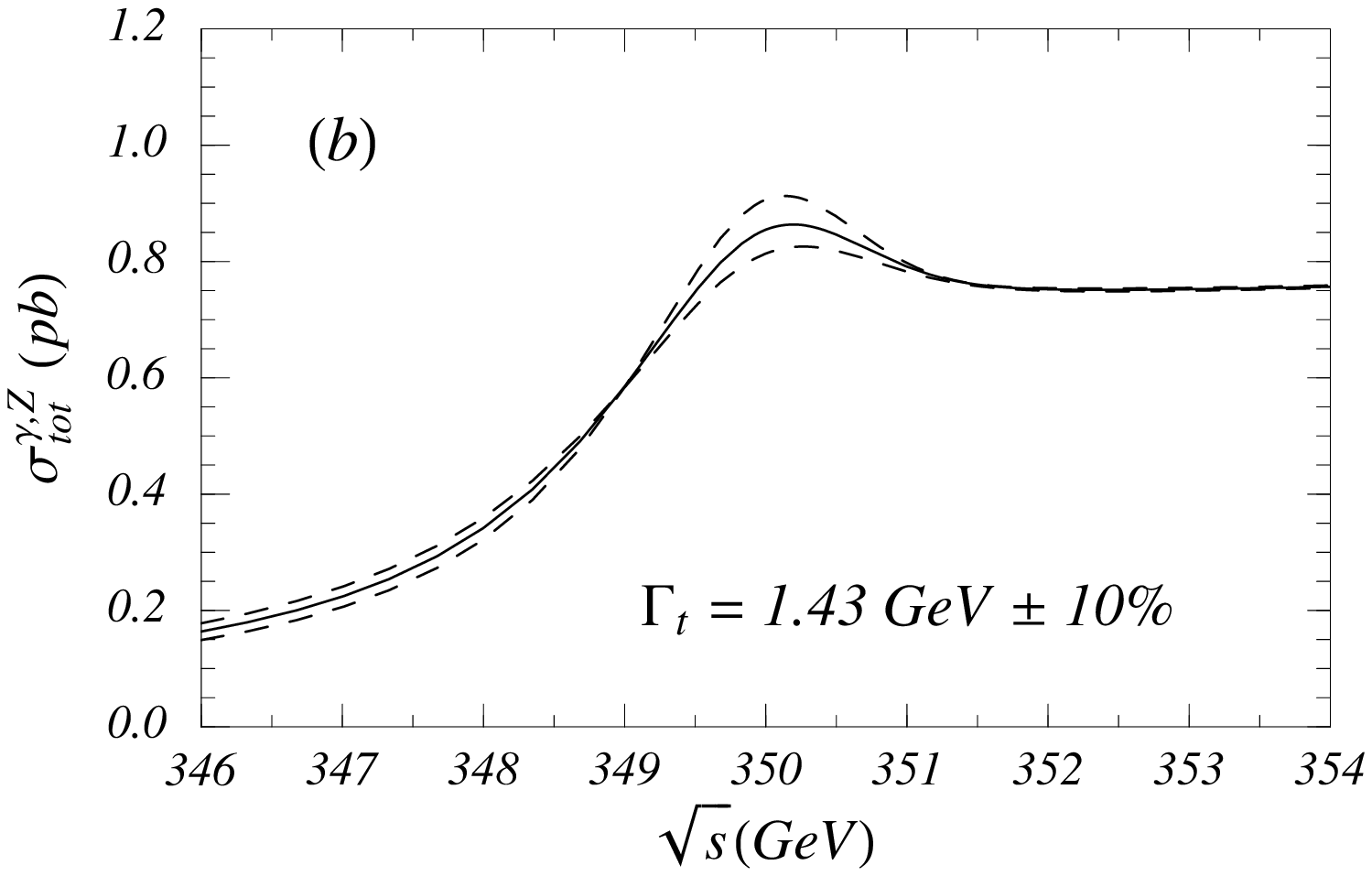}
\\[2.cm]
\leavevmode
\epsfxsize=2.7cm
\leavevmode
\epsffile[220 580 420 710]{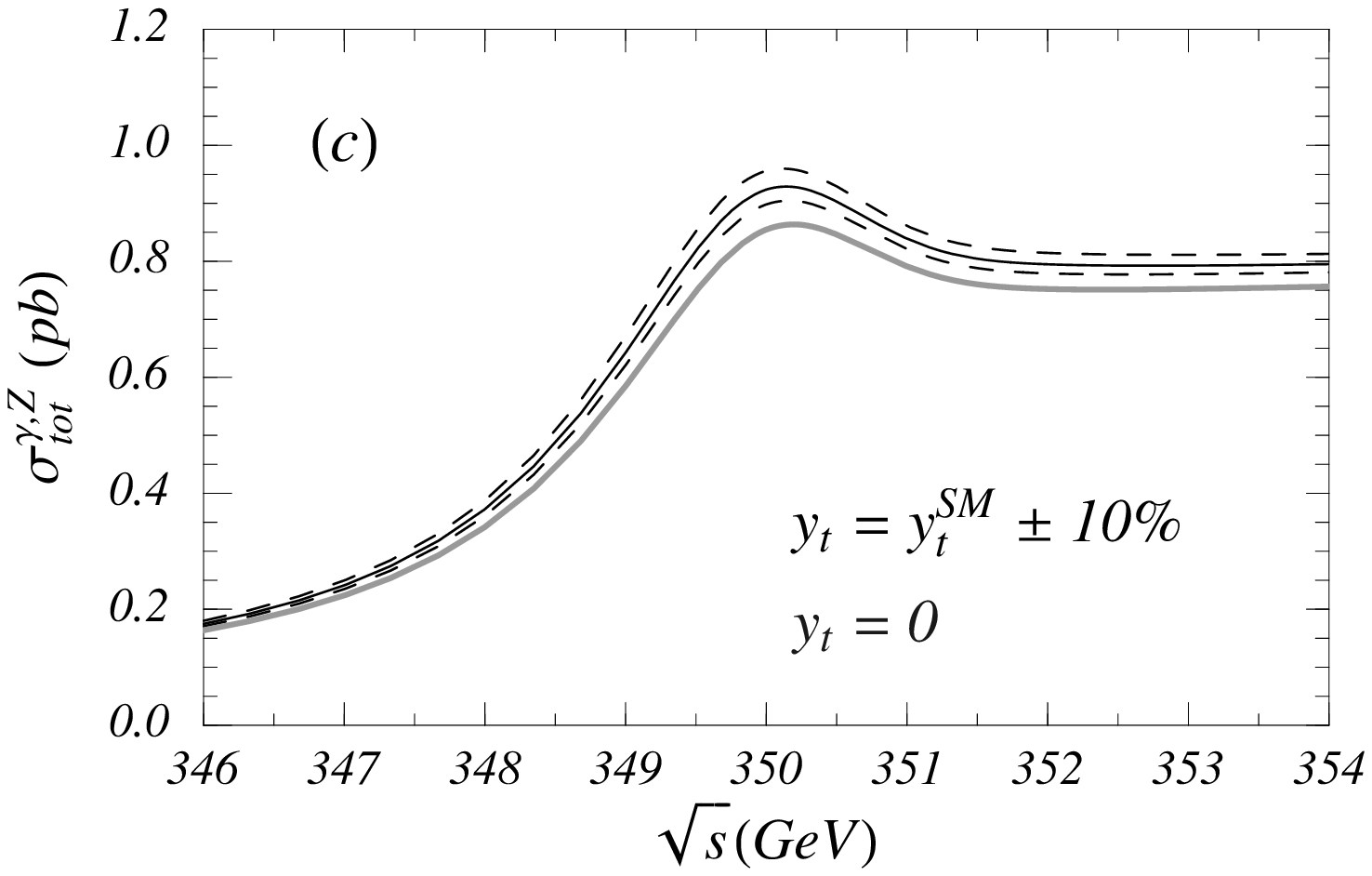}
\vskip 1.5cm
 \caption{   Variation of the NNLL cross section in
   vNRQCD~\protect\citebkcap{Hoang3} 
without beam effects for a) the value of the
strong coupling, b) the top quark width, and c) the inclusion of a Standard
Model (SM) Higgs boson.  Changes relative to the central value (solid lines) are
shown by dashed red lines. In c) there are two solid lines, the lower black
line is the decoupling limit for the Higgs boson, and the upper blue line is
for a SM Higgs with mass $m_H=115\,{\rm GeV}$.
The parameters $M_{\rm 1S}=175$~GeV, $\Gamma_t=1.43$~GeV,
$\alpha_s(M_Z)=0.118$, $y_t=0$ and $\nu=0.15$ have been used unless
stated otherwise.  \label{fig:outlook} }
\end{center}
\end{figure}
In Figs.~\ref{fig:outlook} the variation of the NNLL cross
section $\sigma_{\rm tot}^{\gamma,Z}$ obtained in Ref.~\citebk{Hoang3} is
displayed as a function of $\sqrt{s}$  for different
choices of the input parameters $\alpha_s(M_Z)$, $\Gamma_t$ and $y_t$.  In
all figures the parameters $M_{\rm 1S}^{}=175$~GeV, $\Gamma_t=1.43\,{\rm GeV}$,
$\alpha_s(M_Z)=0.118$, $y_t=0$ and $\nu=0.15$ are chosen, unless
stated otherwise. The upper left panel shows the cross section for
$\alpha_s(M_Z)=0.116$ (lower dashed red line), $0.118$ (solid black line), and
$0.120$ (upper dashed red line).  At the peak one finds a $\pm 2.7\%$
variation when varying $\alpha_s(M_Z)$ by $\pm 0.002$.  The upper
right panel shows the cross section for $\Gamma_t=1.43$~GeV 
(solid black line). The dashed red lines correspond to variations of 
$\Gamma_t$ by $\pm 10\%$, where for a smaller width the peak becomes more
pronounced. At the peak one finds a $(-2.3\%,+2.6\%)$ variation of the cross
section when varying $\Gamma_t$ by $\pm 5\%$.  Finally, the lower panel shows
the cross section for zero Yukawa coupling (solid black line), the Standard
Model value for the Yukawa coupling (solid blue line), and a $\pm 20\%$
variation of the coupling with respect to the SM value (upper/lower dashed red
lines).  The Higgs mass was chosen to be $m_H=115$~GeV.  At the peak one finds a
$(+3.3\%,-2.6\%)$ variation when varying $y_t^{\rm SM}$ by $\pm 20\%$.  For
$m_H=130$ and $150\,{\rm GeV}$ the corresponding variations are
$(+2.9\%,-2.3\%)$ and $(+2.5\%,-2.0\%)$ respectively. Thus, given that
the remaining theoretical uncertainty in the normalization of the
cross section is $3$\%, measurements with theoretical uncertainties
$\delta\alpha_s(M_Z)\sim 0.002$,
$\delta\Gamma_t/\Gamma_t\sim 5\%$ and $\delta y_t/y_t\sim 20\%$ appear
feasible. These uncertainties match quite well the estimated experimental  
uncertainties from a recent simulation study by Martinez\,\cite{Martinez1} (see
end of Sec.\ \ref{subsectionfixedorderttbar}).

\vspace{1cm}

\section{Bottom Mass from $\Upsilon$ Sum rules}
\label{sectionsumrules}

The determination of the Cabibbo-Kobayashi-Maskawa (CKM) matrix
elements is one of the main goals of current B physics experiments. A
sufficiently accurate determination of the size of the CKM matrix
elements and their relative phases will lead to a better understanding
of the origin of CP violation, the structure of the weak interaction,
and, possibly, to the establishment of physics beyond the Standard
Model. For the extraction of the CKM matrix elements from inclusive B
decay rates, such as of $V_{cb}$ from the semileptonic decays
into $D$ mesons or of $V_{ub}$ from semileptonic decays into light
hadrons, an accurate and precise knowledge of the bottom quark 
mass is desirable due to the strong dependence of the total decay
rate on the bottom quark mass parameter 
(Sec.~\ref{subsectionthresholdmasses}).

Non-relativistic sum rules for the masses and electronic decay widths
of $\Upsilon$ mesons, bottom-antibottom quark bound states with
${}^{2s+1}L_j={}^3S_1$, are an ideal tool to determine the bottom
quark mass: using causality and global duality arguments one can
relate integrals over the total cross section for the production of
hadrons containing a bottom-antibottom quark pair in $e^+e^-$
collisions to derivatives of the vacuum polarization function of
bottom quark currents at zero-momentum transfer. The $n$-th derivative
is referred to as the $n$-th moment of the total cross section.
For a particular range of $n$ the moments are dominated by the
experimental data on the $\Upsilon$ mesons and, at the same time, can
be calculated reliably using perturbative QCD in the non-relativistic
expansion. Because the moments have, for dimensional reasons, a strong
dependence on the bottom quark mass, $P_n\sim 1/M^{2n}$, 
these sum rules can be used to determine the bottom quark mass to high
precision. Within the last few years there have been several analyses
at NLO and NNLO in the non-relativistic expansion in the fixed order
approach. Initially, the bottom pole mass was 
extracted,\,\cite{Voloshin2,Kuhn1,Penin1,Hoang14}
whereas later analyses\,\cite{Melnikov2,Hoang15,Beneke6,Hoang8}
determined threshold mass parameters  
and the \ms mass $\overline m(\overline m)$. For an application of
non-relativistic sum rules to the $\psi$ family for an extraction of
the charm quark mass see Ref.~\citebk{Eidemuller1}.

\subsection{Basic Theoretical Issues and Experimental Data}
\label{subsectionsumrulesbasic}
The sum rules for the $\Upsilon$ mesons start from the correlator of
two electromagnetic vector currents of bottom quarks at momentum
transfer $q$,
\begin{eqnarray}
\Pi(q^2) & = &
-\,i \int dx\,e^{i\,q.x}\,
   \langle\, 0\,|\,T\,j^v_\mu(x)\,j^{v\,\mu}(0)\,|\,0\, \rangle
\,.
\label{vacpoldef}
\end{eqnarray}
The $n$-th
moment $P_n$ of the vacuum polarization function is defined as
\begin{equation}
P_n \, \equiv \,
\frac{4\,\pi^2\,Q_b^2}{n!}\,
\bigg(\frac{d}{d q^2}\bigg)^n\,\frac{\Pi(q^2)}{q^2}\bigg|_{q^2=0}
\,,
\label{momentsdef1}
\end{equation}
where $Q_b=-1/3$ is the electric charge of the bottom quark.
Due to causality the $n$-th moment $P_n$ can also be written as a
dispersion integral
\begin{equation}
P_n \, = \,Q_b^2\,
\int\limits^\infty_{s_{\rm min}} \frac{d s}{s^{n+1}}\,R^v(s)
\,,
\label{momentsdef2}
\end{equation}
where
\begin{equation}
Q_b^2\,R^v(s) \, = \, \frac{\sigma(e^+e^-\to\gamma\to \mbox{$b\bar b$+X})}
{\sigma_{\rm pt}}
\, = \,
\frac{4\,\pi\,Q_b^2}{s}\,
\mbox{Im}
\,\Pi(s)
\label{Rdefinitioncovariant}
\end{equation}
is the total normalized photon-induced cross section of bottom
quark--antiquark production in $e^+e^-$ annihilation. 
The lower limit of
the integration in Eq.~(\ref{momentsdef2}) is set by the mass of the
lowest-lying $b\bar b$ resonance. Assuming global duality, the moments $P_n$ can
be either calculated from experimental data on $R$ or theoretically
using perturbative QCD.

The experimental moments $P_n^{\rm ex}$ are determined by using
the data on the $\Upsilon$ meson masses, $M_{\rm kS}$, and $e^+e^-$ partial
widths, $\Gamma_{\rm kS}$, for $k=1,\ldots,6$, see
Ref.~\citebk{PDG}. The formula that is used for the experimental moments
has the typical form 
\begin{eqnarray}
P_n^{\rm ex} & = &  
\frac{9\,\pi}{\tilde\alpha^2_{\rm em}}\,\sum\limits_{k=1}^6\,
\frac{\Gamma_{\rm kS}}{M_{\rm kS}^{2n+1}} 
\, + \,
\int\limits_{s_{{\rm B}\bar{\rm B}}}^\infty
\frac{ds}{s^{n+1}}\,r_{\rm cont}(s)
\,,
\label{Pnexperiment}
\end{eqnarray}
and is based on the narrow width approximation for the known
$\Upsilon$ resonances; $\tilde\alpha_{\rm em}$ is the electromagnetic
coupling at the scale $10$~GeV. Because the difference in the
electromagnetic coupling for the different $\Upsilon$ masses is
negligible, one can chose $10$~GeV as the scale of the electromagnetic
coupling for all resonances. Since
the continuum cross section above the ${\rm B}\bar{\rm B}$
threshold is quite poorly known experimentally a number of different
methods to estimate it has been used in the literature.
Usually it is
approximated by some theoretically motivated model, based on the
perturbative predictions for the continuum region.
For very low values of $n\lsim 4$ the uncertainties in the knowledge of the
continuum cross section are a significant contribution in the
uncertainties of the bottom mass extractions.
For $n\gsim 4$ the continuum contribution is suppressed 
sufficiently so that only a rough estimate of the continuum region
is needed. Here, it is sufficient to assume that $r_c$ is a constant
compatible with the perturbative predictions. 
In Tab.~\ref{tabdata} a compilation is given of the experimental numbers for
the $\Upsilon$ resonances needed
in Eq.~(\ref{Pnexperiment}).
\begin{table}[t!]  
\vskip 0mm
\begin{center}
\begin{tabular}{|c||r@{$.$}l|c|} \hline
 nS & \multicolumn{2}{|c||}{$M_{\rm nS}/[\mbox{GeV}]$} 
  & $\Gamma_{\rm nS}/[\mbox{keV}]$ 
\\ \hline\hline
 1S & $\hspace{0.6cm} 9$&$460$ & $1.32\pm 0.04\pm 0.03$
\\ \hline
 2S & $10$&$023$ & $0.52\pm 0.03\pm 0.01$
\\ \hline
 3S & $10$&$355$ & $0.48\pm 0.03\pm 0.03$
\\ \hline
 4S & $10$&$58$ & $0.25\pm 0.03\pm 0.01$ 
\\ \hline
 5S & $10$&$86$ & $0.31\pm 0.05\pm 0.07$
\\ \hline
 6S & $11$&$02$ & $0.13\pm 0.03\pm 0.03$
\\ \hline
\end{tabular}
\caption{\label{tabdata}    
The experimental numbers for the $\Upsilon$ masses and electronic decay
widths used for the calculation of the experimental moments
$P_n^{\rm ex}$. For the widths, the first error is statistical and the
second one systematic. All errors are estimated from the
numbers presented in Refs.~\protect\citebkcap{PDG,Albrecht1}. The electromagnetic
coupling at $10$~GeV and the ${\rm B}\bar{\rm B}$ threshold point are
$\tilde \alpha_{\rm em}^{-1} = \alpha_{\rm em}^{-1}(10\,\mbox{GeV}) 
= 131.8(1\pm0.005)$ and
$(\sqrt{s})_{{\rm B}\bar{\rm B}} = 2\times 5.279$~GeV.
The errors in the $\Upsilon$ masses and the ${\rm B}\bar{\rm B}$ 
threshold $(\sqrt{s})_{{\rm B}\bar{\rm B}}$
are small and can be neglected.
}
\end{center}
\vskip 3mm
\end{table}

A reliable computation of the theoretical moments $P_n^{\rm th}$ based on 
perturbative QCD is only possible if the effective energy range
contributing to the integration in Eq.~(\ref{momentsdef2}) 
is sufficiently larger than
$\lqcd\sim 300$~MeV.\cite{Poggio1}
For large values of $n$ the theoretical moments are dominated by the
non-relativistic kinematic regime close to the $\Upsilon$ resonances,
and it can be shown that the size of the
energy range is proportional $M/n$. This implies 
that $n$ should be chosen sufficiently smaller than $15$--$20$. 
A practical upper bound for $n$ is $10$, which has been used in
Refs.~\citebk{Hoang14,Hoang15,Beneke6,Hoang8}. Larger values
for $n$ up to $20$ have been used in the analyses of 
Refs.~\citebk{Voloshin2,Kuhn1,Penin1,Melnikov2}. One argument 
in favor of using also values of $n$ larger than $10$ is that the
leading order gluon condensate contribution of the large-$n$ 
theoretical moments\,\cite{Voloshin2,Onishchenko1} (see 
Fig.~\ref{fignonpertG2} and Eq.~(\ref{condensate1})),
\begin{eqnarray}
P_n^{\tiny <G^2>} 
& = & - 
\frac{\pi\,n^3}{72\,M^4}\,
e^{-0.4 C_F \alpha_s\sqrt{n}}\,
\langle 0|\alpha_s\,G_{\mu\nu} G^{\mu\nu}_{}|0\rangle\,
P^{\rm p.t.}_n
\, + \,\ldots
\,,
\qquad
\label{momentsnonperturbative}
\end{eqnarray}
where the $P_n^{\rm p.t.}$ is the $n$'s moment obtained from
perturbation theory, is less then a percent even for $n<20$.
On the other hand, the estimate of non-perturbative effects based on
a local expansion in gluonic and light quark condensates is only reliable if
$M/n\gg\Lambda_{\rm QCD}$, see Ref.~\citebk{Voloshin1}. In addition,
the size of the non-perturbative corrections coming from the
dimension-$6$ (and higher) condensates as well as all perturbative
Wilson coefficients for the condensates are not well known. It has
also been argued recently that the condensates might actually not be
the dominant non-perturbative effects.\cite{Zakharov2}

In order to suppress uncertainties from non-perturbative corrections,
$n$ should be chosen as small as possible. If $n$ is chosen very
small, typically $n\lsim 4$, the dynamics of the $b\bar b$ pair
encoded in the moments is dominated by relativistic energies and
momenta of order $M$, and the usual perturbative expansion
in the number of loops can be employed to determine the theoretical
moments.  At present the normalized photon-induced heavy
quark pair production cross section including the full mass dependence
is known at two-loop order.\cite{Chetyrkin2} Very few analyses
in terms of low-$n$ moments exist in the literature. 
A recent analysis in terms of low-$n$ moments and extracting the
bottom \ms mass as $\overline m(\overline m)= 4.21\pm 0.05$~GeV 
was carried out in Ref.~\citebk{Kuhn2}. 
Since this review concentrates on
non-relativistic dynamics, no discussion on this method can given
here. It should be noted that the result from
Ref.~\citebk{Kuhn2} is consistent with the large-$n$ sum rule analyses
that used threshold masses (see Tab.~\ref{tabbottommasses}).

The strategy that has been employed most frequently in the literature
is to use $n$ as large as possible in order to suppress
the contribution from the $b\bar b$ continuum. Here, $n$ can already
been considered large for $n\gsim 4$. 
The experimental uncertainties
for large-$n$ extractions of the bottom mass are around $15$~MeV.
In this case, theoretical
uncertainties dominate and there is the chance
to reduce the uncertainties in the bottom mass extractions by 
theoretical considerations. In large-$n$ 
moments the $b\bar b$ dynamics is dominated by non-relativistic
physics and it is necessary to sum the Coulomb singularities
$(\alpha_s/v)^i$ in the cross section (see Eqs.\
(\ref{RNNLLorders}) and (\ref{RNNLorders})).  
Because the size of the energy range
contributing to the $n$-th moment is of order $M/n$,
the typical center-of-mass velocity of the bottom quarks in the $n$-th
moment is of order $v_n=1/\sqrt{n}$.
This counting rule allows for a quantitative formulation
of the two requirements for the choice of $n$: theoretical reliability
demands $M v_n^2\gg\lqcd$, and dominance of the
non-relativistic dynamics demands $v_n\ll 1$. If both requirements are
met, the $b\bar b$ dynamics encoded in the moments is non-relativistic
and perturbative, i.e.\ the hierarchy $M\gg M v_n\gg M v_n^2 \gg
\lqcd$ is valid and the moments can be computed with the effective
theory methods described in the previous section. Thus, from the
conceptual point of view there is only a quite narrow window of $n$
values between about $4$ and $10$ that should be used for large-$n$
sum rules. If these considerations are applied to $c\bar c$ systems,
there is no window of $n$ values left for which a sum rule analysis
based on non-relativistic moments can be justified.

At present only fixed order analyses have been carried out for the
large-$n$ $\Upsilon$ sum rules. Initially the bottom pole mass was
determined. After the re\-le\-vance
of the bad high order behavior of the pole mass for $\QQbar$ systems
was fully appreciated and suitable threshold mass schemes were
proposed (Sec.\ \ref{sectionquarkmass}), subsequent analyses extracted
threshold masses, which were in a second step converted to the \ms
mass $\overline m(\overline m)$. In the following I give a historically
ordered compilation of the various sum rule analyses including brief
descriptions of the respective methods. The main results that have
been given by the authors are collected in Tab.\
\ref{tabbottommasses}. In Tab.\ \ref{tabbottommasses} I have also collected
bottom quark mass results from $b\bar b$ sum rules not discussed in
this review and from perturbative computations of the
$\Upsilon(\mbox{1S})$ mass (Sec.\ \ref{subsectionbottomspectrum}).
The results are compatible to recent lattice
analyses (see Refs.~\citebk{latticebquarkmass} and references therein) 
and the measurements from three-jet rates a LEP.\,\cite{LEPbquark}
It should be noted that the uncertainties obtained from recent lattice
analyses have uncertainties that are comparable to the ones obtained
from the perturbative analyses discussed in this
review. However, the lattice numbers for 
$\overline m(\overline m)$ generally tend towards larger
values.\,\cite{latticebquarkmass}

%
%
\begin{table}[t!]  
\vskip 0mm
\begin{center}
\begin{scriptsize}
\begin{tabular}{|l|l|c|l|} \hline
 author & $\overline m_b(\overline m_b)$ & other mass  & comments, Ref.
\\ \hline\hline
  Voloshin  \hfill 95
    &  
    & $m^{}_{\rm pole}=4.83\pm 0.01$
    & {\tiny NLO $\Upsilon$ sum rules, no theo.uncert.}~\protect\citebkcapx{Voloshin2}
\\ \hline
  K\"uhn  \hfill 98 
    & 
    & $m^{}_{\rm pole}=4.78\pm 0.04$
    & NLO $\Upsilon$ sum rules~\protect\citebkcapx{Kuhn1}
\\ \hline
  Penin \hfill 98
    & 
    & $m^{}_{\rm pole}=4.78\pm 0.04$
    & NNLO $\Upsilon$ sum rules~\protect\citebkcapx{Penin1}
\\ \hline
  Hoang \hfill 98
    & 
    & $m^{}_{\rm pole}=4.88\pm 0.13$
    & NLO $\Upsilon$ sum rules~\protect\citebkcapx{Hoang14} 
\\ \hline
  Hoang  \hfill 98
    & $4.26\pm 0.09$~* 
    & $m^{}_{\rm pole}=4.88\pm 0.09$
    & NNLO $\Upsilon$ sum rules~\protect\citebkcapx{Hoang14} 
\\ \hline
  Melnikov \hfill  98
    & $4.20\pm 0.10$
    & $M^{1\mbox{\tiny GeV}}_{\rm kin}=4.56\pm 0.06$
    & NNLO $\Upsilon$ sum rules~\protect\citebkcapx{Melnikov2} 
\\ \hline
  Penin  \hfill  98 
    & $4.21\pm 0.11$~* 
    & $m^{}_{\rm pole}=4.80\pm 0.06$
    & NNLO $\Upsilon$ sum rules~\protect\citebkcapx{Penin1} 
\\ \hline
  Jamin  \hfill 98
    & $4.19\pm 0.06$
    & 
    & $\Upsilon$ sum rules; no exact info~\protect\citebkcapx{Jamin1} 
\\ \hline
  Hoang  \hfill 99
    & $4.20\pm 0.06$ 
    & $M^{}_{\rm 1S}=4.71\pm 0.03$
    & NNLO $\Upsilon$ sum rules~\protect\citebkcapx{Hoang15} 
\\ \hline
  Beneke  \hfill 99
    & $4.26\pm 0.09$ 
    & $M^{2\mbox{\tiny GeV}}_{\rm PS}=4.60\pm 0.11$
    & NNLO $\Upsilon$ sum rules~\protect\citebkcapx{Beneke6}
\\ \hline
  Hoang \hfill  00
    & $4.17\pm 0.05$ 
    & $M^{}_{\rm 1S}=4.69\pm 0.03$
    & {\tiny NNLO $\Upsilon$ sum rules, charm mass eff.}~\protect\citebkcapx{Hoang8}
\\ \hline
  K\"uhn  \hfill 01  
    & $4.21\pm 0.05$ 
    & 
    & low-$n$ $\Upsilon$ sum rules, ${\cal O}(\alpha_s^2)$~\protect\citebkcapx{Kuhn2} 
\\ \hline\hline
  Pineda \hfill  97
    &  
    & $m^{}_{\rm pole}=5.00^{+ 0.10}_{- 0.07}$
    & {\tiny NNLO $\Upsilon(\mbox{1S})$ mass \& non-pert. eff.}~\protect\citebkcapx{Pineda7}
\\ \hline
  Beneke \hfill  99
    & $4.24\pm 0.09$ 
    & $M^{2\mbox{\tiny GeV}}_{\rm PS}=4.58\pm 0.08$
    & {\tiny NNLO $\Upsilon(\mbox{1S})$ mass \& non-pert. eff.}~\protect\citebkcapx{Beneke6} 
\\ \hline
  Hoang  \hfill 99
    & $4.21\pm 0.07$ 
    & \mbox{}\hspace{3.5mm}$M^{}_{\rm 1S}=4.73\pm 0.05$
    & {\tiny NNLO $\Upsilon(\mbox{1S})$ mass \&. non-pert. eff.}~\protect\citebkcapx{Hoang10} 
\\ \hline
  Pineda \hfill 01
    & $4.21\pm 0.09$ 
    & $M_{\rm RS}^{2\mbox{\tiny GeV}} = 4.39\pm 0.11$
    & {\tiny NNLO $\Upsilon(\mbox{1S})$ mass \& non-pert. eff.}~\protect\citebkcapx{Pineda6}
\\ \hline
  Brambilla  01
  & $4.19\pm 0.03$ & 
  & NNLO $\Upsilon(\mbox{1S})$ mass, p.th. only~\protect\citebkcapx{Brambilla5}
\\ \hline
\end{tabular}
\end{scriptsize}
\caption{\label{tabbottommasses}
Collection in historical order of results of bottom quark mass
determinations from 
$\Upsilon$ sum rules (upper part) and 
calculations of the $\Upsilon(\mbox{1S})$ mass (lower part). 
All results are given in units of GeV and rounded to units of
$10$~MeV.
Only results where $\alpha_s$ was taken as an input are shown. The
uncertainties quoted in the respective references have been added
quadratically. For analyses with several authors only the respective
first author is quoted. 
The \ms mass result that are indicated with a star have been
determined with an inconsistent conversion formula.
}
\end{center}
\vskip 3mm
\end{table}

\subsection{Pole Mass Results}
\label{subsectionpolemassbb}

Extractions of the bottom quark pole mass were carried out at NLO by
Voloshin\,\cite{Voloshin2} and by K\"uhn et al.,\,\cite{Kuhn1} and at 
NNLO by Penin and Pivovarov\,\cite{Penin1} and by
Hoang.\cite{Hoang14} In this section the 
methods and results are briefly described. The corresponding 
values for $\overline m(\overline m)$ are computed from the formula for
$m^{}_{\rm pole}/r_m$  in Eq.~(\ref{polemsbar}). The two-loop
relation is used for a NLO sum rule analysis and the three-loop relation 
for a NNLO sum rule    
analysis. This order-dependent conversion is necessary to achieve the
numerical cancellation of the 
large infrared-sensitive contributions contained in the pole mass
definition (Sec.\ \ref{subsectionthresholdmasses}). 
Since in all analyses the information on the exact
correlation of the numerical pole mass value with the set of
theoretical parameters used for the fits was not quoted,
I carry out the conversion treating the pole mass
result as uncorrelated. All uncertainties are added quadratically and 
perturbative conversion uncertainties are included.
The values for the \ms mass obtained
from  the pole mass in this way needs to be interpreted with some
caution, because the full correlation should have been taken into account
(Sec.\ \ref{subsectionpolemassVstat}). 

The NLO analysis by Voloshin\,\cite{Voloshin2} was the first large-$n$
sum rule 
analysis consistent with non-relativistic power counting. Voloshin
used time-independent perturbation theory to determine the NLO
corrections to the correlator ${\cal A}_1$ (Eq.~(\ref{correlatordef})). 
The moment integration of
Eq.\ (\ref{momentsdef2}) was carried out in a strict non-relativistic
expansion. After deforming the contour into the complex energy plane
away from the bound state poles
the integration can be solved using inverse Laplace transformations.
This method has the feature that bound state and continuum
contributions are treated at the same time and in exactly the same
way with respect to the non-relativistic expansion, i.e.\ the
corrections to bound state energies in the energy denominators 
of the resonances are also expanded. For each moment
Voloshin set the scale $\mu$ in the Coulomb potential such that the
NLO correction to the Green 
function vanishes. The scale of $\alpha_s$ in $c_1$
(Eqs.~(\ref{Rveft})) was set to the bottom mass.
He carried out a two parameter $\chi^2$ fit for
$m_{\rm pole}$ and $\alpha_s$ using four moments $P_n$ ($n=8,12,16,20$)
and obtained $m_{\rm pole}=4.827\pm 0.007$~MeV and
$\alpha_s(M_Z)=0.109\pm 0.001$. 
The result does not account for perturbative uncertainties.
Using Voloshin's value for $\alpha_s$ the corresponding \ms mass,
and two-loop conversion, 
gives 
$\overline m(\overline m)=4.33\pm 0.01$~GeV. 
For $\alpha_s(M_Z)=0.118\pm 0.003$ one obtains  
$\overline m(\overline m)=4.27\pm 0.04$~GeV.

Voloshin's NLO analyses was repeated by K\"uhn, Penin and Pivovarov
(KPP) in Ref.~\citebk{Kuhn1} and Hoang in Ref.~\citebk{Hoang14}. 
KPP used time-independent perturbation theory
for ${\cal A}_1$ and solved the moment integration exactly in the form
of Eq.\ (\ref{momentsdef2}). This method treats bound state and
continuum contributions differently, because the bound state energy
corrections in the energy denominators of the resonances are not
expanded out. KPP also improved the convergence of the perturbative
series by absorbing some part of the higher order corrections in the
Coulomb  potential into the LO Coulomb charge.
KPP fitted individual moments for $10\le n\le 20$ taking
$\alpha_s(M_Z)=0.118$ as an input. The scale of $\alpha_s$ in $c_1$
was set equal to $\mu$, the scale in the LO Coulomb potential.
They obtained
$m_{\rm pole}=4.78\pm 0.04$~GeV, where the uncertainty was estimated
from the variation of the result with $n$, $\mu=1.2$--$4.8$~GeV and
when the NNLO corrections to the Coulomb potential in the Schr\"odinger
equation are included. 
For the corresponding \ms mass, using the two-loop relation, one obtains
$\overline m(\overline m)=4.18\pm 0.05$~GeV
for $\alpha_s(M_Z)=0.118\pm 0.003$.

Hoang\,\cite{Hoang14} used Voloshin's method to determine the NLO
moments, but did not fix $\mu$. He carried out $\chi^2$ fits using
moments for various sets of $n$'s in the range between $4$ and $10$
for a given value of $\mu$. Perturbative
uncertainties were estimated by repeating the fit many times
choosing random values for $\mu$ between $1.5$ and $3.5$~GeV and for
the scale of $\alpha_s$ 
in $c_1$ between $2.5$ and $10$~GeV. In a two-parameter fit he
obtained  $m_{\rm pole} = 4.78\pm 0.14$~GeV and
$\alpha_s(M_Z)=0.109\pm 0.023$. 
The result for the pole mass  corresponds to 
$\overline m(\overline m)=4.18\pm 0.13$~GeV
at the two-loop level for $\alpha_s(M_Z)=0.118\pm 0.003$ 
as an input.
Taking $\alpha_s(M_Z)$ as an input Hoang obtained 
$m_{\rm pole} = 4.88\pm 0.13$~GeV for $\alpha_s(M_Z)=0.118\pm 0.004$.
This corresponds to 
$\overline m(\overline m)=4.28\pm 0.13$~GeV
at the two-loop level for $\alpha_s(M_Z)=0.118\pm 0.003$.

In the first NNLO analyses by  Penin and Pivovarov (PP)\,\cite{Penin1}
and the NNLO analysis by 
Hoang\,\cite{Hoang14} time-independent perturbation theory was used to
determine the NNLO corrections to the correlators. The UV-divergences
were regularized with a cutoff and the two-loop
corrections to $c_1$ were determined using the direct matching
method.\cite{Hoang1} The moment integration and the fitting procedure
were equivalent to the respective methods by PP and Hoang in their
NLO analyses. PP improved the convergence as in
Ref.~\citebk{Kuhn1} and fitted individual moments for $10\le n\le 20$
taking $\alpha_s(M_Z)=0.118$ as an input.  The scale of
$\alpha_s$ in 
$c_1$ and the cutoff scale
were set equal to $\mu$, the scale of $\alpha_s$ in the LO Coulomb
potential. They obtained
$m_{\rm pole}=4.78\pm 0.04$~GeV, where the uncertainty was estimated
from the varying $n$ and the scale 
$\mu=m_{\rm pole}\pm 1$~GeV.
In order to determine $\overline m(\overline m)$ from this NNLO
pole mass result one now has to use the three-loop relation for
$n_\ell=4$ massless quarks. For $\alpha_s(M_Z)=0.118\pm 0.003$ one obtains 
$\overline m(\overline m)=4.05\pm 0.07$~GeV.

In his NNLO analysis Hoang\,\cite{Hoang14} carried out $\chi^2$ fits 
using moments for various sets of $n$'s in the range between $4$ and
$10$ for a given value of $\mu$. As in his NLO analysis, perturbative
uncertainties were determined from 
varying independently $\mu$ between $1.5$ and $3.5$~GeV, and the
scale in $\alpha_s$ in $c_1$ and the cutoff between $2.5$ and
$10$~GeV. In a two-parameter fit Hoang obtained  
$m_{\rm pole} = 4.83\pm 0.07$~GeV and $\alpha_s(M_Z)=0.110\pm 0.014$.
The pole mass value corresponds to 
$\overline m(\overline m)=4.10\pm 0.09$~GeV
at the three-loop level for $\alpha_s(M_Z)=0.118\pm 0.003$.
Taking $\alpha_s(M_Z)=0.118\pm 0.004$ as an input he obtained 
$m_{\rm pole} = 4.88\pm 0.09$~GeV,
which corresponds to 
$\overline m(\overline m)=4.14\pm 0.10$~GeV
using the three-loop conversion for $\alpha_s(M_Z)=0.118\pm 0.003$.
For the determination of the \ms mass Hoang used the 
two-loop relation and obtained $\overline m(\overline m)=4.26\pm
0.09$~GeV for the  two-parameter fit and 
$\overline m(\overline m)=4.24\pm 0.08$~GeV, if $\alpha_s$ is taken as
an input. In both cases Hoang accounted for the correlation between
$m_{\rm pole}$ and $\alpha_s(M_Z)$ obtained from the fitting. 

In the second NNLO analysis by PP\,\cite{Penin1} the methods from
their first analysis were used, 
but $\mu$ was varied between $3.5$ and $6.5$~GeV and moments for $n$
between $8$ and $12$ were employed. In this analysis PP obtained 
$m_{\rm pole}=4.80\pm 0.06$~GeV, taking $\alpha_s(M_Z)=0.118$ as an
input. For $\alpha_s(M_Z)=0.118\pm 0.003$ this result corresponds to
$\overline m(\overline m)=4.06\pm 0.08$~GeV
if the proper three-loop relation is used.
PP used the two-loop relation and obtained 
$\overline m(\overline m)=4.21\pm 0.11$~GeV.

\subsection{Threshold Mass Results}

Extractions of threshold masses at NNLO were carried out by Melnikov
and Yelkhovsky,\,\cite{Melnikov2} Hoang\,\cite{Hoang15,Hoang8} and
Beneke and Signer.\cite{Beneke6} In this section the methods and
results of the respective analyses are briefly described. 

In the analysis by Melnikov and Yelkhovsky (MY)\,\cite{Melnikov2} the
kinetic mass $M^{}_{\rm kin}(1~\mbox{GeV})$ was determined. The theoretical
moments were determined in time-independent perturbation theory in the
pole mass scheme using
the same methods as previous NNLO analyses. The moment integration was
carried out in the form Eq.\ (\ref{momentsdef2}) and the energy
denominators of the bound state resonances were not expanded out.
MY fitted individual moments for $n=14,16,18$ taking
$\alpha_s(M_Z)=0.118$ as an input and extracted the bottom pole mass
as a correlated parameter for a given choice of $\mu$ between $2$ and
$4.5$~GeV. The scale of $\alpha_s$ in $c_1$ and the cutoff were fixed
to $5$~GeV. For each value of $\mu$ MY determined subsequently the
kinetic mass using the numerical value for $\mu$ for which the pole
mass value was determined. To achieve the numerical cancellation of
the large infrared-sensitive corrections it was important to keep the
bound state energy denominators unexpanded. In Tab.\
\ref{tabMYbbsum} the results obtained by MY for $\mu$ between $2$ and
$4.5$~GeV are displayed.
%
%
\begin{table}[t]
\begin{center}
$$
\begin{array}{||c||c||c|c||c|c||c|c||}
\hline
&
&
\multicolumn{2}{|c|}{14}&
\multicolumn{2}{|c|}{16}&
\multicolumn{2}{|c|}{18} \\
\cline{3-8}
{\rm Scales}& {\rm Order~of~PT}
&m_{\rm pole} & M_{\rm kin} &  m_{\rm pole} &  M_{\rm kin}  
& m_{\rm pole} & M_{\rm kin} 
\\ \hline \hline
\mu = 4.5 ~{\rm GeV} &{\rm LO}&  4.69 & 4.57 & 4.7 
&  4.57&  4.7&  4.575 \\ 
\cline{2-8}
\mu_{c_1} = 5 ~{\rm GeV} &{\rm NLO}
& 4.74&  4.49&  4.75 &  4.50 & 4.76
& 4.51 
\\ \cline{2-8}
\Lambda = 5 ~{\rm GeV}
&{\rm NNLO} &  4.89 &  4.51 & 4.90 
&  4.52 &  4.91 &  4.53  \\ \hline \hline
\mu = 3.5 ~{\rm GeV} &{\rm LO} & 4.73 & 4.59 
& 4.73& 4.59& 4.73 
& 4.595 
\\ \cline{2-8}
\mu_{c_1} = 5 ~{\rm GeV} &{\rm NLO} & 4.77& 4.50
& 4.78& 4.511
& 4.79& 4.52 
\\ \cline{2-8}
\Lambda = 5 ~{\rm GeV}
&{\rm NNLO}& 4.95& 4.54& 4.96 & 4.55 & 4.96&
4.55
\\ \hline \hline 
\mu = 2.5 ~{\rm GeV}
&{\rm LO} & 4.79& 4.63&
4.79& 4.63 & 4.79& 4.63 
\\ \cline{2-8} 
\mu_{c_1} =5 ~{\rm GeV} &{\rm NLO} & 4.82& 4.505& 4.83&
4.517 & 4.84&  
4.525
\\ \cline{2-8} 
\Lambda = 5 ~{\rm GeV}&{\rm NNLO}& 5.08 &
4.595 &  5.08& 4.595 &  5.09 & 4.6 
\\  \hline \hline
\mu = 2 ~{\rm GeV} &{\rm LO} & 4.84 & 4.67 
& 4.84& 4.666& 4.84 
& 4.66 
\\ \cline{2-8}
\mu_{c_1} = 5 ~{\rm GeV} &{\rm NLO}& 4.84& 4.49
& 4.85& 4.505
& 4.86& 4.514 
\\ \cline{2-8}
\Lambda = 5 ~{\rm GeV}
&{\rm NNLO}& 5.21& 4.66& 5.21 & 4.66 & 5.21&
4.66
\\
\hline
\end{array}
$$
\vspace{-0.3cm}
\caption{\label{tabMYbbsum}   
The bottom quark pole mass and the kinetic mass $M_{\rm kin}(1~\mbox{GeV})$  
in units of GeV
for $\mu$ between $2$ and $4.5$~GeV obtained  by Melnikov and
Yelkhovsky in
Ref.~\protect\citebkcap{Melnikov2} for the moments 
$n=14,16,18$.} 
\end{center}
\end{table}
MY found pole mass values in the
range between $4.5$ and $5.2$~GeV, where the NNLO shifts were larger
than the NLO one. For the kinetic mass the results were more stable
in the range between $4.49$ and $4.67$~GeV. However, the NNLO shifts
were still quite large compared to the NLO ones. For $\mu=4.5$ the
convergence is quite good, but for $\mu=2$~GeV the NNLO and NLO shifts
are of equal size. Since the resulting series for 
$m_{\rm kin}(1~\mbox{GeV})$  is alternating,
MY argued heuristically that the convergence could be improved by
an Euler transformation on the series. In this way they obtained the
result
\[
M^{}_{\rm kin}(1~\mbox{GeV}) \, = \, 4.56 \, \pm \, 0.06~\mbox{GeV}
\,,
\]
which is somewhat between the NLO and NNLO values and has an
uncertainty that is smaller than the range covered by the NLO and NNLO
values obtained in the analysis.
Since the kinetic mass is a short-distance mass there is no conceptual
need to convert the result to the \ms mass at a particular order.
In general, one should go for the best (i.e.\ highest order) conversion.
MY determined the \ms mass at two-loop order and obtained
$\overline m(\overline m)=4.20\pm 0.10$~GeV. 

\begin{figure}[t!] 
\begin{center}
\leavevmode
\epsfxsize=4.2cm
\epsffile[170 460 470 720]{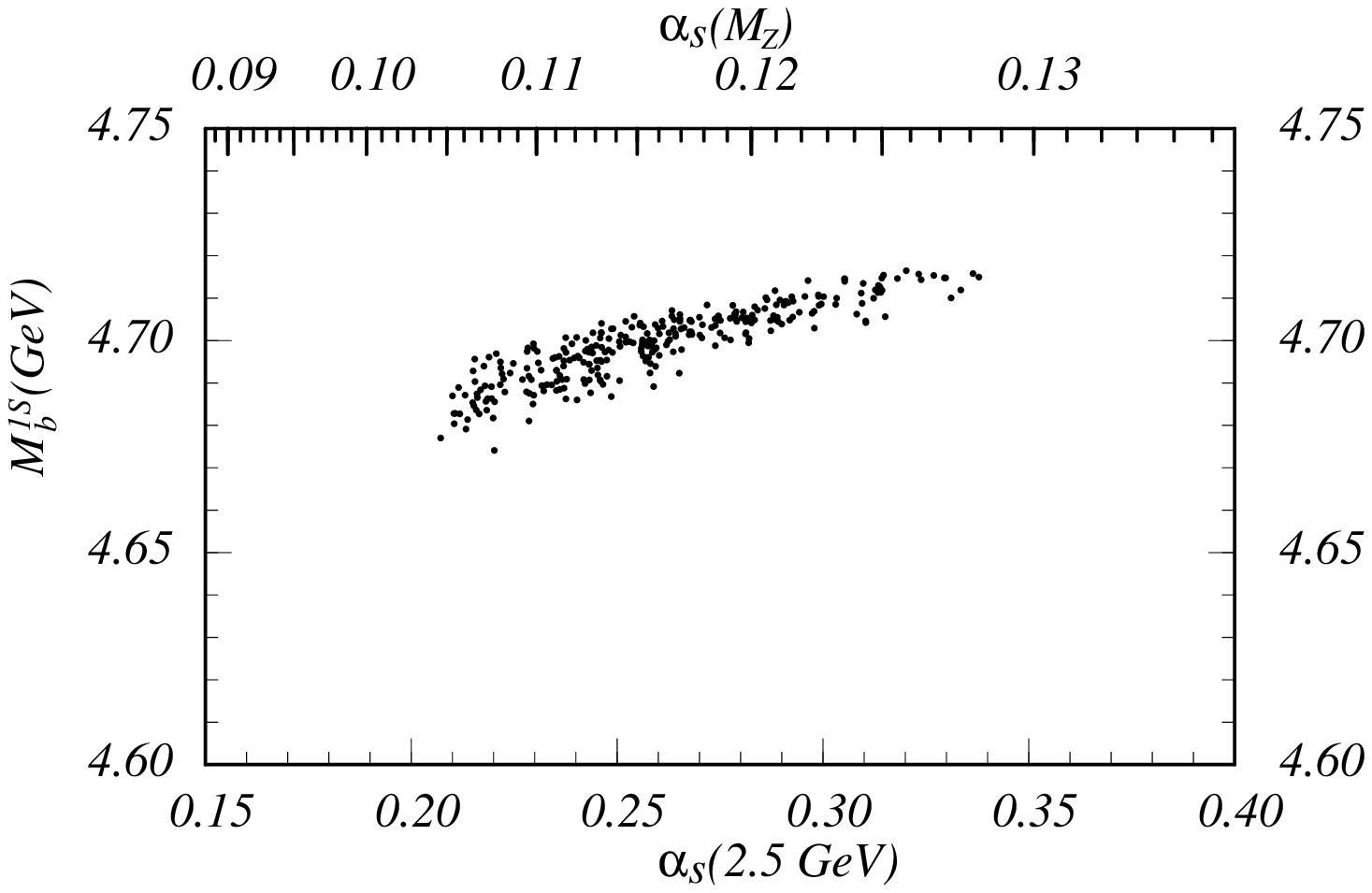}
\hspace{1.7cm}
\leavevmode
\epsfxsize=4.2cm
\epsffile[170 460 470 720]{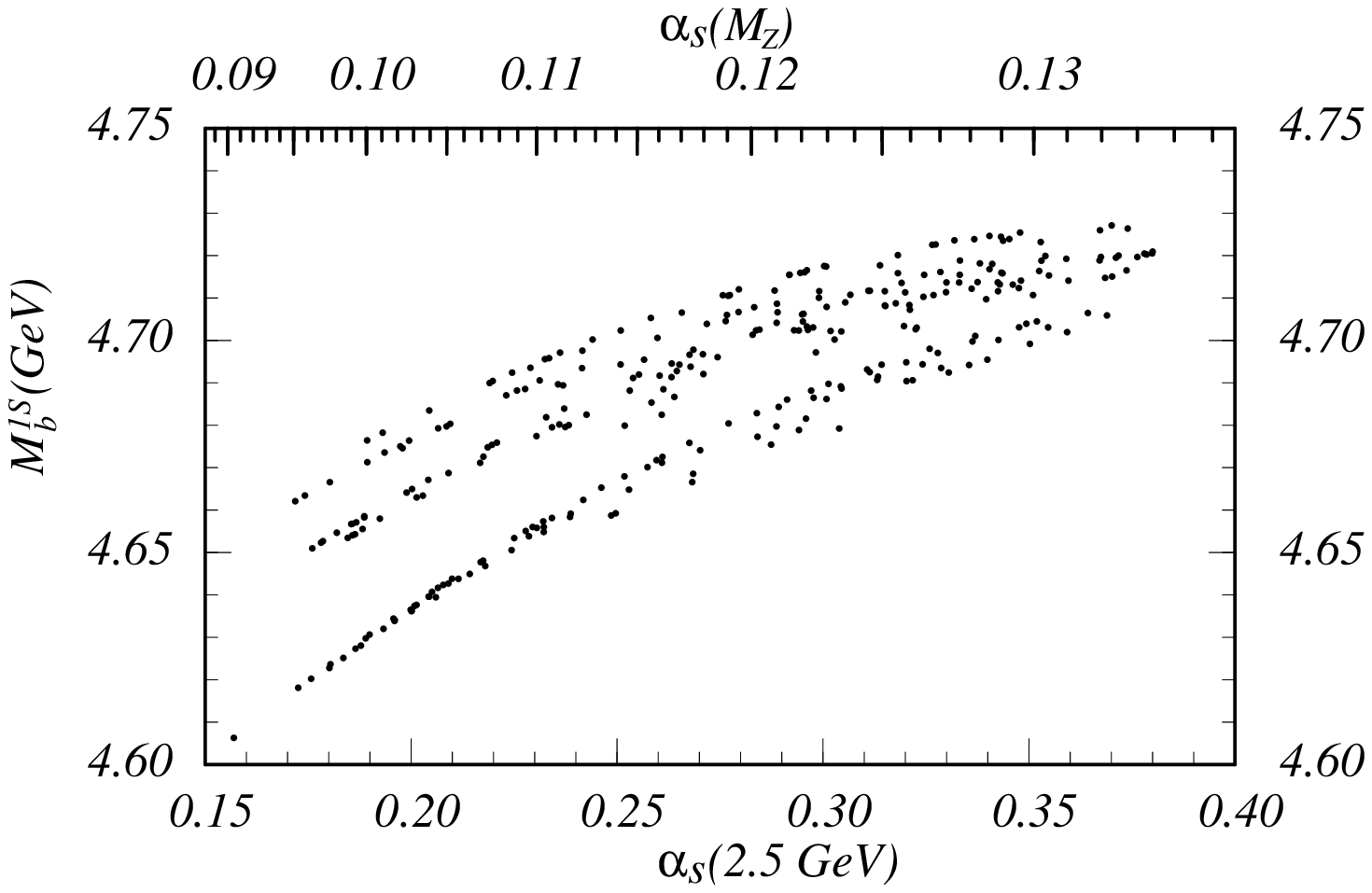}\\[3mm]
\leavevmode
\epsfxsize=4.2cm
\epsffile[170 460 465 720]{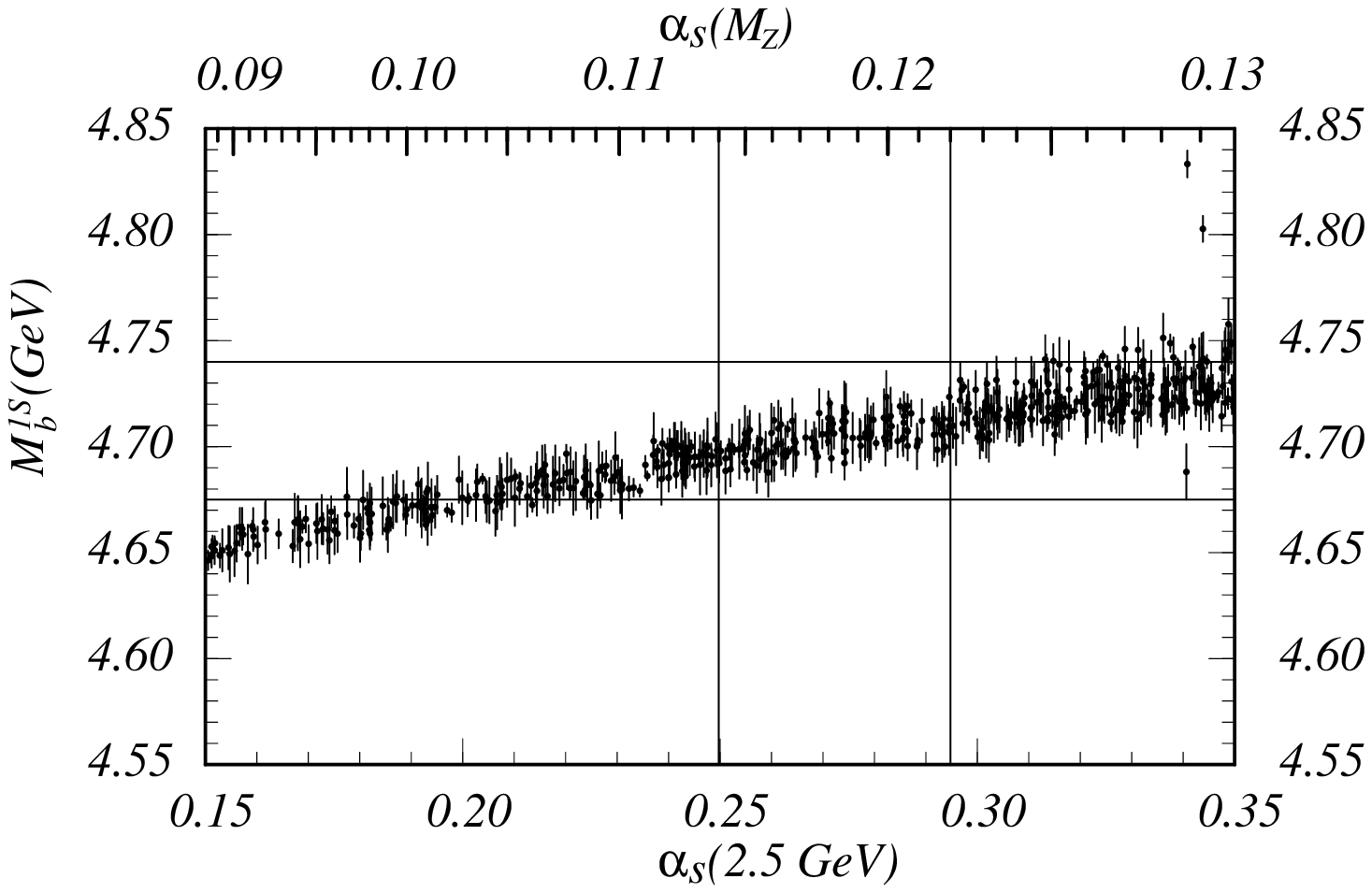}
\hspace{1.7cm}
\leavevmode
\epsfxsize=4.2cm
\epsffile[170 460 465 720]{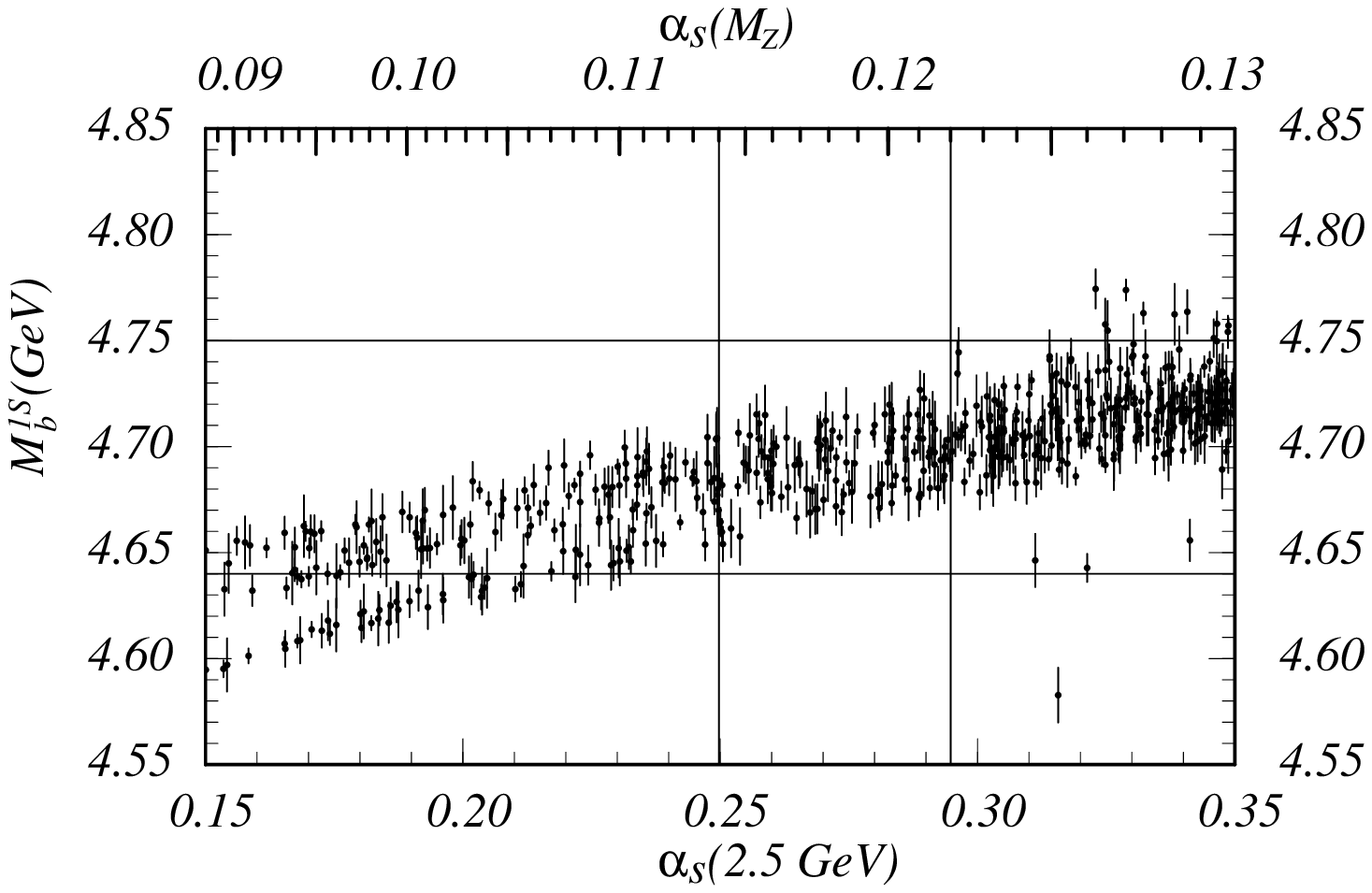}
\begin{picture}(0,0)(1,1)
\put(-288,187){$(a)$}
\put(-113,187){$(b)$}
\put(-288,74){$(c)$}
\put(-113,74){$(d)$}
\end{picture}
%
%
 \caption{\label{fig1Sfits}    
Result for the allowed region in the $M_{\rm 1S}^b$-$\alpha_s$ plane
obtained by Hoang in Ref.~\protect\citebkcap{Hoang15}. 
(a) Two parameter fit at NNLO and (b) NLO. The dots represent points
of minimal 
$\chi^2$ for a large number of random choices for the scales and sets
of $n$'s. Experimental errors are not displayed. 
(c) Single parameter fit at NNLO and (d) NLO taking $\alpha_s$ as an
input. The 
dots represent points of minimal $\chi^2$ for a large number of random
choices for the scales, sets 
of $n$'s and $\alpha_s$ . Experimental errors at 95\% CL are displayed
as vertical bars. 
}
 \end{center}
\end{figure}

In the analysis by Hoang\,\cite{Hoang15} the 1S mass was
determined. The theoretical moments were computed directly in the 1S
mass scheme 
using time-independent perturbation theory and the methods of
Ref.~\citebk{Hoang14} (Sec.\ \ref{subsectionpolemassbb}). All NLO and NNLO
corrections in the Schr\"odinger 
equation in Eq.\ (\ref{NNLLSchroedinger}) were strictly treated
perturbatively, i.e.\ the cancellation of the large infrared-sensitive
corrections associated with the pole mass were eliminated
analytically. For the fitting procedure exactly the same
method was used as for the earlier pole mass analysis in
Ref.~\citebk{Hoang14}. Hoang carried out $\chi^2$ fits 
using various sets of $n$'s in the range between $4$ and
$10$. For all moments in a given set the same value of $\mu$ was chosen.
This fitting procedure is different from fits of individual moments,
because it put highest statistical weight on linear combinations of
moments that are the least sensitive to the experimental uncertainties in
the electronic partial widths of the $\Upsilon$ mesons. Since the
correlation of the experimental moments coming from the $e^+e^-$ partial
widths (see Eq.\ (\ref{Pnexperiment})) is similar to the correlation
of the theoretical moments  
coming from their dependence on the $b\bar b$
wave function at the origin, the resulting $\chi^2$ function has a
smaller sensitivity to variations of $\mu$ than the individual
moments.
In contrast to fits of individual moments, which exclusively use
information on the size of the moments, the $\chi^2$ fitting
procedure also uses information on the relative size of the moments as
a function of $n$. Hoang estimated the perturbative uncertainties by
repeating the fit many times choosing independently random values for
$\mu$ between $1.5$ and $3.5$~GeV, for the scale of $\alpha_s$ in
$c_1$ and for the cutoff between $2.5$ and $10$~GeV. 
In a two-parameter fit Hoang obtained  
$M^{}_{\rm 1S} = 4.67\pm 0.07$~GeV and $\alpha_s(M_Z)=0.112\pm 0.023$
at NLO and
$M^{}_{\rm 1S} = 4.70\pm 0.03$~GeV and $\alpha_s(M_Z)=0.116\pm 0.013$
at NNLO.
A graphical representation of the NLO and NNLO fit results is displayed in
Fig.\ \ref{fig1Sfits}a and b.
In a single parameter fit taking $\alpha_s(M_Z)=0.118\pm 0.004$ as an input 
Hoang obtained
$M^{}_{\rm 1S} = 4.70\pm 0.05$~GeV at NLO. At NNLO he obtained 
\[
M^{}_{\rm 1S} = 4.71\pm 0.03~\mbox{GeV}
\,,
\]
which was considered the main result of the analysis. 
A graphical representation of the NLO and NNLO fit results is
displayed in Fig.~\ref{fig1Sfits}c and d. The results in
Fig.~\ref{fig1Sfits} show that  
the 1S mass has a very weak correlation to $\alpha_s$. Thus the
uncertainty in $\alpha_s$ is practically irrelevant for the
error in the 1S mass.
In contrast to the analysis by Melnikov and Yelkovsky\,\cite{Melnikov2}
and later by Beneke and Signer\,\cite{Beneke6} the results at NLO
and NNLO by Hoang are consistent with each other. The range of the 
NNLO result is contained entirely in the range of the NLO result.
This is a consequence of the fitting method used by Hoang.
Hoang converted the result to the \ms mass at two loops using the
large-$\beta_0$ approximation for the three-loop correction as an
estimate for the perturbative uncertainty. He
obtained $\overline m(\overline m)=4.20\pm 0.06$~GeV for $\alpha_s(M_Z)=0.118\pm
0.004$. The uncertainty is larger than for the 1S mass because of the
perturbative uncertainty in the conversion ($40$~MeV) 
and the uncertainty in $\alpha_s(M_Z)$ ($40$~MeV), which arises from a
relatively large one-loop correction in the relation between 1S and
\ms mass, see Eq.\ (\ref{polemsbar}). 
In Ref.~\citebk{Hoang8} Hoang repeated the analysis taking into
account also the effects of the non-zero charm quark mass. For $\overline
m_c(\overline m_c)=1.4\pm 0.3$~GeV he obtained 
\[
M^{}_{\rm 1S} = 4.69\pm 0.03~\mbox{GeV}
\]
for the 1S mass for the same fitting procedure. The result was
converted to the \ms mass using the full three-loop relation leading to
$\overline m(\overline m)=4.17\pm 0.05$~GeV for $\alpha_s(M_Z)=0.118\pm
0.003$. 
  
\begin{figure}[t]
\begin{center}
\leavevmode
\epsfxsize=1.8cm
\epsffile[250 225 370 565]{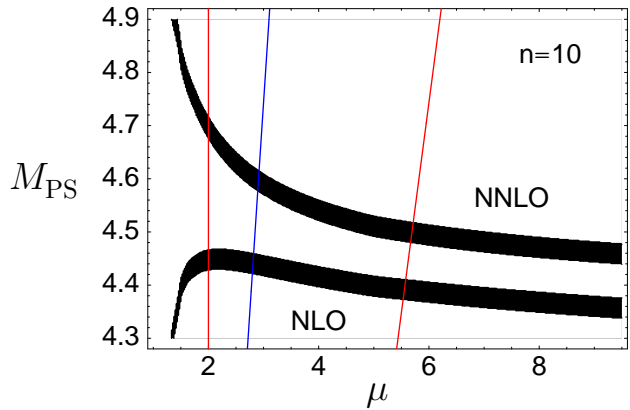}
\end{center}
\begin{picture}(0,0)(1,1)
\put(170,10){\Large $\mu$}
\put(35,90){\Large $M^{}_{\rm PS}$}
\end{picture}
\vspace*{-.3cm}
\caption{\label{figBSbb}   
The value of $M_{\rm PS}(2\,\mbox{GeV})$ in units of GeV 
obtained by Beneke and Signer in Ref.~\protect\citebkcap{Beneke6}
from the 10th moment as a function of the renormalization scale 
in NLO and NNLO and for $\alpha_s(M_Z)=0.118$. 
The dark region specifies the variation due to the 
experimental error on the moment. The two outer lines display
the scale variation from which the perturbative error was computed.}
\end{figure}

In the analysis by Beneke and Signer (BS)\,\cite{Beneke6}
the PS mass $m^{}_{\rm PS}(2~\mbox{GeV})$ was determined. The
theoretical moments were computed in the PS mass scheme using
time-independent perturbation theory to
solve the Schr\"odinger equation of Eq.\ (\ref{NNLLSchroedinger}).
BS used dimensional regularization and identified the form of the
potentials and coefficients using the threshold expansion, i.e.\ they
did not carry out an explicit matching calculation.
The moments of BS depend on $\mu$, the scale of $\alpha_s$ and the
factorization scale, which was set to the bottom quark mass. 
The moment integration was solved in the form of Eq.\
(\ref{momentsdef2}), but all NLO and NNLO corrections were strictly
treated perturbatively by expanding out the resonance delta functions
in $R^v$ in order to achieve an explicit analytical
cancellation of the large infrared-sensitive corrections associated with
the pole mass definition. BS also carried out a 
summation of some NLL order logarithms in $c_1$. The prescription
they used, however, is not consistent
with the known NLL order anomalous dimensions in pNRQCD or vNRQCD.
BS fitted the $n=10$ moment taking $\alpha_s(M_Z)=0.118\pm 0.003$ as
an input and varying $\mu$ between $2$~GeV and 
$2 m^{}_{\rm PS}(2~\mbox{GeV})/\sqrt{10}$. At NNLO they obtained  
\[
M^{}_{\rm PS}(2~\mbox{GeV}) = 4.60\pm 0.11~\mbox{GeV}
\,,
\label{BenekeSignerbottom}
\]
where the scale variation is the dominant source of uncertainty
($100$~MeV). The uncertainty from the strong coupling ($30$~MeV) and  
from the experimental data ($20$~MeV) are small.
At NLO BS obtained $m^{}_{\rm PS}(2~\mbox{GeV}) = 4.44$~GeV as the
central value with a smaller scale variation. 
No explicit uncertainties for the NLO result were quoted.
A graphical representation of the results is shown in Fig.\
\ref{figBSbb}.
%
%
Similar to the analysis of Melnikov and Yelkhovsky, which was
also based on fits of individual moments, the NLO and NNLO results are
not compatible numerically.
BS considered the NNLO result given above as the main result of
the analysis. The discrepancy with the NLO number was not included
in the error estimate.
The PS mass at $2~\mbox{GeV}$ has a stronger
correlation to the input value of $\alpha_s(M_Z)$ than for example the
1S mass. Taking into account this correlation to $\alpha_s$ and $\mu$ 
BS obtained $\overline m(\overline m)=4.26\pm 0.09$~GeV for the \ms
mass using two-loop conversion. 
The mean value also contains a $-10$~MeV
shift from a large-$\beta_0$ estimate for the (at the time of the analysis 
unknown) three-loop correction. The uncertainty from the strong
coupling ($10$~MeV) and from the conversion ($20$~MeV) were small.
The mean value for the \ms mass obtained by BS is higher than the one
by MY. This is because BS did not take into account their NLO results
in the determination of the mean value.

The analyses by MY, Hoang and BS are conceptually equivalent
with respect to the computations of the moments in the framework of
NRQCD. The differences in the results are a consequence of differences
in the fitting methods and the treatment of perturbative
uncertainties. Because the uncertainties are dominated by
theory, the error estimate is, naturally, a quite delicate task.
As such none of the analyses can a priori be considered better or more
realistic than another one. Nevertheless, the quoted uncertainties
have been subject to controversial discussions,
which shall be not repeated in this review. From a (at least for me)
conservative point of view it is fair to say all three analyses
contain elements that allow to consider the quoted uncertainties as
optimistic. 
The analysis by MY and BS showed that the results for the
threshold masses at NLO and NNLO obtained from fits of individual
moments are not compatible. MY dealt with this problem by using
heuristic arguments on alternating series and BS by ignoring the NLO
results in the final result. The origin of the difference between
NLO and NNLO results is that the R-ratio 
that is contained in the moments of Eq.\ (\ref{momentsdef2}) has 
quite large NNLO corrections. We face this problem also in fixed
order computations of the $t\bar t$ production cross section close to
threshold, which is discussed in Sec.\
\ref{subsectionfixedorderttbar}, see Fig.~\ref{figthresholdmasses}.
In the fitting procedure by Hoang the normalization problem of the
R-ratio was treated by also using information of the relative size of
the moments for different $n$, similar to taking ratios of the
moments. 
   
Clearly, the situation is not completely satisfactory and further
improvements in the computation of theoretical moments will be
necessary to settle the question of the uncertainties in the bottom
quark determination from $\Upsilon$ sum rules. One promising path
might be to use renormalization-group-improved perturbation theory
for the computation of the moments. In the case of top production
this has indeed improved the stability of the R-ratio (Sec.\
\ref{subsectionttbarcrossvNRQCD}). At the present time a
renormalization-group-improved sum rule analysis has not yet been
carried out.

\vspace{1cm}

\section{Heavy Quarkonium Spectrum} 
\label{sectionspectrum}

Theoretical descriptions of the heavy quarkonium spectrum represent
the classic application of non-relativistic quarkonium physics. Early
work in QCD-inspired potential models provided very precise
determinations of charmonium and bottomonium spectra
and physical insights into the long-distance aspects of interquark
forces.\cite{Buchmuller1} 
However, potential models cannot be derived quantitatively 
from first principles in QCD, which makes them less suitable for
determinations of QCD parameters, most notably the heavy quark
masses. With the
advent of effective theories for heavy quarkonium systems (Secs.\
\ref{sectionNRQCD}, \ref{sectionpNRQCD}, \ref{sectionvNRQCD})
a first principle description of the spectrum has become possible
conceptually, but precise computations are difficult,
because a reliable technology only exists for systems where the
hierarchy $m\gg mv\gg mv^2\gg\lqcd$ is valid. This is the case that
is discussed generically in Sec.\
\ref{subsectioncalculatesprectrum}. In this case the
predictions of the effective theories can be computed perturbatively
supplemented by local condensate contributions, which incorporate
averaged non-perturbative effects. 
 As a matter of principle, at most the
bottomonium ground states can be expected to be described with this
technology. Determinations of the bottom quark mass from the mass of the
$\Upsilon(\mbox{1S})$ state are described in Sec.\
\ref{subsectionbottomspectrum}. Considerations of the toponium
spectrum are irrelevant, due to the large top decay width 
$\Gamma_t\approx 1.5~\mbox{GeV}\gg\lqcd$ (see Sec.\
\ref{sectionttbar}).

For most quarkonium systems $mv^2$ is of order $\lqcd$
or even smaller. In the framework of pNRQCD
an approach for these systems has been roughly
outlined\,\cite{Brambilla1} 
(Sec.\ \ref{subsectionlqcdlarger}), but no systematic prescription to
carry out quantitative computations of the spectrum has yet been
devised. In the framework of vNRQCD such systems can a priori not be
treated because ultrasoft modes are by construction already present at
the hard matching scale and because soft and ultrasoft scales are
correlated at all times (Sec.\ \ref{sectionvNRQCD}). In the framework
of NRQCD (Sec.\ \ref{sectionNRQCD}), on the other hand, a systematic
treatment of the spectrum of these systems is unknown with analytic
methods, but is possible with lattice
simulations.\cite{Latticereviews} On the 
lattice,  theoretical precision is limited by systematic
uncertainties, such as from unquenching and extrapolating and matching
to the continuum limit. No further discussion on lattice methods for
heavy quarkonium spectra is given in this review. 
In a recent analysis Brambilla et al.\,\cite{Brambilla4,Brambilla5}
examined the 
charmonium, bottomonium and $B_c$ spectrum using only the perturbative
contributions for the case  $m\gg mv\gg mv^2\gg\lqcd$. Setting the
renormalization scale using a minimal sensitivity prescription they
found remarkable agreement of the bottomonium sprectrum up to the
$n=3$ radial excitation without
accounting for any non-perturbative effects. Brambilla et
al. concluded that perturbative methods devised for the case 
$m\gg mv\gg mv^2\gg\lqcd$ might have a wider range of applicability
than previously expected. For subsequent discussions following the same
interpretation see also  Ref.~\citebk{Kiyo2}. The results of Brambilla
et al., however, appear to be in sharp contradiction to results on the
size of non-perturbative corrections obtained for higher bottomonium
excitations in Ref.~\citebk{Pineda6}.

\subsection{Calculation of the $\QQbar$ Spectrum}
\label{subsectioncalculatesprectrum}

For the case $m\gg mv\gg mv^2\gg\lqcd$ the computation of the heavy
quarkonium spectrum follows the presentation given in Sec.\
\ref{sectionQQbarproduction} on heavy quark production close to
threshold. At NNLL order in the renormalization-group-improved
approach of pNRQCD (Sec.\ \ref{sectionpNRQCD}) or vNRQCD (Sec.\
\ref{sectionvNRQCD}), or at NNLO in the fixed order NRQCD (Sec.\
\ref{sectionNRQCD}) approach the perturbative contribution of $\QQbar$
dynamics can be 
described by a common text-book two-body Schr\"odinger equation. In
momentum space representation and in a threshold mass scheme the
Schr\"odinger equation has the generic form 
\begin{eqnarray} 
\lefteqn{
 \bigg[\, \frac{\bmp^2}{M}\bigg(1+\frac{\delta M}{M}\bigg) -
 \frac{\bmp^4}{4M^3} - (E +2\, \delta M) \, \bigg]
 \, \tilde\Psi({\bmp})
}
\nonumber\\[2mm] & &\quad\qquad\qquad\qquad\qquad
 + \int \frac{d^3{\bmk}}{(2\pi)^3}\,\tilde V({\bmp},{\bmk})\, 
 \tilde\Psi({\bmk})
 = \, 0
\,.
\label{NNLLSchroedingerwave}
\end{eqnarray}
The perturbative contribution of the masses of the $\QQbar$ states
are given by the energy eigenvalues of Eq.\
(\ref{NNLLSchroedingerwave}),
\begin{eqnarray}
M_{\QQbar}^{\rm\scriptsize pert} & = &
2M + E
\,.
\end{eqnarray}
The classification of the solutions according to quantum numbers is 
standard text-book knowledge and shall not be repeated here.
The renormalization-group-improved computation for the energy
eigenvalues at NNLL order takes the generic form
\begin{eqnarray}
E & = &
M\alpha_s^2 \,
 \sum\limits_i \left(\alpha_s\ln\alpha_s \right)^i\,
 \bigg\{1\,\mbox{(LL)}; \alpha_s,\,\mbox{(NLL)}; 
 \alpha_s^2\,\mbox{(NNLL)}\bigg\}
 \,.
 \label{ENNLLorders}
\end{eqnarray}
In the fixed order approach  logarithms of $\alpha_s$ are not
summed and the energy eigenvalues at NNLO have the generic form
\begin{eqnarray}
E & = &
M\alpha_s^2 \,
 \bigg\{1\,\mbox{(LO)}; \alpha_s,\,\mbox{(NLO)}; 
 \alpha_s^2\,\mbox{(NNLO)}\bigg\}
 \,.
 \label{ENNLLO}
\end{eqnarray}
In Eq.\ (\ref{NNLLSchroedingerwave}) the difference between
renormalization-group-improved and fixed order computation is just the
form of the coefficients of the potential $\tilde V$ (see Secs.\
\ref{subsectioncrosssectionvnrqcd} and
\ref{subsectionQQbarfixedorder}). From the technical point of view the
computations needed to solve Eq.\ (\ref{NNLLSchroedingerwave}) are
equivalent in both approaches. In the fixed order approach the first
computation of the NNLO $\QQbar$ spectrum was carried out by Pineda
and Yndurain\,\cite{Pineda7} (see also
Refs.~\citebk{Titard1,Melnikov2}). Renormalization-group-improved
computations at NNLL 
order were carried out by Hoang et al. in Ref.~\citebk{Hoang4} and by 
Pineda in Ref.~\citebk{Pineda3}. Technically, the energy eigenvalues
are determined 
using Rayleigh-Schr\"odinger time-independent perturbation theory
starting from the well known leading order non-relativistic Coulomb
solution and including successively higher order terms from the
potential and the kinetic energy term. To achieve an explicit
cancellation of the unphysical large-order behavior associated with the
pole mass it is mandatory to consistently include the contributions of
$\delta M$ at each order of perturbation theory (Sec.\
\ref{subsectionthresholdmasses}). Since at NNLL order (or NNLO) the
corresponding integrals do not contain any UV divergences, no
regularization prescription for Eq.\ (\ref{NNLLSchroedingerwave})
needs to be specified at this order.

At N$^3$LL order (or N$^3$LO) the computations of the perturbative part
of the $\QQbar$ spectrum becomes more complicated. Apart from N$^3$LL
order (or N$^3$LO) corrections to the potential $\tilde V$, which
include for example two-loop corrections to the $1/m|{\bmk}|$
potential or three-loop corrections to the Coulomb potential and the
corresponding three- and four-loop anomalous dimensions, also
non-logarithmic corrections from ultrasoft gluons need to be
considered (see for example Figs.\ \ref{figpnrqcdselfenergy} and 
\ref{figCoulombusoft}). The latter corrections involve UV divergences
and depend on the regularization scheme. The corresponding UV
divergences are canceled by conterterms of the $1/m^2$, $1/m|{\bmk}|$
and $1/{\bmk^2}$ potentials (Secs.\ \ref{subsectionpNRQCDrunning},  
\ref{subsectionvNRQCDrunning} and \ref{sectionvNRQCDvspNRQCD}), which
become renormalized. The sum of the ultrasoft corrections and the
corrections from the potentials are scheme-independent. 
The ultrasoft corrections are the QCD analogue of
the Bethe-log corrections known from QED and cannot be represented by
a two-body Schr\"odinger equation, because they correspond to a higher
Fock $\QQbar$-Gluon state. Some corrections 
at N$^3$LO in the fixed order approach are already known. The
non-logarithmic corrections from ultrasoft gluons were determined in 
Ref.~\citebk{Kniehl2}, and two-loop corrections to the
$1/m|{\bmk}|$ potential were determined in Ref.~\citebk{Kniehl1}. 
The N$^3$LO corrections in the large-$\beta_0$ approximation were
determined in Refs.~\citebk{Kiyo1,Hoang8}. From the NNLL order
renormalization-group-improved computations\,\cite{Hoang4,Pineda3} (see
also Ref.~\citebk{Brambilla3}) all N$^3$LO corrections proportional to
powers of $\ln\alpha_s$ are also known.
In Ref.~\citebk{Kiyo3} Kiyo and Sumino claimed that additional 
NNLO corrections $\propto M\alpha_s^4$ and $M\alpha_s^4\ln\alpha_s$ are
found when Bethe-Salpeter rather than effective theory techniques are
employed with dimensional regularization. The contributions found in
Ref.~\citebk{Kiyo3} originate
from an incomplete expansion of the small $k_0$ component of the
one-loop corrections to the Coulomb potential in the potential region
(Eq.\ (\ref{momentumregions}))
and are unphysical. 

In the case $m\gg mv\gg mv^2\gg\lqcd$ the non-perturbative
contributions to the spectrum can be expressed as an
expansion in terms of local gluon and light quark
condensates.\cite{Voloshin1,Leutwyler1} At leading order in the
multipole expansion the 
first non-perturbative correction arises from radiation and absorption
of a gluon with energy of order $\lqcd$ from the $\QQbar$
pair. Conceptually this contribution is closely related to the
ultrasoft correction mentioned before. For $mv^2\gg\lqcd$, both
contributions can be separated by an expansion in
$\lqcd/mv^2$, see Refs.~\citebk{Voloshin1,Brambilla3}. Physically, this
expansion means that the time span between emission and absorption of
the gluon is much smaller than the non-perturbative correlation time.
In configuration space representation 
the coupling of the gluon to the $\QQbar$ pair is just the
${\bmr}.{\bmE}$ interaction in pNRQCD. In momentum space
representation the interaction arises from the diagrams in
Fig. \ref{figpnrqcdradiation}. The first non-perturbative correction
is in analogy to Eq.\ (\ref{condensate1}) and
reads\,\cite{Voloshin1,Leutwyler1} 
\begin{eqnarray}
E_{nl}^{\rm\scriptsize n.p.} & = &
-\frac{\pi}{18}\,\langle 0|\alpha_s G_{\mu\nu} G^{\mu\nu}_{}|0\rangle
\,
\int\! d^3{\bmx}\,\int\! d^3{\bmy}\,{\bmx}.{\bmy}\,
 \Psi_{nl}({\bmx})\, G_o({\bmx},{\bmy})\, \Psi_{nl}({\bmy})
\nonumber\\ & = &
\frac{8\,\pi\,n^6\,M\,a_{nl}}{9\,(M C_F\alpha_s)^4}\,
\langle 0|\alpha_s G_{\mu\nu} G^{\mu\nu}_{}|0\rangle
\,,
\label{condensate2}
\end{eqnarray}
where $a_{1S}=1.652$, $a_{2S}=1.783$, $a_{3S}=1.869$, $a_{2P}=1.123$,
etc..
The next order terms in the $\lqcd/mv^2$ expansion correspond to
dimension-six condensates and have been determined by Pineda in
Ref.~\citebk{Pineda8}. It should be noted that the coefficients of the
operator expansion in local condensates are at present only known at
LO (or LL order) in the non-relativistic expansion and still have a
rather large scheme-dependence. 

The summation of all orders in $\lqcd/mv^2$ leads 
formally to non-local condensate contributions (see
e.g.\ Refs.~\citebk{Brambilla1,Dosch1}) 
describing the non-perturbative corrections for $\lqcd\sim
mv^2$. A reliable quantitative treatment of non-perturbative
corrections for this case in quarkonium systems has not yet
been developed. However, for $\lqcd\sim mv^2$ also the perturbative
contributions are affected  because gluons with ultrasoft energies
and energies of order $\lqcd$ are identical and cannot be separated,
and because $\alpha_s(mv^2)$ has to be
considered of order one.
This means that the non-perturbative/ultrasoft corrections
have to be counted as NNLO contributions.

\subsection{Bottom Mass Determinations from the $\Upsilon(\mbox{1S})$ Mass}
\label{subsectionbottomspectrum}

There are a number of analyses where the bottom mass was
determined from the $\Upsilon(\mbox{1S})$ mass using fixed order
calculations at NNLO as described in the previous section. From the
technical point of view, in all analyses it was assumed that the
hierarchy $m\gg mv\gg mv^2\gg\lqcd$ is valid and that non-perturbative
corrections can be expressed as an operator expansion in terms of
local condensates. The expression for the perturbative contribution of
the $\Upsilon(\mbox{1S})$ mass at NNLO in a threshold mass scheme
(Sec.\ \ref{subsectionthresholdmasses})
with $\delta M=M-m^{}_{\rm pole}=\delta M^{}_{\rm LO}+
\delta M^{}_{\rm NLO}+\delta M^{}_{\rm NNLO}$
reads ($a_s=\alpha_s^{(n_\ell=4)}(\mu)$)
\begin{eqnarray}
M_{\Upsilon(\mbox{\scriptsize 1S})}^{\mbox{\scriptsize pert}} 
& = &
2\, M
\,- \,2\,\Big[\,
    M \Delta^{\rm LO} + \delta M^{}_{\rm LO}
  \,\Big]
\,-\,2\,\Big[\,
    M \Delta^{\rm NLO} + \delta M^{}_{\rm NLO} 
  \,\Big]
\nonumber
\\[2mm] & & \hspace{1cm}
\,-\,2\,\Big[\,
    M \Delta^{\rm NNLO} + \delta M^{}_{\rm NNLO} 
    - \Delta^{\rm LO} \delta M^{}_{\rm LO}
 \,\Big]
\,,
\label{upsilon1S}
\end{eqnarray}
where, assuming massless light quarks,
\begin{eqnarray}
\Delta^{\rm LO} & = &
 \frac{C_F^2\,a_s^2}{8}
\,,\qquad 
\Delta^{\rm NLO} \, = \,
\frac{C_F^2\,a_s^2}{8}\, 
\Big(\frac{a_s}{\pi}\Big)\,\bigg[\,
\beta_0\,\bigg( L + 1 \,\bigg) + \frac{a_1}{2} 
\,\bigg]
\,,
\nonumber
\\[2mm] 
\Delta^{\rm NNLO} & = &
\frac{C_F^2\,a_s^2}{8}\, \Big(\frac{a_s}{\pi}\Big)^2\,
\bigg[\,
\beta_0^2\,\bigg(\, \frac{3}{4} L^2 +  L + 
                             \frac{\zeta_3}{2} + \frac{\pi^2}{24} +
                             \frac{1}{4} 
\,\bigg) + 
\beta_0\,\frac{a_1}{2}\,\bigg(\, \frac{3}{2}\,L + 1
\,\bigg)
\nonumber\\[3mm]
& & \hspace{0.5cm} +
\frac{\beta_1}{4}\,\bigg(\, L + 1
\,\bigg) +
\frac{a_1^2}{16} + \frac{a_2}{8} + 
\bigg(\, C_A - \frac{C_F}{48} \,\bigg)\, C_F \pi^2 
\,\bigg]
\,,
\nonumber
\\[2mm] 
L & \equiv & 
\ln\Big(\frac{\mu}{C_F\,a_s\,M}\Big)
\,.
\end{eqnarray}
The one- and two-loop coefficients of the beta-function, 
$\beta_0$ and $\beta_1$, and the
constants $a_1$ and $a_2$ are from the corrections to the Coulomb
potential in Eq.~(\ref{staticpotentialschroeder}).
In the 1S mass scheme 
$\delta M^{}_{\rm LO,NLO}=-M \Delta^{\rm LO,NLO}$ and
$\delta M^{}_{\rm NNLO}=-M \Delta^{\rm NNLO}-M (\Delta^{{\rm LO}})^2$,
and, by construction, perturbative corrections are zero. The
non-perturbative contribution from Eq.\ (\ref{condensate2}) reads
\begin{eqnarray}
E_{10}^{\rm\scriptsize n.p.}
& = &
\frac{0.465\,\pi}{M^3\,\alpha_s^4}\,
\langle 0|\alpha_s G_{\mu\nu} G^{\mu\nu}_{}|0\rangle
\,.
\label{upsilon1Scondensate}
\end{eqnarray}
Typical values for the gluon condensate used in the
literature are in the range 
$\langle 0|\alpha_s G_{\mu\nu} G^{\mu\nu}_{}|0\rangle =
0.05\pm 0.03\,\mbox{GeV}^4$.
The threshold mass is extracted from fitting 
$M_{\Upsilon(\mbox{\scriptsize 1S})}^{\mbox{\scriptsize pert}}+
E_{10}^{\rm\scriptsize n.p.}$
to the experimental value\,\cite{PDG} 
$M_{\Upsilon(\mbox{\scriptsize 1S})}^{\mbox{\scriptsize exp}}
=9460.30\pm 0.26$~MeV.

In the following I review the bottom mass determinations based on
fixed order NNLO calculations in historical order. The results
obtained in the analyses are collected in Tab.\ \ref{tabbottommasses}.
To simplify the presentation on the determination
of uncertainties I discuss only the dominant source of uncertainties
as obtained in the different analyses.

Pineda and Yndurain (PY)\,\cite{Pineda7} determined the pole mass. For the
strong coupling PY used a scheme where part of the corrections to the
Coulomb potential are absorbed into $\alpha_s$. In this scheme the
size of the non-perturbative contribution was at the level of
$10$~MeV. PY obtained 
$m^{}_{\rm pole}=5.00^{+0.10}_{-0.07}$~GeV, where the uncertainty was
dominated by the error in the strong coupling,
$\alpha_s(M_Z)=0.114^{+0.006}_{-0.004}$. PY also determined the \ms
mass using two-loop conversion, 
$\overline m(\overline m)=4.44^{+0.04}_{-0.03}$~GeV. Converting to the
\ms mass at three loops, which is necessary to achieve the numerical
cancellation of the large unphysical corrections associated with the
pole mass, the central value reads 
$\overline m(\overline m)=4.32$~GeV
for $\alpha_s(M_Z)=0.114$, using the formula for $m^{}_{\rm pole}/r_m$ 
in Eq.~(\ref{polemsbar}). 

Hoang\,\cite{Hoang10} determined the 1S mass. In the 1S scheme there is, by
construction, no correlation to $\alpha_s$. Perturbative uncertainties
cannot be estimated in the usual way by changing the renormalization
scale $\mu$. A conservative estimate would be, for example, to account
for the difference in the result when 
$\Delta_{\rm NNLO}$ is replaced by its large-$\beta_0$
approximation. Hoang assigned $100$~MeV for the size of the
non-perturbative contributions and treated them as an overall
uncertainty without including them in the determination of the
central value. He obtained $M^{}_{\rm 1S}=4.73\pm
0.05$~GeV. Perturbative uncertainties were not included. Using
three-loop conversion to the \ms mass he obtained
$\overline m(\overline m)=4.21\pm 0.07$~GeV for 
$\alpha_s(M_Z)=0.118\pm 0.004$. The uncertainty is larger
than the one of the 1S mass because the \ms-1S mass relation has a
rather strong dependence on $\alpha_s$ coming from the one-loop
correction in the \ms-pole mass relation.
Since the 1S mass (as
well as the PS and the RS mass mentioned below) are short-distance
masses the conversion to the \ms mass does not have to be carried out
at a specific order. However, one should go for the highest order
conversion to keep perturbative uncertainties small. A compact
approximation of the three-loop 
conversion formula for the \ms-1S mass relation, which also accounts
for charm mass effects and the scale $\mu$ in the strong coupling, was
given in Ref.~\citebk{Hoang8},
\begin{eqnarray}
\overline m(\overline m)
& = &
\Big[\,
4.169\,\mbox{GeV} 
 -  0.01\,\Big(
  \overline m_c(\overline m_c)-1.4\,\mbox{GeV} \Big)
 +  0.925\,\Big(
  M^{}_{\rm 1S}
   -4.69\,\mbox{GeV} \Big)
\nonumber
\\ & & 
\, - 9.1\,\Big( 
  \alpha_s^{(5)}(M_Z) - 0.118
 \Big)\,\mbox{GeV}
 +  0.0057\,\Big( 
   \mu - 4.69
 \Big)\,\mbox{GeV}
\,\Big]
\,.
\label{msbarapproximation}
\end{eqnarray}
For $\overline m_c(\overline m_c)>0.4$~GeV and $\mu>2.5$~GeV the
difference between this approximation formula and the exact result is
less than $3$~MeV. 

Beneke and Signer (BS)\,\cite{Beneke6} determined the PS mass. They estimated
the uncertainty in the PS mass coming from the non-perturbative
contributions as $\pm 70$~MeV and obtained
$M^{}_{\rm PS}(2~\mbox{GeV})=4.58\pm 0.08$~GeV, where non-perturbative
contributions were not included in the central value and
$\alpha_s(M_Z)=0.118\pm 0.004$ was taken as an input.
For the \ms mass
BS used three-loop conversion, and they obtained
$\overline m(\overline m)=4.24\pm 0.09$~GeV. 

Pineda\,\cite{Pineda6} determined the RS and RS$^\prime$ masses
assuming a $75$~MeV 
uncertainty from the non-perturbative and N$^3$LO ultrasoft
contributions. He obtained 
$M^{}_{\rm RS}(2~\mbox{GeV})=4.39\pm 0.11$~GeV and
$M^{}_{\rm RS^\prime}(2~\mbox{GeV})=4.78\pm 0.08$~GeV. The error in
the RS mass is significantly larger than the error from the
non-perturbative and N$^3$LO ultrasoft contributions due to a
theoretical ambiguity in the 
subtraction term that defines the RS mass. For the \ms mass
he used three-loop conversion, and obtained a combined result of
$\overline m(\overline m)=4.21\pm 0.09$~GeV. 
 
Brambilla, Sumino and Vairo (BSV)\,\cite{Brambilla5} determined the \ms mass
directly from the $\Upsilon(\mbox{1S})$ mass without taking into
account non-perturbative contributions, which technically means that
they used Eq.~(\ref{msbarapproximation}) inserting half the
$\Upsilon(\mbox{1S})$ mass as the 1S mass value. They fixed the
renormalization scale $\mu$ using a minimal sensitivity requirement,
which corresponds to $\mu\approx 2.5$~GeV in
Eq.~(\ref{msbarapproximation}). BSV obtained  
$\overline m(\overline m)=4.19\pm 0.03$~GeV taking
$\alpha_s(M_Z)=0.118\pm 0.002$ as an input.

\vspace{1cm}

\section{Conclusions} 
\label{sectionconclusions}

In the last few years the conceptual understanding of the perturbative
treatment of heavy quarkonium systems with $m\gg mv\gg mv^2\gg\lqcd$,
$m$ and $v$ being the heavy quark mass and velocity, respectively,
has improved considerably. New effective theories have been developed
that are more sophisticated than NRQCD and that allow
systematic QCD calculations using fixed order
or renormalization-group-improved perturbation theory.
Starting from
solutions of the non-relativistic Schr\"odinger equation, 
perturbative corrections can now be computed in
dimensional regularization to any order using a fixed set of rules.
At the
same time, the required technology has been developed to carry out
computations at next-to-next-to-leading and higher orders, and the
role of the infrared sensitivity of quark mass definitions in the
non-relativistic context has been understood. In this review I have
tried to give a comprehensive overview of the developments and the
applications of the new technologies to bottom quark mass
determinations and top quark threshold physics. What comes next?

The main field of application of the new technologies is top quark  
pair production close to threshold, which is a major part of the top
physics program at the Linear Collider. This is because the top mass
and the top quark widths are so much larger than the typical
hardonization scale. 
Practically all aspects of the non-relativistic top-antitop dynamics
can be described  predominantly with perturbative methods based on the 
new developments. Thus, top threshold measurements will very likely
provide the insights into the interplay of perturbative and
non-perturbative aspects of QCD one hoped for already after the
discovery of $J/\psi$. Still, a lot of work remains to be done to
establish the theoretical tools that will make this possible. At
present, the new effective theory methods have exclusively been
applied to computations of the total cross section, where
non-perturbative and electroweak effects are small. The most
important outcome of these calculations is that top mass measurements
from a threshold scan will have a precision at the level of $100$~MeV,
which will be unsurpassed by any other method. 
But calculations should also be attempted with the same theoretical
precision for differential observables. At certain kinematic endpoints
the impact of non-perturbative and also electroweak effects can be
enhanced. At present, systematic studies of such effects using the
new technologies have not yet been carried out. 
Measured by the progress that already happened in the last few years I
am optimistic that developments in this direction will take place in
short time. 

\vspace{1cm} 

\section*{Acknowledgments}  
I would like to express my gratitude to Stan Brodsky, Johann K\"uhn,
Patrick Labelle, Zoltan Ligeti, Aneesh Manohar, Martin Smith, Tim Stelzer,
Iain Stewart, Thomas Teubner, Scott Willenbrock and Mohammed Zebarjad
for pleasant and lively collaborations in projects on perturbative  
heavy quarkonium physics. 
I particularly thank Aneesh Manohar, Antonio Pineda and Iain Stewart 
for discussions on non-relativistic effective theories,
and Ralf Hofmann, Thomas Teubner,  
and Iain Stewart for comments on the manuscript.

\vspace{1cm}

\sloppy
\raggedright

\end{document}